\documentclass[11pt]{article}
\usepackage{pgfplots}
\pgfplotsset{compat=1.7}
\usepackage{subcaption}
\usepackage[margin=1in,letterpaper]{geometry}
\usepackage[T1]{fontenc}
\usepackage{lmodern}
\usepackage{amsmath}
\usepackage{amssymb}
\usepackage{amsthm}
\usepackage{color,soul}
\usepackage{xcolor}
\usepackage{graphicx}
\usepackage{algorithm}
\usepackage{tcolorbox}
\usepackage[normalem]{ulem}
\usepackage{tikz}
\usetikzlibrary{patterns}
\usepackage[noend]{algpseudocode}
\usepackage[colorlinks=true, linkcolor=red, urlcolor=blue, citecolor=gray,hyperindex,breaklinks]{hyperref}
\PassOptionsToPackage{linktocpage}{hyperref}

\newcommand\blfootnote[1]{%
	\begingroup
	\renewcommand\thefootnote{}\footnote{#1}%
	\addtocounter{footnote}{-1}%
	\endgroup
}
\usepackage{comment}
\theoremstyle{plain}
\newtheorem{thm}{\protect\theoremname}
\theoremstyle{plain}
\newtheorem{claim}[thm]{\protect\claimname}
\theoremstyle{plain}
\newtheorem{prop}[thm]{\protect\propositionname}
\theoremstyle{plain}
\newtheorem{lem}[thm]{\protect\lemmaname}
\theoremstyle{plain}
\newtheorem{cor}[thm]{\protect\corollaryname}
\theoremstyle{definition}
\newtheorem{defn}[thm]{\protect\definitionname}
\theoremstyle{definition}
\newtheorem{assump}{\protect\assumptionname}
\theoremstyle{definition}
\newtheorem{rem}{\protect\remarkname}
\theoremstyle{plain}

\makeatother

\providecommand{\claimname}{Claim}
\providecommand{\lemmaname}{Lemma}
\providecommand{\propositionname}{Proposition}
\providecommand{\theoremname}{Theorem}
\providecommand{\corollaryname}{Corollary}
\providecommand{\definitionname}{Definition}
\providecommand{\assumptionname}{Assumption}
\providecommand{\remarkname}{Remark}

\global\long\def\RR{\mathbb{R}}

\global\long\def\R{{\cal R}}

\global\long\def\e{{\mathbf{e}}}

\global\long\def\p{{\mathbf{p}}}

\emergencystretch=\maxdimen
\newcommand{\wt}{\widetilde}
\title{LSH}
\newcommand{\muu}[1]{\exp_{\mu}\left({#1}\right)}

{\makeatletter
	\gdef\xxxmark{%
		\expandafter\ifx\csname @mpargs\endcsname\relax 
		\expandafter\ifx\csname @captype\endcsname\relax 
		\marginpar{xxx}
		\else
		xxx 
		\fi
		\else
		xxx 
		\fi}
	\gdef\xxx{\@ifnextchar[\xxx@lab\xxx@nolab}
	\long\gdef\xxx@lab[#1]#2{{\bf [\xxxmark #2 ---{\sc #1}]}}
	\long\gdef\xxx@nolab#1{{\bf [\xxxmark #1]}}
	\long\gdef\xxx@lab[#1]#2{}\long\gdef\xxx@nolab#1{}%
}

\newcommand{\E}{\mathbb{E}}
\renewcommand{\e}{\epsilon}

 \newenvironment{proofof}[1]{\noindent{\bf Proof of #1:}}{$\qed$\par}

\newcommand{\expm}{\exp_\mu}
\renewcommand{\R}{\mathbb{R}}

\renewcommand{\p}{\mathbf{p}}

\begin{document}
\title{Kernel Density Estimation through Density Constrained Near Neighbor Search}
\author{
  Moses Charikar\\Stanford University
  \and
  Michael Kapralov\\EPFL
  \and
  Navid Nouri\\EPFL
  \and 
  Paris Siminelakis\\UC Berkeley
  }

\maketitle

\blfootnote{Moses Charikar was supported by a Simons Investigator Award, a Google Faculty Research Award and an Amazon Research Award.}
\blfootnote{This project has received funding from the European Research Council (ERC) under the European Union’s Horizon 2020 research and innovation programme (grant agreement No 759471).}
\blfootnote{Paris Siminelakis was supported by ONR DORECG award N00014-17-1-2127.}
\begin{abstract}
In this paper we revisit the kernel density estimation problem: given a kernel $K(x, y)$ and a dataset of $n$ points in high dimensional Euclidean space, prepare a data structure that can quickly output, given a query $q$, a $(1+\e)$-approximation to $\mu:=\frac1{|P|}\sum_{p\in P} K(p, q)$. First, we give a single data structure based on classical near neighbor search techniques that improves upon or essentially matches the query time and space complexity for all radial kernels considered in the literature so far. We then show how to improve both the query complexity and runtime by using recent advances in data-dependent near neighbor search. 

We achieve our results by giving a new implementation of the natural importance sampling scheme. Unlike previous approaches, our algorithm first samples the dataset uniformly (considering a geometric sequence of sampling rates), and then uses existing approximate near neighbor search techniques on the resulting smaller dataset to retrieve the sampled points that lie at an appropriate distance from the query. We show that the resulting sampled dataset has strong geometric structure, making approximate near neighbor search return the required samples much more efficiently than for worst case datasets of the same size. As an example application, we show that this approach yields a data structure that achieves query time $\mu^{-(1+o(1))/4}$ and space complexity $\mu^{-(1+o(1))}$ for the Gaussian kernel. Our data dependent approach achieves query time $\mu^{-0.173-o(1)}$ and space $\mu^{-(1+o(1))}$ for the Gaussian kernel.  The data dependent analysis relies on new techniques for tracking the geometric structure of the input datasets in a recursive hashing process that we hope will be of interest in other applications in near neighbor search.

\end{abstract}

\newpage
\tableofcontents
\newpage
\section{Introduction}

Kernel density estimation is a fundamental problem with numerous applications in machine learning, statistics and data analysis~\cite{fan1996local,scholkopf2001learning,joshi2011comparing,schubert2014generalized,genovese2014nonparametric,arias2015estimation,gan2017scalable}. Formally, the Kernel Density Estimation (KDE) problem is: preprocess a dataset $P$ of $n$ points $\mathbf{p}_1,\ldots, \mathbf{p}_n\in \R^d$ into a small space data structure that allows one to quickly approximate, given a query $\mathbf{q}\in \R^d$, the quantity
\begin{align}\label{eq:kde-intro}
K(P,\mathbf{q}):=\frac{1}{|P|}\sum_{\mathbf{p}\in P}K(\mathbf{p},\mathbf{q}).
\end{align}
where $K(\mathbf{p},\mathbf{q})$ is the kernel function. The Gaussian kernel
\begin{align*}
K(\mathbf{p},\mathbf{q}):=\exp(-||\mathbf{p}-\mathbf{q}||_2^2/2)
\end{align*}
is a prominent example, although many other kernels (e.g., Laplace, exponential, polynomial etc) are the method of choice in many applications~\cite{shawe2004kernel,RasmussenW06}.

In the rest of the paper, we use the notation $\mu^*$ defined as $\mu^*:=K(P,\mathbf{q})$, and $\mu$ is a quantity that satisfies $\mu^*\le \mu \le 4\mu^*$.\footnote{We have replaced $\mu^*$ with $\mu$ in the abstract for the ease of notation in the abstract.} Moreover, in the statement of the main results, we assume that a constant factor lower bound to the actual kernel density, $\mu^*$, is known. In general, if we only know that $\mu^* \ge \tau$ for some $\tau$, then the $\mu^*$ terms in the space should be replaced by $\tau$ (similar to prior results in the literature). However, the query time can always be stated in terms of $\mu^*$.

The kernel density estimation problem has received a lot of attention over the years, with very strong results available for low dimensional datasets. For example, the celebrated fast multipole method~\cite{fast-multipole} and the related Fast Gauss Transform can be used to obtain efficient data structure for KDE (and in fact solves the more general problem of multiplying by a kernel matrix). However, this approach suffers from an exponential dependence on the dimension of the input data points, a deficiency that it shares with  other tree-based methods~\cite{gray2001n,gray2003nonparametric,yang2003improved,lee2006dual,ram2009linear}. A recent line of work~\cite{charikar2017hashing,charikar2019multi-resolution,backurs2018efficient,backurs2019space} designed sublinear query algorithms for kernel density estimation in high dimensions using variants of the Locality Sensitive Hashing~\cite{charikar2017hashing} framework of Indyk and Motwani~\cite{IndykM98}. 

Most of these works constructed estimators based on locality sensitive hashing, and then bounded the variance of these estimators to show that a small number of repetitions suffices for a good estimate. Bounding the variance of LSH-based estimators is nontrivial  due to correlations inherent in sampling processes based on LSH, and the actual variance turns out to be nontrivially high.

In this work we take a different approach to implementing importance sampling for KDE using LSH-based near neighbor search techniques.  At a high level, our approach consists of first performing independent sampling on the dataset, and then using using LSH-based near neighbor search primitives to extract relevant data points from this sample\footnote{The approach of~\cite{backurs2018efficient} also used near neighbor search techniques, but was only using $c$-ANN primitives as a black box, which turns out to be constraining -- this only leads to strong results for slowly varying kernels (i.e., polynomial kernels).  Our data-independent result recovers the results of~\cite{backurs2018efficient}, up to a $\mu^{-o(1)}$ loss, as a special case.}.  The key observation  is that the sampled dataset in the KDE problem has nice geometric structure: the number of data points around a given query cannot grow too fast as a function of distance and the actual KDE value $\mu$ (we refer to these constraints as density constraints -- see Section~\ref{sec:tech-overview} for more details). The fact that our approach departs from the idea of constructing unbiased estimators of KDE directly from LSH buckets turns out to have two benefits: first, we immediately get a simple algorithm that uses classical LSH-based near neighbor search primitives (Euclidean LSH of Andoni and Indyk~\cite{DBLP:conf/focs/AndoniI06})  to improve on or essentially matches all prior work on kernel density estimation for radial kernels. The result is formally stated as Theorem~\ref{thm:main-data-indep} for the Gaussian kernel below, and its rather compact analysis in a more general form that extends to other kernels is presented in Section~\ref{sec:data-independent}.  The second benefit of our approach is that it distills a clean near neighbor search problem, which we think of as near neighbor search under density constraints, and improved algorithms for that problem immediately yield improvements for the KDE problem itself. This clean separation allows us to use the recent exciting data-depending techniques pioneered by~\cite{DBLP:conf/soda/AndoniINR14,AR15,DBLP:conf/soda/AndoniLRW17} in our setting. It turns out that while it seems plausible that data-dependent techniques can improve performance in our setting, actually designing an analyzing a data-dependent algorithm for density constrained near neighbor search is quite nontrivial. The key difficulty here lies in the fact that one needs to design tools for tracking the evolution of the density of the dataset around a given query through a sequence of recursive partitioning steps (such evolution turns out to be quite involved, and in particular governed by a solution to an integral equation involving the log density of the kernel and properties of Spherical  LSH).  The design of such tools is our main technical contribution and is presented in Section~\ref{sec:dd}. The final result for the Gaussian kernel is given below as Theorem~\ref{thm:main}, and  extensions to other kernels are presented in Section~\ref{sec:dd}.

\subsection{Our results}

We instantiate our results for the Gaussian kernel as an illustration, and then discuss extensions to more general settings. We assume that $\mu^*=n^{-\Theta(1)}$, since this is the interesting regime for this problem. For $\mu^{*}=n^{-\omega(1)}$ under the Orthogonal Vectors Conjecture (e.g. \cite{rubinstein2018hardness}), the problem cannot be solved faster than $n^{1-o(1)}$ using space $n^{2-o(1)}$~\cite{charikar2019multi-resolution}, and for larger values $\mu^{*}=n^{-o(1)}$ random sampling solves the problem in $n^{o(1)}/\epsilon^{2}$ time and space.  
\paragraph{Data-Independent LSH} Our first result uses data-independent LSH of Andoni-Indyk~\cite{DBLP:conf/focs/AndoniI06} to improve upon the previously best known result \cite{charikar2017hashing} and follow up works that required query time $\wt{O}(\mu^{-0.5-o(1)}/\epsilon^{2})$ if only polynomial space in $1/\mu$ is available.

\begin{thm}\label{thm:main-data-indep}
Given a kernel $K(\mathbf{p},\mathbf{q}):=e^{-a||\mathbf{p}-\mathbf{q}||_2^2}$ for any $a>0$, $\epsilon = \Omega\left(\frac{1}{\mathrm{polylog} n}\right)$, $\mu^*=n^{-\Theta(1)}$ and a data set of points $P$, there exists an algorithm for preprocessing and an algorithm for query procedure such that after receiving query $\mathbf{q}$ one can approximate $\mu^*:=K(P,\mathbf{q})$ (see Definition~\ref{def:kernel}) up to $(1\pm \epsilon)$ multiplicative factor, in time $\wt{O}\left(\epsilon^{-2}\left(\frac{1}{\mu^*}\right)^{0.25+o(1)}\right)$, and the space consumption of the data structure is $$\min\left\{ \epsilon^{-2} n\left(\frac{1}{\mu^*}\right)^{0.25+o(1)},\epsilon^{-2}\left(\frac{1}{\mu^*}\right)^{1+o(1)}\right\}.$$
\end{thm}
\begin{rem}
	In Theorem~\ref{thm:main-data-indep} (and similar theorems in the rest of the paper), we assumed that $\epsilon= \Omega\left(\frac{1}{\mathrm{polylog} n}\right)$ and $\mu^*=n^{-\Theta(1)}$, so that we can assume  $d=\wt{O}(1)$ (and ignore the dependencies on dimension in the statements). The reason (for $d=\wt{O}(1)$) is that in this case the contribution of far points (points at distance $\Omega(\log n)$) is negligible and for close points, we can use Johnson-Lindenstrauss (JL) lemma to reduce the dimension to $O(\mathrm{polylog } n)$,  without distorting the kernel value by a more than $1\pm o(1)$ multiplicative factor. If we remove these assumptions, we need to multiply the query-time and space bounds by dimension $d$. 
\end{rem}
This theorem is stated and proved as Theorem~\ref{thm:Gaus4} in Section~\ref{sec:data-independent}. To get a sense of the improvement, the result of \cite{charikar2017hashing} exhibited query time that is roughly a square root of the query time of uniform random sampling. Our result uses the same LSH family as in \cite{charikar2017hashing} but achieves  query time that is itself roughly the square root of that of \cite{charikar2017hashing}! 

\paragraph{Data-Dependent LSH}  Our main technical contribution is a collection of techniques for using data dependent hashing introduced by~\cite{DBLP:conf/soda/AndoniINR14,AR15,DBLP:conf/soda/AndoniLRW17} in the context of kernel density estimation. Unlike these works, however, who had no assumptions on the input data set, we show how to obtain refined bounds on the efficiency of near neighbor search under density constraints imposed by assumptions on KDE value as a function of the kernel. This turns out to be significantly more challenging: while in approximate near neighbor search, as in~\cite{DBLP:conf/soda/AndoniLRW17}, it essentially suffices to track the size of the dataset in recursive iterations of locality sensitive hashing and partitioning into spheres, in the case of density constrained range search problems arising from KDE one must keep track of the distribution of points across different distance scales in the hash buckets, i.e. track evolution of functions as opposed to numbers. This leads to a natural linear programming relaxation that bounds the performance of our algorithm that forms the core of our analysis\footnote{The actual optimal evolution is described by an integral equation involving the log density of the kernel function and collision probabilities of LSH on the Euclidean sphere, but we do not make the limiting claim formal here since the ultimate integral equation appears to not have a closed form solution, and hence would not be useful for analysis purposes.}.  Our ultimate result for the Gaussian kernel is:

\begin{thm}\label{thm:main}
	For Gaussian kernel $K$, any data set of points $P$ and any $\epsilon = \Omega\left(\frac{1}{\mathrm{polylog} n}\right)$, $\mu^*=n^{-\Theta(1)}$, using Algorithm~\ref{alg:4} for preprocessing and Algorithm~\ref{alg:query4} for the query procedure, one can approximate $\mu^*:=K(P,\mathbf{q})$ (see Definition~\ref{def:kernel}) up to $(1\pm \epsilon)$ multiplicative factor, in time $\wt{O}(\mu^{-0.173-o(1)}/\epsilon^2)$. The space complexity of the algorithm is also bounded by $$\min\left\{O(n\cdot \mu^{-(0.173+o(1))}/\epsilon^{2}),O\left(\mu^{-(1+c+o(1))}/\epsilon^2\right)\right\},$$ for $c=10^{-3}$.\footnote{This $c$ can be set to any small constant that one desires. For our setting of parameters $c=10^{-3}$.}
\end{thm}
The proof of Theorem~\ref{thm:main} is given in Section~\ref{sec:dd}.

Our techniques extend to other kernels  -- the extensions are presented in Section~\ref{sec:dd}.

\subsection{Related Work}

For $d\gg 1$,  KDE was studied extensively in the 2000's with the works of~\cite{gray2001n,gray2003nonparametric,yang2003improved,lee2006dual,ram2009linear} that employed hierarchical space partitions (e.g. kd-trees, cover-trees) to obtain sub-linear query time  for datasets with low intrinsic dimensionality~\cite{karger2002finding}. Nevertheless, until recently~\cite{charikar2017hashing},  in the regime of $d=\Omega(\log n)$ and  under worst case assumptions,  the best known algorithm was simple random sampling that for constant $\delta>0$ requires $O(\min\{1/\epsilon^{2}\mu, n\})$  evaluations of the kernel function to provably approximate the density at any query point $q$.

\cite{charikar2017hashing} revisited the problem and introduced a technique, called Hashing-Based-Estimators (HBE), to   implement low-variance Importance Sampling (IS) efficiently for any query through Locality Sensitive Hashing (LSH). For the Gaussian $f(r)=e^{-r^{2}}$, Exponential $f(r)=e^{-r}$, and $t$-Student kernels $f(r)=(1+r^{t})^{-1}$ the authors gave the first sub-linear algorithms that require $O(\min\{1/\epsilon^{2}\sqrt{\mu},n\})$ kernel evaluations. Using ideas from Harmonic Analysis, the technique was later extended in~\cite{charikar2019multi-resolution},  to apply to more general kernels resulting in the first data structures that require $O(\min\{1/\epsilon^{2}\sqrt{\mu}, n\})$ kernel evaluations to approximate the density for log-convex kernels $e^{\phi(\langle x,y\rangle)}$.  Furthermore, under the Orthogonal Vectors Conjecture it was shown that  there does not exist a data structure that solves the KDE problem under the Gaussian kernel in time $n^{1-o(1)}/{\mu^{o(1)}}$ and space $n^{2-o(1)}/\mu^{o(1)}$.

The work  most closely related to ours is that of \cite{backurs2018efficient}.   \cite{backurs2018efficient} introduced a technique, called Spherical Integration, that uses black-box calls to $c$-ANN data structures (constructed on sub-sampled versions of the data set) to sample points from ``spherical annuli'' $(r,cr)$ around the query, for all annuli that had non-negligible contribution to the density of the query. For kernels with polynomial tails of degree $t$, their approach required $\wt{O}(c^{5t})$ calls to such data-structures (without counting the query time  required for each such call) to estimate the density. Unfortunately, this approach turns out to be constraining due to its reliance on {\bf black-box} $c$-ANN calls, and in particular only applies to polynomial kernels. Our techniques in this paper recover the result of~\cite{backurs2018efficient} up to $\mu^{-o(1)}$ factors as a special case (see Section~\ref{sec:data-independent}). Furthermore, the $\mu^{-o(1)}$ factor loss that we incur is only due to the fact that we are using the powerful Euclidean LSH family in order to achieve strong bounds for kernels that exhibit fast decay (e.g., Gaussian, exponential and others) using the same algorithm. For polynomial kernels the dependence on $\mu$ in our approach can be reduced to polylogarithmic in $1/\mu$ by using an easier hash family (e.g., the hash family of ~\cite{datar2004locality}; see~\cite[Chapter 10]{siminelakis2019} for details). 

\paragraph{Scalable approaches to KDE and Applications}
Recent works \cite{siminelakis2019rehashing,backurs2019space} also address scalability issues of the original approach of \cite{charikar2017hashing}.  \cite{siminelakis2019rehashing}  designed a more efficient adaptive procedure that can be used along with  Euclidean LSH~\cite{datar2004locality} to solve KDE for a variety of power-exponential kernels, most prominently the Gaussian. Their algorithm is the first practical algorithm for Gaussian KDE with worst case guarantees that improve upon random sampling in high dimensions.  Experiments in real-world data sets show~\cite{siminelakis2019rehashing} that the method of \cite{charikar2017hashing}, yields practical improvements for many real world datasets. \cite{backurs2019space} introduced a way to sparsify hash tables and showed that in order to estimate densities $\mu^{*}\geq \tau \geq \frac{1}{n}$ one can reduce the space usage of the data structures~\cite{siminelakis2019rehashing} from $O(1/\tau^{3/2}\epsilon^{2})$ to $O(1/\tau\epsilon^{2})$. The authors also evaluated their approach on real world data for the Exponential $e^{-\|x-y\|_{2}}$ and Laplace $e^{-\|x-y\|_{1}}$ kernels showing improvements compared to \cite{charikar2017hashing} and uniform random sampling. A related approach of Locality Sensitive Samplers~\cite{spring2017new} has also been applied to obtain practical procedures in the contexts of Outlier detection~\cite{luo2018arrays}, Gradient Estimation~\cite{chen2019lsh} and Clustering~\cite{luo2019scaling}. Finally, \cite{wu2018local} uses similar ideas to address the problem of approximate range counting on the unit sphere.

\paragraph{Core-sets and Kernel sketching} The problem of KDE is phrased in terms of guarantees for any single query $\mathbf{q}\in \R^{d}$. A related problem is that of Core-sets for kernels~\cite{phillips2013varepsilon}, where the goal is to find a (small) set $S\subset P$ such that the kernel density estimate on $P$ is close to the one on $S$. After  recent flurry of research efforts ~\cite{phillips2018improved,phillips2018near} has resulted in near optimal~\cite{phillips2018near} unweighted $|S|=O(\sqrt{d\log(1/\epsilon)} /\epsilon)$  and optimal~\cite{DBLP:conf/colt/KarninL19}  weighted core-sets $|S|=O(\sqrt{d}/\epsilon)$ for positive definite kernels. Somewhat related to this problem is the problem of oblivious sub-space embeddings for polynomial kernels \cite{NIPS2014_5240,pham2013fast,avron2017random,ahle2019oblivious}.

\subsection{Outline}
We start by giving a technical overview of the paper in Section~\ref{sec:tech-overview}. Preliminary definitions and results are presented in Section \ref{sec:prelims}. In Section \ref{sec:data-independent},  we present our data-independent result for Gaussian KDE and state a general version of our result for other decreasing kernels. We present our data structure based on Data-Dependent LSH for Gaussian KDE in Section \ref{sec:dd} and its analysis in Sections \ref{sec:query} (Query time), \ref{sec:exepath} (Valid execution path analysis), \ref{sec:main-tech-lemma} (Linear Program analysis), and  \ref{sec:lp-sol} (Primal-Dual solution). 


\section{Technical overview}\label{sec:tech-overview}

In this section we give an overview of our results and the main ideas behind them. For simplicity we use the Gaussian kernel, even though both our results extend to more general settings. Thus, for the purposes of this overview our problem is: preprocess a dataset $P$ of $n$ points $\mathbf{p}_1,\ldots, \mathbf{p}_n\in \R^d$ into a small space data structure that allows fast KDE queries, i.e. can quickly approximate, given $\mathbf{q}\in \R^d$, the quantity
\begin{align}\label{eq:kde-tech-overview}
K(P,\mathbf{q}):=\frac{1}{|P|}\sum_{\mathbf{p}\in P}K(\mathbf{p},\mathbf{q}),
\end{align}
where 
\begin{align*}
K(\mathbf{p},\mathbf{q}):=\exp(-||\mathbf{p}-\mathbf{q}||_2^2/2).
\end{align*}
We present two schemes based on ideas from data independent and data dependent LSH schemes. Both schemes employ the strategy of first sampling the dataset at a sequence of geometric levels, and then using near neighbor search algorithms to retrieve all points at an appropriate distance from the query from the sample. The difference between the two approaches lies in the implementation and analysis of the near neighbor search primitive used for this retrieval. In what follows we first overview our approach to implementing importance sampling for KDE using near neighbor search primitives, and then instantiate this scheme with data-independent (Section~\ref{sec:overviewDI}) and data-dependent (Section~\ref{sec:overviewDD}) schemes.

\subsection{Data-independent algorithm (Section~\ref{sec:data-independent})}\label{sec:overviewDI}
We start by showing a new application of data-independent locality sensitive hashing to KDE that results in a simple scheme that provides the following result.

\begin{thm}[Informal version of Theorem~\ref{thm:Gaus4}] \label{thm:informal14}
	If  $\mu^*:=K(P,\mathbf{q})$, then there exists an algorithm that can approximate $\mu^*$ up to $(1\pm \epsilon)$ multiplicative factor, in time $\left(\frac{1}{\mu^*}\right)^{0.25+o(1)}$, using a data structure of size  $$\min\left\{ \epsilon^{-2} n\left(\frac{1}{\mu^*}\right)^{0.25+o(1)},\epsilon^{-2}\left(\frac{1}{\mu^*}\right)^{1+o(1)}\right\}.$$
\end{thm}

We remark that the actual non-adaptive algorithm that we present in Section~\ref{sec:data-independent} is more general than the above and applies to a wide class of kernels. In particular, it simultaneously improves upon all prior work on radial kernels that exhibit fast tail decay (such as the exponential and the Gaussian kernels) ~\cite{charikar2017hashing} as well as matches the result of ~\cite{backurs2018efficient} on kernels with only inverse polynomial rate of decay up to $\mu^{-o(1)}$ factors.

We now outline the algorithm and the analysis. The main idea is simple: we note that in order to approximate the sum on the right hand side of~\eqref{eq:kde-tech-overview}, ideally we would like to do importance sampling, i.e. pick every point with probability proportional to its contribution to the KDE value. It is of course not immediate how to do this, since the contribution depends on the query, which we do not know at the preprocessing stage. However, we show that it is possible to simply prepare sampled versions of the input dataset using a fixed geometric sequence of sampling rates, and then use locality sensitive hashing to retrieve the points relevant to the given query from this sample efficiently. Below, we present an overview of our algorithm. 
 
\paragraph{Geometric weight levels:} Let $J:=\lceil \log \frac{1}{\mu}\rceil$ and partition the points in the data set into $J$ sets, such that the contribution of any point in the $j$'th set to the kernel density is $\approx 2^{-j}$. If $w_i:=K(\mathbf{p}_i,\mathbf{q})$, then we define (see Definition~\ref{def:geom}) level sets $$L_j:=\left\{\mathbf{p}_i \in P: w_i \approx 2^{-j} \right\}.$$
The kernel density can be expressed in terms of the level sets as
\begin{align*}
K(P,\mathbf{q})\approx \frac{1}{n} \sum_{j=1}^J |L_j|\cdot 2^{-j},
\end{align*}
which implies size upper bounds for $L_j$, namely:
\begin{align}\label{eq:lj-size}
|L_j| \lesssim 2^jn\mu.
\end{align}
This means that for every query $\mathbf{q}$ such that the KDE value at $\mathbf{q}$ equals $\mu$ to within constant factors one can place an upper bound of $2^jn\mu$ on the number of points at distance corresponding to level $L_j$ -- these are exactly the geometric weight constraints that make our near neighbor search primitives very efficient.   Note that we are only considering level sets $L_j$ for $j$ at most $J=\lceil \log \frac{1}{\mu}\rceil$. We describe our implementation of importance sampling now. 

\paragraph{Importance sampling:}
Suppose that one designs a sampling procedure that samples each point $\mathbf{p}_i$ with probability $p_i$ and calculates the following estimator
\begin{align*}
Z = \sum_{i} \frac{\chi_i}{p_i}w_i
\end{align*}
where $\chi_i=1$ if $\mathbf{p}_i$ is sampled and $\chi_i=0$ otherwise. Obviously, this estimator is an unbiased estimator for $n\mu^*$. So, if we can prove that this estimator has a relatively low variance, then by known techniques (repeating many times, averaging and taking the median) one can approximate $\mu^*$, efficiently. It can be shown (see Claim~\ref{claim:variance})  that if $p_i$'s are proportional to $w_i$'s (more specifically, we set $p_i\approx \frac{w_i}{ n \mu}$) then the variance is low. This approach is known as {\bf importance sampling}. In other words, we need to sample points with higher contribution, with higher probability.

If $L_j$'s were known to the algorithm in the preprocessing phase, then for each $j$, one could have sampled points in $L_j$ with probability $\approx \frac{1}{2^jn\mu}$. However, the query is not known in the preprocessing phase and hence geometric weight levels are not known beforehand. 

Our approach is the following: for each $j$ we sample the data set $P$ with probability $\frac{1}{2^jn\mu}$. Then, we prepare a data structure (for this sampled data set) that can {\bf recover} any sampled point with contribution $\approx 2^{-j}$ in the query procedure, efficiently and with high probability. Note that the number of points with contribution $\ge 2^{-j}$ is upper bounded by $2^{j}n\mu$. So, on average after the sub-sampling we expect to have at most $O(1)$ point from $L_1\cup\ldots\cup L_j$. On the other hand, since Gaussian kernel is a decreasing function of distance, points in $L_{j+1}\cup\ldots\cup L_J$ are actually further than the query. Thus, our {\bf recovery} problem can be seen as an instance of {\bf near neighbor} problem. Therefore, we use the {\bf locality sensitive hashing (LSH)} approach, which has been used in the literature for solving the approximate near neighbor problem.
\paragraph{Using Euclidean LSH for recovery:} Now, we explain how one can use Euclidean LSH scheme to design a data structure to recover points from $L_j$ in the corresponding sub-sampled data set. 

We first present an informal and over-simplified version of LSH function used in \cite{DBLP:conf/focs/AndoniI06}. Roughly speaking \cite{DBLP:conf/focs/AndoniI06} presents the following result (see Lemma~\ref{lem:andoni-indyk} for the formal statement): 
\begin{lem}[Informal version of Lemma~\ref{lem:andoni-indyk}]\label{lm:euclidean-lsh-informal}
	For every $r$ there exists a (locality sensitive) hash family such that, if $\mathbf{p}$ (a `close' point) and $\mathbf{p}'$ (a `far' point) are at distance $r$ and $\ge c\cdot r$ (for some $c\ge1$) of some point $\mathbf{q}$, respectively, then if 
	$$
	p:=\Pr[h(\mathbf{p})=h(\mathbf{q})],
	$$ then 
	\begin{align*}
	\Pr[h(\mathbf{p'})=h(\mathbf{q})] \leq p^{(1-o(1))c^2}.
	\end{align*} 
\end{lem}

Now given a query $\mathbf{q}$, for every $j$ we use Euclidean LSH to retrieve the points in $L_j$ from a sample of the dataset where every point is included with probability $\frac{1}{2^jn\mu}$. We repeat the hashing process multiple times to ensure high probability of recovery overall,  as in the original approach of~\cite{IndykM98}. However, the parameter setting and the analysis are different, since in the context of KDE we can exploit the geometric structure of the sampled dataset, namely upper bounds on the sizes of level sets $L_j$ given in \eqref{eq:lj-size} above -- we outline the parameter setting and analysis now.

\begin{figure}
	\centering
	\tikzstyle{vertex}=[circle, fill=black!70, minimum size=3,inner sep=1pt]
	\tikzstyle{svertex}=[circle, fill=black!100, minimum size=5,inner sep=1pt]
	\tikzstyle{gvertex}=[circle, fill=green!80, minimum size=7,inner sep=1pt]
	
	\tikzstyle{evertex}=[circle,draw=none, minimum size=25pt,inner sep=1pt]
	\tikzstyle{edge} = [draw,-, color=red!100, very  thick]
	\tikzstyle{bedge} = [draw,-, color=green2!100, very  thick]
	\begin{tikzpicture}[scale=1.7, auto,swap]

	\draw[pattern=north east lines, pattern color=blue] (0,0) circle (2cm);
	\fill[fill=white] (0,0) circle (1.73cm);
	\draw[pattern=north west lines, pattern color=brown] (0,0) circle (1.73cm);
	\fill[fill=white] (0,0) circle (1.414cm);
	\draw[pattern=north east lines, pattern color=green] (0,0) circle (1.414cm);
\fill[fill=white] (0,0) circle (1cm);
	\draw[pattern=north west lines, pattern color=red] (0,0) circle (1cm);

	\draw (0,0) circle (1cm);
\draw (0,0) circle (1.414cm);

\draw (0,0) circle (1.73cm);
\draw (0,0) circle (2cm);

	\draw (0,0)[above] node {{$\mathbf{q}$}};

	\node[vertex](v1) at (0, 0) {};  

	\end{tikzpicture}
	\caption{Illustration of distance levels induced by geometric weight levels. Areas marked with colors red, green, brown, blue and so on correspond to geometric weight levels $L_1,\ldots,L_4$ and so on.} \label{fig:inducedistlevels}
\end{figure}
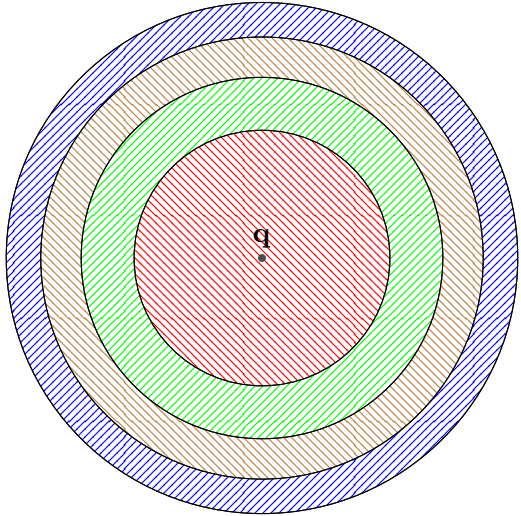

On the other hand, geometric weight levels induce distance levels (see Definition~\ref{def:geom} and Figure~\ref{fig:inducedistlevels}). Roughly speaking, for the Gaussian kernel if $\mathbf{p} \in L_j$ and $\mathbf{p}'\in L_i$, then
\begin{align*}
\frac{\|\mathbf{p}'-\mathbf{q}\|_2}{\|\mathbf{p}-\mathbf{q}\|_2}\approx \sqrt{\frac{i}{j}}=: c_{i,j}.
\end{align*}

Recall that since we sampled the data set with probability $\frac{1}{2^jn\mu}$ then for every $i$ we will have at most $\approx 2^{i-j}$ points from $L_i$ in the sampled set, in expectation. In particular, most likely the sample does not contain points from level sets $i<j$.
We instantiate Euclidean LSH from Lemma~\ref{lm:euclidean-lsh-informal} with the `near' distance $r$ being the distance to the target level set $L_j$. Let $p$ denote the probability that the query collides with a point in $L_j$. Now by Lemma~\ref{lm:euclidean-lsh-informal} we upper bound the expected number of points from level sets $L_i, i>j,$ in the bucket of the query:
\begin{align*}
\sum_{i> j} 2^{i-j}\cdot p^{c_{i,j}^2}
\end{align*}
We now select $p$ (note that Lemma~\ref{lm:euclidean-lsh-informal} allows flexibility in selecting $p$, which is achieved by concatenating hash functions; see Section~\ref{sec:data-independent} for the detailed analysis). We set $p$ such that the number of points from each $L_i$ in the bucket of the query is at most $1$ for all $i>j$. For every such $i$, 
$2^{i-j}\cdot p^{\frac{i}{j}} \le 1$ implies $p \le \left(\frac{1}{2}\right)^{j-\frac{j^2}{i}}$, and hence we let

$$p=p_j=\min_{i>j} \left(\frac{1}{2}\right)^{j-\frac{j^2}{i}},$$
where we give the probability a subscript $j$ to underscore that this is the setting for level set $L_j$.

On the other hand, note that since the point that we want to recover will be present in the query's bucket with probability $p_j$, we need to repeat this procedure $\wt{O}\left(\frac{1}{p_j}\right)$ times, to recover the point with high probability. This means that for every $j$ the contribution of level set $L_j$ to the query time will be $\wt{O}\left(\frac{1}{p_j}\right)$. Now, note that 
\begin{equation}\label{eq:di-opt}
\begin{split}
\max_{j\in [J]} \log_2 \frac1{p_j}&=\max_{j\in[J]}\max_{i\in(j,J]}\left( j-\frac{j^2}{i}\right)\\
&=J\cdot \max_{j\in[J]} ~~\frac{j}{J}\cdot \left(1-\frac{j}{J}\right)\\
&=\frac{J}{4}\\
&=\frac{1}{4}\log \frac{1}{\mu},
\end{split}
\end{equation}
implying a $(1/\mu)^{0.25}$ upper bound on the query time. This (informally) recovers the result mentioned in Theorem~\ref{thm:informal14}. Note that the space complexity of our data structure is no larger than the number of data points times the query time, i.e., $\approx n (1/\mu)^{0.25}$, since at every sampling rate we hash at most the entire dataset about $(1/\mu)^{0.25}$ times independently.  The space complexity can also be bounded by $\wt{O}(1/\mu)$ by noting that the datasets for which we have the highest query time and hence many repetitions are in fact heavily subsampled versions of the input dataset. These bounds are incomparable, and the latter is preferable for large values of KDE value $\mu$. 

We used the Gaussian kernel in the informal description above to illustrate our main ideas, but the approach extends to a very general class of kernels. In particular, it gives improvements over all prior work on the KDE problem for shift invariant kernels (with the only exception that our results essentially match the results of ~\cite{backurs2018efficient}, where an already very efficient algorithm with a polylogarithmic dependence on $1/\mu$ is presented). We present the detailed analysis of this approach in Section~\ref{sec:data-independent}.

\subsection{Data dependent algorithm (Section~\ref{sec:dd})}\label{sec:overviewDD}

We note that the efficiency of our implementation of importance sampling relies heavily on the efficiency of near neighbor search primitive under density constraints.  In this section we show how to use data-dependent techniques, i.e. data partitioning followed by the use of the more efficient Spherical LSH, to achieve significantly better results. Our approach builds on the exciting recent line of work on data-dependent near neighbor search~\cite{DBLP:conf/soda/AndoniINR14,AR15,DBLP:conf/soda/AndoniLRW17}, but the fact that we would like to optimally use the assumptions on the density of various spherical ranges that follow from assumptions on KDE value, the analysis turns out to be significantly more challenging. In particular, the core of our approach is a linear program that allows one to analyze the worst case evolution of densities during the hashing process. The analysis is presented in Section~\ref{sec:dd}, Section~\ref{sec:main-tech-lemma} and Section~\ref{sec:lp-sol}. Since the analysis is somewhat involved, we present it for the case of the Gaussian kernel to simplify notation. We then provide a version of the key lemma for other kernels and state the corresponding results.
 
 \begin{thm}[Informal version of Theorem~\ref{thm:Gaus5}] \label{thm:informal-dd}
	There exists an algorithm that, when $K$ is the Gaussian kernel and $\mu^*:=K(P,\mathbf{q})$, for $\e\in (0, 1)$ approximates $\mu^*$ to within a $(1\pm \epsilon)$ multiplicative factor, in expected time $\left(\frac{1}{\mu^*}\right)^{0.173+o(1)}$ and space $\min\{n \left(\frac{1}{\mu^*}\right)^{0.173+o(1)}, \left(\frac{1}{\mu^*}\right)^{1+o(1)}\}$.
\end{thm}

Our techniques extend to kernels beyond the Gaussian kernel (e.g., the exponential kernel, for which we obtain query time $\left(\frac{1}{\mu^*}\right)^{0.1+o(1)}$ and space $n \left(\frac{1}{\mu^*}\right)^{0.1+o(1)}$). We outline the extension in Section~\ref{sec:dd}. 

Recall that we need to preprocess a dataset $P$ of $n$ points $\mathbf{p}_1,\ldots, \mathbf{p}_n\in \R^d$ into a small space data structure that allows fast KDE queries, i.e., can quickly approximate, given $\mathbf{q}\in \R^d$, the quantity
\begin{align}\label{eq:kde-tech-overview-dd-1}
\mu^*=K(P,\mathbf{q})=\frac{1}{|P|}\sum_{\mathbf{p}\in P}\exp(-||\mathbf{p}-\mathbf{q}||_2^2/2).
\end{align}

Recall also that we assume knowledge of a quantity $\mu$ such that
\begin{equation}\label{eq:mu-assumption-overview}
\mu^*\leq \mu\leq 4\mu^*.
\end{equation}
This is without loss of generality by a standard reduction -- see Section~\ref{sec:dd}, Remark~\ref{rm:wlog}.
For simplicity of presentation, in this section we use a convenient rescaling of points so that 
\begin{align}\label{eq:kde-tech-overview-dd}
\mu^*=K(P,\mathbf{q})=\frac{1}{|P|}\sum_{\mathbf{p}\in P}(1/\mu)^{-||\mathbf{p}-\mathbf{q}||_2^2/2}.
\end{align}
Note that this is simply a rescaling of the input points, namely multiplying every coordinate by $(\log (1/\mu))^{-1/2}$. This is for analysis purposes only, and the algorithm does not need to perform such a rescaling explicitly.  We fix the query $\mathbf{q}$ for the rest of this section.

\paragraph{Densities of balls around query.} Upper bounds on the number of points at various distances from the query point in dataset (i.e., densities of balls around the query) play a central part in our analysis. For any $x\in (0,\sqrt{2})$  let
 \begin{equation}\label{eq:dx-def-overview}
 D_x(\mathbf{q}):= \{ ||\mathbf{p}-\mathbf{q}||:~\mathbf{p}\in P, ||\mathbf{p}-\mathbf{q}||\gtrsim x\},
 \end{equation}
denote the set of possible distances from $\mathbf{q}$ to points in the dataset $P$. Note that we are ignoring distances that are too close to $x$ -- this is for technical reasons that let us introduce some simplifications with respect to the analysis of ~\cite{DBLP:conf/soda/AndoniLRW17} at the expense of a small constant loss in the exponent of the ultimate query time (see Section~\ref{sec:query-dd} for more discussion of this). When there is no ambiguity we drop $\mathbf{q}$ and $x$ and we simply call it $D$. For any $y\in D$ we let 
 \begin{equation}\label{eq:p-y-def-overview}
 P_y(\mathbf{q}):=\{ \mathbf{p}\in P: ||\mathbf{p}-\mathbf{q}||\le y\}
 \end{equation}
 be the set of points at distance $y$ from $\mathbf{q}$. Since for every $y>0$
\begin{equation*}
\begin{split}
\mu^*=K(P, \mathbf{q})&=\frac1{n}\sum_{\mathbf{p}\in P} \mu^{||\mathbf{p}-\mathbf{q}||_2^2/2}\\
&\geq \frac{\mu^{y^2/2}}{n} |P_y(\mathbf{q})|
\end{split}
\end{equation*}
we get
\begin{align}\label{eq:density-constraints-overview}
|P_y(\mathbf{q})| \leq n \mu^*\cdot \left(\frac1{\mu}\right)^{\frac{y^2}{2}}\leq n\cdot \left(\frac1{\mu}\right)^{\frac{y^2}{2}-1},
\end{align}
since $\mu^*\leq \mu$ by assumption.

We implement the same importance sampling strategy as in Section~\ref{sec:overviewDI}: sample the dataset at a geometric sequence of sampling rates, and for each such sampling rate use approximate near neighbor search primitives (in this case data dependent ones) to retrieve the relevant points (which are generally a few closest points to the query) from the sample. The rescaling of the input space~\eqref{eq:kde-tech-overview-dd} together with the assumption~\eqref{eq:mu-assumption-overview}  implies that one essentially only needs to care about points $\p\in P$ such that 
$$
||\mathbf{p}-\mathbf{q}||_2\approx x\text{~for some $x\in [0, \sqrt{2})$}.
$$
This is because every $\mathbf{p}\in P$ such that $||\mathbf{p}-\mathbf{q}||_2\geq \sqrt{2}$ contributes at most $(1/\mu)^{-||\mathbf{p}-\mathbf{q}||_2^2/2}\leq \mu\leq 4\mu^*$ by~\eqref{eq:mu-assumption-overview}. This means that the contribution of such points can be approximated well by simply sampling every point with probability $\approx 1/n=1/|P|$ and examining the entire sample -- see Section~\ref{sec:dd-kde} for details. Therefore in the rest of this section (and similarly in its formal version, namely Section~\ref{sec:dd}) we focus on the following single scale recovery problem:

\vspace{0.05in}
\fbox{
\parbox{0.9\textwidth}
{
\begin{center}
Given $x\in (0, \sqrt{2})$ and a sample $\wt{P}$ of the dataset $P$ that includes every point with probability $\frac1{n}\cdot \left(\frac1{\mu}\right)^{1-\frac{x^2}{2}}$, recover all sampled points at distance at most $\approx x$ from the query.
\end{center}
}
}
\vspace{0.05in}

Fix $x\in (0, \sqrt{2})$, and recall that $\wt{P}$ contains every point in $P$ independently with probability $\frac1{n}\cdot \left(\frac1{\mu}\right)^{1-\frac{x^2}{2}}$. Note that by~\eqref{eq:density-constraints-overview} for every $y\in (x, \sqrt{2}]$ the expected number of points at distance at most $y$ from $q$ that are included in $\wt{P}$ is upper bounded by 
\begin{equation}\label{eq:density-ub-11}
n\cdot \left(\frac1{\mu}\right)^{\frac{y^2}{2}-1}\cdot \frac1{n}\cdot \left(\frac1{\mu}\right)^{1-\frac{x^2}{2}}\approx \left(\frac1{\mu}\right)^{\frac{y^2-x^2}{2}}.
\end{equation}

What we defined so far is of course just a reformulation of our approach from Section~\ref{sec:overviewDI}, and indeed our data-dependent result follows the overall uniform sampling scheme. The difference comes in a much more powerful primitive for recovering data points at distance $\approx x$ from the query from the uniform sample. We describe this primitive now.  In this development we start with the observation that underlies the work of~\cite{DBLP:conf/soda/AndoniLRW17} on data-dependent near neighbor search. Namely, one first observes that if the points in the sampled dataset $\wt{P}$ were uniformly random on the sphere (except of course for the actual points at distance $\approx x$ from the query $\mathbf{q}$), then instead of Euclidean LSH one could use random spherical caps to partition the dataset, leading to significantly improved performance.  In order to leverage this observation, the work of~\cite{DBLP:conf/soda/AndoniLRW17} introduces the definition of a pseudo-random dataset (see Definition~\ref{def:psrs} below), gives an efficient procedure for decomposing any dataset into pseudorandom components and shows that the pseudorandom property is sufficiently strong to allow for about the same improvements as a random dataset does. Then their algorithm is a recursive process that partitions a given input dataset using random spherical caps, decomposes the resulting smaller datasets into pseudorandom components and recurses. Our algorithm follows this recipe, but the analysis turns out to be significantly more challenging due to the fact that we need to track the evolution of the densities of balls around the query during this recursive process. In what follows we state the necessary definitions and outline our algorithm.

The work of~\cite{DBLP:conf/soda/AndoniLRW17} introduces a key definition of a pseudorandom dataset (see Definition~\ref{def:psrs}), which we reuse in our analysis and state here for convenience of the reader:

\noindent{\em {\bf Definition~\ref{def:psrs}} (Restated)
Let $P$ be a set of points lying on $\mathcal{S}^{d-1}(o,r)$  for some $o \in \RR^d$ and $r\in \RR_+$. We call this sphere a \emph{pseudo-random} sphere\footnote{Whenever we say \emph{pseudo-random sphere}, we implicitly associate it with parameter $\tau, \gamma$ which are fixed throughout the paper.}, if $\nexists \mathbf{u}^*\in \mathcal{S}^{d-1}(o,r)$ such that 
\begin{align*}
\left| \left\{  \mathbf{u}\in P : ||\mathbf{u}-\mathbf{u}^*||\le r(\sqrt{2}-\gamma)       \right\}\right| \ge \tau\cdot |P|.
\end{align*}
}

In other words, a dataset is pseudorandom on a sphere if at most a small fraction of this dataset can be captured by a spherical cap of nontrivially small volume. It turns out~\cite{DBLP:conf/soda/AndoniLRW17} that every dataset can be partitioned into pseudorandom components efficiently, so one can assume that the input dataset is pseudorandom. The significance of this lies in the fact that the power of Spherical LSH manifests itself on the points $\mathbf{p}$ at distance $\sqrt{2}-\gamma$ from the query essentially as well as on uniformly random points. Thus, if the fraction $\tau$ of `violating' points is small, one now use Spherical LSH to partition the dataset into hash buckets and then recursive on the hash buckets, partition them into pseudorandom components and proceed recursively in this manner. Our algorithms follows this recipe, but the analysis introduces new techniques, as we describe below. We start by fixing some notation. Our algorithm (Algorithm~\ref{alg:DD-KDE-Prep2}) recursively constructs a tree $\mathcal{T}$ with alternating levels of \textsc{SphericalLSH} nodes and \textsc{Pseudorandomify} nodes, which correspond to partitioning the dataset using locality sensitive hashing and extraction of dense components as per Definition~\ref{def:psrs} respectively.  At every \textsc{SphericalLSH} node (Algorithm~\ref{alg:DD-KDE-spherical}) we repeatedly generate subsets $P'$ of the dataset $P$ by sampling a Gaussian vector $g\sim N(0, 1)^d$
and letting
$$
P' \gets \left\{\mathbf{p}\in P : \left\langle \frac{p-o}{R},g\right\rangle \ge \eta \right\},
$$
where $R$ and $o$ are the radius and center of the sphere that dataset $P$ resides on, and $\eta=\omega(1)$ is an appropriately chosen parameter -- we choose $\eta$ to ensure that the collision probability of the query with a point at distance $x$ from it is exactly $\mu^{1/T}$ for a parameter $T$ (see line~\ref{line:1T} of Algorithm~\ref{alg:DD-KDE-spherical}). Crucially, we chose the parameter $\eta$ to ensure that the size of the spherical cap is not too large. Specifically, for a parameter $T=\omega(1)$  that governs the depth of our recursive process we choose $\eta$ to ensure that for every $\mathbf{p}\in P$ such that $||\mathbf{p}-\mathbf{q}||_2\approx x$ one has
$$
\Pr_{g\sim N(0, I)^d}\left[\left\langle \frac{\mathbf{p}-o}{R},g\right\rangle \ge \eta|\left\langle \frac{\mathbf{q}-o}{R},g\right\rangle \ge \eta\right]\approx \mu^{1/T},
$$
where we assume for simplicity of presentation here that the query is on the sphere. The number of datasets $\wt{P}$ is chosen to be such that the query $\mathbf{q}$ collides with any given point $\mathbf{p}$ at distance $\approx x$ with high constant probability over all $O(T)$ levels of the tree $\mathcal{T}$.  This means (see Section~\ref{sec:query}) that the expected number of datasets that the query $\mathbf{q}$ will be exploring is $(1/\mu)^{1/T}$. We limit the depth of the exploration process to $\approx 0.172\cdot T$ (see line~\ref{line:Jstop} of Algorithm~\ref{alg:DD-KDE-spherical}), so that the \textsc{Query} algorithm (see Algorithm~\ref{alg:DD-KDE-Query}) explores at most $((1/\mu)^{1/T})^{0.172\cdot T}=(1/\mu)^{0.172}$ leaf datasets in the tree $\mathcal{T}$. The main challenge lies in showing that these leaf datasets have small (nearly constant) expected size. In other words, we need to bound the effect of such a filtering process on the density of balls of various radius $y$ around the query $\mathbf{q}$. Generally, the densities along any root to leaf path are decreasing because of two effects:
\paragraph{Truncation due to pseudorandom spheres:} First effect that we consider is the condition that pseudo-randomness of spheres imply over the densities. Consider any query $\mathbf{q}$ and any pseudo-random sphere with radius $r$, and let $\ell$ be the distance from $\mathbf{q}$ to the center of the sphere. Let $\mathbf{q}'$ be the projection of the query on the sphere. Then, by pseudo-randomness of the sphere, we know that most of the points are orthogonal to $\mathbf{q}'$, i.e., have distance $\approx \sqrt{2}r$ from $\mathbf{q}'$ (see Lemma~\ref{claim:psrs}). However, we are interested in the condition that implies over the densities. Roughly speaking, the orthogonal points are at distance $c:=\sqrt{\ell^2+r^2}$. So we expect that the number of points at distance $\approx c$ will dominate the densities.
\begin{claim}[Informal version of Claim~\ref{claim:roundedpseudorand}]\label{claim:trunc-overview}
	Suppose that a sphere with center $o$ and radius $r$ is pseudo-random. Then, if $\ell\approx ||\mathbf{q}-o||$, $c:=\sqrt{\ell^2+r^2}$ and for all $y$ we let $B_{y}$ be the number of points at distance $y$ from $\mathbf{q}$ in the sphere. Then, the following conditions hold.
	\begin{align*}
	\sum_{y\leq c-r \psi } B_{y}\leq \frac{\tau}{1-2\tau}\cdot\sum_{y\in (c-r\psi, c+r\psi )} B_{y},
	\end{align*}
	and 
	\begin{align*}
	\sum_{y\ge c+r \psi } B_{y}\leq \frac{\tau}{1-2\tau}\cdot\sum_{y\in (c-r\psi, c+r\psi )} B_{y},
	\end{align*}
	where $\psi=o(1)$ is small factor.
\end{claim} 
\paragraph{Removing points due to Spherical LSH:} The second phenomenon that reduces the densities is spherical LSH rounds. We set the size of the spherical cap as described above. Under this setting of size of spherical cap, the probability that a spherical cap conditioned on capturing the query, captures $\mathbf{p}$, which is at distance $y$ from $\mathbf{q}$, is given by Claim~\ref{cl:nonadapt}, which is restated informally below. 

\begin{claim} [Informal version of Claim~\ref{cl:nonadapt}] \label{cl:density-change} Consider a sphere of radius $r$ around point $o$, and let $\ell\approx||\mathbf{q}-o||$. Also let $\mathbf{p}$ be a point on the sphere such that $y=||\mathbf{p}-\mathbf{q}||$. Now, suppose that one generates a Gaussian vector $g$ as in Algorithm~\ref{alg:DD-KDE-spherical}. Then, we have
	\begin{align*}
	\Pr_{g\sim N(0,1)^d}\left[\langle g,\frac{\mathbf{p}-o}{||\mathbf{p}-o||}\rangle \ge \eta| \langle g,\frac{\mathbf{q}-o}{||\mathbf{q}-o||}\rangle \ge \eta \right]\lesssim \muu{-  \frac{4(r/x')^2-1}{4(r/y')^2-1}\cdot \frac{1}{T}}.
	\end{align*}
	where 
	\begin{itemize}
		\item $\eta$ is such that $\frac{F(\eta)}{G(x'/r,\eta)}\approx \left(\frac{1}{\mu}\right)^{\frac{1}{T}}$ (see line~\ref{line:1T} of Algorithm~\ref{alg:DD-KDE-spherical}).
		\item $x':=\textsc{Project}(x,\ell,r)$ (see Definition~\ref{def:projection}). 
		\item $y' := \textsc{Project}(y,\ell,r)$.
	\end{itemize}
\end{claim}

We use Claim~\ref{claim:trunc-overview} and Claim~\ref{cl:density-change} to bound the evolution of the density of various balls around the query $\mathbf{q}$ in the datasets constructed on the way from the root of the tree $\mathcal{T}$ down to a leaf. 

 Formally, we gather all necessary information about such a path in the definition of a {\em valid execution path} below:

\begin{center}
    \fbox{
      \begin{minipage}{0.99\textwidth}
\noindent{\bf Definition~\ref{def:validpath}} (Valid execution path; slightly informal version)
		Let $R:=(r_j)_{j=1}^{J}$ and $L:=(\ell_j)_{j=1}^{J}$ for some positive values $r_j$'s and $\ell_j$'s such that for all $j\in[J]$, $x \gtrsim |\ell_j-r_j|$. Also let $D$ be as defined in \eqref{eq:dx-def-overview}. Then, for
		\begin{equation*}
		\begin{split}
		A&:=(a_{y,j}), ~~y\in D, j\in[J]\cup \{0\}\text{~~~~~~~~~~(Intermediate densities)}\\
		B&:=(b_{y,j}), ~~y\in D, j\in[J+1]\cup \{0\}\text{~~~~~(Truncated intermediate densities)}
		\end{split}
		\end{equation*}
		$(L,R,A,B)$ is called a \emph{valid execution path}, if the conditions below are satisfied for $\psi:=o(1)$ and $c_j:= \sqrt{r_j^2+\ell_j^2}$ for convenience.
		\begin{description}
			\item[(1) Initial densities condition.] The $a_{y, 0}$ and $b_{y,0}$ variables are upper-bounded by the initial expected densities in the sampled dataset: for all $y\in D$
			\begin{equation*}
			\sum_{y'\in[0,y]\cap D}a_{y',0}\le   \min\left\{ \muu{\frac{y^2-x^2}{2}},\muu{\frac{1-x^2}{2}}\right\}
			\end{equation*}
			and
			\begin{equation*}
			\sum_{y'\in[0,y]\cap D}b_{y',0}\le  \min\left\{ \muu{\frac{y^2-x^2}{2}},\muu{\frac{1-x^2}{2}}\right\}
			\end{equation*}
			\item[(2) Truncation conditions (effect of \textsc{PseudoRandomify}).]  For any $j\in[J]$, for all $y\in D\setminus [\ell_j-r_j,\ell_j+r_j]$ one has $b_{y,j}=0$ (density is zero outside of the range corresponding to the $j$-th sphere on the path; condition {\bf (2a)}), 
			for all $y\in D\cap [\ell_j-r_j,\ell_j+r_j]$ one has $b_{y,j}\le a_{y,j-1}$ (removing points arbitrarily {\bf (2b)}) and 
			\begin{equation*}
			\sum_{y\in [0,c_j-\psi r_j]\cap D}~b_{y,j}\leq \frac{\tau}{1-2\tau}\cdot \sum_{y\in(c_j-\psi r_j,c_j+\psi r_j)\cap D}b_{y,j} ~~~~~\text{(condition {\bf (2c)})}
			\end{equation*}
			
			\item[(3) LSH conditions.] For every $j\in [J]$ and all $y\in [\ell_j-r_j,\ell_j+r_j]\cap D$
			\begin{equation*}
			\begin{split}
			a_{y,j} \le  {b_{y,j}}\cdot\muu{-\frac{4\left(\frac{r_j}{x'}\right)^2-1}{4\left(\frac{r_j}{y'}\right)^2-1}\cdot \frac{1}{T}}
			\end{split}
			\end{equation*}
			where $x':=\textsc{Project}(x+\Delta,\ell_j,r_j)$ and $y':=\textsc{Project}(y-\Delta/2,\ell_j,r_j)$. See Remark~\ref{rem:Delta} below for a discussion about $\Delta$ factors. 
			\item [(4) Terminal density condition.] For any $y$ such that $a_{y,J}$ is defined, $b_{y,J+1}\le a_{y,J}$. 
		\end{description}
	
\end{minipage}
    }
\end{center}

Thus, our main goal is to show that 

\vspace{0.05in}
\fbox{
\parbox{0.9\textwidth}
{
\begin{center}
For every valid execution path $(L, R, A, B)$ one has $\sum_{y} b_{y,J+1}=n^{o(1)}$.
\end{center}
}
}
\vspace{0.05in}

The main challenge here is optimizing over sequences $(\ell_j, r_j)_{j=1}^J$ (distance to center of the sphere from $\mathbf{q}$ and the radius of the sphere). We perform this optimization in two steps, which we describe below.

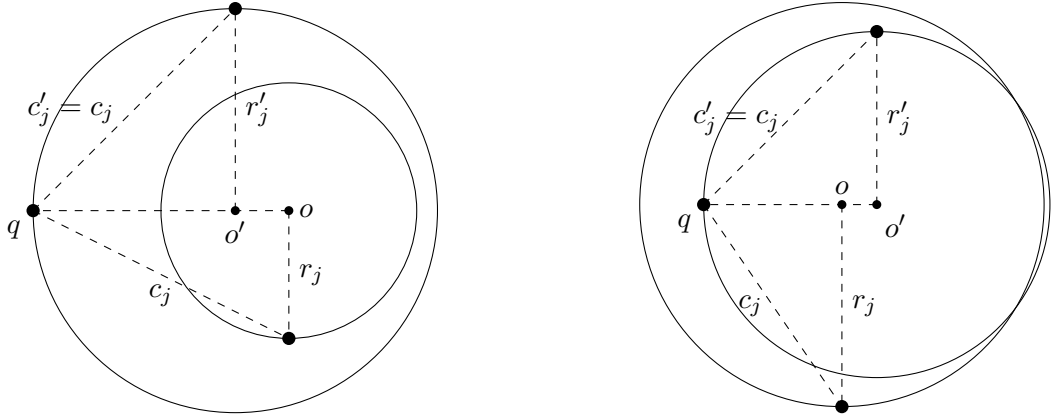
\begin{figure}
\begin{subfigure}[b]{0.5\textwidth}
	\centering
	\tikzstyle{vertex}=[circle, fill=black!70, minimum size=3,inner sep=1pt]
	\tikzstyle{svertex}=[circle, fill=black!100, minimum size=5,inner sep=1pt]
	\tikzstyle{gvertex}=[circle, fill=green!80, minimum size=7,inner sep=1pt]
	
	\tikzstyle{evertex}=[circle,draw=none, minimum size=25pt,inner sep=1pt]
	\tikzstyle{edge} = [draw,-, color=red!100, very  thick]
	\tikzstyle{bedge} = [draw,-, color=green2!100, very  thick]
	\begin{tikzpicture}[scale=1.7, auto,swap]

	\draw (0,0) circle (1cm);

	\fill[fill=black] (0,0) circle (1pt);
	\draw[dashed] (0,0 ) -- node[right]{$r_j$} (0,-1);

	\node[svertex](v1) at (-2, 0) {};
	\draw (-2.15,-0.15) node {{$q$}};
	
	\draw[dashed] (-2,0) --node[below]{$c_j$}(0,-1);  
	\draw[dashed] (0,0 ) --node[below]{} (-2,0);
	\node[svertex](v1) at (0, -1) {};   
	\draw (0,0)[right] node {{$o$}};

	\draw (-0.41886,0)[below] node {{$o'$}};
	\fill[fill=black] (-0.41886,0) circle (1pt);
	\draw (-0.41886,0) circle (1.581138cm);

	\draw[dashed] (-0.41886,0 ) -- node[right]{$r'_j$} (-0.41886,+1.581138);
	
	\node[svertex](v5) at (-0.41886,+1.581138) {};  
	\draw[dashed] (-2,0 ) -- node[left]{$c'_j=c_j~$} (-0.41886,+1.581138);
	\end{tikzpicture}
	\subcaption{When the query is outside of the sphere.}
\end{subfigure}
\begin{subfigure}[b]{0.5\textwidth}
	\centering
	\tikzstyle{vertex}=[circle, fill=black!70, minimum size=3,inner sep=1pt]
	\tikzstyle{svertex}=[circle, fill=black!100, minimum size=5,inner sep=1pt]
	\tikzstyle{gvertex}=[circle, fill=green!80, minimum size=7,inner sep=1pt]
	
	\tikzstyle{evertex}=[circle,draw=none, minimum size=25pt,inner sep=1pt]
	\tikzstyle{edge} = [draw,-, color=red!100, very  thick]
	\tikzstyle{bedge} = [draw,-, color=green2!100, very  thick]
	\begin{tikzpicture}[scale=1.7, auto,swap]
	
	\draw (-1.5+1.354,0) circle (1.354cm);

	\fill[fill=black] (-1.5+1.354,0) circle (1pt);
	\draw[dashed] (-1.5+1.354,0 ) -- node[right]{$r'_j$} (-1.5+1.354,1.354);

	\node[svertex](v1) at (-1.5, 0) {};
	\draw (-1.65,-0.15) node {{$q$}};
	
	\draw[dashed] (-1.5,0) --node[left]{$c'_j=c_j$}(-1.5+1.354,1.354);  
	\draw[dashed] (-1.5+1.354,0 ) --node[below]{} (-1.5,0);
	\node[svertex](v1) at (-1.5+1.354, +1.354) {};   
	\draw (0,0)[below] node {{$o'$}};

	\draw (-0.41886,0)[above] node {{$o$}};
	\fill[fill=black] (-0.41886,0) circle (1pt);
	\draw (-0.41886,0) circle (1.581138cm);

	\draw[dashed] (-0.41886,0 ) -- node[right]{$r_j$} (-0.41886,-1.581138);
	
	\node[svertex](v5) at (-0.41886,-1.581138) {};  
	\draw[dashed] (-1.5,0 ) -- node[left]{$c_j$} (-0.41886,-1.581138);
	\end{tikzpicture}
	\subcaption{When the query is inside the sphere}
\end{subfigure}
\caption{Converting a (non-zero-distance) sphere to its corresponding zero-distance sphere}\label{fig:zerodist} 
\end{figure}

\begin{figure}[H]
\begin{subfigure}[b]{0.5\textwidth}
	\centering
	\tikzstyle{vertex}=[circle, fill=black!70, minimum size=3,inner sep=1pt]
	\tikzstyle{svertex}=[circle, fill=black!100, minimum size=5,inner sep=1pt]
	\tikzstyle{gvertex}=[circle, fill=green!80, minimum size=7,inner sep=1pt]
	
	\tikzstyle{evertex}=[circle,draw=none, minimum size=25pt,inner sep=1pt]
	\tikzstyle{edge} = [draw,-, color=red!100, very  thick]
	\tikzstyle{bedge} = [draw,-, color=green2!100, very  thick]
	\begin{tikzpicture}[scale=1.7, auto,swap]

	\draw (0,0) circle (1cm);

	\fill[fill=black] (0,0) circle (1pt);

	\node[svertex](v1) at (-2, 0) {};
	\draw (-2.15,-0.15) node {{$q$}};

	\draw[dashed] (0,0 ) --node[below]{} (-2,0);
	
	\draw (0,0)[right] node {{$o$}};

	\draw (-0.41886,0)[below] node {{$o'$}};
	\fill[fill=black] (-0.41886,0) circle (1pt);
	\draw (-0.41886,0) circle (1.581138cm);

	\node[svertex](v1) at (-0.4, 0.916){};
	\draw (-0.4, 0.916)[above] node {{$p$}};
	\node[svertex](v1) at (-0.925123, 1.4979){};
	\draw (-0.925123, 1.4979)[above] node {{$p'$}};

	\draw[dashed,red] (-2,-1.8436) arc (-90:90:1.8436cm);

	\end{tikzpicture}
	\subcaption{When the query is outside of the sphere}
\end{subfigure}
\begin{subfigure}[b]{0.5\textwidth}
	\centering
	\tikzstyle{vertex}=[circle, fill=black!70, minimum size=3,inner sep=1pt]
	\tikzstyle{svertex}=[circle, fill=black!100, minimum size=5,inner sep=1pt]
	\tikzstyle{gvertex}=[circle, fill=green!80, minimum size=7,inner sep=1pt]
	
	\tikzstyle{evertex}=[circle,draw=none, minimum size=25pt,inner sep=1pt]
	\tikzstyle{edge} = [draw,-, color=red!100, very  thick]
	\tikzstyle{bedge} = [draw,-, color=green2!100, very  thick]
	\begin{tikzpicture}[scale=1.7, auto,swap]

	\draw (-1.5+1.354,0) circle (1.354cm);

	\fill[fill=black] (-1.5+1.354,0) circle (1pt);

	\node[svertex](v1) at (-1.5, 0) {};
	\draw (-1.65,-0.15) node {{$q$}};
	
	\node[svertex](v1) at (-0.25, 1.35) {};
 	\draw (-0.25, 1.15) node {{$p'$}};
 	
 	\node[svertex](v1) at (-0.55, 1.58) {};
 	\draw (-0.55, 1.78) node {{$p$}};

	\draw[dashed] (-1.5+1.354,0 ) --node[below]{} (-1.5,0);

	\draw (0,0)[below] node {{$o'$}};

	\draw (-0.41886,0)[above] node {{$o$}};
	\fill[fill=black] (-0.41886,0) circle (1pt);
	\draw (-0.41886,0) circle (1.581138cm);
	\draw[dashed,red] (-1.5,-1.8436) arc (-90:90:1.8436cm);

	\end{tikzpicture}
	\subcaption{When the query is inside the sphere}
\end{subfigure}
\caption{Mapping points from a sphere to its corresponding zero-distance sphere.} \label{fig:ppp}
\end{figure}
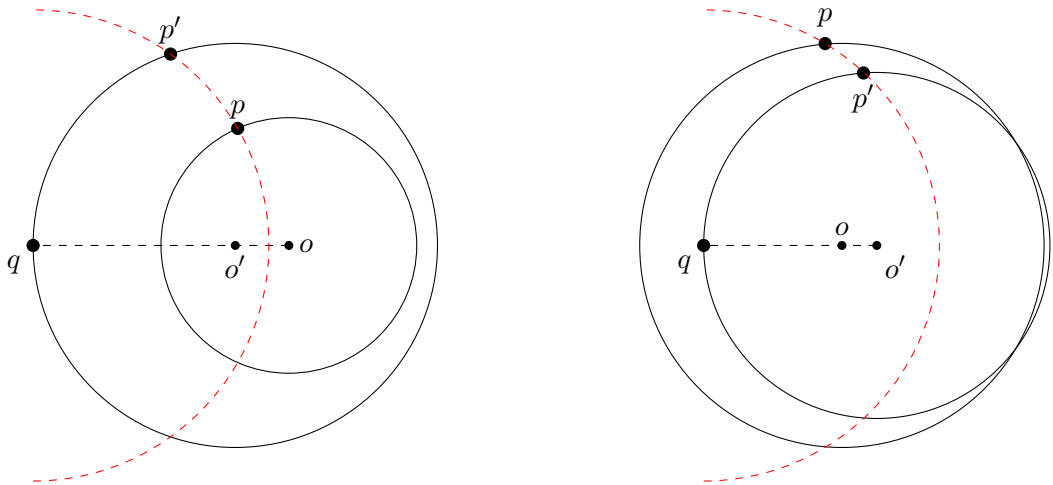

\noindent{\bf Step 1.}Suppose that there are two spheres such that the distance from the query to the orthogonal points for these spheres are the same. Also, assume that for the first sphere the query is not on the sphere, but for the second sphere the query is on the sphere (see Figure~\ref{fig:zerodist}). Now, let $\mathbf{p}$ and $\mathbf{p}'$ lie on the first and the second spheres, respectively. Moreover, assume that they have the same distance from the query (see Figure~\ref{fig:ppp}). Comparing the spherical LSH effect on these two spheres, we prove that $\mathbf{p}$ is removed with higher probability compared to $\mathbf{p}'$ (see Claim~\ref{lm:zd-reduction}). So, when the query is on the sphere, the densities are shrinking with a lower rate.

{\bf Step 2.} Now consider two spheres with different radii, and assume that query $\mathbf{q}$ lies on them at the same time. Due to less curvature on the larger sphere, after one round of spherical LSH on these two spheres, the densities are shrinking with a lower rate on this sphere compared to the smaller sphere (see Claim~\ref{claim:oneswap}). 

The following definition enables us to state our claims more efficiently.

\noindent{\em {\bf Definition~\ref{defn:zerodist}} (Zero-distance and monotone path)
	Let $(L,R,A,B)$ be an execution path defined in Definition~\ref{def:validpath}. If for $R=(r_j)_{j=1}^{J}$, $r_j$'s are non-increasing in $j$, and $L=R$, then we say that $(L,R,A,B)$ is a \emph{zero-distance and monotone} execution path. When $L=R$, we usually drop $L$, and simply write $(R,A,B)$.
}

Now, using the two steps above, we can argue that for any valid execution path, we can find a zero-distance monotone execution path, with the same terminal densities and the same length (see Lemma~\ref{lm:monotone-path} restated below for the convenience of the reader).

 \noindent{\em {\bf Lemma~\ref{lm:monotone-path}} (Zero-distance and monotone path)
 	For every valid execution path $(L, R, A, B)$ (see Definition~\ref{def:validpath}), there exists a \emph{zero-distance and monotone} valid execution path $(R', A', B')$ (see Definition~\ref{defn:zerodist}) such that $b'_{y,J+1}=b_{y,J+1}$ for all $y\in D$\footnote{We need the final condition to argue that we have the same number of points remaining at the end.} and $|R'|=|R|$ (i.e., the length of the paths are equal). 
 }

The proof of the lemma (the formal version of the two steps mentioned above) is given in Section~\ref{sec:exepath}. 

As mentioned before, we analyze the evolution of density of points in various distances. First, we define a grid of distances around query, which we use to properly round the distances of real spheres in the execution of algorithm. Second, instead of analyzing continuous densities, we define a new notion, called \emph{discretized log-densities} (see Definition~\ref{def:fy} below), for which we round densities to the discretized distances in a natural way, and for simplicity of calculations we take the log of these densities. 

\noindent{\em {\bf Definition~\ref{def:zx}} ($x$-centered grid $Z_x$; restated)
For every $x\in (0, R_{max})$ define the grid 
	$Z_x=\{z_I,z_{I-1},\ldots,z_0\}$ by letting $z_I=x$, letting $z_{I-i}:=\left(1+\delta_z\right)^i\cdot z_I$ for all  $i\in[I]$ and choosing the smallest integer $I$ such that  $z_0\geq R_{max}\sqrt{2}$.
}

\noindent{{\em {\bf Definition~\ref{def:fy}} (Discretized log-densities $f_{z_i, j}$; restated)
	For any zero-distance monotone valid execution path $(R,A,B)$ (as per Definition~\ref{def:validpath}) with radii bounded by $R_{max}$ and $J=|R|$,  for all $j \in \left[J\right]$ let $k_j$ be the index of the largest grid element which is not bigger than $r_j\cdot (\sqrt{2}+\psi)$, i.e.,
	\begin{align}\label{eq:zkj-ov}
	r_j \cdot (\sqrt{2}+ \psi) \in   [z_{k_j} , z_{k_j-1} )
	\end{align}
	and for every integer $i \in \{k_j,\ldots, I\}$ define
	\begin{align}
	f_{z_i,j} &:= \log_{1/\mu} \left(\sum_{y\in D \cap [z_{i+1},z_{i-1}) } b_{y,j}\right)  \label{eq:f-def-ov}
	\end{align}
	Note that  the variables $b_{y, j}$ on the right hand side of~\eqref{eq:f-def-ov} are the $b_{y, j}$ variables of the execution path $(R, A, B)$.
}

These two steps, allow us to analyze the evolution of densities over the course of time. In this section, we present an LP (see \eqref{def:lp}) that its optimal cost bounds the query time of our algorithm. The main idea behind the linear program is to relax the notion of a zero-distance monotone path, which may involve only a small number of decreasing sphere radii, to a process that uses a grid $Z=Z_x$ of decreasing radii and possibly applies locality sensitive hashing at every such point (see the spherical LSH constraint in \eqref{def:lp-overview} below), and applies pseudorandmification, i.e. ensures that the dataset is dominated by points at distance $z_j\approx \sqrt{2}r_j$ from the query (see the truncation constraints in~\eqref{def:lp-overview} below). We note that the grid $Z_x$ represents distances to points on the $j$-the sphere that are nearly orthogonal to the query, i.e. whose at distance $\approx \sqrt{2} r_j$, as opposed to the radii themselves. It is also important to note that the linear program is parameterized by two quantities: the target distance $x\in [0, \sqrt{2}]$  and a parameter $j^*$ that indexes a point $z_{j^*}$ in the grid $Z_x$. The quantity $z_{j^*}$ should be thought of as the distance scale that contributes the most to query time, i.e. the band that the has the most number of points in the final densities (see non-empty range constraints in the linear program~\eqref{def:lp-overview}, as well as the similar calculation~\eqref{eq:di-opt} in Section~\ref{sec:overviewDI}). To obtain our final bound on the query time, we enumerate over all $x$ and $j^*\in Z_x$, upper bound the value of the corresponding LP$(x, j^*)$ and take the maximum. Finally, we note that the intended LP solution is as follows. Consider a root to leaf path in the tree $\mathcal{T}$ constructed by \textsc{PreProcess}$(\wt{P}, x, \mu)$ that an invocation of \textsc{Query}$(\mathbf{q},\mathcal{T}, x)$, and suppose that the sequence of radii of spheres traversed by \textsc{Query} is exactly $Z_x$. Then letting $\alpha_j$ denote the number of LSH nodes that correspond to sphere with radius $r_j=z_j/\sqrt{2}$, divided by $T$, should intuitively give a feasible solution\footnote{This statement is somewhat imprecise, and in fact is quite nontrivial to make fully formal -- this is exactly what our algorithm achieves by introducing the notion of valid execution paths.}.

In Section~\ref{sec:main-tech-lemma} we will show in details why this LP formulation is enough to analyze the query time. Informally, this LP considers all possible root to leaf paths, and applies corresponding \emph{truncation} and \emph{spherical LSH} functions on the density and its cost is related to the length of root to leaf paths. We show in Section~\ref{sec:main-tech-lemma} that any execution path with large enough final densities gives a feasible solution to the linear program whose cost is (almost) equal to the length of the path divided by $T$. Thus, if we take any path with length more than $T\cdot \text{OPT(LP)}$, the final densities are small. 

Letting $Z:=Z_x$ to simplify notation,  we will consider $I$ linear programs defined below in~\eqref{def:lp}, enumerating over all $j^*\in [I]$, where we let $x'=x+\Delta$:

\begin{center}
	\fbox{
		\begin{minipage}{0.99\textwidth}
			\begin{align}\label{def:lp-overview} 
			\text{LP}(x, j^*): ~~~~~~~\max_{\alpha\ge 0}&~\sum_{j=1}^{j^*-1}  \alpha_j\\
			\forall y\in Z&: g_{y,1}\le \min\left\{ {\frac{ y^2-x^2}{2}}, {1-\frac{x^2}{2}}\right\}&&\text{Density constraints}\nonumber\\
			\text{~for all~} &j<j^*,  y \in Z, y<z_j:\nonumber\\
			&~ g_{y,j}\le {g_{z_j,j}} &&\text{Truncation}\nonumber\\
			& ~~~g_{y,{j+1}} \le g_{y,{j}}-\frac{2\left(z_j/x\right)^2-1}{2\left(z_j/y\right)^2-1}\cdot \alpha_j &&\text{Spherical LSH}\nonumber\\
			& ~~~g_{z_{j^*},j^*}\geq 0&&\text{Non-empty range constraint}\nonumber
			\end{align}
		\end{minipage}
	}
\end{center}

The following claim is the main technical claim relating zero-distance monotone execution paths and the linear program~\eqref{def:lp-overview}:

\noindent{\em {\bf Claim~\ref{claim:feasible}}(Feasible LP solution from an execution path; Restated)
	If integer $J$ is such that $J> \frac{T}{1-10^{-4}}\mathrm{OPT(LP)}$ then, for all $y\le z_{j^*-1}$ , $f_{y,J+1}< 7\delta_z$ for $j^*= k_{J}+1$ (see Definition~\ref{def:fy} for the definition of $k_J$).
}

The proof of Claim~\ref{claim:feasible} is somewhat delicate, and exploits specific properties of the (negative) log-density of the Gaussian kernel. In fact, one can construct rather simple kernels with non-decreasing log-density for which Claim~\ref{claim:feasible} is false -- we give an example in Figure~\ref{fig:badexample}.  
Informally, we call a kernel well-behaved, if the log-densities after applying a few rounds of LSH (and corresponding truncations), are increasing up to some point and then they are decreasing. More formal description is given after the following paragraph. 
 
Intuitively, the reason is the difference between how the LP works and how the algorithm works. In the algorithm if we are running LSH on some sphere $z$ we apply truncations based on distance $z$ after each round of LSH (except the last step, for the intuition we can ignore this fact) and when we move to the next sphere $z'$, the algorithm applies truncation to log-densities with respect to log-density at $z'$. However, the LP applies all the LSH rounds at once and then does truncation with respect to all bands from $z$ to $z'$. Now, if some kernel is not well-behaved, say like the kernel depicted in Figure~\ref{fig:badexample} then when the LP wants to move from $z$ to $z'$ it also truncates the log-densities with respect to the log-density at any $\eta\in(z',z)$. Then, for some $\eta$ as shown in Figure~\ref{fig:badexample} the log-density at some $\eta\in (z',z)$ is lower compared to the density at $z$ and $z'$. Thus the log-densities in the LP shrink faster than the algorithm, which makes this approach not applicable to these set of kernels. However, for instance in the case of Gaussian kernel, the truncation with respect to log-densities at $\eta\in (z',z)$, do not impose a problem since the log-density at any $\eta\in(z',z)$ is larger than the minimum of densities at $z$ and $z'$. This informally suggests that the evolution of the LP, can be seen as evolution of log-densities for well-behaved kernels, and thus can be used to analyze the run-time of the algorithm. 

Now, we present a relatively more formally definition of well-behaved kernels. We say that a kernel $k(\mathbf{p}, \mathbf{q})=\exp(-h(||\mathbf{p}-\mathbf{q}||_2))$ with the input space scaled so that $\exp(-h(\sqrt{2}))=\mu$ is well-behaved if for every integer $t\ge1$, $x\in(0,\sqrt{2})$ and any sequence $c_1\ge c_2\ge \ldots\ge c_t\ge x$, such that 
	\begin{equation*}
	f(y)=\frac{y^2-x^2}{2}-\sum_{s=1}^{t} \frac{2(c_s/x)^2-1}{2(c_s/y)^2-1}\cdot \frac{1}{T}
	\end{equation*}
	satisfies $f(\sqrt{2}c_t)>0$, the following conditions hold. There exists $y^*\in (x, \sqrt{2}c_t]$ such that the function 
	satisfies $f(y^*)=0$ is monotone increasing on the interval $[y^*, \eta]$, where $\eta$ is where the (unique) maximum of $f$ on $(y^*, \sqrt{2}c_t]$ happens.  See Fig.~\ref{fig:gexample} for an illustration. 
Intuitively, a log-density $h$ is well-behaved if the result of applying any amount of LSH on any collection of spheres to $h$ results in a function with at most one maximum.  This lets us control the structure of log-densities that arise after several iterations of LSH and truncation primitives in a valid execution path (and thus in a root to leaf path in $\mathcal{T}$ that a query $\mathbf{q}$ traverses).

\begin{figure}
	\begin{subfigure}[b]{0.5\textwidth}
		\begin{tikzpicture}
		\begin{axis}[
		restrict y to domain=-10:10,
		samples=1000,
		width=7cm, height=7cm,
		ymin=-0.2 ,ymax=0.6,
		xmin=0, xmax=3.5,
		xtick={1.2,2,3},
		xticklabels={$z'$,$\eta$,$z$},
		ytick={},
		yticklabels={},
		xlabel={Distance from the query},
		ylabel={Log-density},
		axis x line=center,
		axis y line=left,
		every axis x label/.style={
			at={(ticklabel* cs:1)},
			anchor=south,}
		]

		\addplot [
		domain=0:1, 
		samples=10, 
		color=red,
		]
		{0};
		\addplot [
		domain=1:1.001, 
		samples=10, 
		color=red,
		]
		{(x-1)*200};
		
		\addplot [
		domain=1:2, 
		samples=10, 
		color=red,
		]
		{0.2};
			\addplot [
		domain=2:2.001, 
		samples=10, 
		color=red,
		]
		{(x-2)*200+0.2};
					
		\addplot [
		domain=2:3, 
		samples=10, 
		color=red,
		]
		{0.4};
		
			\addplot [
		domain=0:1, 
		samples=10, 
		color=blue,
		]
		{-0.05*x^2};
		
			\addplot [
		domain=1:1.001, 
		samples=10, 
		color=blue,
		]
		{(x-1)*200-0.05};
		
		\addplot [
		domain=1:2, 
		samples=10, 
		color=blue,
		]
		{-0.05*(x-1)^2+0.2-0.05};
		
		\addplot [
		domain=2:2.001, 
		samples=10, 
		color=blue,
		]
		{-0.05+0.2-0.05+(x-2)*200};
		
		\addplot [
		domain=2:3, 
		samples=10, 
		color=blue,
		]
		{0.4-0.1-0.05*(x-2)^2};

		\end{axis}
		\end{tikzpicture}

\subcaption{A kernel, which is not well-behaved}\label{fig:badexample}
\end{subfigure}
\begin{subfigure}[b]{0.5\textwidth}
\begin{tikzpicture}
\begin{axis}[
restrict y to domain=-10:10,
samples=1000,
width=7cm, height=7cm,
ymin=-0.2 ,ymax=1,
xmin=0, xmax=2.5,
xtick={1,1.5,2},
xticklabels={$z'$,$\eta$,$z$},
ytick={},
yticklabels={},
xlabel={Distance from the query},
ylabel={Log-density},
axis x line=center,
axis y line=left,
every axis x label/.style={
	at={(ticklabel* cs:1)},
	anchor=south,}
]

\addplot+[dashed,gray, const plot, no marks] coordinates { (2,0) (2 , 0.437665) } node[above,pos=.57,black] {};

\addplot+[dashed,gray, const plot, no marks] coordinates { (1,0) (1 , 0.288 ) } node[above,pos=.57,black] {};

\addplot [
domain=0.2:1.3, 
samples=100, 
color=red,
]
{(x^2/2.4-0.04/2.4)};

\addplot [
domain=1.3:2, 
samples=100, 
color=red,
]
{  0.6875 };

\addplot [
domain=0.27:1.175, 
samples=100, 
color=blue,
]
{  (x-0.24)^2/2 
};

\addplot [
domain=1.175:1.257, 
samples=100, 
color=blue,
]
{ (x-0.24)^2/2};

\addplot [
domain=1.257:2, 
samples=100, 
color=blue,
]
{ 0.51781-(x-1.2)^2/8   };

\end{axis}

\end{tikzpicture}
\subcaption{A well-behaved kernel}\label{fig:gexample}
\end{subfigure}
\caption{In both figures, the red curve and the blue curve represents the densities before and after running LSH rounds on sphere $z$, respectively. In the case of well-behaved kernels the density at any $\eta\in (z',z)$ is lower-bounded by the minimum of densities at $z$ and $z'$. However, for a kernel which is not well-behaved, for instance for the $\eta$ shown in the left figure, the density is lower than the density at $z$ and $z'$.}	\label{fig:goodexample}
\end{figure}
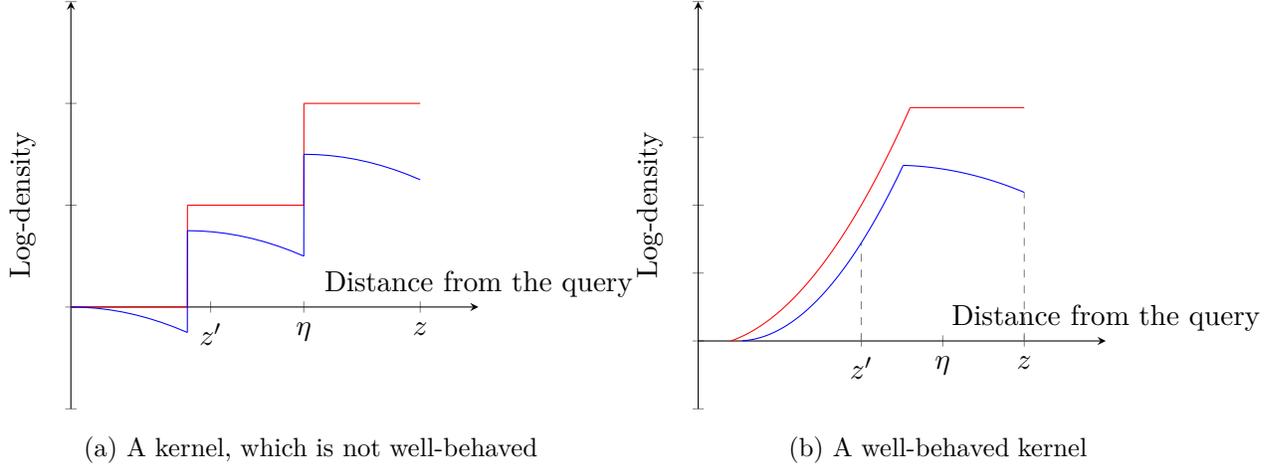

We show in Section~\ref{sec:main-tech-lemma} (see Claim~\ref{claim:monotone}) that the Gaussian kernel is well behaved, and use this fact that prove Claim~\ref{claim:feasible}. We also show a similar claim for the class of kernels whose negative log density is concave (the exponential kernel is one example).  This lets us extend our result to kernels beyond Gaussian (see Remark~\ref{rem:generalkernels} in Section~\ref{sec:dd}).

On the other hand, we  show numerically that the solution of the LP in~\eqref{def:lp-overview} is upper bounded by $0.1718$ for the Gaussian kernel. This is done in Section~\ref{sec:lp-sol} by formulating the dual LP 
\begin{align}
\min&~\sum_{y\in Z} \left\{\frac{y^2-x^2}{2},1-\frac{x^2}{2}\right\} r_{y, 0}\label{eq:dual-overview}\\
\text{such that}&:\nonumber\\
\forall  j\in [j^*-1], y\in Z, y<z_j&: r_{y, j-1}-r_{y, j}+q_{y, j}=0&(g_{y,j})&~~~\text{Mass transportation}\nonumber\\
\forall  j\in [j^*-1]&: r_{z_j, j-1}-\sum_{x\in Z, x<z_j} q_{x, j}=0&(g_{z_j,j})&~~~\text{Max tracking}\nonumber\\
\forall y\in Z, y<z_{j^*}&: r_{y, j^*-1}=0&(g_{y,j^*})&~~~\text{Sink}\nonumber\\
&-\eta+r_{z_{j^*}, j^*-1}=0&(g_{z_{j^*},j^*})&~~~\text{Terminal flow}\nonumber\\
j\in[j^*-1]&:\sum_{y\in Z: y<z_j} \frac{2\left(z_j/x\right)^2-1}{2\left(z_j/y\right)^2-1}r_{y, j}\geq 1&(\alpha_j)&\nonumber\\
&r_{y, j}, q_{y, j}\geq 0&&\nonumber\\
&\eta\geq 0&&\nonumber
\end{align}
and exhibiting a dual feasible solution of value $\approx 0.1716$ for a fine grid of points $Z_x$ and every $x$ in a fine grid over $[0, \sqrt{2}]$. We also give an analytic upper bound of $\frac{x^2}{2}(1-\frac{x^2}{2})+0.001$ on the value of the LP~\eqref{def:lp-overview}.

\newpage
\section{Preliminaries}\label{sec:prelims}

We let $\mu^*\in (0, 1]$ denote the kernel density of a dataset $P$ in $\R^d$ at point $\mathbf{q}\in \R^d$:
\begin{align*}
\mu^*=K(P,\mathbf{q}):=\frac{1}{|P|}\sum_{\mathbf{p}\in P}K(\mathbf{p},\mathbf{q}).
\end{align*}

\subsection{Basic notation}

	Throughout the paper we assume that the points lie in a $d$-dimensional Euclidean space, $\RR^d$. We let $\mathcal{S}^{d-1}$ denote the set of points on the unit radius sphere around the origin in $\RR^d$. Also, for any $o\in\RR^d$ and $R>0$, we let $\mathcal{S}^{d-1}(o,R)$ to be the set of points on the sphere centered at $o$ and radius $R$, and for any point $\mathbf{q}\in \RR^d\setminus\{o\}$, the projection of $\mathbf{q}$ onto $\mathcal{S}^{d-1}(o,R)$ is defined as the closest point in $\mathcal{S}^{d-1}(o,R)$ to $\mathbf{q}$. For any pair of points $\mathbf{u}, \mathbf{v}\in\RR^d$, we let $||\mathbf{u}-\mathbf{v}||$ to be the Euclidean distance of $\mathbf{u}$ and $\mathbf{v}$. 
	
	For any integer $J$ we define $[J]:=\{1, 2,\ldots, J\}$.
	For ease of notation in the rest of the paper, we let $\muu{a}:=\left(\frac{1}{\mu}\right)^a$ and (abusing notation somewhat) let $\exp_2(a)=2^a$  for any $a\in \RR$. 

\subsection{$F(\eta)$ and $G(s,\eta,\sigma)$}
In this section, we define notations and present results, which we later use to analyze the collision probability of spherical-LSH. 
\begin{lem}[Lemma 3.1, \cite{DBLP:conf/soda/AndoniLRW17}]\label{lm:f-eta} If for any $u\in \mathcal{S}^{d-1}$ we define 
	\begin{align*}
	F(\eta):=\Pr_{z\sim N(0,1)^d}\left[ \langle z,u \rangle\ge \eta \right],
	\end{align*}
	then, for $\eta \rightarrow \infty$ 
	\begin{align*}
	F(\eta)=e^{-(1+o(1))\cdot\frac{\eta^2}{2}}.
	\end{align*}
\end{lem}
\begin{lem}[Lemma 3.2, \cite{DBLP:conf/soda/AndoniLRW17}]\label{lm:g-eta} If for any $u,v \in \mathcal{S}^{d-1}$ such that $s:=||u-v||$, we define
	\begin{align*}
	G(s,\eta,\sigma):=\Pr_{z\sim N(0,1)^d}\left[ \langle z,u \rangle\ge \eta \text{ and } \langle z,v\rangle \ge \sigma \right],
	\end{align*}
	then if $\sigma, \eta \rightarrow \infty$, and $\frac{\max\{\sigma,\eta\}}{\min\{\sigma,\eta\}}\ge \alpha(s)$, then one has
	\begin{align*}
	G(s,\eta,\sigma)=e^{-(1+o(1))\cdot \frac{\eta^2+\sigma^2-2\alpha(s)\eta\sigma}{2\beta^2(s)}},
	\end{align*}
	where $\alpha(s):=1-\frac{s^2}{2}$ and $\beta(s):=\sqrt{1-\alpha^2(s)}$.
\end{lem}

\begin{defn}\label{def:g-eta}
	For ease of notation we also define 
	\begin{align*}
	G(s,\eta) := G(s,\eta,\eta).
	\end{align*}
\end{defn}

\subsection{Projection}
\begin{defn}\label{def:projection}
	Let $\mathbf{q}$ be a point on $\mathcal{S}^{d-1}(o,R_1)$ and $\mathbf{p}$ be a point on $\mathcal{S}^{d-1}(o,R_2)$, such that $y:=||q-p||$. Now, if we define $\mathbf{q}'$ as the projection of $\mathbf{q}$ on $\mathcal{S}^{d-1}(o,R_2)$. Then, we define the following
	$$\textsc{Project}(y,R_1,R_2):= ||\mathbf{q}'-\mathbf{p}||.$$
\end{defn}
\begin{lem}\label{lem:projection}
	For any $R_1,R_2\in \RR_{+}$ and $o \in \RR^d$ assume that we have points $\mathbf{q},\mathbf{p}$ on spheres $\mathcal{S}_1:=\mathcal{S}^{d-1}(o,R_1)$ and $\mathcal{S}_2:=\mathcal{S}^{d-1}(o,R_2)$, respectively. Also, let $x:=||\mathbf{p}-\mathbf{q}||$ and let $\mathbf{q}'$ be the projection of point $\mathbf{q}$ on $\mathcal{S}_2$. Then we have the following
	\begin{align*}
	\textsc{Project}(x,R_1,R_2)=||\mathbf{q}'-\mathbf{p}||=\sqrt{\frac{R_2}{R_1}\left(x^2-\left(R_2-R_1\right)^2\right)}.
	\end{align*}
\end{lem}
The proof is deferred to Appendix~\ref{app:prelim}.
\subsection{Pseudo-Random Spheres}
\begin{defn}(Pseudo-random spheres)\label{def:psrs}
	Let $P$ be a set of points lying on $\mathcal{S}^{d-1}(o,r)$  for some $o \in \RR^d$ and $r\in \RR_+$. We call this sphere a \emph{pseudo-random} sphere\footnote{Whenever we say \emph{pseudo-random sphere}, we implicitly associate it with parameter $\tau, \gamma$ which are fixed throughout the paper.}, if $\nexists \mathbf{u}^*\in \mathcal{S}^{d-1}(o,r)$ such that 
	\begin{align*}
	\left| \left\{  \mathbf{u}\in P : ||\mathbf{u}-\mathbf{u}^*||\le r(\sqrt{2}-\gamma)       \right\}\right| \ge \tau\cdot |P|.
	\end{align*}
\end{defn}

As shown in~\cite[Section 6]{AR15}, it is possible to decompose a dataset $P$ into pseudo-random components in time $\mathrm{poly}\left(d, \gamma^{-1},\tau^{-1},\log|P|\right)\cdot |P|$. We present a slight modification of their argument using our notation in Appendix~\ref{sec:ball_carving}.   The following claim summarizes the properties of a pseudo-random sphere that we use later:

\begin{claim}\label{claim:psrs}
	If $P$ is a set of points lying on $\mathcal{S}^{d-1}(o,r)$ for some $o\in\RR^d$ and $r\in \RR_+$, and $P$ is a pseudo-random sphere (see Definition~\ref{def:psrs}) then for any point $\mathbf{q}'$ on the sphere we have the following property
	\begin{align*}
	\left| \left\{  \mathbf{u}\in P : ||\mathbf{u}-\mathbf{q}'||\le r(\sqrt{2}-\gamma)       \right\}\right| \le \frac{\tau}{1-2\tau}\cdot \left| \left\{  \mathbf{u}\in P :||\mathbf{u}-\mathbf{q}'||\in\left(r\left(\sqrt{2}-\gamma\right), r (\sqrt{2}+\gamma)\right)    \right\}\right|,
	\end{align*}
and consequently,
	\begin{align*}
	\left| \left\{  \mathbf{w}\in P :||\mathbf{w}-\mathbf{q}'||\in\left(r\left(\sqrt{2}-\gamma\right), r \left(\sqrt{2}+\gamma\right)\right)    \right\}\right|=\Omega(|P|).
	\end{align*}
\end{claim}
The proof is deferred to Appendix~\ref{app:prelim}.


 \section{Kernel Density Estimation Using Andoni-Indyk LSH}\label{sec:data-independent}
 In this section, we present an algorithm for estimating KDE, using the Andoni-Indyk LSH framework. In order to state the main result of this section for general kernels, we need to define a few notions first. Thus, we state the main result for Gaussian kernel in the following theorem, and then state the general result, Theorem~\ref{thm:query-time-gen}, after presenting the necessary definitions.
 \begin{thm}\label{thm:Gaus4}
 	Given a kernel $K(\mathbf{p},\mathbf{q}):=e^{-a||\mathbf{p}-\mathbf{q}||_2^2}$ for any $a>0$, $\epsilon = \Omega\left(\frac{1}{\mathrm{polylog} n}\right)$, $\mu^*=n^{-\Theta(1)}$ and a data set of points $P$, using Algorithm~\ref{alg:4} for preprocessing and Algorithm~\ref{alg:query4} for the query procedure, one can approximate $\mu^*:=K(P,\mathbf{q})$ (see Definition~\ref{def:kernel}) up to $(1\pm \epsilon)$ multiplicative factor, in time $\wt{O}\left(\epsilon^{-2}\left(\frac{1}{\mu^*}\right)^{0.25+o(1)}\right)$, for any query point $\mathbf{q}$. Additionally, the space consumption of the data structure is $$\min\left\{ \epsilon^{-2} n\left(\frac{1}{\mu^*}\right)^{0.25+o(1)},\epsilon^{-2}\left(\frac{1}{\mu^*}\right)^{1+o(1)}\right\}.$$
 \end{thm}

 Throughout this section, we refer to Andoni-Indyk LSH's main result stated in the following lemma.

 \begin{lem}[\cite{DBLP:conf/focs/AndoniI06}]\label{lem:andoni-indyk}
 	Let $\mathbf{p}$ and $\mathbf{q}$ be any pair of points in $\mathbb{R}^d$. Then, for any fixed $r>0$, there exists a hash family $\mathcal{H}$ such that, if $p_{\mathrm{near}}:=p_1(r):=\Pr_{h\sim\mathcal{H}}[h(\mathbf{p})=h(\mathbf{q}) \mid ||\mathbf{p}-\mathbf{q}||\le r]$ and $p_{\mathrm{far}}:=p_2(r,c):=\Pr_{h\sim\mathcal{H}}[h(\mathbf{p})=h(\mathbf{q}) \mid ||\mathbf{p}-\mathbf{q}||\ge cr]$ for any $c\ge1$, then $$\rho:=\frac{\log 1/p_{\mathrm{near}}}{\log 1/p_{\mathrm{far}}}\le \frac{1}{c^2}+O\left(\frac{\log t}{t^{1/2}}\right),$$ for some $t$, where $p_{\mathrm{near}}\ge e^{-O(\sqrt{t})}$ and each evaluation takes $d t^{O(t)}$ time.
 \end{lem}
 \begin{rem}\label{rem:AI}
 	From now on, we use $t=\log^{2/3}n$, which results in $n^{o(1)}$ evaluation time and $\rho=\frac{1}{c^2}+o(1)$. In that case, note that if $c=O\left(\log^{1/7}n\right)$, then 
 	\begin{align*}
 	\frac{1}{\frac{1}{c^2}+O\left(\frac{\log t}{t^{1/2}}\right)}=c^2(1-o(1)).
 	\end{align*}
 \end{rem}
 \begin{defn}\label{def:kernel}
 	For a query $\mathbf{q}$, and dataset $P=\{\mathbf{p}_1,\ldots,\mathbf{p}_n\}$, we define 
 	\begin{align*}
 	\mu^*:=K(P,\mathbf{q}):=\frac{1}{|P|}\sum_{\mathbf{p}\in P}K(\mathbf{p},\mathbf{q})
 	\end{align*}
 	where for any $\mathbf{p}\in P$, $K(\mathbf{p},\mathbf{q})$ is a monotone decreasing function of $||\mathbf{q}-\mathbf{p}||$. Also, we define $$w_i:=K(\mathbf{p}_i,\mathbf{q}).$$
 \end{defn}
From now on, we assume that $\mu$ is a quantity such that 
\begin{align}
\mu^*\le \mu
\end{align}
We also use variable $J:=\left\lceil\log_2 \frac{1}{\mu}\right\rceil$.
 \begin{defn}[Geometric weight levels]\label{def:geom}
 	For any $j\in[J]$
 	\begin{align*}
 	L_j:=\left\{\mathbf{p}_i \in P: w_i \in \left(2^{-j},2^{-j+1}\right] \right\}.
 	\end{align*}
 	This implies corresponding distance levels (see Figure~\ref{fig:inducedistlevels} and Figure~\ref{fig:defgeom}), which we define as follows
 	\begin{align*}
 	\forall j\in [J]:~r_j:=\max_{\text{s.t. }f(r) \in \left(2^{-j},2^{-j+1}\right]} r.
 	\end{align*}
where $f(r):=K(\mathbf{p},\mathbf{p}')$ for $r=||\mathbf{p}-\mathbf{p}'||$. Also define $L_{J+1}:=P \setminus \cup_{j\in[J]}L_j$.\footnote{One can see that $L_{J+1}=\{\mathbf{p}_i\in P: w_i\le 2^{-J}\}$.}
 \end{defn}
 	\begin{figure}[H]
 	\centering
 	\tikzstyle{vertex}=[circle, fill=black!70, minimum size=3,inner sep=1pt]
 	\tikzstyle{svertex}=[circle, fill=black!100, minimum size=5,inner sep=1pt]
 	\tikzstyle{gvertex}=[circle, fill=green!80, minimum size=7,inner sep=1pt]
 	
 	\tikzstyle{evertex}=[circle,draw=none, minimum size=25pt,inner sep=1pt]
 	\tikzstyle{edge} = [draw,-, color=red!100, very  thick]
 	\tikzstyle{bedge} = [draw,-, color=green2!100, very  thick]
 	\begin{tikzpicture}[scale=1, auto,swap]
 	
 	\draw (0,0) circle (2cm);
 	\draw (0,0) circle (3cm);
 	
 	 \draw[pattern=north west lines, pattern color=gray] (0,0) circle (3cm);
 	 \fill[fill=white] (0,0) circle (2cm);
 	
 	\fill[fill=black] (0,0) circle (1pt);
 	\draw[dashed] (0,0 ) -- node[above]{$r_{j-1}$} (-2,0);
 	\draw[dashed] (0,0 ) -- node[right]{$r_j$} (-2.4,-1.8);
 \node (A) at (-2.5, 2.5) {$L_j$};
\node (B) at (-1,1) {};
\draw [-> ] (-2.3,2.3) -- (-1.8,1.8);
 	\draw (0,0)[above] node {{$\mathbf{q}$}};
 	\end{tikzpicture}
 	\caption{Illustration of definition of $r_j$'s based on $L_j$'s.} \label{fig:defgeom}
 \end{figure}
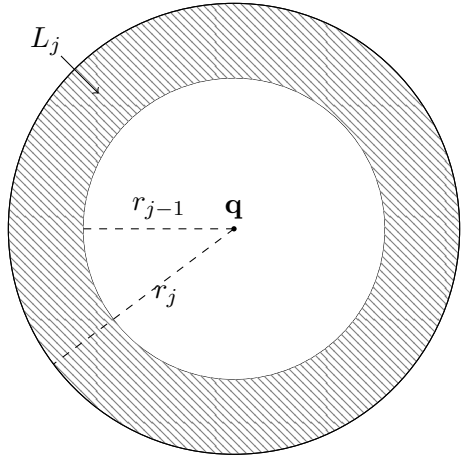
 
 We start by stating basic bounds on collision probabilities under the Andoni-Indyk LSH functions in terms of the definition of geometric weight levels $L_j$ (Definition~\ref{def:geom}):
 \begin{claim}\label{claim:collprob}
 	Assume that kernel  $K$ induces weight level sets, $L_j$'s, and corresponding distance levels, $r_j$'s (as per Definition~\ref{def:geom}). Also, for any query $\mathbf{q}$, any integers $i\in[J+1],j \in [J]$ such that $i>j$, let $\mathbf{p}\in L_j$ and $\mathbf{p}'\in L_i$. And assume that $\mathcal{H}$ is an Andoni-Indyk LSH family designed for near distance $r_j$ (see Lemma~\ref{lem:andoni-indyk}). Then, for any integer $k\ge 1$, we have the following conditions:
 	\begin{enumerate}
 		\item $\Pr_{h^*\sim\mathcal{H}^k}\left[ h^*(\mathbf{p})=h^*(\mathbf{q})\right] \ge p_{\mathrm{near},j}^k$,
 		\item $\Pr_{h^*\sim\mathcal{H}^k}\left[ h^*(\mathbf{p}')=h^*(\mathbf{q})\right] \le p_{\mathrm{near},j}^{kc^2(1-o(1))}$,
 	\end{enumerate}
 where $c:=c_{i,j}:=\min\left\{\frac{r_{i-1}}{r_j},\log^{1/7}n\right\}$ (see Remark~\ref{rem:AI}) and $p_{\mathrm{near},j}:=p_1(r_j)$ in Lemma~\ref{lem:andoni-indyk}.
 \end{claim}
\begin{proof}
	If $\mathbf{p} \in L_j$ by Definition~\ref{def:geom}, we have
	\begin{align*}
	||\mathbf{q}-\mathbf{p}|| \le r_j.
	\end{align*}
	Similarly using the fact that the kernel is decaying, for $\mathbf{p}' \in L_i$ we have
	\begin{align*}
	||\mathbf{q}-\mathbf{p}'|| \ge  r_{i-1} \ge c\cdot r_j.
	\end{align*}
	So, by Lemma~\ref{lem:andoni-indyk} and Remark~\ref{rem:AI} the claim holds. Figure~\ref{fig:AIgeom} shows an instance of this claim. 
\end{proof}
 	\begin{figure}
	\centering
	\tikzstyle{vertex}=[circle, fill=black!70, minimum size=3,inner sep=1pt]
	\tikzstyle{svertex}=[circle, fill=black!100, minimum size=5,inner sep=1pt]
	\tikzstyle{gvertex}=[circle, fill=green!80, minimum size=7,inner sep=1pt]
	
	\tikzstyle{evertex}=[circle,draw=none, minimum size=25pt,inner sep=1pt]
	\tikzstyle{edge} = [draw,-, color=red!100, very  thick]
	\tikzstyle{bedge} = [draw,-, color=green2!100, very  thick]
	\begin{tikzpicture}[scale=0.8, auto,swap]

			\draw (0,0) circle (4cm);
	\draw (0,0) circle (5cm);

	\draw[pattern=north west lines, pattern color=gray] (0,0) circle (5cm);
	\fill[fill=white] (0,0) circle (4cm);
	
	\draw (0,0) circle (1.3cm);
	\draw (0,0) circle (1.8cm);
	
	\draw[pattern=north east lines, pattern color=gray] (0,0) circle (1.8cm);
	\fill[fill=white] (0,0) circle (1.3cm);
	
	\fill[fill=black] (0,0) circle (1pt);

	\draw[->] (0,0 ) -- node[above]{$r_{i-1}$} (-4,0);

	\draw[->] (0,0 ) -- node[left]{$r_j$} (0,-1.8);
	
	\node (A) at (-1.8, 1.8) {$L_j$};
	\node (B) at (-1,1) {};
	\draw [-> ] (-1.6,1.6) -- (-1.1,1.1);
	
		\node (C) at (-4, 4) {$L_i$};
	
	\draw [-> ] (-3.8,3.8) -- (-3.3,3.3);
	\draw (0,0)[above] node {{$\mathbf{q}$}};
	
	\draw (1.3,1.1)[above] node {{$\mathbf{p}$}};
\node[vertex](v1) at (1, 1.3) {};  
	
	\node[vertex](v1) at (0, 0) {};  
		\draw (3,3.4)[above] node {{$\mathbf{p'}$}};
	\node[vertex](v1) at (2.6, 3.6) {};  
	\end{tikzpicture}
	\caption{Illustration of $r_j$ and $r_{i-1}$ in terms of $L_j$ and $L_{i}$.} \label{fig:AIgeom}
\end{figure}
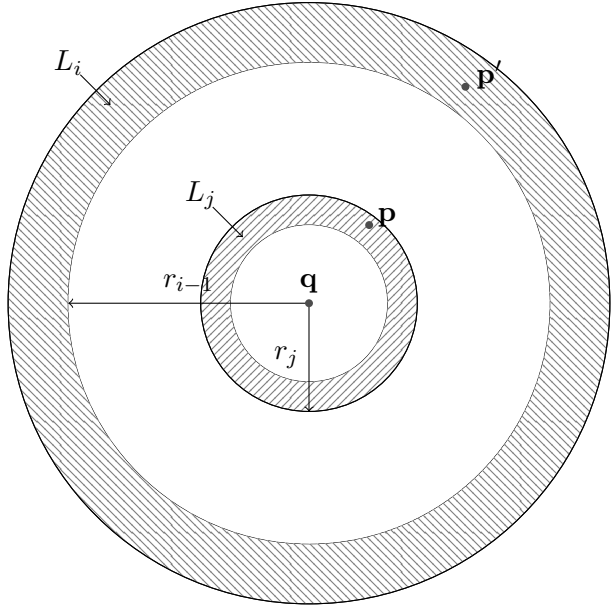

Now, we prove an upper-bound on sizes of the geometric weight levels, i.e., $L_j$'s (see Definition~\ref{def:geom}).
 \begin{lem}[Upper bounds on sizes of geometric weight levels]\label{lem:sizeL}
 	For any $j\in [J]$, we have  $$|L_j|\le 2^{j}n \mu^*\le 2^{j}n\mu.$$
 \end{lem}
 \begin{proof}
 	For any $j\in [J]$ we have 
 	\begin{align*}
 	n\mu\ge n \mu^*&=\sum_{\mathbf{p}\in P}K(\mathbf{p},\mathbf{q})&&\text{By Definition~\ref{def:kernel}}\\
 	&\ge \sum_{i\in[J]}\sum_{\mathbf{p}\in L_i} K(\mathbf{p},\mathbf{q})\\
 	&\ge \sum_{\mathbf{p}\in L_j} K(\mathbf{p},\mathbf{q})\\
 	&\ge  |L_j|\cdot  2^{-j}
 	\end{align*}
 	which proves the claim.
 \end{proof}

\begin{defn}[Cost of a kernel]\label{def:costK}
	Suppose that a kernel $K$ induces geometric weight levels, $L_j$'s, and corresponding distance levels, $r_j$'s (see Definition~\ref{def:geom}). For any $j\in[J]$ we define \emph{cost} of kernel $K$ for weight level $L_j$ as 
		\begin{align*}
	\mathrm{cost}(K,j)&:= \exp_2\left(\max_{i=j+1,\ldots,J+1}  \left\lceil  \frac{i-j}{c_{i,j}^2(1-o(1))}     \right\rceil\right),
	\end{align*}
	where $c_{i,j}:=\min\left\{\frac{r_{i-1}}{r_j},\log^{1/7}n\right\}$. Also, we define the general \emph{cost} of a kernel $K$ as
	\begin{align*}
	\mathrm{cost}(K):= \max_{j\in[J]} \mathrm{cost}(K,j).
	\end{align*}
\end{defn}

\paragraph{Description of algorithm:}
The algorithm runs 	in $J$ phases. For any $j\in [J]$, in the $j$'th phase, we want to estimate the contribution of points in $L_j$ to $K(P,\mathbf{q})$. We show that it suffices to have an estimation of the number of points in $L_j$. One can see that if we sub-sample the data set with probability $\min\{\frac{1}{2^jn\mu},1\}$, then in expectation we get at most $O(1)$ points from $L_{i}$ for any $i\le j$. Now, assume that a point $\mathbf{p}\in L_j$ gets sampled by sub-sampling, then we want to use Andoni-Indyk LSH to distinguish this point from other sub-sampled points, efficiently. Thus, we want to find the appropriate choice of $k$ for the repetitions of Andoni-Indyk LSH (see Claim~\ref{claim:collprob}). Suppose that we call Claim~\ref{claim:collprob} with some $k$ (which we calculate later in \eqref{eq:kj}). Then we have 
\begin{align*}
\Pr_{h^*\sim\mathcal{H}^k}\left[ h^*(\mathbf{p})=h^*(\mathbf{q})\right] \ge p_{\mathrm{near},j}^k,
\end{align*}
which implies that in order to recover point $\mathbf{p}$ with high probability, we need to repeat the procedure $\wt{O}\left( p_{\mathrm{near},j}^{-k}\right)$ times. Another factor that affects the run-time of the algorithm is the number of points that we need to check in order to find $\mathbf{p}$. Basically, we need to calculate the number of points that hash to the same bucket as $\mathbf{q}$ under $h^*$'s. For this purpose, we use the second part of Claim~\ref{claim:collprob}, which bounds the collision probability of far points, i.e., points such as $\mathbf{p}'\in L_i$ for any $i>j$. Intuitively, for any point $\mathbf{p}'\in L_i$ for any $i>j$, by Claim~\ref{claim:collprob} we have 
\begin{align*}
\Pr_{h^*\sim\mathcal{H}^k}\left[ h^*(\mathbf{p}')=h^*(\mathbf{q})\right] \le p^{kc^2(1-o(1))}
\end{align*}
where $c:=c_{i,j}:=\min\left\{ \frac{r_{i-1}}{r_j},\log^{1/7}n\right\}$ and $p:=p_{\mathrm{near},j}$\footnote{The indices are dropped for $c_{i,j}$ and $p_{\mathrm{near},j}$ for ease of notation.}. On the other hand, by Lemma~\ref{lem:sizeL}, for $i=j+1,\ldots,J$ we have
\begin{align*}
|L_{i}|\leq 2^{i}n\mu^{*}\leq 2^{i}n\mu.
\end{align*}
Then, one has the following bound,
\begin{align}
&\nonumber\mathbb{E}\left[\left| \{ \mathbf{p}' \in L_i: h^*(\mathbf{p}')=h^*(\mathbf{q}) \}   \right|\right] \\
&\nonumber\le 2^{i}n\mu\cdot \frac{1}{2^jn\mu}\cdot p^{kc^2(1-o(1))}&&\text{Sub-sampling and then applying LSH}\\
&=2^{i-j}\cdot p^{kc^2(1-o(1))}.\label{eq:hashsubsample}
\end{align}
Since we have $O\left(\log \frac{1}{\mu}\right)$ geometric weight levels, then the expression in \eqref{eq:hashsubsample} for the worst $i$, bounds the run-time up to $O\left(\log \frac{1}{\mu}\right)$ multiplicative factor. In order to optimize the run-time up to $\wt{O}(1)$ multiplicative factors, we need to set $k$ such that the expression in \eqref{eq:hashsubsample} gets upper-bounded by $O(1)$ for all $i>j$.  So, in summary, for any fixed $j\in[J]$, we choose $k$ such that any weight level $L_i$ for $i\ge j$ contributes at most $\wt{O}(1)$ points in expectation to the hash bucket of the query, i.e., $h^*(\mathbf{q})$. One can see that we can choose $k$ as follows
\begin{align}\label{eq:kj} 
k:=k_j:=\frac{-1}{ \log p} \cdot\max_{i=j+1,\ldots,J+1}  \left\lceil  \frac{i-j}{c_{i,j}^2(1-o(1))}     \right\rceil.
\end{align}

For sampling the points in $L_{J+1}$, it suffices to sample points in the data set with probability $\frac{1}{n}$ (see line~\ref{line:beyondJsample} in Algorithm~\ref{alg:4}), since the size of the sampled data set is small and there is no need to apply LSH. One can basically scan the sub-sampled data set.
 \begin{algorithm}[H]
	\caption{Preprocessing}  
	\label{alg:4} 
	\begin{algorithmic}[1]
		
		\Procedure{$\textsc{PreProcess}(P,\epsilon )$}{}
	
		\State {} \Comment{$P$ represents the set of data points}
		\State{} \Comment{$\epsilon$ represents the precision of estimation}
		\State $K_1 \gets \frac{C\log n}{\epsilon^2}\cdot \mu^{-o(1)}$\Comment{$C$ is a universal constant}\label{line:q-K1}
		\State $J \gets  \left\lceil\log {\frac{1}{\mu}}\right\rceil$\Comment{We use geometric weight levels with base $2$, see~Definition~\ref{def:geom}}
		\For {$a=1,2,\dots,K_1$}\Comment{$O(\log n/\epsilon^2)$ independent repetitions}
		\For {$j=1,2,\dots,J$}\Comment{$J=\left\lceil\log \frac{1}{\mu}\right\rceil$ geometric weight levels}
		\State $K_2\gets 100\log n \cdot p_{\mathrm{near},j}^{-k_j}$\label{line:q-K2}\\ \Comment{See Claim~\ref{claim:collprob} and \eqref{eq:kj} for definition of $p_{\mathrm{near},j}$ and $k_j$}
		\State $p_{\text{sampling}} \gets \min\{\frac{1}{2^jn\mu},1\}$ \label{line:setpalg1}
		\State {$\wt{P} \gets $~sample each element in $P$ with probability $p_{\text{sampling}}$.} \label{line:sub-sample4} 
		\For {$\ell=1,2,\dots,K_2$}
		\State Draw a hash function from hash family $\mathcal{H}^{k_j}$ as per Claim~\ref{claim:collprob} and call it $H_{a,j,\ell}$
		\State Run $H_{a,j,\ell}$ on $\wt{P}$ and store non-empty buckets
		\EndFor
		\EndFor
		\State {$\wt{P}_a \gets $~sample each element in $P$ with probability $\frac{1}{n}$}\label{line:beyondJsample}
		\State Store $\wt{P}_a$\Comment{Set $\wt{P}_a$ will be used to recover points beyond $L_{J+1}$}
		\EndFor
		\EndProcedure
	\end{algorithmic}
\end{algorithm}
 \begin{algorithm}[H]
	\caption{Query procedure}  
	\label{alg:query4} 
	\begin{algorithmic}[1]
		
		\Procedure{$\textsc{Query}(P,\mathbf{q},\epsilon,\mu)$}{}
		\State {} \Comment{$P$ represents the set of data points}
		\State{} \Comment{$\epsilon$ represents the precision of estimation}
		\State $K_1 \gets \frac{C\log n}{\epsilon^2}\cdot \mu^{-o(1)}$\Comment{$C$ is a universal constant}
		\State $J \gets \left\lceil\log {\frac{1}{\mu}}\right\rceil$ \Comment{We use geometric weight levels with base $2$, see~Definition~\ref{def:geom}}
		\For {$a=1,2,\dots,K_1$}\Comment{$O(\log n/\epsilon^2)$ independent repetitions}
		\For {$j=1,2,\dots,J$}\Comment{$J=\left\lceil\log \frac{1}{\mu}\right\rceil$ geometric weight levels}
		\State $K_2\gets 100\log n \cdot p_{\mathrm{near},j}^{-k_j}$ \Comment{See Claim~\ref{claim:collprob} and \eqref{eq:kj} for definition of $p_{\mathrm{near},j}$ and $k_j$}
		\For {$\ell=1,2,\dots,K_2$}
		\State Scan $H_{a,j,\ell}(q)$ and recover points in $L_j$
		\EndFor
		\EndFor

		\State Recover points from $L_{J+1}$ in the sub-sampled dataset, $\wt{P}_a$.
		\State $S \gets $ set of all recovered points in this iteration
		\For{$\mathbf{p}_i\in S$}
		\State $w_i \gets K(\mathbf{p}_i,\mathbf{q})$
		\If {$\mathbf{p}_i\in L_j$ for some $j\in[J]$}
		\State $p_i\gets\min\{\frac{1}{2^jn\mu},1\}$,
		\ElsIf {$\mathbf{p}_i\in P\setminus \cup_{j\in[J]}L_j$} 
		\State $p_i\gets \frac{1}{n}$ 
		\EndIf
		\EndFor
		\State $Z_a\gets\sum_{\mathbf{p}_i\in S}\frac{w_i}{p_i}$\label{line:est4}
		\EndFor
		\EndProcedure
	\end{algorithmic}
\end{algorithm}
Now, we present the main result of this section. 
\begin{thm}[Query time]\label{thm:query-time-gen}
	For any kernel $K$, the expected query-time of the algorithm is equal to $\wt{O}\left(\epsilon^{-2}n^{o(1)}\cdot \mathrm{cost}(K)\right)$.
\end{thm}
Assuming Theorem~\ref{thm:query-time-gen}, we prove Theorem~\ref{thm:Gaus4}.

\begin{proofof}{Theorem~\ref{thm:Gaus4}}
	We first start by proving the query time bound and then we prove the space consumption of the data structure, and the guarantee over the precision of the estimator is given in Claim~\ref{claim:variance}.
\paragraph{Proof of the query time bound:}
We calculate the cost of Gaussian kernel $e^{-a||\mathbf{x}-\mathbf{y}||_2^2}$. First, we present the weight levels and distance levels induced by this kernel. As per Definition~\ref{def:kernel}, let 
\begin{align*}
	\mu^*:= K(P,\mathbf{q})=\sum_{\mathbf{p}\in P}e^{-a||\mathbf{p}-\mathbf{q}||_2^2}.
\end{align*}
By Definition~\ref{def:geom}, one has
\begin{align*}
L_j&:=\left\{\mathbf{p}_i\in P: w_i\in \left(2^{-j},2^{-j+1}\right]\right\}\\
&=\left\{   \mathbf{p}_i\in P : ||\mathbf{p}_i-\mathbf{q}||_2 \in \left[\sqrt{\frac{(j-1)\ln 2}{a}}, \sqrt{\frac{j\ln 2}{a}}\right)   \right\},
\end{align*}
which immediately translates to $r_j:= \sqrt{\frac{j\ln 2}{a}}$ for all $j\in[J]$. Also, we for all $i\in[J+1],j \in [J]$ such that $i>j$, we have 

\begin{align*}
c_{i,j}&:=\min\left\{ \frac{r_{i-1}}{r_j},\log^{1/7}n\right\}\\
&=\min\left\{ \sqrt{\frac{{i-1}}{j}},\log^{1/7}n\right\}
\end{align*}
At this point, one can check that 
\begin{align*}
\max_{j\in[J]}\max_{i=j+1,\ldots,J+1}  \left\lceil  \frac{i-j}{c_{i,j}^2(1-o(1))}     \right\rceil=
(1+o(1))\frac{1}{4}\log\frac{1}{\mu},
\end{align*}
Therefore, the cost of Gaussian kernel is 
\begin{align*}
\mathrm{cost}(K)=\left(\frac{1}{\mu}\right)^{(1+o(1))\frac{1}{4}}.
\end{align*}
Now, invoking Theorem~\ref{thm:query-time-gen}, the statement of the claim about the query time holds. 
\paragraph{Proof of the space bound:} First, since the query time is bounded by $\epsilon^{-2}\left(\frac{1}{\mu^*}\right)^{0.25+o(1)}$, then the number of hash functions used is also bounded by the same quantity. This implies that the expected size of the space needed to store the data structure prepared by the preprocessing algorithm is $\epsilon^{-2}n\left(\frac{1}{\mu^*}\right)^{0.25+o(1)}$, since for each hash function we are hashing at most $n$ points (number of points in the dataset). 

For the other bound, we need to consider the effect of sub-sampling the data set. Fix $j\in [J]$. In the phase when we are preparing the data structure to recover points from $L_j$, we sub-sample the data set with probability $\min\{\frac{1}{2^jn\mu},1\}$, and then we apply $\wt{O}\left(p_{\text{near},j}^{-k_j}\right)$ hash functions to this sub-sampled data set. Since 
$$
k_j=\frac{-1}{ \log p} \cdot\max_{i=j+1,\ldots,J+1}  \left\lceil  \frac{i-j}{c_{i,j}^2(1-o(1))}     \right\rceil,
$$
by~\eqref{eq:kj}, where $p=p_{\text{near},j}$, we have 
\begin{equation}
p_{\text{near},j}^{-k_j}=\exp_2\left(\max_{i=j+1,\ldots,J+1}  \left\lceil  \frac{i-j}{c_{i,j}^2(1-o(1))}    \right\rceil-j\right).
\end{equation}
At the same time, the expected size of the sampled dataset is bounded by $n\cdot \min\{\frac{1}{2^jn\mu},1\}\leq \frac1{\mu}\cdot 2^{-j}$. Putting this together with the equation above, we get that the expected size of the dataset constructed for level $L_j$ is upper bounded by 
\begin{equation}\label{eq:size-bound-92h3tg}
\frac{1}{\mu} \exp_2\left(\max_{i=j+1,\ldots,J+1}  \left\lceil  \frac{i-j}{c_{i,j}^2(1-o(1))}    \right\rceil-j\right).
\end{equation}
Now for every $i=j+1,\ldots, J$ such that $c_{i, j}=\sqrt{\frac{i-1}{j}}$ one has 
$$
\max_{i=j+1,\ldots,J+1}  \left\lceil  \frac{i-j}{c_{i,j}^2(1-o(1))}    \right\rceil-j=\max_{i=j+1,\ldots,J+1}  \left\lceil  j\cdot \frac{i-j}{(i-1)(1-o(1))}    \right\rceil-j\leq o(J),
$$
and for the other values of $i$ we have $\max_{i=j+1,\ldots,J+1}  \left\lceil  \frac{i-j}{\log^{1/7} n (1-o(1))}    \right\rceil-j\leq o(J)$ as well.
Putting this together with~\eqref{eq:size-bound-92h3tg} and multiplying by $J=O(\log(1/\mu))=\mu^{-o(1)}$ to account for the number of choices $j\in [J]$, we get the second bound for the expected size of the data structure
$\epsilon^{-2}\left(\frac{1}{\mu^*}\right)^{1+o(1)}$.
\paragraph{Proof of the precision of the estimator:} First, we prove the following claim, which guarantees high success probability  for recovery procedure.
\begin{claim}[Lower bound on probability of recovering a sampled point]\label{claim:AIhighprob}
	Suppose that we invoke Algorithm~\ref{alg:4} with $(P,\epsilon)$. Suppose that in line~\ref{line:sub-sample4} of Algorithm~\ref{alg:4}, when $k=k^*$ and $j=j^*$, we sample some point $\mathbf{p} \in L_{j^*}$. We claim that with probability at least $1-\frac{1}{n^{10}}$, there exists $\ell^* \in [K_2]$ such that $H_{k^*,j^*,\ell^*}(\mathbf{p})=H_{k^*,j^*,\ell^*}(\mathbf{q})$.
\end{claim}
\begin{proof}
	By Claim~\ref{claim:collprob} we have 
	\begin{align*}
	\Pr_{h^*\sim\mathcal{H}^k}\left[ h^*(\mathbf{p})=h^*(\mathbf{q})\right] \ge p_{\mathrm{near},j}^{k_{j}}.
	\end{align*}
	Now note that we repeat this process for $K_2=100\log n\cdot p_{\mathrm{near},j}^{-k_j}$ times. So any point $\mathbf{p}$ which is sampled from band $L_{j^*}$ is recovered in at least one of the repetitions of phase $j=j^*$, with high probability.
\end{proof}
Now, we argue that the estimators are unbiased (up to small inverse polynomial factors)
\begin{claim}[Unbiasedness of the estimator]\label{claim:unbias4}
	For every $\mu^*\in (0,1)$, every $\mu \ge \mu^*$, every $\epsilon\in (\mu^{10},1)$, every $\mathbf{q}\in \R^d$, estimator $Z_a$ for any $a\in [K_1]$ constructed in $\textsc{Query}(P,\mathbf{q},\epsilon,\mu)$ (Algorithm~\ref{alg:query4}) satisfies the following:
	\begin{align*}
	(1-n^{-9})n\mu^{*}\leq \E[Z_a] \leq n\mu^{*}
	\end{align*}
\end{claim}
\begin{proof}
	Let $\mathcal{E}$ be the event that every sampled point is recovered and let $Z:=Z_a$ (see line~\ref{line:est4} in Algorithm~\ref{alg:query4}). By Claim~\ref{claim:AIhighprob} and union bound, we have 
	\begin{align*}
	\Pr[\mathcal{E}] \ge 1-n^{-9}
	\end{align*}
	We have that $\E[Z] = \sum_{i=1}^{n}\frac{\E[\chi_{i}]}{p_{i}}w_{i}$ with $(1-n^{-9}) p_{i}\leq \E[\chi_{i}]\leq p_{i}$, where we now define $\chi_i=1$ if point $\mathbf{p}_i$ is sampled and recovered in the phase corresponding to its weight level, and $\chi_i=0$ otherwise. Thus
	\begin{equation}
	(1-n^{-9})n\mu^{*}\leq \E[Z] \leq n\mu^{*}.
	\end{equation}
\end{proof}

\begin{rem}\label{rm:wlog}
We proved that our estimator is unbiased\footnote{Up to some small inverse polynomial error.} for {\bf any choice} of $\mu\ge \mu^*$. Therefore if $\mu\ge4\mu^*$,  by Markov's inequality the estimator outputs a value larger than $\mu$ at most with probability $1/4$. We perform $O(\log n)$ independent estimates, and conclude that $\mu$ is higher than $\mu^*$ if the median of the estimated values is below $\mu$. This estimate is correct with high probability, which suffices to ensure that we find a value of $\mu$ that satisfies $\mu/4< \mu^*\le \mu$ with high probability by starting with some $\mu=n^{-\Theta(1)}$ (since our analysis assumes $\mu^*=n^{-\Theta(1)}$) and repeatedly halving our estimate (the number of times that we need to halve the estimate is $O(\log n)$ assuming that $\mu$ is lower bounded by a polynomial in $n$, an assumption that we make).
\end{rem}
\begin{claim}[Variance bounds]\label{claim:variance}
	For every $\mu^*\in (0, 1)$, every $\epsilon\in (\mu^{10}, 1)$, every $\mathbf{q}\in \mathbb{R}^d$, using estimators $Z_a$, for $a \in [K_1]$ constructed in \textsc{Query}($P,\mathbf{q},\epsilon,\mu$) (Algorithm~\ref{alg:query4}), where $ \mu/4\le \mu^*\le \mu$, one can output a $(1\pm \epsilon)$-factor approximation to $\mu^*$. 
\end{claim}
\begin{proof}
	By Claim~\ref{claim:unbias4} and noting that $Z \le n^2 \mu^*$, where the worst case (equality) happens when all the points are sampled and all of them are recovered in the phase of their weight levels. Therefore, 
	\begin{align*}
	\mathbb{E}\left[ Z|\mathcal{E}\right]\cdot \Pr[\mathcal{E}] + n^2\mu^* (1-\Pr[\mathcal{E}]) \ge \mathbb{E}[Z].
	\end{align*}
	Also, since $Z$ is a non-negative random variable, we have  
	\begin{align*}
	\mathbb{E}\left[Z| \mathcal{E} \right]\le 	\frac{\mathbb{E}\left[Z\right]}{\Pr[\mathcal{E}]}\le \frac{n\mu^*}{\Pr[\mathcal{E}]}=n\mu^* (1+o(1/n^9))
	\end{align*}
	Then, we have
	\begin{align*}
	\mathbb{E}[Z^2]
	&=\mathbb{E}\left[\left(\sum_{\mathbf{p}_i\in P}\chi_i \frac{w_i}{p_i}\right)^2\right] \\
	&=\sum_{i\ne j} \mathbb{E}\left[\chi_i\chi_j \frac{w_iw_j}{p_ip_j}\right]+\sum_{i\in [n]}\mathbb{E}\left[\chi_i \frac{w_i^2}{p_i^2}\right]\\
	&\le \sum_{i\ne j}w_iw_j + \sum_{i\in [n]}\frac{w_i^2}{p_i}\mathbb{I}[p_i=1]+ \sum_{i\in [n]}\frac{w_i^2}{p_i}\mathbb{I}[p_i\ne1]\\
	&\le \left(\sum_{i}w_i\right)^2  +\sum_{i\in[n]}w_i^2+\max_i\left\{\frac{w_i}{p_i}\mathbb{I}[p_i\ne 1]\right\}\sum_{i\in [n]}w_i\\
	&\le 2n^2(\mu^*)^2+ \max_{j\in [J],\mathbf{p}_i\in L_{j}}\{w_{i}2^{j+1}\}n\mu \cdot n \mu^{*} \\
	&\le 4n^2\mu^2 &&\text{Since $\mu^*\le \mu$}
	\end{align*}
	and 
	\begin{align*}
	\mathbb{E}[Z^2|\mathcal{E}]\le \frac{\mathbb{E}[Z^2]}{\Pr[\mathcal{E}]}\le n^2 \mu^{2-o(1)}(1+o(1/n^9))
	\end{align*}
	
	Now, since $\mu \le 4\mu^*$, in order to get a $(1\pm \epsilon)$-factor approximation to $\mu^*$, with high probability, it suffices to repeat the whole process $K_1=\frac{C\log n}{\epsilon^2}\cdot \mu^{-o(1)}$ times, where $C$ is a universal constant.
	
	Suppose we repeat this process $m$ times and $\bar{Z}$ be the empirical mean, then:
	\begin{align*}
	\Pr[|\bar{Z}-\mu^{*}|\geq \epsilon n\mu^{*}] &\le \Pr[|\bar{Z}-\E[Z]|\geq \epsilon \mu^{*}-|\E[Z]-n\mu^{*}|]\\
	&\le \Pr[|\bar{Z}-\E[Z]|\geq (\epsilon-n^{-9})n\mu^{*}]\\
	&\le \frac{\E[\bar{Z}^{2}]}{(\epsilon-n^{-9})^{2}(n^{2}\mu^{*})^{2}}\\
	&\le \frac{1}{m}\frac{16n^{2}(\mu^{*})^{2}}{(\epsilon-n^{-9})^{2}(n^{2}\mu^{*})^{2}}
	\end{align*}
	Thus by picking $m=O(\frac{1}{\epsilon^{2}})$ and taking the median of $O(\log(1/\delta))$ such means we get a $(1\pm \epsilon)$-approximation with probability at least $1-\delta$ per query.\end{proof}
All in all, we proved the expected query time bound, the expected space consumption and the precision guarantee in the statement of the theorem.
\end{proofof}
Now, we calculate the cost of kernel for $t$-student kernel. 
\paragraph{$t$-student kernel ($\frac{1}{1+||\mathbf{x}-\mathbf{y}||_2^t}$):} 
We directly calculate distance levels induced by this kernel as follows
\begin{align*}
r_j=\sqrt[t]{2^j-1}
\end{align*}
which implies that for all $i\in[J+1],j\in[J]$ such that $i>j$,
\begin{align*}
c_{i,j}&:=\min\left\{  \frac{r_{i-1}}{r_j},\log^{1/7}n \right\}\\
&=\min\left\{  \sqrt[t]{\frac{2^{i-1}-1}{2^j-1}},\log^{1/7}n \right\}.
\end{align*}
Now, one can check that 
\begin{align*}
\max_{j\in[J]}\max_{i=j+1,\ldots,J+1}  \left\lceil  \frac{i-j}{c_{i,j}^2(1-o(1))}     \right\rceil=\frac{\log\frac{1}{\mu}}{\log^{2/7}n}(1+o(1)).
\end{align*}
Thus, we have 
\begin{align*}
\mathrm{cost}(K)=\mu^{-o(1)}.
\end{align*}
We note that this matches the result of~\cite{backurs2018efficient} up to the difference between $\mu^{-o(1)}$ and $\log (1/\mu)$ terms. The $\mu^{-o(1)}$ dependence comes from the fact that we used the LSH of~\cite{DBLP:conf/focs/AndoniI06}, and the dependence can be improved to $\log(1/\mu)$ by using the hash family of~\cite{datar2004locality}, for instance.

\paragraph{Exponential Kernel ($e^{-\|x-y\|_{2}}$)}

The distance levels induced by the kernel are given by $r_{j}=j\log(2)$ for $j\in [J]$. Hence, we get that $c_{ij}=\min\left\{\frac{r_{i-1}}{r_{j}},\log^{1/7}n\right\}=\min\{\frac{i-1}{j},\log^{1/7}n\}$. If $i> j\log^{1/7}n+1$ then the cost is increasing in $i>0$ becomes:
\begin{equation*}
\mathrm{cost}(K,j)=\exp_2\left( \frac{J+1-j}{\log^{2/7}n}\right) \leq \exp_2\left(\frac{J}{\log^{2/7}n}\right) = \mu^{-o(1)}.
\end{equation*}
Thus, for the rest we will assume that $i\leq j \log^{1/7}n+1$, and we need to find the maximum over $j$ of 
\begin{align*}
(1+o(1))\max_{i=j+1,\ldots,J+1}\left\lceil \frac{j^{2}((i-1)-(j-1))}{(i-1)^{2}}\right\rceil
\end{align*} 

Setting $x=i-1$ and $A=j-1$, we optimize the function $\frac{(x-A)}{x^{2}}$ for $x\geq A+1$. We get that the optimal value is attained for  $i^{*}(j)=\max\{\min\{2j-1,J+1\},j+1\}$. We distinguish three cases:
\begin{enumerate}
	\item   $j=1$: then $i^{*}=2$ and we get $\mathrm{cost}(K,1)=\mu^{-o(1)}$
	\item  $j>\frac{J+2}{2}$: then the maximum over $i$ is $\frac{j^{2}(J+1-j)}{J^{2}}(1+o(1))$, and the optimal choice of $j$ is $j^{*}=\frac{2(J+1)}{3}$. We thus get
	\begin{equation*}
	\max_{j>\frac{J+2}{2}}\{\mathrm{cost}(K,j)\} =\mu^{-(1+o(1)) \frac{4}{27}}
	\end{equation*}
	\item $j\leq \frac{J+2}{2}$: then the maximum over $i$ is $\frac{j^{2} }{4(j-1)}(1+o(1))$  and the optimal choice for $j$ is $j^{*}=\frac{J+2}{2}$. We thus get
	\begin{equation*}
	\max_{j\leq \frac{J+2}{2}}\{\mathrm{cost}(K,j)\} =\mu^{-(1+o(1)) \frac{1}{8}}.
	\end{equation*}
\end{enumerate}
Overall, the worst-case cost is attained for $i^{*}=J$ and $j^{*}=\frac{2J}{3}$ and yields
\begin{equation*}
\mathrm{cost}(K) = \mu^{-(1+o(1)) \frac{4}{27}}.
\end{equation*}

\begin{proofof}{Theorem~\ref{thm:query-time-gen}}
	One should note that the query time of our approach depends on the number of times that we hash the query and the number of points that we check, i.e., the number of points that collide with the query. First, we analyze the number of points colliding with the query. We  Fix $j\in[J]$, so, we want to estimate the contribution of points in $L_i$ to $K(P,\mathbf{q})$. We consider 3 cases:
	\paragraph{Case 1. $i\le j$:} Note that we have $|L_i| \le 2^{i}n\mu$ and note that in $j$'th phase, we sample the data set with rate $\min\{\frac{1}{2^jn\mu},1\}$. Thus, we have at most $1=O(1)$ sampled points from $L_i$ in expectation.
	\paragraph{Case 2. $i=j+1,\dots,J$:} Again, note that by Lemma~\ref{lem:sizeL}, $|L_i|\le 2^{i}n\mu$, and the sampling rate is $\min\{\frac{1}{2^jn\mu},1\}$. Thus, we have at most $2^{i-j}$ sampled points from $L_i$ in expectation. Now, we need to analyze the effect of LSH. Note that we choose LSH function such that the near distance is $r_j$ (see Claim~\ref{claim:collprob}). Also, note that as per \eqref{eq:kj}, we use
	\begin{align*}
		k:=k_j :=\frac{-1}{ \log p_{\mathrm{near},j}} \cdot\max_{i=j+1,\ldots,J+1}  \left\lceil  \frac{i-j}{c_{i,j}^2(1-o(1))}     \right\rceil.
	\end{align*}
	as the number of concatenations. 
	Now, we have the following collision probability for $\mathbf{p}\in L_i$ using Claim~\ref{claim:collprob}
	\begin{align*}
	\Pr_{h^*\in \mathcal{H}^{k}}\left[  h^*(\mathbf{p})=h^*(\mathbf{q})\right] \le p^{kc^2(1-o(1))},
	\end{align*}
	where $c:=c_{i,j}:=\min\left\{   \frac{r_{i-1}}{r_j},\log^{1/7}n\right\}$ and $p:=p_{\mathrm{near},j}$ for ease of notation. This implies that the expected number of points from weight level $L_i$ in the query hash bucket is at most
	\begin{align*}
	2^{i-j}\cdot p^{kc^2(1-o(1))}=\wt{O}(1)
	\end{align*}
	by the choice of $k$. 
	\paragraph{Case 3. points in $L_{J+1}$:} We know that we have $n$ points, so after sub-sampling, we have at most $\frac{1}{2^j\mu}$ points from this range, remaining in expectation. For any $\mathbf{p}\in L_{J+1}$, note that $||\mathbf{p}-\mathbf{q}|| \ge c\cdot r_j$ for $c:=c_{J+1,j}:=\min\left\{   \frac{r_{J}}{r_j},\log^{1/7}n\right\}$. Then, 
	\begin{align*}
	\Pr_{h^*\in \mathcal{H}^k}\left[  h^*(\mathbf{p})=h^*(\mathbf{q})\right] \le p^{kc^2(1-o(1))},
	\end{align*}
	which implies that the expected number of points form this range in the query hash bucket is at most
	\begin{align*}
	\frac{1}{2^j\mu}\cdot p^{kc^2(1-o(1))}=2^{J-j}\cdot p^{kc^2(1-o(1))}=\wt{O}(1)
	\end{align*}
	by the choice of $k_j$.
	
	All in all, we prove that each weight level $L_i$ for $i\in[J+1]$ contribute at most $\wt{O}(1)$ points to the hash bucket of query. Now, we need to prove a bound on the number of times we evaluate our hash function. One should note that by the choice of $k_j$ in \eqref{eq:kj} we have
	\begin{align*}
	k_j=\wt{O}(1)
	\end{align*}
	which basically means that we only concatenate $\wt{O}(1)$ LSH functions. Thus, we the evaluation time of $h^*(q)$ for any $h^*\in \mathcal{H}^k$ is $\wt{O}(n^{o(1)})$, by Remark~\ref{rem:AI}. On the other hand, note that for recovering the points in $L_{J+1}$ we just sub-sampled the data set with probability $\frac{1}{n}$ so in expectation we only scan $1$ point. So in total, since we repeat this for all $j\in[J]$ and $J=\lceil \log \frac{1}{\mu}\rceil$, by the choice of $K_1$ and $K_2$ assigned in lines~\ref{line:q-K1} and \ref{line:q-K2} of Algorithm~\ref{alg:4}, respectively, the claim holds.  
\end{proofof}

\newpage 
\section{Improved algorithm via data dependent LSH}\label{sec:dd}

In this section, we improve the algorithm presented in the previous section using data dependent LSH approach for the Gaussian kernel. 
Consider a data set $P\subset \R^d$, a positive real number $a$, and a query $\mathbf{q}\in \R^d$. Let 
$$
\mu^*:=K(P, \mathbf{q})=\sum_{\mathbf{p}\in P} e^{-a||\mathbf{p}-\mathbf{q}||_2^2}
$$
denote the KDE value at the query $\mathbf{q}\in \R^d$ of interest, and for the rest of the paper suppose that the algorithm is given a parameter $\mu$ that satisfies the following property 
\begin{equation}\label{eq:mu-def}
 \mu^*\le\mu.
\end{equation}
We prove the following main result in the rest of the paper.
\begin{thm}\label{thm:Gaus5}
	Given a kernel $K(\mathbf{p},\mathbf{q}):=e^{-a||\mathbf{p}-\mathbf{q}||_2^2}$ for any $a>0$, $\epsilon = \Omega\left(\frac{1}{\mathrm{polylog} n}\right)$, $\mu^*=n^{-\Theta(1)}$ and a data set of points $P$, there exists a preprocessing algorithm and a corresponding query algorithm that one can approximate $\mu^*:=K(P,\mathbf{q})$ (see Definition~\ref{def:kernel}) up to $(1\pm \epsilon)$ multiplicative factor, in time $\wt{O}\left(\epsilon^{-2}\left(\frac{1}{\mu^*}\right)^{0.173+o(1)}\right)$, for any query point $\mathbf{q}$. Additionally, the space consumption of the data structure is $$\min\left\{ \epsilon^{-2} n\left(\frac{1}{\mu^*}\right)^{0.173+o(1)},\epsilon^{-2}\left(\frac{1}{\mu^*}\right)^{1+c+o(1)}\right\}.$$
	for a small constant $c=10^{-3}$.
\end{thm}
\begin{proof}
	First, in Section~\ref{sec:prepro-dd} we present the main primitives in the preprocessing phase (Algorithms~\ref{alg:DD-KDE-Prep2}, \ref{alg:DD-KDE-spherical} and \ref{alg:pseudorandomify}) and prove the space bound in Lemma~\ref{lm:space}. The standard outer algorithm is presented in Appendix~\ref{sec:dd-kde} for completeness. The main query primitive in query algorithm is presented in Section~\ref{sec:query-dd}, and the query time is proved in Section~\ref{sec:query} in Lemma~\ref{lm:query}. The correctness proof (precision of the estimator) is rather standard and similar to the correctness proof in Section~\ref{sec:data-independent} and is given in Appendix~\ref{sec:dd-kde} for completeness. 
\end{proof}

\begin{rem}\label{rem:generalkernels}
Although we present the analysis for the Gaussian kernel, our techniques can be used for other kernels such as the exponential  kernel as well.  We do not present the full analysis to simplify presentation of our main result for the Gaussian kernel, but provide proofs of key lemmas in Appendix~\ref{app:generalkernels}. Specifically, we present the equivalent of Claims~\ref{claim:monotone} and~\ref{cl:mina}, which underly our LP analysis, for kernels whose negative log density is concave (this includes the exponential kernel $\exp(-||x||_2)$). Our dual solution presented in Section~\ref{sec:lp-sol} gives an upper bound of $\approx 0.1$ on the value of the corresponding LP. Replacing the parameter $\alpha^*$ in the algorithms presented in this section with $0.1$ thus yield an data structure for KDE with the exponential kernel with query time $\wt{O}\left(\epsilon^{-2}\left(\frac{1}{\mu^*}\right)^{0.1+o(1)}\right)$ and space consumption $\epsilon^{-2} n\left(\frac{1}{\mu^*}\right)^{0.1+o(1)}$ for the exponential kernel.
\end{rem}

In order to simplify notation we apply the following normalization without loss of generality: For any point in $\mathbf{p} \in P\cup \{\mathbf{q}\}$,  let $\mathbf{p}':=\sigma \mathbf{p}$ and $\sigma:=\sqrt{\frac{2a}{\log(1/\mu)}}$ such that a point $\mathbf{p}'$ at distance $\sqrt{2}$ from the query $\mathbf{q}'$ contributes exactly $\mu$ to the kernel. In other words, we assume by convenient scaling that 
$$
K(P, \mathbf{q})=\frac1{n}\sum_{\mathbf{p}'} (\mu)^{||\mathbf{p}'-\mathbf{q}'||_2^2/2}.
$$
and to lighten notation we will assume that $\sigma=1$, i.e. points are already properly scaled.
 For $x\in (0, \sqrt{2})$, let $\wt{P}$ be the dataset obtained from $P$ by including every point independently  with probability $\min\left\{\frac1{n}\cdot \left(\frac1{\mu}\right)^{1-\frac{x^2}{2}},1\right\}$. We state these conditions in a compact way as follows and use them in the rest of the paper.
 \begin{assump}
 	We have the followings
 	\begin{itemize}\label{assump:1}
 		\item $P\subset \RR^d$ and $|P| = n $.
 		\item $\mathbf{q} \in \RR^d$.
 		\item $\mu^*:=K(P, \mathbf{q})$
 		\item $\frac{1}{\mu^*}=n^{\Omega(1)}$
 		\item $\mu$ is such that $\mu^* \le \mu$.
 		\item $\mu=n^{-\Theta(1)}$.
 		\item The points are scaled so that $K(\mathbf{p},\mathbf{q})=\mu^{\frac{||\mathbf{p}-\mathbf{q}||^2}{2}}$.
 		\item $\wt{P}$ is obtained by independently sub-sampling elements of $P$ with probability $\min\left\{\frac1{n}\cdot \left(\frac1{\mu}\right)^{1-\frac{x^2}{2}},1\right\}$, for some $x\in (0,\sqrt{2})$, which is clear from the context.
 	\end{itemize}
 \end{assump}
In this section we design a data structure that allows preprocessing $\wt{P}$ as above using small space such that every point at distance at most $x$ from any query $\mathbf{q}$ is recovered with probability at least $0.8$ (see Lemma~\ref{lem:tree-correctness}).

In what follows we present our preprocessing algorithm (Algorithm~\ref{alg:DD-KDE-Prep2}) in Section~\ref{sec:prepro-dd}, the query algorithm (Algorithm~\ref{alg:DD-KDE-Query}) in Section~\ref{sec:query-dd} as well as proof of basic bounds on their performance in the same sections. Our main technical contribution is the proof of the query time bound. This proof relies on a novel linear programming formulation that lets us bound the evolution of the density of points around the query $q$ as the query percolates does the tree $\mathcal{T}$ of hash buckets produced by \textsc{PreProcess}. This analysis is given in Section~\ref{sec:query}, with the main supporting technical claims presented in Section~\ref{sec:main-tech-lemma}.

\subsection{Preprocessing algorithm and its analysis}\label{sec:prepro-dd}

Our preprocessing algorithm is recursive. At the outer level, given the sampled dataset $\wt{P}$ as input, the algorithm hashes $\wt{P}$ into buckets using Andoni-Indyk Locality sensitive hashing. The goal of this is to ensure that with high probability all hash buckets that a given query explores are of bounded diameter, while at the same time ensuring that any close point $\mathbf{p}$ hashes together with $\mathbf{q}$ in at least one of the hash buckets with high constant probability. The corresponding analysis is presented in Sections~\ref{sec:query-dd} and \ref{sec:query}.

Our main tool in partitioning the data set into (mostly) low diameter subsets is an Andoni-Indyk Locality Sensitive Hash family.
Such a family is provided by Lemma~\ref{lem:andoni-indyk}, which was our main tool in obtaining the non-adaptive KDE primitives in Section~\ref{sec:data-independent}, and Corollary~\ref{cor:AI} below. We restate the lemma below for convenience of the reader:

\noindent{\em {\bf Lemma~\ref{lem:andoni-indyk}}  (\cite{DBLP:conf/focs/AndoniI06}) (Restated)
	Let $\mathbf{p}$ and $\mathbf{q}$ be any pair of points in $\mathbb{R}^d$. Then, for any fixed $r>0$, there exists a hash family $\mathcal{H}$ such that, if $p_{\mathrm{near}}:=p_1(r):=\Pr_{h\sim\mathcal{H}}[h(\mathbf{p})=h(\mathbf{q}) \mid ||\mathbf{p}-\mathbf{q}||\le r]$ and $p_{\mathrm{far}}:=p_2(r,c):=\Pr_{h\sim\mathcal{H}}[h(\mathbf{p})=h(\mathbf{q}) \mid ||\mathbf{p}-\mathbf{q}||\ge cr]$ for any $c>1$, then $$\rho:=\frac{\log 1/p_{\mathrm{near}}}{\log 1/p_{\mathrm{far}}}\le \frac{1}{c^2}+O\left(\frac{\log t}{t^{1/2}}\right),$$for some $t$, where $p_{\mathrm{near}}\ge e^{-O(\sqrt{t})}$ and each evaluation takes $d t^{O(t)}$ time.
}
One should also recall Remark~\ref{rem:AI}, which ensures $n^{o(1)}$ evaluation time, with appropriate choice of $t$ in Lemma~\ref{lem:andoni-indyk}. 

\begin{cor}\label{cor:AI}
	Let $\alpha$ be a constant, and let $x\in (0,\sqrt{2})$ and $y$ be  such that $y\ge x$. Then, there exists a hash family $\mathcal{H}$ such that for any points $\mathbf{q}\in \R^d$, $\mathbf{p}$ and $\mathbf{p}'$, where $||\mathbf{p}-\mathbf{q}||\le x$ and $||\mathbf{p}'-\mathbf{q}||\ge y$, we have the following conditions
	\begin{itemize}
		\item $\Pr_{h\sim \mathcal{H}}\left[h(\mathbf{q})=h(\mathbf{p})\right] \ge \mu^{\alpha}$
		\item $\Pr_{h\sim \mathcal{H}}\left[h(\mathbf{q})=h(\mathbf{p}')\right] \le \mu^{\alpha  c^2 (1-o(1))}$
	\end{itemize}
where $c:= \min\left\{\frac{y}{x},\log^{1/7}n\right\}$, and we call such a hash family a $(\alpha,x,\mu)$-AI hash family.
\end{cor}

Our preprocessing algorithm is given below. It simply hashes the dataset several times independently using an Andoni-Indyk LSH family and calls \textsc{Spherical-LSH} (Algorithm~\ref{alg:DD-KDE-spherical} below) on the buckets. The hashing is repeated several times to ensure that the query collides with any given close point with high probability in at least one of the hashings. Overall \textsc{PreProcess} simply reduces the the diameter of the dataset, whereas most of the work is done by \textsc{Spherical-LSH}, defined below.

\begin{algorithm}[H]
	\caption{\textsc{PreProcess}: $\wt{P}$ is the subsampled data-set, $x$ is the target distance to recover}  
	\label{alg:DD-KDE-Prep2}
	\begin{algorithmic}[1]
		\Procedure{$\textsc{PreProcess}(\wt{P},x, \mu)$}{}
		\State Add a root $w_0$ to the recursion tree $\mathcal{T}$
		\State $w_0.P \gets \wt{P},~w_0.level\gets 0,~w_0.g\gets 0$ \Comment{$g=0$ since this node uses Euclidean LSH}

		\If {$x>\sqrt{2}$}
		\Return $\wt{P}$ \Comment{In that case the expected size of $\wt{P}$ is small}
		\EndIf
		\State $\alpha \gets 10^{-4}$ \label{line:ai-alpha}\Comment{Choice of $\alpha$ affects hash bucket diameter, see Lemma~\ref{lm:diam}}
		\State $K_1=100\left(\frac{1}{\mu}\right)^\alpha$\label{line:const-c} \Comment{ The number of repetitions of the first round of hashing }
		\For {$j = 1,2,\dots, \lceil K_1 \rceil$}
		\State Pick $h_{j}$ from a $(\alpha,x,\mu)$-AI hash family, $\mathcal{H}$\Comment{See Definition~\ref{cor:AI}} \label{line:h12}
		\State $B \gets $ set of non-empty hash buckets by hashing points in $P$ using $h_{j}$.
		\For {each $b\in B$}
		\State Add a node $v$ as a child of $w_0$ in recursion tree $\mathcal{T}$
		\State  $v.P\gets b,~v.level\gets 0,~v.g\gets 0$
		\State $v.o \gets$ any point in bucket $b$ 
		\State $\textsc{Spherical-LSH}(v,x, \mu)$
		\EndFor
		\EndFor
		\State \Return $\mathcal{T}$
		\EndProcedure
	\end{algorithmic}
\end{algorithm}

We will use the following basic upper bound on the Euclidean diameter of LSH buckets:
\begin{lem}[Diameter bound for Andoni-Indyk LSH buckets]\label{lm:diam}
	Under Assumption~\ref{assump:1}, suppose that $\mathcal {H}$ is a $(\alpha,x,\mu)$-AI hash family (see Corollary~\ref{cor:AI}), for some constant $\alpha$ and let $c:= \min\left\{\frac{R_{\mathrm{diam}}}{x},\log^{1/7}n\right\}$ for some $R_{\mathrm{diam}}\geq \sqrt{2}$, then if $\alpha c^2 = 2+\Omega(1)$ then one has 
\begin{description}
\item[(a)]	$\mathbb{E}_{h\sim \mathcal{H}, \wt{P}}\left[\left|\left\{\mathbf{p}\in \wt{P}:  ||\mathbf{p}-\mathbf{q}||\geq R_{\mathrm{diam}}\text{~and~}h(\mathbf{p})=h(\mathbf{q})\right\}\right|\right] \le \left(\frac{1}{\mu}\right)^{2-\alpha c^2(1-o(1))}$
\item[(b)] 	and consequently 
	\begin{equation*}
	\Pr_{h\sim \mathcal{H}, \wt{P}}[\text{diameter of $h^{-1}(\mathbf{q})\cap \wt{P}$ is larger than $R_{\mathrm{diam}}$}]\leq \left(\frac{1}{\mu}\right)^{2-\alpha c^2(1-o(1))}.
	\end{equation*}
\end{description}	
\end{lem}
\begin{proof}
	One has for every $R_{\mathrm{diam}}\geq \sqrt{2}$
	\begin{align*}
	\mathbb{E}_{\wt{P}}\left[\left|\left\{\mathbf{p}\in \wt{P}: ||\mathbf{p}-\mathbf{q}||\geq R_{\mathrm{diam}}\right\}\right|\right] &\le \mathbb{E}_{\wt{P}}\left[\left|\wt{P}\right|\right]= n\cdot\frac{1}{n} \left(\frac{1}{\mu}\right)^{1-\frac{x^2}{2}}=\left(\frac{1}{\mu}\right)^{1-\frac{x^2}{2}}
	\end{align*}
	
	Taking the expectation with respect to the hash function $h$, we get, 
	\begin{equation*}
	\begin{split}
	\mathbb{E}_{h\sim \mathcal{H}, \wt{P}}\left[\left|\left\{\mathbf{p}\in \wt{P}:  ||\mathbf{p}-\mathbf{q}||\geq R_{\mathrm{diam}}\text{~and~}h(\mathbf{p})=h(\mathbf{q})\right\}\right|\right]  &\le \left(\frac{1}{\mu}\right)^{1-\frac{x^2}{2}-\alpha c^2(1-o(1))}\\
	&\le\left(\frac{1}{\mu}\right)^{2-\alpha c^2(1-o(1))},
	\end{split}
	\end{equation*}
	establishing {\bf (a)}.
	Claim {\bf (b)} now follows by applying Markov's inequality since $\alpha c^2=2 +\Omega(1)$.
\end{proof}
Now, we establish a constant upper bound on the diameter of data set after the Andoni-Indyk LSH round. Since we have $\mu=n^{-\Theta(1)}$, and $\alpha=10^{-4}$ as per  line~\ref{line:ai-alpha} of Algorithm~\ref{alg:DD-KDE-Prep2}, by Lemma~\ref{lm:diam} if we let $R_{\mathrm{diam}}$ be a large enough constant, then one has 
\begin{equation}\label{eq:92y3t9gdf}
\begin{split}
\Pr_h[\text{diameter of $h^{-1}(\mathbf{q})\cap \wt{P}$ is larger than $R_{diam}$}]&\le n^{-20}
\end{split}
\end{equation}
Let $\mathcal{E}_{diam}$ denote the event that all Andoni-Indyk hash buckets that the query hashes to have diameter bounded by $R_{diam}$. We have, combining the failure event over sampling of $\wt{P}$ (over-sampling by a factor more than $O(\log n)$) with~\eqref{eq:92y3t9gdf} that $\Pr[\bar{\mathcal{E}}]\leq 2n^{-20}\le n^{-19} $. Conditioned on ${\mathcal{E}}$ buckets that the query hashes have diameter bounded by $R_{\mathrm{diam}}$. Now, if we take any point in the data set and consider a ball of radius $R_{\text{max}}:=2R_{\text{diam}}$, using the triangle inequality, it contains all the points of this hash bucket. This ensures that all the spheres in the recursion tree have radius bounded by $R_{\text{max}}=\Theta(1)$. 
\begin{cor}[Bounded diameter spheres]
	 All the spheres that the query scan in the algorithm have radius bounded by $R_{\text{max}}=\Theta(1)$.\footnote{Since we did not use any density constraints other than the upper bound of $n$ on the number of points, this corollary applies for all spheres in the Algorithm.} 
\end{cor}

We are now ready to present our main preprocessing primitive \textsc{Spherical-LSH}, given as Algorithm~\ref{alg:DD-KDE-spherical} below. The input to the algorithm is a node in the recursion tree $\mathcal{T}$ created by recursive invocations of \textsc{Spherical-LSH}. Every such node $v$ is annotated with a dataset $v.P$, a radius $v.r$ of a ball enclosing the dataset, the center $v.o$ of that ball and a level, $v.level$, initially set to $0$ for the root of the tree $\mathcal{T}$ that is created by \textsc{PreProcess}.  \textsc{Spherical-LSH} then proceeds as follows. First it calls the \textsc{Pseudorandomify} procedure (Algorithm~\ref{alg:pseudorandomify} below). This procedure partitions the input dataset $v.P$ into subsets that are pseudorandom as per Definition~\ref{def:psrs}. A similar procedure was used in the work of~\cite{DBLP:conf/soda/AndoniLRW17} on space/query time tradeoffs for nearest neighbor search. Intuitively, a dataset is pseudorandom if the points belong to a thin spherical shell and furthermore do not concentrate on any spherical cap in this shell (appropriately defined). These pseudorandom datasets are added to the recursion tree $\mathcal{T}$ as children of $v$. \textsc{Spherical-LSH} then generates random subsets of these pseudorandom spheres defined by random spherical caps, adds these datasets to the recursion tree $\mathcal{T}$ and recursively calls itself until a depth budget $T$ (see line~\ref{line:t-def} below) is exhausted. Note that the radius of spherical caps generated depends on the distance $x'$ from the projected query point to the target near point (which is assumed to be at distance $x$ from the query). Note that since the query is not available at the preprocessing stage, the algorithm prepares data structures for all possible values of $x'$ (see line~\ref{line:xprime-enumeration} in Algorithm~\ref{alg:DD-KDE-spherical} below). Note that the value of $x'$ is passed down the recursion tree. In the section below, we set the parameters that we use in the algorithms. 

\subsection{Parameter settings}\label{sec:param}
\begin{itemize}
	
	\item $\gamma = \frac{1}{\log\log\log n}$ and $\tau=\frac{1}{10}$ are the parameters used for pseudo-random spheres (see Definition~\ref{def:psrs}) in Algorithm~\ref{alg:pseudorandomify}.
	\item $\alpha^*=0.172$ (see Section~\ref{sec:lp-sol}), $T = \sqrt{\log n}$ and $J= \min\left\{\alpha^*\cdot T, \left(\frac{x^2}{2}\left(1-\frac{x^2}{2}\right)+10^{-4}\right)\cdot T\right\}$ are parameters to bound the depth of the recursion tree. 
	\item $\delta = \exp(-(\log\log n)^C)$ for some large enough constant $C$, is a parameter used for partitioning point in a ball to discrete spheres of radii multiplies of $\delta$ (see Algorithm~\ref{alg:pseudorandomify})
	\item $\delta' = \exp(-(\log\log n)^C)$ for some large enough constant $C$, is a parameter for rounding $x'$'s to $x''$'s (see lines~\ref{line:x-prime} and \ref{line:x-tilde} in Algorithm~\ref{alg:DD-KDE-Query})
	\item $R_{\mathrm{min}}=10^{-5}$ is a lower bound on the radius of spheres that we process further, i.e., we stop whenever the radius becomes less than $R_{\mathrm{min}}$.
	
	\item $\Delta = 10^{-20}$ is a tiny constant. For a discussion about $\Delta$ see Remark~\ref{rem:Delta}.
	\item $\alpha=10^{-4}$ is a parameter used for the Andoni-Indyk LSH round (see Algorithm~\ref{alg:DD-KDE-Prep2})
	\item $\delta_z=10^{-6}$ is a parameter used in discretizing continuous densities in Definition~\ref{def:fy}. 
	\item $\delta_x=10^{-8}$ is used for defining a grid over $(0,\sqrt{2})$, such that for any $x$ from this grid we prepare the data structure to recover points from $[x-\delta,x)$ (see Algorithm~\ref{alg:DD-KDE-Prep}).
\end{itemize}

\begin{algorithm}[H]
	\caption{\textsc{Spherical-LSH}: $x$ is the target distance to recover, $v$ is the node in recursion tree (corresponds to a subset of the dataset)}  
	\label{alg:DD-KDE-spherical} 
	\begin{algorithmic}[1]
		\Procedure{$\textsc{Spherical-LSH}(v,x, \mu )$
		}{}
		
		\State $\gamma \gets \frac{1}{\log \log \log n}$\label{line:epsilon}
		\State $T\gets \sqrt{\log n}$\label{line:t-def}

		\State  $U\gets \textsc{PseudoRandomify}(v,\gamma)$
		\For {$w\in U$}
		\State Add $w$ as a child of $v$ in recursion tree $\mathcal{T}$
		\State $P\gets w.P$ \Comment{The dataset of $w$}
		\State $R\gets w.r$\Comment{Radius of the sphere of $w$}
		\State {\bf if~}$R<R_{min}$ {\bf continue}\label{line:radius-dmin-lsh}
		\State $\delta' \gets \exp(-(\log\log n)^C)$ \label{line:delta-prime-def}
		\State $o\gets w.o$\Comment{Center of sphere of $w$}

		\State $W\gets \left\{\lfloor\frac{\Delta-\delta}{\delta'}\rfloor\cdot \delta',\left(\lfloor\frac{\Delta-\delta}{\delta'}\rfloor+1\right)\delta',\ldots\right\}\cap (0,R(\sqrt{2}+\gamma)]$ \label{line:lowx}
		\\
		\Comment{The smallest element in $W$ is $\Theta(1)$ by the setting of parameters. See \ref{sec:param}}
		\For {$x'' \in W$}	\label{line:xprime-enumeration}\Comment{Enumerate over potential target distances}

			\State {\bf if~}$x''>R(\sqrt{2}+\gamma)$ {\bf continue} \label{line:xsq2}
			\State Choose $\eta$ such that $\frac{F(\eta)}{G(x''/R,\eta)}=\left(\frac{1}{\mu}\right)^{\frac{1}{T}}$ \label{line:1T} \\
			 \Comment{Choose $\eta$ such that a query explores $\left(\frac{1}{\mu}\right)^{\frac{1}{T}}$ children in expectation}

			\For {$i = 1 ,\dots ,\left\lceil  \frac{100}{G(x''/R,\eta)}   \right\rceil$}\label{line:all-gaussians}
				\State Sample a Gaussian vector $g\sim N(0, 1)^d$
				\State $P' \gets \left\{\mathbf{p}\in P : \left\langle \frac{\mathbf{p}.new-o}{R},g\right\rangle \ge \eta \right\}$\\
				 \Comment{$\mathbf{p}.new$ is the rounded $\mathbf{p}$ to the surface of the sphere (see line~\ref{line:p-new} of Algorithm~\ref{alg:pseudorandomify})}
				\If {$P'\ne \emptyset$}
				\State Add a node $v'$  as child of $w$ in $\mathcal{T}$
				\State $v'.P\gets P',~v'.level\gets v.level+1,~v'.g\gets g,~v'.r\gets R,~v'.o\gets o,~v'.x\gets x''$
				\State $v'.\eta\gets \eta$
				\If {$v'.level \ne J$}\\
				\Comment{Stop whenever the level becomes $J$ (see Section~\ref{sec:param} for the value of $J$.)}\label{line:Jstop}\label{line:termination}
				\State $\textsc{Spherical-LSH}(v',x, \mu)$ \Comment{Recurse unless budget has been exhausted}
				\EndIf
				\EndIf
			\EndFor
		\EndFor
		\EndFor

		\EndProcedure
	\end{algorithmic}
\end{algorithm}

Algorithm~\ref{alg:DD-KDE-Prep} is the standard (similar to Section~\ref{sec:data-independent}) outer algorithm and is presented in Appendix~\ref{sec:dd-kde}. It simply calls \textsc{PreProcess} (Algorithm~\ref{alg:DD-KDE-Prep2}) presented in this section. The following lemma bounds the space complexity of the preprocessing algorithm:
\begin{lem}\label{lm:space}
Under Assumption~\ref{assump:1}, the expected space consumption of the datastructure generated by \textsc{PreProcess-KDE}($\wt{P}, \mu$) (Algorithm~\ref{alg:DD-KDE-Prep}) is bounded by 
$$\min\left\{  n\muu{0.173},\muu{1+c+o(1)}      \right\},$$
for small constant $c=10^{-3}$. 
\end{lem}
\begin{proof}
First, we calculate the expected size of the data structure created by $\textsc{PreProcess}(\wt{P},x,\mu)$ for any $x\in\{\delta_x,2\delta_x,\ldots\}\cap (0,\sqrt{2})$ (see line~\ref{line:xingrid} in Algorithm~\ref{alg:DD-KDE-Prep}). Note that the expected size of the sampled dataset is $$\E[|\wt{P}|]\leq \min\left\{ \muu{1-\frac{x^2}{2}},n\right\}.$$ Since \textsc{PseudoRandomify} does not duplicate points, every point in the dataset is duplicated (due to their presence in different spherical caps) at most
\begin{align*}
\muu{\frac{1}{T}} \cdot |W| =\muu{\frac{1}{T}} \cdot \exp\left((\log\log n)^{O(1)}\right)
\end{align*}  
times in expectation each time we increase the level. So, in total every point is duplicated at most
\begin{align*}
\muu{\frac{J}{T}}\cdot |W|^{J}=\muu{\frac{J}{T}}\cdot |W|^{J}
\end{align*} in expectation. Indeed, in every level we enumerate over at most $|W|= \exp((\log\log n)^{O(1)})$ possibilities for $x''$, amounting to at most a factor of $|W|^{J}=\exp\left((\log\log n)^{O(1)}\cdot O(\sqrt{\log n})\right)=n^{o(1)}$ duplication due to the termination condition in line~\ref{line:termination} of Algorithm~\ref{alg:DD-KDE-spherical}. Finally, \textsc{PreProcess} itself hashes every point $100\muu{\alpha}\leq 100\muu{10^{-4}}$ times (see line~\ref{line:const-c} and line~\ref{line:ai-alpha} of Algorithm~\ref{alg:DD-KDE-Prep2}). Putting these bounds together yields that the space consumption of $\textsc{PreProcess}(\wt{P},x,\mu)$  is at most
	\begin{align*}
	&\min\left\{ \muu{1-\frac{x^2}{2}},n\right\}\cdot\muu{10^{-4}+ \min\left\{\alpha^*,\frac{x^2}{2}\left(1-\frac{x^2}{2}\right)+10^{-4}\right\}+o(1)}
	\end{align*}
	in expectation, where $c:=10^{-4}$. Now, note that we also repeat this procedure $\left(\frac{1}{\mu}\right)^{4\delta_x+o(1)}$ times (see Algorithm~\ref{alg:DD-KDE-Prep}), which results in the following bound on the total space consumption
	\begin{align*}
	&\max_{x\in (0,\sqrt{2})}\left(\min\left\{ \muu{1-\frac{x^2}{2}},n\right\}\cdot\muu{10^{-4}+ \min\left\{\alpha^*,\frac{x^2}{2}\left(1-\frac{x^2}{2}\right)+10^{-4}\right\}+4\delta_x+o(1)} \right)\\ &\le \min\left\{  n\muu{0.173},\muu{1+c+o(1)}      \right\},
	\end{align*}
	for $c=10^{-3}$.
\end{proof}

Finally, we introduce the procedure \textsc{PseudoRandomify} (Algorithm~\ref{alg:pseudorandomify} below) used in \textsc{Spherical-LSH}.  This procedure is quite similar to the corresponding primitive in~\cite{DBLP:conf/soda/AndoniLRW17} and is guaranteed to output pseudo-random spheres with parameters $\tau$ and $\gamma$ (See Definition~\ref{def:psrs}).

\begin{algorithm}[H]
	\caption{PseudoRandomify}  
	\label{alg:pseudorandomify} 
	\begin{algorithmic}[1]
		\Procedure{$\textsc{PseudoRandomify}(v,\gamma)$ }{}
		\State $\delta\gets \exp(-(\log\log n)^C)$ \label{line:delta-ps}
		\State $R_{min}\gets$ sufficiently small constant larger than $\Delta$ and $\delta_x$ (see Section~\ref{sec:param})  \label{line:d-min}
		\State $P \gets v.P$ \Comment{Dataset of node $v$}
		\State $R\gets v.r$ \Comment{Radius of sphere of node $v$}
		\State $o \gets v.o$ \Comment{Center of sphere of node $v$}
		\State $\tau \gets \frac{1}{10}$
		\State {\bf if~}$R<R_{min}$ {\bf return}\label{line:radius-dmin-ps}
		\For {$\mathbf{p}\in P$} 
		\State $\mathbf{p}.new\gets o+\delta \lceil \frac{||\mathbf{p}-o||}{\delta}\rceil \cdot \frac{\mathbf{p}-o}{||\mathbf{p}-o||}$\label{line:p-new} \Comment{$\mathbf{p}$ represents the initial coordinates of point $\mathbf{p}$}
		\EndFor
		\State $V\gets \emptyset$				
		\For {$i \gets 1 \dots \lceil \frac{R}{\delta }\rceil$}\Comment{Process all resulting spheres}
		\State $\wt{P}\gets \{\mathbf{p}\in P~:||\mathbf{p}.new-o||=\delta i \}$
		\If {$\wt{P}\ne \emptyset$}
		\State $\hat{R}\gets (\sqrt{2}-\gamma)R$
		\State $m \gets |\wt{P}|$
		\State $m'\gets 0$
		\While {$m' \le \frac{m}{2} $}
		\State $m \gets |\wt{P}|$
		\While {$\exists \hat{o} \in \partial B(o, \delta i): |B(\hat{o},\hat{R})\cap \wt{P}|\ge \frac{1}{2}\cdot \tau\cdot m$}\label{line:pseudorandom}
		\State \Comment{Using {\bf rounded} $p.new$ coordinates (see line~\ref{line:p-new}) in line above} 
		\State $P' \gets \wt{P}\cap B(\hat{o},\hat{R})$
		\State $B(o',R')\gets \textsc{SEB}(P')$ \Comment{SEB=smallest enclosing ball}

		\State Create a node $tmp$
		\State $tmp.P \gets P',~tmp.level\gets v.level,~tmp.g\gets 0,~tmp.r\gets R',~tmp.o\gets o'$
		\State $V\gets V\cup \textsc{PseudoRandomify}(tmp,\gamma)$
		\State $\wt{P}\gets \wt{P} \setminus B(\hat{o},\hat{R})$
		
		\EndWhile

		\State $m'\gets |\wt{P}|$
		\EndWhile
		\State Create a node $w$
		\State $w.P\gets \wt{P},~w.level \gets v.level,~w.g\gets 0,~w.r\gets \delta i, w.o\gets o$
		\State $V \gets V\cup \{w\}$
		\EndIf
		\EndFor
		\State\Return $V$	
		\EndProcedure
	\end{algorithmic}
\end{algorithm}

\subsection{Query procedure}\label{sec:query-dd}
We now present our query procedure (Algorithm~\ref{alg:DD-KDE-Query} below). The procedure simply traverses the recursion tree $\mathcal{T}$ from the root, exploring leaves that the query is mapped to according to line~\ref{line:g-capture}. Since every node $u$ of the tree $\mathcal{T}$ corresponds to a pseudorandom dataset $u.P$ residing (essentially) on a sphere of radius $u.r$ centered at $u.o$, the query is projected onto the sphere, after which one recursively explores the children of $u$ in $\mathcal{T}$ whose Gaussian vectors (see line~\ref{line:g-capture}) are sufficiently correlated with the projected query. One notable feature in comparison to the corresponding procedure in~\cite{DBLP:conf/soda/AndoniLRW17} is the follows. Note that the procedure of~\cite{DBLP:conf/soda/AndoniLRW17} recurses on a sphere even if the intersection of a sphere of radius $x$ around the query (i.e. the range in which we would like to report points) barely touches the sphere that the dataset resides on. Our data structure, however, uses an increased search range $x+\Delta$ (see Figure~\ref{fig:Deltaaa}), which results in somewhat higher runtime, but allows one to only recurse when the extended search range has nontrivial overlap with the sphere in question -- see lower bound on $x'$ in line~\ref{line:lowx} of Algorithm~\ref{alg:DD-KDE-spherical}. This additive $\Delta$ technique, can also be used to simplify the technical proofs of \cite{DBLP:conf/soda/AndoniLRW17}, by not allowing their algorithm to recurse on tiny spheres at distance roughly $x$ (i.e., when the distance $x$ barely touches the sphere). The reason is that all the points on these small spheres has distance at most $x+2R_{min}$ from the query, and we have small number of such points in expectation, by sub-sampling and density constraints.
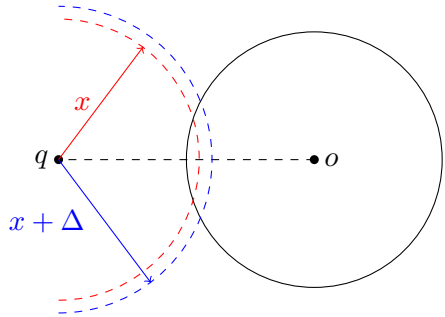
\begin{figure}[H]
	\centering
	\tikzstyle{vertex}=[circle, fill=black!70, minimum size=3,inner sep=1pt]
	\tikzstyle{svertex}=[circle, fill=black!100, minimum size=5,inner sep=1pt]
	\tikzstyle{gvertex}=[circle, fill=green!80, minimum size=7,inner sep=1pt]
	
	\tikzstyle{evertex}=[circle,draw=none, minimum size=25pt,inner sep=1pt]
	\tikzstyle{edge} = [draw,-, color=red!100, very  thick]
	\tikzstyle{bedge} = [draw,-, color=green2!100, very  thick]
	\begin{tikzpicture}[scale=1.7, auto,swap]
	
	\draw (0,0) circle (1cm);

	\fill[fill=black] (0,0) circle (1pt);

	\draw (-2,0)[left] node {{$q$}};
	\fill[fill=black] (-2,0) circle (1pt);
	
	\draw[dashed] (0,0 ) --node[below]{} (-2,0);
	
	\draw (0,0)[right] node {{$o$}};

	\draw [->,red ] (-2,0) -- node[left]{$x$}(-2+0.6*1.1,0.8*1.1);
	\draw [->,blue ] (-2,0) -- node[left]{$x+\Delta~$}(-2+0.6*1.2,-0.8*1.2);

	\draw[dashed,red] (-2,-1.1) arc (-90:90:1.1cm);
	\draw[dashed,blue] (-2,-1.2) arc (-90:90:1.2cm);

	\end{tikzpicture}
	\caption{An (exaggerated) illustration of $x$ and $x+\Delta$.} \label{fig:Deltaaa}
\end{figure}
\begin{algorithm}[H]
	\caption{Query }  
	\label{alg:DD-KDE-Query} 
	\begin{algorithmic}[1]
		\Procedure{$\textsc{Query}(\mathbf{q},\mathcal{T},x )$	
		}{}	
			\State $P_x\gets \emptyset$
			\State $\delta\gets \exp(-(\log\log n)^C)$ \label{line:delta-def}
			\State $v \gets $ root of $\mathcal{T}$
			\If {$v.level = 0$}\label{line:ai-query}
					\State $K_1=100 \left(\frac{1}{\mu}\right)^\alpha$ \Comment{ $\alpha$ is the constant from line~\ref{line:ai-alpha} in Algorithm~\ref{alg:DD-KDE-Prep2}}
					\For {$j = 1,2,\dots, \lceil K_1 \rceil$}
					\State Locate $\mathbf{q}$ in $h_{j}$ and $u\gets $ the corresponding node in $\mathcal{T}$
					\For {each $w$ child of $u$}
					\State $\mathcal{T}_w \gets $ sub-tree of $w$ and its descendants. 
					\State $P_x \gets P_x \cup \textsc{Query}(\mathbf{q},\mathcal{T}_w,x)$
					\EndFor
					\EndFor
					\Return $P_x$
			\ElsIf {$\mathcal{T}$ is just one node, without any children}
			\State\Return $v.P$\label{line:depth1}
			\Else
			\State $o \gets v.o$ 
			\State $R \gets v.r$
			\State $R_2\gets ||\mathbf{q}-o||$
				\State $x' \gets \textsc{Project}(x+\Delta, R_2,R)$\label{line:x-prime}
			\State$ x''\gets$ smallest element in the grid $W$ (line~\ref{line:lowx} of Algorithm~\ref{alg:DD-KDE-spherical}) which is not less than $x'$\label{line:x-tilde}
			\If {$ x+\delta < |R-R_2|$} {\bf return}\label{line:xclose} \\
			\Comment{Then no point from distance $x$ can be on this sphere}\EndIf 
			\If{$\nexists u$ child of $v$, such that $u.x=x''$} \label{line:deadendscan1}
			\State \Return $v.P$ 
			\EndIf
			\For {each $u$ child of $v$}
				
				\State $\Delta \gets 10^{-20}$  \label{line:Delta-def}

					\If {$u.x =x''$}			
					\State $g\gets u.g$
					\State $\eta \gets u.\eta$
					\If {$\langle g, \frac{\mathbf{q}-o}{||\mathbf{q}-o||} \rangle \ge \eta$}\label{line:g-capture}
									\For {each $w$ child of $u$}
					
								\State $\mathcal{T}_w\gets $ sub-tree of $w$ and its descendants.
					\State $P_x\gets P_x  \cup \textsc{Query}(\mathbf{q},\mathcal{T}_w,x)$
					
					\EndFor
					
					\EndIf	
										\EndIf

			\EndFor
			
					\State\Return $P_x$
			\EndIf

		\EndProcedure
	\end{algorithmic}
\end{algorithm}

\renewcommand{\P}{{\mathcal P}}
Since the correctness analysis of this procedure is standard and similar to \cite{DBLP:conf/soda/AndoniLRW17}, we present it in Appendix~\ref{sec:dd-kde} (Lemma~\ref{lem:tree-correctness}), for completeness. Basically, we prove the query procedure outputs any given point within distance $x$ with high constant probability.

\section{Query time analysis}\label{sec:query}

The main result of this section is the following lemma which bounds the expected query time of the algorithm. 

\begin{lem}\label{lm:query}
	The expected query time is bounded by $O\left(\left(\frac{1}{\mu}\right)^{0.173+o(1)}\right)$.
\end{lem}

Throughout this section we consider the setting where one is given a query $\mathbf{q}\in \R^d$ and a parameter $\mu\in (0, 1]$  with the promise that 
\begin{equation}\label{eq:mu-mustar}
\mu^*\le \mu,
\end{equation}
where 
$$
\mu^*=K(P, \mathbf{q})
$$
is the true kernel density value. We assume that $\mu^*=n^{-\Theta(1)}$, since this is the interesting regime for this problem. For $\mu^{*}=n^{-\omega(1)}$ under the Orthogonal Vectors Conjecture (e.g. \cite{rubinstein2018hardness}), the problem cannot be solved faster than $n^{1-o(1)}$ using space $n^{2-o(1)}$~\cite{charikar2019multi-resolution}, and for larger values $\mu^{*}=n^{-o(1)}$ random sampling solves the problem in $n^{o(1)}/\epsilon^{2}$ time and space.  

\paragraph{Densities of balls around query.} Upper bounds on the number of points at various distances from the query point in dataset (i.e., densities of balls around the query) play a central part in our analysis. The core of our query time bound amounts to tracking the evolution of such densities in the recursion tree $\mathcal{T}$. In order to analyze the evolution of these upper bounds we let, for a query $\mathbf{q}\in \mathbb{R}^d$ (which we consider fixed throughout this section) and any $x\in (0,\sqrt{2})$ let
 \begin{equation}\label{eq:dx-def}
 D_x(\mathbf{q}):= \{ ||\mathbf{p}-\mathbf{q}||:~\mathbf{p}\in P, ||\mathbf{p}-\mathbf{q}||\ge x+1.5\Delta\},
 \end{equation}
 denote the set of possible distances from the query to the points in the dataset which are further that $x+1.5\Delta$ from the query.
  When there is no ambiguity we drop $q$ and $x$ and we simply call it $D$. For any $y\in D$ we let 
 \begin{equation}\label{eq:p-y-def}
 P_y(\mathbf{q}):=\{ \mathbf{p}\in P: ||\mathbf{p}-\mathbf{q}||\le y\}
 \end{equation}
 be the set of points at distance $y$ from $\mathbf{q}$. Since for every $y>0$
\begin{equation*}
\begin{split}
 \mu^*=K(P, \mathbf{q})&=\frac1{n}\sum_{\mathbf{p}\in P} \mu^{||\mathbf{p}-\mathbf{q}||_2^2/2}\\
 &\geq \frac{\mu^{y^2/2}}{n} |P_y(\mathbf{q})|
\end{split}
\end{equation*}
 we get
 $$
|P_y(\mathbf{q})| \leq n \mu^*\cdot \left(\frac1{\mu}\right)^{\frac{y^2}{2}}\leq n\cdot \left(\frac1{\mu}\right)^{\frac{y^2}{2}-1},
 $$
 since $\mu^*\leq 4\mu$ by assumption.

\paragraph{Densities in the subsampled dataset.} Fix $x\in (0, \sqrt{2})$, and recall that $\wt{P}$ contains every point in $P$ independently with probability $\frac1{n}\cdot \left(\frac1{\mu}\right)^{1-\frac{x^2}{2}}$. Note that for every $y$ the expected number of points at distance at most $y$ from query $\mathbf{q}$ that are included in $\wt{P}$ is upper bounded by 
\begin{equation}\label{eq:density-ub}
\min\left\{n\cdot \left(\frac1{\mu}\right)^{\frac{y^2}{2}-1},n\right\}\cdot \frac1{n}\cdot \left(\frac1{\mu}\right)^{1-\frac{x^2}{2}}\leq  \min\left\{ \muu{\frac{y^2-x^2}{2}},\muu{\frac{1-x^2}{2}}\right\},
\end{equation}
and Figure~\ref{fig:density-ub} illustrates this. 
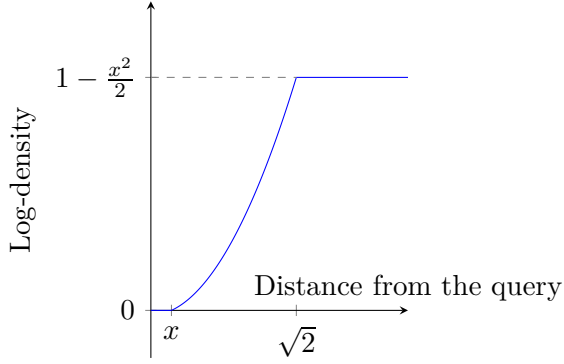
\begin{figure}[H]
	\begin{center}
		\begin{tikzpicture}
		\begin{axis}[
		restrict y to domain=-10:10,
		samples=1000,
		width=5cm, height=180pt,
		ymin=-0.2 ,ymax=1.3,
		xmin=0, xmax=2.5,
		xtick={0,0.2,1.41421},
		xticklabels={$0$,$x$,$\sqrt{2}$},
		ytick={0,0.98},
		yticklabels={$0$,$1-\frac{x^2}{2}$},
		xlabel={Distance from the query},
		ylabel={Log-density},
		axis x line=center,
		axis y line=left,
		every axis x label/.style={
			at={(ticklabel* cs:1)},
			anchor=south,}
		]
		
		]
		Below the red parabola is defined
		\addplot [
		domain=0:0.2, 
		samples=100, 
		color=blue,
		]
		{0};
		
		\addplot [
		domain=0.2:1.41421, 
		samples=100, 
		color=blue,
		]
		{(x^2/2-0.02)};
		\addplot [
		domain=1.41421:2.5, 
		samples=100, 
		color=blue,
		]
		{0.98};
		
		\addplot [
		domain=0:1.41421, 
		samples=100, 
		color=gray,
		style=dashed,
		]
		{0.98};
		\end{axis}
		\end{tikzpicture}
		\caption{Upper-bound on log-densities after sub-sampling.}	\label{fig:density-ub}
	\end{center}
	
\end{figure}
Our main goal is to track the progress of the query $\mathbf{q}$ and any $\mathbf{p}$, for which we have $||\mathbf{p}-\mathbf{q}||_2\le x$, that was included in the set $\wt{P}$, and exploit the upper bounds~\eqref{eq:density-ub} on the number of points at various distances from $\mathbf{q}$ in $\mathbf{q}$'s `hash bucket' to show that the process quickly converges to a constant size data set at a leaf of $\mathcal{T}$. It is not hard to see (Lemma~\ref{lm:numleaves-explored} below)  that the number of nodes in $\mathcal{T}$ that the query explores is low. The main challenge is to show that the expected size of a leaf data set in $\mathcal{T}$ is small, since for that one needs to prove strong upper bounds on the number of points at various distances from the query in dataset that the query traverses on its path to a leaf in $\mathcal{T}$. We exploit two effects:
\begin{description}
\item[(Removal of points due to truncation)] The \textsc{Pseudorandomify} procedure, which is crucial to ensuring that spheres at nodes on $\mathcal{T}$ are pseudorandom, essentially acts as a trunction primitive on the density curve. See conditions {\bf (2)} in Definition~\ref{def:validpath} below.
\item[(Removal of points due to LSH)] As the query explores the children of an LSH node $v\in \mathcal{T}$ the probability that a given point $\mathbf{p}\in v.P$ belongs to the same spherical cap as $\mathbf{q}$ depends on the distance between $\mathbf{p}$ and $\mathbf{q}$. This implies bounds on the evolution of the density of points at various distances $y$ from $\mathbf{q}$ in the datasets that $\mathbf{q}$ explores on its path towards a leaf in $\mathcal{T}$. See conditions {\bf (3)} in Definition~\ref{def:validpath} below.
\end{description}

The bulk of our analysis is devoted to understanding the worst case sequence of geometric configurations, i.e. spheres, that the query encounters on its path towards a leaf in $\mathcal{T}$.  

\subsection{Path geometries }\label{sec:pathgeometry}
We start by defining the path geometries in the recursion tree. Assume an invocation of \textsc{PreProcess} algorithm (Algorithm~\ref{alg:DD-KDE-Prep2}) and let $\mathcal{T}$  be the sub-tree that the query explores. Let $$\mathcal{P}:= (w_0,v_0,w_1,v_1,\ldots, w_J,v_J)$$ be any path from root to a LSH leaf at level $J$.

For any $j\in[J]$, suppose that given $x'':= v_j.x$ and $r:=v_i.r$, we are interested in the distance from the query to the center of the sphere ($v_j.o$). For simplicity of notation let $\wt{\ell}=||\mathbf{q}-v_j.o||$. Recall that $x''$ is the {\emph rounded} value for $x'=\textsc{Project}(x+\Delta,\wt{\ell},r)$ (see lines~\ref{line:x-prime} and \ref{line:x-tilde} of Algorithm~\ref{alg:DD-KDE-Query}). However, this equation is a degree two polynomial in $\wt{\ell}$, so it has at most two solutions. For intuition, Figure~\ref{fig:b} shows these two solutions with an example. The solutions to the equation correspond to the points that the dashed circle intersects with the dashed line, i.e., position of $\mathbf{q}$. Now, recall that $x''$ is the rounded $x'$ (see line~\ref{line:x-tilde} of Algorithm~\ref{alg:DD-KDE-Query}). So, $x'$ can change in a small interval. This corresponds to moving the center of the dashed circle over the red arc. This changes the position of intersections, however, they still belong to a relatively small interval (shown in blue in Figure~\ref{fig:b}), we denote this intervals by \emph{left interval} and \emph{right interval}. Now, given query $\mathbf{q}$, we check weather it corresponds to the left interval or the right interval, and based on that we set $b_j$ to be $1$ or $2$, respectively. We also let $\ell$ be the distance of the leftmost point in the interval of the query, from the center of the sphere. And we call $\ell$ the distance induced by $(x'',r)$ and $\mathbf{q}$. In appendix~\ref{app:b} we formally argue this procedure.
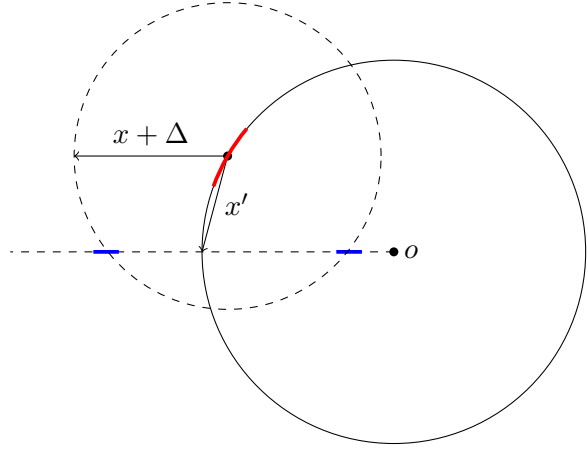
\begin{figure}[H]
	\centering
	\tikzstyle{vertex}=[circle, fill=black!70, minimum size=3,inner sep=1pt]
	\tikzstyle{svertex}=[circle, fill=black!100, minimum size=5,inner sep=1pt]
	\tikzstyle{gvertex}=[circle, fill=green!80, minimum size=7,inner sep=1pt]
	
	\tikzstyle{evertex}=[circle,draw=none, minimum size=25pt,inner sep=1pt]
	\tikzstyle{edge} = [draw,-, color=red!100, very  thick]
	\tikzstyle{bedge} = [draw,-, color=green2!100, very  thick]
	\begin{tikzpicture}[scale=1.7, auto,swap]

	\draw (0,0) circle (1.5cm);

	\fill[fill=black] (0,0) circle (1pt);

	\draw[dashed] (0,0 ) --node[below]{} (-3,0);
	\draw[line width=0.5mm, blue ] (-0.25,0 ) --node[below]{} (-0.45,0);
	
	\draw[line width=0.5mm, blue ] (-2.15,0 ) --node[below]{} (-2.35,0);
	\draw (0,0)[right] node {{$o$}};

	\draw[->] (-1.3,0.75 ) --node[above]{$x+\Delta$} (-2.5,0.75);
	\draw[->] (-1.3,0.75 ) --node[right]{$x'$} (-1.5,0);
	\draw[dashed] (-1.3,0.75) circle (1.2cm);
	\fill[fill=black] (-1.3,0.75) circle (1pt);

	\draw[line width=0.5mm,red] (-1.15,0.964) arc (140:160:1.5cm);

	\end{tikzpicture}
	\caption{Geometric illustration of equation $x'=\textsc{Project}(x+\Delta,\wt{\ell},r)$ when we have access to an approximation of $x'$ (red arc).} \label{fig:b}
\end{figure}
\begin{defn}[Path geometry and induced distances]\label{def:pathgeom}
For any query $\mathbf{q}$ and tree $\mathcal{T}$ (as described above) for any root to leaf path $$\mathcal{P}= (w_0,v_0,w_1,v_1,\ldots, w_J,v_J),$$ we call $$G(\mathcal{P}):=((x''_1,r_1,b_1),\ldots,(x''_J,r_J,b_J))$$ the geometry of path $\mathcal{P}$ where for all $i\in[J]$,
\begin{enumerate}
	\item $x''_i:=v_i.x$,
	\item  $r_i:=v_i.r$,
	\item  $b_i$ is as described above (formally defined in Appendix~\ref{app:b}). 
\end{enumerate} 
Additionally, we call $L(\mathcal{P}):=(\ell_1,\ldots,\ell_J)$ the induced distances of path $\mathcal{P}$, where for all $i\in[J]$, $\ell_i$ is induced by $(x''_i,r_i)$ as explained above and formally defined in Appendix~\ref{app:b}.
\end{defn}

\begin{defn}[Sphere geometries]
	For any query $\mathbf{q}$ and tree $\mathcal{T}$ (as described above) for any root to leaf path $$\mathcal{P}= (w_0,v_0,w_1,v_1,\ldots, w_J,v_J),$$ if the geometry of this path is defined as $$G(\mathcal{P}):=((x''_1,r_1,b_1),\ldots,(x''_J,r_J,b_J))$$ then for any $j\in[J]$ we say that $w_j$ and $v_j$ has geometry $(x''_j,r_j,b_j)$.
\end{defn}

Recall from Definition~\ref{def:psrs} that the \textsc{Pseudoranomify} procedure (Algorithm~\ref{alg:pseudorandomify}) ensures that most of the points on any pseudorandom sphere $w$ are nearly orthogonal to $\mathbf{q}-w.o$. We want to know, how the fact that a sphere is pseudorandom translates to densities. For the first step, we need to understand if a point on the sphere is almost orthogonal to the projection of the query on the sphere, then what the range of possible distances of these points from the query is. We define the $c:=\sqrt{\ell^2+r^2}$ which simplifies the notation. As Figure~\ref{fig:cj} suggests, we expect the orthogonal points to be at distance $\approx c$. The following claim formally argues how pseudorandomness of a sphere translates to a condition on the densities.

	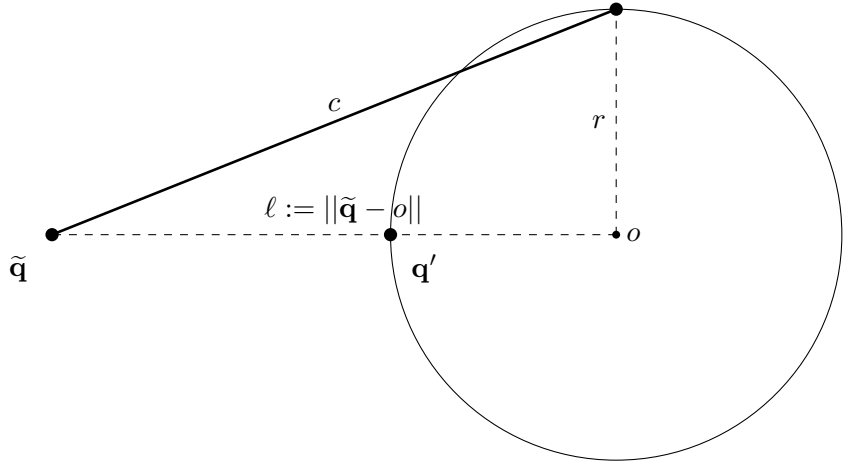
\begin{figure}[H]
	\centering
	\tikzstyle{vertex}=[circle, fill=black!70, minimum size=3,inner sep=1pt]
	\tikzstyle{svertex}=[circle, fill=black!100, minimum size=5,inner sep=1pt]
	\tikzstyle{gvertex}=[circle, fill=green!80, minimum size=7,inner sep=1pt]
	
	\tikzstyle{evertex}=[circle,draw=none, minimum size=25pt,inner sep=1pt]
	\tikzstyle{edge} = [draw,-, color=red!100, very  thick]
	\tikzstyle{bedge} = [draw,-, color=green2!100, very  thick]
	\begin{tikzpicture}[scale=1.5, auto,swap]
	
	\draw (0,0) circle (2cm);

	\fill[fill=black] (0,0) circle (1pt);
	\draw[dashed] (0,0 ) -- node[left]{$r$} (0,2);

	\node[svertex](v1) at (-5, 0) {};
	\draw (-5.3,-0.3) node {{$\wt{\mathbf{q}}$}};
	\node[svertex](v1) at (-2, 0) {};
	\draw (-1.7,-0.3) node {{$\mathbf{q}'$}};
	\path[draw, line width=1pt, -, color=black!100, line width=1pt] (-5,0) --node[above]{$c$}(0,2);  
	\draw[dashed] (0,0 ) --node[above]{$~~\ell:=||\wt{\mathbf{q}}-o||$} (-5,0);
	\node[svertex](v1) at (0, 2) {};   
	\draw (0,0)[right] node {{$o$}};
	\end{tikzpicture}
	\caption{Illustration of the definition of $c$, the distance from the query to a `typical' point on the sphere.} \label{fig:cj}
\end{figure}
\begin{claim}[Truncation claim]\label{claim:roundedpseudorand}
Given query $\mathbf{q}$, let $w$ be a pseudo random sphere with geometry $(x'',r,b)$ which induces distance $\ell$. Let $w.P$ be the set of points on this sphere, i.e., for any $\mathbf{p}\in w.P$, $\mathbf{p}.new$ is on the sphere. For all $y$ let $B_{y}$ be the number of points at distance $y$ from $\mathbf{q}$ in $w.P$. Then, the following conditions hold.
\begin{align*}
\sum_{y\leq c-r \psi } B_{y}\leq \frac{\tau}{1-2\tau}\cdot\sum_{y\in (c-r\psi, c+r\psi )} B_{y},
\end{align*}
and 
\begin{align*}
\sum_{y\ge c+r \psi } B_{y}\leq \frac{\tau}{1-2\tau}\cdot\sum_{y\in (c-r\psi, c+r\psi )} B_{y},
\end{align*}
where $\psi=\gamma^{1/3}+\delta'^{1/4}+\delta^{1/4}$ (the same as in Claim~\ref{claim:roundedpsrsph}) and $c:=\sqrt{\ell^2+r^2}$.
\end{claim}
\begin{proof}
	The proof is just a simple application of Claim~\ref{claim:roundedpsrsph} to this sphere.
\end{proof}
 
Suppose that one has two points on the sphere at some distance from each other, we can use Lemma~\ref{lm:f-eta} and Lemma~\ref{lm:g-eta}, to find collision probabilities under a spherical cap of size $\eta$. However, in general the query is not on the sphere, so we need to translate distance from $\mathbf{q}$ to any point $\mathbf{p}$ to distance from $\mathbf{q}'$ (projection of $\mathbf{q}$ on the sphere) to $\mathbf{p}$, using a function called \textsc{Project} (formally defined in Definition~\ref{def:projection} and its formula is given in Lemma~\ref{lem:projection}). Also, there are some rounding steps, such as rounding the points to the sphere and rounding of the distance from the query to the center of the sphere (rounding of $\wt{\ell}$ to $\ell$). Considering all these issues, the following claim illustrates the effect of spherical LSH on the points based on their distance from the query.  

\begin{claim}[Spherical LSH claim]\label{cl:nonadapt} Suppose that there is a sphere with geometry $(x'', r,b)$ and induced distance $\ell$ (see Section~\ref{sec:pathgeometry} and Definition~\ref{def:pathgeom}) for some $x''\in W$, $r\in \left[\left\lceil\frac{R_{\text{max}}}{\delta}\right\rceil\right] $ and $b\in\{1,2\}$. Let $o$ be the center of the sphere. Also, let $\mathbf{p}$ be a point such that $y=||\mathbf{p}-\mathbf{q}||$ and $\mathbf{p}.new$ is on the sphere (see line~\ref{line:p-new} of Algorithm~\ref{alg:pseudorandomify}). Now, suppose that one generates a Gaussian vector $g$ as in Algorithm~\ref{alg:DD-KDE-spherical}. Then, we have
	\begin{align*}
	\Pr_{g\sim N(0,1)^d}\left[\langle g,\frac{\mathbf{p}.new-o}{||\mathbf{p}.new-o||}\rangle \ge \eta| \langle g,\frac{\mathbf{q}-o}{||\mathbf{q}-o||}\rangle \ge \eta \right]\le \muu{-  \frac{4(r/x')^2-1}{4(r/y')^2-1}\cdot \frac{1}{T}}.
	\end{align*}
	where 
	\begin{itemize}
		\item $\eta$ is such that $\frac{F(\eta)}{G(x''/r,\eta)}=\left(\frac{1}{\mu}\right)^{\frac{1}{T}}$ (see line~\ref{line:1T} of Algorithm~\ref{alg:DD-KDE-spherical}).
		\item $x':=\textsc{Project}(x+\Delta,\ell,r)$.
		\item $y' := \textsc{Project}(y-\Delta/2,\ell,r)$.
	\end{itemize}
\end{claim}
\begin{proof}
	
	Let $o$ be the center of the sphere. Let $\wt{\ell}:=||\mathbf{q}-o||$. Recall by the discussion in Section~\ref{sec:pathgeometry} and Definition~\ref{def:pathgeom} that any sphere geometry $(x'',r,b)$ induces a distance $\ell$. Now, suppose that we move the query in the direction of  the vector from $o$ to $\mathbf{q}$, such that for the new point $\wt{\mathbf{q}}$, we get $||\wt{\mathbf{q}}-o||=\ell$. Now, one should note that the geometry of the sphere with respect to $\mathbf{q}$ and $\wt{\mathbf{q}}$ is the same. Also, the projections of $\mathbf{q}$ and $\wt{\mathbf{q}}$ on the sphere are identical. Also, for point $\mathbf{p}$ at distance $y$ from $\mathbf{q}$, by the triangle inequality for $(\mathbf{q},\wt{\mathbf{q}},\mathbf{p})$, since $\wt{\ell} \in [\ell-\delta'^{1/3},\ell]$ we get  
	\begin{align}
	||\mathbf{p}-\wt{\mathbf{q}}|| \in [y-\delta'^{1/3},y+\delta'^{1/3}].
	\end{align}
	Now, if we let point $\mathbf{q}'$ be the projection of $\wt{\mathbf{q}}$ on the sphere, and let $\mathbf{p}.new$ be the rounded $\mathbf{p}$ on the sphere, then $||\mathbf{p}.new-\wt{\mathbf{q}}|| \in [y-\delta-\delta'^{1/3},y+\delta+\delta'^{1/3}]$, which implies
	\begin{align*}
	y'':=||\mathbf{q}'-\mathbf{p}.new|| \in [\textsc{Project}(y-\delta-\delta'^{1/3},\ell,r),\textsc{Project}(y+\delta+\delta'^{1/3},\ell,r)].
	\end{align*}
	Note that with this definition of $y''$ one has
	\begin{align}
	\Pr_{g\sim N(0,1)^d}\left[\langle g,\frac{\mathbf{p}.new-o}{||\mathbf{p}.new-o||}\rangle \ge \eta| \langle g,\frac{\mathbf{q}-o}{||\mathbf{q}-o||}\rangle \ge \eta \right]= \frac{G(y''/r,\eta)}{F(\eta)}.
	\end{align}
	Now, by invoking Claim~\ref{cl:collision-prob}, {\bf (b)}
	\begin{equation}\label{eq:collision-prob}
	\Pr_{g\sim N(0,1)^d}\left[\langle g,\frac{\mathbf{p}.new-o}{||\mathbf{p}.new-o||}\rangle \ge \eta| \langle g,\frac{\mathbf{q}-o}{||\mathbf{q}-o||}\rangle \ge \eta \right]\le \muu{-\frac{4(r_j/x')^2-1}{4(r_j/y')^2-1}\cdot \frac{1}{T}}
	\end{equation}
	Now, we verify the preconditions of Claim~\ref{cl:collision-prob}, {\bf (b)}. Condition {\bf (p1)} of Claim~\ref{cl:collision-prob} is satisfied by setting of $\delta$ as $\delta+\delta'^{1/3}$.\footnote{To be more clear, we set the $\delta$ of claim~\ref{cl:collision-prob} as $\delta+\delta'^{1/3}$ where $\delta$ and $\delta'$ are the parameters of the algorithm.} Condition {\bf (p2)} is satisfied by line~\ref{line:delta-prime-def} in Algorithm~\ref{alg:DD-KDE-spherical}. Condition {\bf (p3)} is satisfied by setting of $\Delta$ in line~\ref{line:Delta-def} of Algorithm~\ref{alg:DD-KDE-Query}. Finally, condition {\bf (p4)} is satisfied due to line~\ref{line:xsq2} in Algorithm~\ref{alg:DD-KDE-spherical} that ensures that a nontrivial data structure is only prepared for $x'\leq R(\sqrt{2}+\gamma)$, and no recursion is performed otherwise. 
	
	Conditioned on event $\mathcal{E}_{diam}$ (which ensures constant upper-bound on the radii of spheres, see the discussion in Section~\ref{sec:prepro-dd}), $r=O(1)$. Thus, we can invoke part {\bf (b)} of Claim~\ref{cl:collision-prob} applies and gives \eqref{eq:collision-prob}. 
\end{proof}

In the following definition we summarize the effect of sub-sampling the dataset, the truncation rounds and the spherical LSH rounds on densities along the path.

\begin{center}
    \fbox{
      \begin{minipage}{0.99\textwidth}
\begin{defn}[Valid execution path]\label{def:validpath}
Let $R:=(r_j)_{j=1}^{J}$ and $L:=(\ell_j)_{j=1}^{J}$ for some positive values $r_j$'s and $\ell_j$'s such that for all $j\in[J]$, $x+\delta \ge |\ell_j-r_j|$. Also let $D$ be as defined in \eqref{eq:dx-def}. Then, for
\begin{equation*}
\begin{split}
A&:=(a_{y,j}), ~~y\in D, j\in[J]\cup \{0\}\text{~~~~~~~~~~(Intermediate densities)}\\
 B&:=(b_{y,j}), ~~y\in D, j\in[J+1]\cup \{0\}\text{~~~~~(Truncated intermediate densities)}
 \end{split}
 \end{equation*}
$(L,R,A,B)$ is called a \emph{valid execution path}, if the conditions below are satisfied. We define $\psi:=\gamma^{1/3}+\delta'^{1/4}+\delta^{1/4}$ and $c_j:= \sqrt{r_j^2+\ell_j^2}$ for convenience.
\begin{description}
\item[(1) Initial densities condition.] The $a_{y, 0}$ and $b_{y,0}$ variables are upper-bounded by the initial expected densities in the sampled dataset: for all $y\in D$
\begin{equation*}
\sum_{y'\in[0,y]\cap D}a_{y',0}\le   \min\left\{ \muu{\frac{y^2-x^2}{2}},\muu{\frac{1-x^2}{2}}\right\}
\end{equation*}
and
\begin{equation*}
\sum_{y'\in[0,y]\cap D}b_{y',0}\le  \min\left\{ \muu{\frac{y^2-x^2}{2}},\muu{\frac{1-x^2}{2}}\right\}
\end{equation*}
\item[(2) Truncation conditions (effect of \textsc{PseudoRandomify}).]  For any $j\in[J]$, for all $y\in D\setminus [\ell_j-r_j,\ell_j+r_j]$ one has $b_{y,j}=0$ (density is zero outside of the range corresponding to the $j$-th sphere on the path; condition {\bf (2a)}), 
for all $y\in D\cap [\ell_j-r_j,\ell_j+r_j]$ one has $b_{y,j}\le a_{y,j-1}$ (removing points arbitrarily {\bf (2b)}) and 
\begin{equation*}
\sum_{y\in [0,c_j-\psi r_j]\cap D}~b_{y,j}\leq \frac{\tau}{1-2\tau}\cdot \sum_{y\in(c_j-\psi r_j,c_j+\psi r_j)\cap D}b_{y,j} ~~~~~\text{(condition {\bf (2c)})}
\end{equation*}

\item[(3) LSH conditions.] For every $j\in [J]$ and all $y\in [\ell_j-r_j,\ell_j+r_j]\cap D$
\begin{equation*}
\begin{split}
a_{y,j} \le  {b_{y,j}}\cdot\muu{-\frac{4\left(\frac{r_j}{x'}\right)^2-1}{4\left(\frac{r_j}{y'}\right)^2-1}\cdot \frac{1}{T}}
\end{split}
\end{equation*}
where $x':=\textsc{Project}(x+\Delta,\ell_j,r_j)$ and $y':=\textsc{Project}(y-\Delta/2,\ell_j,r_j)$. See Remark~\ref{rem:Delta} below for a discussion about $\Delta$ factors. 
\item [(4) Terminal density condition.] For any $y$ such that $a_{y,J}$ is defined, $b_{y,J+1}\le a_{y,J}$. 

\end{description}
\end{defn}
      
      \end{minipage}
    }
\end{center}
\begin{rem}\label{rem:Delta}
	Throughout the paper we need good bounds on the probability that a random spherical cap encompasses a data point $\mathbf{p}$, given that the spherical cap captures the projection of the query. The expression in condition {\bf (3)} of Definition~\ref{def:validpath} is a convenient upper bound for this quantity when the distance from $\mathbf{p}$ to $\mathbf{q}$ is equal to $y$. Exact expressions for such collision probabilities are unstable with respect to perturbations of the point $\mathbf{p}$ when $\mathbf{p}$ is antipodal to $\mathbf{q}$ on the sphere, and because of this it is more convenient to work with upper bounds.  Specifically, we upper bound this probability by imagining that the point is slightly closer (by $\Delta/2$) than the actual distance $y$, for a small positive constant $\Delta$ that affects our query time bounds. The advantage is that such probabilities are more stable under small perturbations of the data point $\mathbf{p}$ -- see the proof of Claim~\ref{cl:collision-prob} for more details.  One notes that the expression in condition {\bf (3)} also depends on $x$. This is because we select spherical cap sizes based on $x$ -- see line~\ref{line:1T} of Algorithm~\ref{alg:DD-KDE-spherical}. 
\end{rem}

We introduce the notion of the length of an execution path $(L, R,A,B)$.
\begin{defn}\label{def:cost}
	We define the length of an execution path $(L, R,A,B)$ by $\text{ Length}\left(\left(L, R,A,B\right)\right):=|R|=J$.
\end{defn}
A special class of execution paths that we refer to as zero-distance monotone paths will be central to our analysis:
\begin{defn}(Zero-distance and monotone path)\label{defn:zerodist}
	Let $(L,R,A,B)$ be an execution path defined in Definition~\ref{def:validpath}. If for $R=(r_j)_{j=1}^{J}$, $r_j$'s are non-increasing in $j$, and $L=R$, then we say that $(L,R,A,B)$ is a \emph{zero-distance and monotone} execution path. When $L=R$, we usually drop $L$, and simply write $(R,A,B)$.
\end{defn}

The following crucial lemma allows our LP based analysis of the query time:
\begin{lem}(Reduction to zero-distance monotone execution paths)\label{lm:monotone-path}
	For every valid execution path $(L, R, A, B)$ (see Definition~\ref{def:validpath}), there exists a \emph{zero-distance and monotone} valid execution path $(R', A', B')$ (see Definition~\ref{defn:zerodist}) such that $b'_{y,J+1}=b_{y,J+1}$ for all $y\in D$\footnote{We need the final condition to argue that we have the same number of points remaining at the end.} and $|R'|=|R|$ (i.e., the length of the paths are equal). 
\end{lem}
The proof of this lemma is given in Section~\ref{sec:exepath}.
\subsubsection{Linear programming formulation}\label{sec:linprog}
As we prove in Lemma~\ref{lm:monotone-path}, for any execution path there exists a zero-distance monotone path (see Definition~\ref{defn:zerodist}) with the same length and the same final densities. This means that if we prove that for any zero-distance monotone path, the final densities are small, then this generalizes to all possible execution paths. So, from now on we only consider zero-distance monotone paths.

As mentioned before, we analyze the evolution of density of points at various distances. Instead of analyzing continuous densities, we define a new notion, called \emph{discretized log-densities} (see Definition~\ref{def:fy}), for which we round densities to the discretized distances in a natural way, and for simplicity of calculations we take the log of these densities. These two steps allow us to analyze the evolution of densities over the course of time. More specifically, we define an LP (see~\eqref{def:lp}) such that any zero-distance monotone execution path with large enough final densities, imposes a feasible solution to the LP, with cost (almost) equal to the length of the execution path divided by $T$. Thus, if the length of the execution path is large, final densities cannot be too large (see Section~\ref{sec:main-tech-lemma} and Claim~\ref{claim:feasible} for the formal statement), which means that we managed to reduce the densities to a small amount. 

In section~\ref{sec:main-tech-lemma} we formally describe the procedure for constructing a feasible solution based on discretized densities. 

We start by defining a convenient discretization of the distances on a valid execution path:

\begin{defn}[$x$-centered grid $Z_x$]\label{def:zx} 
	For every $x\in (0, R_{max})$ define the grid 
	$Z_x=\{z_I,z_{I-1},\ldots,z_0\}$ by letting $z_I=x$, letting $z_{I-i}:=\left(1+\delta_z\right)^i\cdot z_I$ for all  $i\in[I]$ and choosing the smallest integer $I$ such that  $z_0\geq R_{max}\sqrt{2}$.
\end{defn}

\begin{defn}[Discretized log-densities $f_{z_i, j}$]\label{def:fy}
	For any zero-distance monotone valid execution path $(R,A,B)$ (as per Definition~\ref{def:validpath}) with radii bounded by $R_{max}$ and $J=|R|$,  for all $j \in \left[J\right]$ let $k_j$ be the index of the largest grid element which is not bigger than $r_j\cdot (\sqrt{2}+\psi)$, i.e.,
	\begin{align}\label{eq:zkj}
	r_j \cdot (\sqrt{2}+ \psi) \in   [z_{k_j} , z_{k_j-1} )
	\end{align}
	and for every integer $i \in \{k_j,\ldots, I\}$ define
	\begin{align}
	f_{z_i,j} &:= \log_{1/\mu} \left(\sum_{y\in D \cap [z_{i+1},z_{i-1}) } b_{y,j}\right)  \label{eq:f-def}
	\end{align}
	Note that  the variables $b_{y, j}$ on the right hand side of~\eqref{eq:f-def} are the $b_{y, j}$ variables of the execution path $(R, A, B)$.
\end{defn}

Letting $Z:=Z_x$ to simplify notation,  we will consider $I$ linear programs defined below in~\eqref{def:lp}, enumerating over all $j^*\in [I]$, where we let $x'=x+\Delta$:

\begin{center}
	\fbox{
		\begin{minipage}{0.99\textwidth}
			\begin{align}\label{def:lp} 
			\text{LP}(x, j^*): ~~~~~~~\max_{\alpha\ge 0}&~\sum_{j=1}^{j^*-1}  \alpha_j\\
			\forall y\in Z&: g_{y,1}\le \min\left\{ {\frac{ y^2-x^2}{2}}, {1-\frac{x^2}{2}}\right\}&&\text{Density constraints}\nonumber\\
			\text{~for all~} &j<j^*,  y \in Z, y<z_j:\nonumber\\
			&~ g_{y,j}\le {g_{z_j,j}} &&\text{Truncation}\nonumber\\
			& ~~~g_{y,{j+1}} \le g_{y,{j}}-\frac{2\left(z_j/x\right)^2-1}{2\left(z_j/y\right)^2-1}\cdot \alpha_j &&\text{Spherical LSH}\nonumber\\
			& ~~~g_{z_{j^*},j^*}\geq 0&&\text{Non-empty range constraint}\nonumber
			\end{align}
		\end{minipage}
	}
\end{center}

Intuitively, LP~\eqref{def:lp} captures the evolution of the density of points at different distances from the query throughout the hashing process. Our main technical claim connecting the LP~\eqref{def:lp} and execution paths in the query process is Claim~\ref{claim:feasible} in Section~\ref{sec:main-tech-lemma}. \xxx[MK]{We should add an explanation of the `intended solution' as in the tech overview}

\subsection{Upper-bounding the expected number of points examined by the query}
In this section we bound the expected number of points that the query examines in the query procedure. Let $\mathcal{T}$ be the tree that the query traverses. Note that the query only examines the points that it sees in the leaves that it visits. One should note that some leaves (which are LSH nodes for this case) in the tree have level $J$ (see line~\ref{line:Jstop} of Algorithm~\ref{alg:DD-KDE-spherical}). However, they are other leaves in tree $\mathcal{T}$, due to two cases:
\begin{enumerate}
	\item Path termination due to $x''>R(\sqrt{2}+\gamma)$. This case happens when query $\mathbf{q}$ is such that it needs to recover points at distance $x''$ on the sphere, but this distance corresponds to points beyond orthogonal. Note that in the preprocessing phase we did not prepare any child with this $x''$ (see line~\ref{line:xsq2} in Algorithm~\ref{alg:DD-KDE-spherical}), so the query will stop at this node and scan the points (see line~\ref{line:deadendscan1} of Algorithm~\ref{alg:DD-KDE-Query}). 
	Roughly speaking, since we only expect $O(1)$ number of points at distance $x$, and since the number of points on the sphere is dominated by the number of points in the orthogonal band, then we expect to see small number of points on this sphere. We formally prove this in Claim~\ref{claim:imcompsize}.
	\item Path termination due to small sphere radius. This simple case corresponds to the cases when \textsc{Pseudorandomify} does not process a ball further due to line~\ref{line:radius-dmin-ps} of Algorithm~\ref{alg:pseudorandomify} or \textsc{SphericalLSH} does not partition the dataset further due to line~\ref{line:radius-dmin-lsh} of Algorithm~\ref{alg:DD-KDE-spherical}. Note that in that case the entire ball is necessarily at distance at most $x+2R_{min}$, and hence the total number of points in the ball is small. We formally argue and prove this in Claim~\ref{claim:imcompsize}.
\end{enumerate}
\begin{claim}\label{claim:imcompsize}
	For any tree $\mathcal{T}$ that the query $\mathbf{q}$ explores, the expected total number of points in the leaves with level less than $J$ is bounded by 
	$$\left(\frac{1}{\mu}\right)^{\alpha+\alpha^*+c},$$
	for $c=10^{-4}$.
\end{claim}
\begin{proof}
	We investigate the two cases mentioned above separately:
	
	\paragraph{Path termination due to $x''>R(\sqrt{2}+\gamma)$.} First, suppose that the exploration process terminates at node $u\in \mathcal{T}$ because of line~\ref{line:xsq2} in Algorithm~\ref{alg:DD-KDE-spherical} . In that case one has by invoking Claim~\ref{claim:psrs} for two diametral points on the sphere, since the current dataset $u.P$ is pseudorandom as per Definition~\ref{def:psrs} and $\tau= 1/10$,
	\begin{align*}
	\left| \left\{  \mathbf{p}\in u.P :||\mathbf{p}-\mathbf{q}'||\in\left(R(\sqrt{2}-\gamma), R(\sqrt{2}+\gamma)\right)    \right\}\right|=\Omega\left(|u.P|\right).
	\end{align*}
	Note that the expected number of points at distance at most $R(\sqrt{2}+\gamma)$ from the query is upper-bounded by the expected number of points at distance at most $x+\Delta+\delta'$, since $x''>R(\sqrt{2}+\gamma)$ and by rounding of $x'$ to $x''$ (see line~\ref{line:x-tilde} in Algorithm~\ref{alg:DD-KDE-Query}). So, after sub-sampling the data set and using the density constraints, we have at most 
	\begin{align*}
	\frac1{n}\cdot \left(\frac1{\mu}\right)^{1-x^2/2}\cdot 4n\cdot \mu^{1-\frac{(x+\Delta+\delta')^2}{2}}
	&= 4\muu{\frac{\left(x+\Delta+\delta'\right)^2}{2}-\frac{x^2}{2}}\\
	&\le 4\muu{\frac{\left(x+2\Delta\right)^2}{2}-\frac{x^2}{2}}\\
	&\le      4\muu{\frac{1}{2}\cdot (4\Delta x +4\Delta^2) }\\
	&\le 		\muu{   			5\Delta				}				   &&\text{Since $x\le \sqrt{2}$ and $\Delta=10^{-20}$}
	\end{align*}
	points. 
	\paragraph{Path termination due to small sphere radius.} As we discussed above for this case, the entire ball is necessarily at distance at most $x+2R_{min}$, since this sphere passed the condition in line~\ref{line:xclose} of Algorithm~\ref{alg:DD-KDE-Query}, and hence on expectation the total number of points in this ball is bounded by 
	$$
	\frac1{n}\cdot \left(\frac1{\mu}\right)^{1-x^2/2}\cdot n\cdot \mu^{1-(x+2R_{min})^2/2}\le \expm\left(4R_{min}\right)
	$$
	where the last line is by our choice of parameters, and since $x\le \sqrt{2}$.
	
	Also, by Lemma~\ref{lm:numleaves-explored} we know that the query explores at most  $\left(\frac{1}{\mu}\right)^{\alpha^*+\alpha+o(1)}$ leaves. Now, by setting of parameters, the claim holds. 
\end{proof}

\begin{lem}\label{lm:leaf-bound}
	Under Assumption~\ref{assump:1}, there exists an event $\mathcal{E}$ that depends on the choice of the hash function in \textsc{PreProcess} only and occurs with probability at least $1-(1/\mu)^{-4}$ such that conditioned on $\mathcal{E}$, the following holds. The query examines at most 
	\begin{equation}\label{eq:leaf-dataset-bound}
\left(\frac{1}{\mu}\right)^{0.173}
	\end{equation}
	number of points in expectation.
\end{lem}
\begin{proof}
First, we just calculate the expected size of the data set examined by the query in invocations of $\textsc{Query}$ (Algorithm~\ref{alg:DD-KDE-Query}), and then we bound the expected total number of points of $\textsc{Query-KDE}$ (Algorithm~\ref{alg:DD-KDE-Q}).
Note that the goal is to prove an upper-bound on the expected number of points that the query examines.

Consider an invocation \textsc{PreProcess} and let $\mathcal{T}$ be the sub-tree of the recursion tree that the query explores. Now, we define processes on this tree that output a subset of leaves of this tree. Suppose that 
$$\mathcal{H}=W\times \left[\left\lceil\frac{R_{\text{max}}}{\delta}\right\rceil\right] \times \{1,2\}$$
And let $J$ be the maximum number of times that we applied spherical LSH. Let $\mathbf{q}$ be the query. Let $M := |\mathcal{H}^J|$ and enumerate elements in $\mathcal{H}^J$. For any leaf in $\mathcal{T}$ if one looks at the path to the root from this leaf, this corresponds to one element in $\mathcal{H}^J$ (See the discussion in Section~\ref{sec:pathgeometry} and Definition~\ref{def:pathgeom}). For $i$'th element of $\mathcal{H}^J$, $h_i=(h_i(j))_{j=1}^{J}$, the procedure $\mathcal{P}_i(\mathcal{T})$ outputs set $E_i$, which is the set of output(s) of $\textsc{Sample}(\mathcal{T},h_i,0)$.\footnote{Also, for the purpose of consistency define $h_i(0)=(0,0,0)$ and let $h_i \gets (h_i(j))_{j=0}^{J}$ and assume that every Andoni-Indyk LSH bucket is consistent with $h_i(0)$.} Note that Algorithm~\ref{alg:procedureDelta} outputs a set of leaves in the tree. 
\begin{algorithm}[H]
	\caption{ }  
	\label{alg:procedureDelta} 
	\begin{algorithmic}[1]
		\Procedure{$\textsc{Sample}(\mathcal{T},h_i,k)$}{}
		
		\State $v \gets$ a uniformly random child of the root of $\mathcal{T}$ which is consistent with $h_i(k)$.  \label{line:vchoice}
		\If {$k=J$}
		\State Return $v$
		\EndIf
		\For {all $w$ in the set of the childern of $v$}
		\If {$w$ is consistant with $h_i(k)$ }\label{line:ifw}
		\State $\mathcal{T'} \gets $ the sub-tree of tree where the root is $w$.
		\State $\textsc{Sample}(\mathcal{T}',h,k+1)$.
		\EndIf
		\EndFor
		\EndProcedure
	\end{algorithmic}
\end{algorithm}

	Also, for any pseudo-random node on the tree that the query visits, since $\mu=n^{-\Omega(1)}$ by assumption, using a simple Chernoff bound argument, we have that it explores at most $$m:=O(1)\left( \frac{1}{\mu}\right)^{\frac{1}{T}}$$ children of this node, with high probability. 

Let $V$ be the set of leaves in $\mathcal{T}$, with level $J$. Partition $V$ into $V_1,\ldots,V_{M}$, such that for all $i\in[M]$, the leaves in $V_i$ admit the geometry defined by $h_i$. 
\begin{claim}
	For any $u\in U_i$ we have the following
	\begin{align*}
	\Pr[u\in E_i | \mathcal{T}] \ge \left(\frac{1}{m}\right)^J\left(\frac{1}{100\left(\frac{1}{\mu}\right)^\alpha}\right).
	\end{align*}
\end{claim}
\begin{proof}
	There is exactly one path from root to $u$. So, $u \in E_i$  if in all choices in line~\ref{line:vchoice} of Algorithm~\ref{alg:procedureDelta}, the algorithm chooses the correct child. This happens with probability at least $\left(\frac{1}{m}\right)^{J}\left(\frac{1}{100\left(\frac{1}{\mu}\right)^\alpha}\right)$. To be more clear, with probability $\left(\frac{1}{100\left(\frac{1}{\mu}\right)^\alpha}\right)$ the correct child of the root is chosen, and the other term correspond to the success probability in $J$ steps.
\end{proof}

Now, we have the following:
\begin{align}
\sum_{i \in [M]}\mathbb{E}\left[\sum_{v\in E_i}|v.P|~| \mathcal{T}\right]&=\sum_{i \in [M]}\mathbb{E}\left[\sum_{u\in V_i}\mathbb{I}\{ u\in E_i\}|u.P|  ~| \mathcal{T}\right]\nonumber\\
&=\sum_{i \in [M]}\sum_{u\in V_i} \Pr[u \in E_i|\mathcal{T}] \cdot |u.P|\nonumber\\
&\ge \left(\frac{1}{m}\right)^J\sum_{i\in[M]}\sum_{u\in V_i} |u.P|\nonumber\\
&=\left(\frac{1}{m}\right)^J\sum_{u\in V} |u.P| \label{eq:expsum}
\end{align}
where expectations are over the random choices of line~\ref{line:vchoice} of Algorithm~\ref{alg:procedureDelta}.

Let $V'$ be the leaves with level $\ne J$. Note that $\sum_{u\in V}|u.P|+\sum_{u\in V'} |u.P|$ is equal to the number of points that the query examines in the leaves of $\mathcal{T}$. Note that Claim~\ref{claim:imcompsize} proves that 
\begin{align}\label{eq:shortpaths}
\mathbb{E}_{\mathcal{T}}\left[ \sum_{u\in V'} |u.P|\right] \le \left(\frac{1}{\mu}\right)^{\alpha+\alpha^*+0.0001}
\end{align}
Now, we need to take expectation over the tree $\mathcal{T}$. 
	From now on, the goal is to prove an upper-bound on 
	\begin{align*}
	\mathbb{E}_{\mathcal{T}}\left[  \mathbb{E}\left[\sum_{v\in E_i}|v.P| ~| \mathcal{T}\right] \right]
	\end{align*}
	where the outer expectation is over the randomness of trees, and the inner expectation is over the randomness of choices in line~\ref{line:vchoice} of Algorithm~\ref{alg:procedureDelta}.
	
For any $\mathcal{T}$, define $W^{(0,\mathcal{T})}$, as the root of $T$. For all $j\in[J]\cup \{0\}$ let $V^{(j,\mathcal{T})}$ be the nodes in the tree selected by line~\ref{line:vchoice} of Algorithm~\ref{alg:procedureDelta}, when $k=j$. Also, for all $j\in[J]$ let $W^{(j,\mathcal{T})}$ be the children of $V^{(j-1,\mathcal{T})}$ which are consistent with $h_{i}(j)$, .i.e., nodes satisfying the condition in line~\ref{line:ifw} of Algorithm~\ref{alg:procedureDelta} when $k=j$. We drop superscripts for the tree, when it is clear from the context.

For all $j\in[J]\cup\{0\}$, $A_{y,j}$ denote the number of points at distance $y$ for all $y \ge x+1.5\Delta$ from the query in $\cup_{u\in V^{(j)}} u.P$. And similarly,
for all $j\in[J]\cup \{0\}$ define $B_{y,j}$ as the number of points at distance $y$ from the query in $\cup_{u\in W^{(j)}} u.P$. 

Also, let $L=(\ell_j)_{j=1}^{J}$ be the distances induced by the geometry $h_i$. Now, define $x'_j:=\text{Project}(x+\Delta,\ell_j,r_j)$ and $y'_j=\text{Project}(y-\Delta/2,\ell_j,r_j)$.
Now, Claim~\ref{cl:nonadapt} implies that for all $j\in[J]$
\begin{align}\label{eq:AyjByjpyj}
\mathbb{E}\left[ A_{y,j} | B_{y,j}, \mathcal{T}_{<j}\right] \le p_{y,j} \cdot B_{y,j},
\end{align}
where the expectation is over the randomness of the tree and the random choice of line~\ref{line:vchoice} of Algorithm~\ref{alg:procedureDelta}, and 
\begin{align}\label{eq:defpyj}
p_{y,j}:=\muu{-  \frac{4(r_j/x'_j)^2-1}{4(r_j/y'_j)^2-1}\cdot \frac{1}{T}}.
\end{align} On the other hand, since $B_{y,j}$ variables correspond to pseudo-random spheres, using Claim~\ref{claim:roundedpseudorand} they should satisfy the following:
\begin{equation}\label{eq:flattening-b2}
\sum_{y\leq c_j-\psi R_j} B_{y, j}\leq \frac{\tau}{1-2\tau}\cdot\sum_{y\in (c_j-\psi R_j, c_j+\psi R_j)} B_{y, j},
\end{equation}
and 
\begin{align}\label{eq:beyondorthogonalpts2}
\sum_{y\ge c_j+\psi R_j} B_{y, j}\leq \frac{\tau}{1-2\tau}\cdot\sum_{y\in (c_j-\psi R_j, c_j+\psi R_j)} B_{y, j}.
\end{align} 
Also, since $\cup_{u\in V^{(j)}} u.P\subseteq \cup_{u\in W^{(j)}} u.P$, then $B_{y,j}\le A_{y,j-1}$. At this point, define 
Now, for all $j\in[J]$ define 

\begin{align*}
\wt{B}_{y,j}:=\mathbb{E}\left[B_{y,j}\right]
\end{align*}
and 
\begin{align}\label{eq:wtABp}
\wt{A}_{y,j} := \wt{B}_{y,j} \cdot p_{y,j}
\end{align}
and define 
\begin{align}\label{eq:terminalwtAB}
\wt{B}_{y,J+1}:=\wt{A}_{y,J}.
\end{align}
Therefore, if $A:=(\wt{A}_{y,j})_{j=1}^{J}$ and $B:=(\wt{B}_{y,j})_{j=1}^{J+1}$ and $L$ is the ordered set of distances induced by the path geometry $h_i$ (see Section~\ref{sec:pathgeometry} and Definition~\ref{def:pathgeom}) and $R$ is the set of radii of the spheres, then we can argue that $(L,R,A,B)$ is a valid execution path by Definition~\ref{def:validpath}. Checking the conditions of Definition~\ref{def:validpath}:
\begin{itemize}
	\item Initial conditions: They are satisfied by the expectation of sub-sampling (see \eqref{eq:density-ub}), i.e., $$\sum_{y' \in [0,y]\cup D}\wt{A}_{y',0}\le \min\left\{ \muu{\frac{y^2-x^2}{2}},\muu{\frac{1-x^2}{2}}\right\}$$
	and 
	$$\sum_{y' \in [0,y]\cup D}\wt{B}_{y',0}\le \min\left\{ \muu{\frac{y^2-x^2}{2}},\muu{\frac{1-x^2}{2}}\right\}.$$
	\item Truncation conditions: {\bf (2a)} is satisfied since if a point is on the sphere, its distance to the query can be in interval $[x+1.5\Delta,\ell_j+r_j]$ which is a sub-interval of $[\ell_j-r_j,\ell_j+r_j]$, by the definition of induced distances and setting of parameters.  {\bf (2b)} holds, since the number of points in each distance is non-increasing from root to leaf. {\bf (2c)} is satisfied by \eqref{eq:flattening-b2}.
	\item LSH conditions: They are satisfied by \eqref{eq:wtABp} and the definition of $p_{y,j}$ in \eqref{eq:defpyj}.
	\item Terminal density condition: It holds by \eqref{eq:terminalwtAB}.
\end{itemize}
we conclude that $(L,R,A,B)$ is a valid execution path. 

Now by Lemma~\ref{lm:monotone-path} there exists a zero distance monotone execution path $(R',A',B')$ such that $A'=(a'_{y,j})$, $B'=(b'_{y,j})$ and $b'_{y,J+1}=\wt{B}_{y,J+1}$. Let $f_{y,j}$'s be defined based on $b'_{y,j}$'s using Definition~\ref{def:fy}. More specifically, 
for every integer $i \in \{k_j,\ldots, I\}$ (see Definition~\ref{def:fy} for the definition of $k_j$) define
\begin{align}
f_{z_i,j} &:= \log_{1/\mu} \left(\sum_{y\in D \cap [z_{i+1},z_{i-1}) } b'_{y,j}\right)  
\end{align}

 Now, by Claim~\ref{claim:feasible} and our setting of $J$ (see Section~\ref{sec:param}), which ensures that $J>\frac{T}{1-10^{-4}}\text{OPT(LP)}$, for all $y\le z_{j^*-1}$ we have $f_{y,J+1}< 7\delta_z$  for $j^*=k_{J}+1$. Now, we need to prove that this implies that $\sum_{y}\wt{A}_{y, J}$ is small:

\begin{claim}\label{cl:datasetsizef}
	If for all $y\le z_{j^*-1}$ we have $f_{y,J+1}< 7\delta_z$ for $j^*=k_{J}+1$, then we have the following bound 
	\begin{align*}
	 \sum_{y}\wt{A}_{y, J}\le \muu{7\delta_z+o(1)}.
	\end{align*}
\end{claim}
The proof is deferred to Appendix~\ref{app:query}. 

We just proved that for any fixed $i\in[M]$, $\sum_{y} \wt{A}_{y,J}$ (which bounds the expected number of points at distance $\ge x+1.5\Delta$ (see \eqref{eq:dx-def}) that the query examines in buckets with geometry $h_i$) is bounded by $\muu{7\delta_z+o(1)}$. Moreover, recall that in this process we only considered points at distance $\ge x+1.5\Delta$. We should also add the contribution of points at distance $<x+1.5\Delta$. For this, just recall that after sub-sampling (even without considering any LSH effect on these points) in expectation we have
\begin{align}
4\left(\frac{1}{\mu}\right)^{\frac{(x+1.5\Delta)^2-x^2}{2}}\le \left(\frac{1}{\mu}\right)^{10\Delta}.
\end{align}  Now, in order to argue that the expected number of points examined by the query is bounded, we need to multiply by $M$, which results in the following bound
\begin{align}
M\cdot\left( \muu{7\delta_z+o(1)}+\muu{10\Delta}\right)
\end{align}
which by the setting of parameters, combining with \eqref{eq:expsum} and summing with \eqref{eq:shortpaths}, and considering the we call $\textsc{Query}$ (Algorithm~\ref{alg:DD-KDE-Query}) at most $\left(\frac{1}{\mu}\right)^{4\delta_x + o(1)}$, gives the following bound on the expected number of points scanned by the query
\begin{align*}
\left(\frac{1}{\mu}\right)^{0.173}.
\end{align*}

\end{proof}

\subsection{Proof of Lemma~\ref{lm:query}}

Before we present the proof of Lemma~\ref{lm:query}, we need to show another auxiliary claim that helps us establish an upper bound on the expected number of leaves that a query explores, which helps upper bound the work done to reach a leaf (recall that Lemma~\ref{lm:leaf-bound} shows that the expected size of the dataset corresponding to a leaves of $\mathcal{T}$ that the query scans is bounded, so combining these two bounds will give us the final result).  
 \begin{lem}\label{lm:numleaves-explored}
For every $\mathbf{q}\in \R^d$, every $x>0$, every $\mu\in (0, 1)$ under Assumption~\ref{assump:1}, if $\mathcal{T}$ is the tree generated by \textsc{PreProcess}$(P, x, \mu)$, then the expected number of leaves explored by a query $q$ in a call to \textsc{Query}$(\mathbf{q}, x, \mathcal{T})$ (Algorithm~\ref{alg:DD-KDE-Query}) is bounded by $\expm(\alpha^*+\alpha+o(1))$.
 \end{lem}
The proof is given in Appendix~\ref{app:query}. We will also need the following technical claim, which we also prove in Appendix~\ref{app:query}.

\begin{claim}\label{cl:fg}
For every $R\ge R_{min}$, for every $x'\in (\Delta, R(\sqrt{2}+\gamma))$ and sufficiently large $\eta$ we have that
$\frac1{G(x'/R, \eta)}=\left(\frac{F(\eta)}{G(x'/R, \eta)}\right)^{O(1/\Delta^2)}$.
\end{claim}
\begin{proofof}{Lemma~\ref{lm:query}}
By Lemma~\ref{lm:numleaves-explored} the expected number of leaves that the query explores is bounded by 
\begin{equation}\label{eq:923hg2g2fg}
\left(\frac1{\mu}\right)^{\alpha^*+\alpha+o(1)}
\end{equation}
 The expected size of the dataset that the query scans is bounded by $\left(\frac{1}{\mu}\right)^{0.173} $ with high probability by Lemma~\ref{lm:leaf-bound}.  Now by an application of Markov's  inequality to~\eqref{eq:923hg2g2fg} we have that the query explores at most $\left(\frac{1}{\mu}\right)^{0.173+o(1)}$ leaves with high probability, and hence the total work is bounded by $(\frac1{\mu})^{0.173}\cdot n^{o(1)}$, as required. 
Finally, we bound  the work done in line~\ref{line:all-gaussians} of Algorithm~\ref{alg:DD-KDE-spherical}. Indeed, recall that $x'<R(\sqrt{2}+\gamma)$ by line~\ref{line:xsq2} of Algorithm~\ref{alg:DD-KDE-spherical}, and at the same time by Claim~\ref{cl:fg} we have 
$$
\frac1{G(x'/R, \eta)}=\left(\frac{F(\eta)}{G(x'/R, \eta)}\right)^{O(1/\Delta^2)}.
$$
Equipped with this observation, we can now finish the proof.  We get using the choice of $T$ in line~\ref{line:1T} of Algorithm~\ref{alg:DD-KDE-spherical}
$$
\frac{100}{G(x'/R, \eta)}=100\cdot \left(\frac{F(\eta)}{G(x'/R, \eta)}\right)^{O(1/\Delta^2)}=100\cdot (1/\mu)^{O(1/(\Delta^2 \cdot T))}=n^{o(1)}
$$
by choice of $\Delta=\Omega(1)$ and $T=\sqrt{\log n}$ in line~\ref{line:t-def} of Algorithm~\ref{alg:DD-KDE-spherical}. This completes the proof.
\end{proofof}

\section{Reduction to zero-distance monotone execution paths}\label{sec:exepath}

In this section, we prove Lemma~\ref{lm:monotone-path}, which proves that for any valid execution path, there exists a zero-distance valid execution path such that the final densities are identical and both have the same length. First, we state the following claims, and then assuming these claims, we prove Lemma~\ref{lm:monotone-path}. Then, we present the proof of these claims. 
\begin{claim}[Reduction to zero distance paths]\label{lm:zd-reduction}
	For any $L$, $R$, $A$ and $B$ such that $(L,R,A,B)$ is a valid execution path (see Definition~\ref{def:validpath}), there exists $R'$ and $A'$ such that $(R',R', A', B)$ is a valid execution path for some $A'$. 
\end{claim}

\begin{claim}[Local improvement towards monotonicity]\label{claim:oneswap}	
	For every valid zero-distance execution path $(R, A, B)$, if for some $i\in [J-1]$ one has $r_i\le r_{i+1}$, then for $R':=(r_1,\ldots, r_{i-1}, r_{i+1},r_{i+1},\ldots,r_{J})$, there exist $A', B'$ such that the path $(R',A',B')$ is a valid execution path and $b'_{y,J+1}=b_{y,J+1}$ for all $y\in D$ (see \eqref{eq:dx-def} for the definition of $D$).
\end{claim}
Now, assuming the correctness of Claim~\ref{lm:zd-reduction} and Claim~\ref{claim:oneswap} we present the proof of Lemma~\ref{lm:monotone-path}.

\begin{proofof}{Lemma~\ref{lm:monotone-path}}
	First, using Claim~\ref{lm:zd-reduction}, we find a zero-distance valid execution path $(L'',R'',A'',B)$.  Now, we repeat the procedure described in Claim~\ref{claim:oneswap} on $(L'',R'',A'',B)$, until it becomes a zero-distance monotone execution path, $(R',A',B')$, which satisfies the conditions of the lemma. 
\end{proofof}

Now we present the proof of Claim~\ref{lm:zd-reduction} and Claim~\ref{claim:oneswap}.

\begin{proofof}{Claim~\ref{lm:zd-reduction}}
	Let $(\ell_j)_{j=1}^{J}=L$ and $(r_j)_{j=1}^{J}=R$. Then $\forall j \in [J]$ we define: \footnote{We define $\ell'_j$'s for the convenience of the reader, otherwise it is clear that $\ell'_j=r'_j$.}
	\begin{align}
	r'_j&:=\sqrt{\frac{\ell_j^2+r_j^2}{2}}\label{eq:rll}\\
	\ell'_j&:=\sqrt{\frac{\ell_j^2+r_j^2}{2}}
	\end{align}
	and we let $R':=(\ell'_j)_{j=1}^{J}=(r'_j)_{j=1}^{J}$. The same as Definition~\ref{def:validpath} for all $j\in [J]$, we define
	\begin{align*}
	c'_j:=\sqrt{(\ell'_j)^2+(r'_j)^2}=\sqrt{2}\cdot r'_j
	\end{align*}
	which translates to $c'_j=c_j$. 
	First, we need to show that 
	\begin{align*}
	[0,\ell_j+r_j] \subseteq [0,\ell'_j+r'_j].
	\end{align*}
	Note that 
	\begin{align*}
	(\ell_j'+r_j')^2-(\ell_j+r_j)^2 = 2c_j^2 - c_j^2-2\ell_jr_j \ge 0
	\end{align*}
	where the last inequality is due to $c_j=\sqrt{r_j^2+\ell_j^2}$. 
	One can see that since we can set $a'_{y,0}=a_{y,0}$ for all $y\in D$, it suffices to show that for all $j\in[J]$, $x\in [|\ell_j-r_j|-\delta,\ell_j+r_j]$ (see line~\ref{line:xclose} of Algorithm~\ref{alg:DD-KDE-Query} and Definition~\ref{def:validpath}) and $y\in[\ell_j-r_j,\ell_j+r_j]$ such that $y-\Delta/2\ge x+\Delta$:
	\begin{align}\label{eq:zero}
	\frac{4\left(\frac{r_j}{\textsc{Proj}(x+\Delta,\ell_j,r_j)}\right)^2-1}{4\left(\frac{r_j}{\textsc{Proj}(y-\Delta/2,\ell_j,r_j)}\right)^2-1}\ge \frac{4(r'_j/(x+\Delta))^2-1}{4(r'_j/(y-\Delta/2))^2-1}
	\end{align}
	We drop the indices $j$ for ease of notation, and let $\alpha:=x+\Delta$ and $\beta:=y-\Delta/2$.
	Note that using the formula for $\textsc{Project}$ (see Lemma~\ref{lem:projection}) have
	\begin{align*}
	&\frac{4\left(\frac{r}{\textsc{Project}(\alpha,\ell,r)}\right)^2-1}{4\left(\frac{r}{\textsc{Project}(\beta,\ell,r)}\right)^2-1}\cdot  \frac{4(r'/\beta)^2-1}{4(r'/\alpha)^2-1}\\
	&=\frac{4\left(\frac{r^2}{\frac{r}{\ell}\left(\alpha^2-(\ell-r)^2\right)}\right)-1}{4\left(\frac{r^2}{\frac{r}{\ell}\left(\beta^2-(\ell-r)^2\right)}\right)-1}\cdot \frac{\frac{4r'^2-\beta^2}{\beta^2}}{\frac{4r'^2-\alpha^2}{\alpha^2}}\\
	&=\frac{(\ell+r)^2-\alpha^2}{\alpha^2-(\ell-r)^2}\cdot \frac{\beta^2-(\ell-r)^2}{(\ell+r)^2-\beta^2}\cdot \frac{\alpha^2}{\beta^2}\cdot \frac{4r'^2-\beta^2}{4r'^2-\alpha^2}\\
	\end{align*}
	where in the second transition above we used the fact that 
	\begin{equation*}
	4\frac{r^2}{\frac{r}{\ell}\left(\alpha^2-(\ell-r)^2\right)}-1=\frac{4r^2 \ell^2-\left(\alpha^2-(\ell-r)^2\right)}{\alpha^2-(\ell-r)^2}=\frac{(\ell+r)^2-\alpha^2}{\alpha^2-(\ell-r)^2}
	\end{equation*}
	and 
	\begin{equation*}	
	4\frac{r^2}{\frac{r}{\ell}\left(\beta^2-(\ell-r)^2\right)}-1=\frac{4r \ell-(\beta^2-(\ell-r)^2)}{\beta^2-(\ell-r)^2}=\frac{(\ell+r)^2-\beta^2}{\beta^2-(\ell-r)^2}.
	\end{equation*}	
	
	Now, by re-ordering the factors, and the fact that $4r'^2=2(\ell^2+r^2)$ by \eqref{eq:rll}
	\begin{align*}
	&\frac{(\ell+r)^2-\alpha^2}{\alpha^2-(\ell-r)^2}\cdot \frac{\beta^2-(\ell-r)^2}{(\ell+r)^2-\beta^2}\cdot \frac{\alpha^2}{\beta^2}\cdot \frac{4r'^2-\beta^2}{4r'^2-\alpha^2}\\
	&=\left(\frac{2(\ell^2+r^2)-\beta^2}{(r+\ell)^2-\beta^2}\cdot \frac{(r+\ell)^2-\alpha^2}{2(\ell^2+r^2)-\alpha^2}\right)\cdot\left( \frac{\alpha^2}{\alpha^2-(r-\ell)^2}\cdot \frac{\beta^2-(r-\ell)^2}{\beta^2}\right) 
	\end{align*}
	We bound the two terms above separately. For the first term we have
	\begin{align*}
	\frac{2(\ell^2+r^2)-\beta^2}{(r+\ell)^2-\beta^2}\cdot \frac{(r+\ell)^2-\alpha^2}{2(\ell^2+r^2)-\alpha^2} = 	\frac{2(\ell^2+r^2)-\beta^2}{\left(2(r^2+\ell^2)-\beta^2\right)-(\ell-r)^2}\cdot \frac{\left(2(r^2+\ell^2)-\alpha^2\right)-(\ell-r)^2}{2(\ell^2+r^2)-\alpha^2}\ge 1
	\end{align*}
where the inequality follow since for any $0<d<a\le b$ one has $\frac{a}{a-d}\frac{b-d}{b}\ge 1$. Set $a=2(r^2+\ell^2)-\beta^2$, $b=2(r^2+\ell^2)-\alpha^2$ and $d=(\ell-r)^2$. Note that, $a\le b$ since $x+\Delta \le y-\Delta/2$, and $d<a$ since $y-\Delta/2 < \ell_j +r_j$.
	
	Now, we bound the second term
	\begin{align*}
	\left( \frac{\alpha^2}{\alpha^2-(r-\ell)^2}\cdot \frac{\beta^2-(r-\ell)^2}{\beta^2}\right) \ge 1
	\end{align*}
	again by the same argument as above, by setting $d=(r-\ell)^2$, $a=\alpha^2$ and $b=\beta^2$. Again, $a\le b$ since $x+\Delta\le y-\Delta/2$, and $d<a$ since $ x+\Delta > |r-\ell|$ (since $\Delta> \delta$ by the setting of parameters).
	
	Now, combining these two facts \eqref{eq:zero} holds. 
\end{proofof}
\begin{rem}
	One should note that in some cases, the radius of a sphere may decrease when converting it to a zero distance sphere and it means that the size of the band corresponding to orthogonal bands may decrease and this may cause the sphere not being pseudo-random anymore. However, one should note that in our algorithm the radius of the sphere is always $\Theta(1)$, meaning that the radius may change by a constant multiplicative factor, so one can re-scale the size of the orthogonal band in the definition (Definition~\ref{def:validpath}) to cover the previously covered distances. 
\end{rem}

\begin{proofof}{Claim~\ref{claim:oneswap}}
	First note that for $r'\ge r$, we have
	\begin{align}\label{eq:monotonecj}
	\frac{4\left(\frac{r}{x+\Delta}\right)^2-1}{4\left(\frac{r}{y-\Delta/2}\right)^2-1}\ge\frac{4\left(\frac{r'}{x+\Delta}\right)^2-1}{4\left(\frac{r'}{y-\Delta/2}\right)^2-1},
	\end{align}
	since $f(c)=	\frac{4\left(\frac{r}{x+\Delta}\right)^2-1}{4\left(\frac{r}{y-\Delta/2}\right)^2-1}$ is a decreasing function in $r$, assuming $y-\Delta/2> x+\Delta$. 
	For the rest of the proof, let $x':=x+\Delta$ and $y':=y-\Delta/2$.
	\paragraph{Defining $A'$ and $B'$.} We now construct the sequence of intermediate densities $A'$ that satisfies the conditions in Definition~\ref{def:validpath} by modifying the original sequence $A$ on position $j$ (the position where non-monotonicity occurs in the original sequence). 
	Let $a'_{y,i}:=a_{y,i}$ and $b'_{y,i}:=b_{y,i}$ for all $y\in D$  and $i \in \left([J]\cup\{0\}\right)\setminus \{j\}$. Also, let $b'_{y,J+1}:=b_{y,J+1}$ for all $y\in D$. Now, let 
	\begin{equation}\label{eq:ayj-def}
	\forall y\in D:~a'_{y,j} := b_{y,j+1}
	\end{equation}
	and also set 
	\begin{align}\label{eq:h3j3k4h}
	b'_{y,j}:=a'_{y,j}\cdot \muu{\frac{4(r_{j+1}/x')^2-1}{4(r_{j+1}/y')^2-1}\cdot\frac{1}{T}}=b_{y,j+1}\cdot \muu{\frac{4(r_{j+1}/x')^2-1}{4(r_{j+1}/y')^2-1}\cdot\frac{1}{T}}
	\end{align} 
	since $r'_j=r_{j+1}$. 
	
	We now prove that our choice of $A'$ and $B'$ above satisfies the conditions of Definition~\ref{def:validpath}, i.e. yields a valid execution path. Initial density condition (condition {\bf (1)}) and the terminal density condition (condition {\bf (4)}) are satisfied since they were satisfied by the original execution path $(R, A, B)$, and we did not modify the path on the first and last coordinates. The LSH condition (condition {\bf (3)}) is also satisfied by \eqref{eq:h3j3k4h} and tha fact that the original execution path satisfied it. We now verify condition {\bf (2)}. Condition {\bf (2a)} follows since $r_{j+1}> r_j$. 
	
	\paragraph{Verifying condition {\bf (2b)}.}	One has, using the assumption that $(L,R,A,B)$ is a valid execution path,
	\begin{align*}
	b'_{y,j} &=  b_{y,j+1}\cdot \muu{\frac{4(r_{j+1}/x')^2-1}{4(r_{j+1}/y')^2-1}\cdot\frac{1}{T}}&&\text{(by \eqref{eq:h3j3k4h})}\\
	&\le a_{y,j}\cdot \muu{\frac{4(r_{j+1}/x')^2-1}{4(r_{j+1}/y')^2-1}\cdot\frac{1}{T}}&& \text{(property {\bf (2b)} for $(R,A,B)$)}\\
	&\le b_{y,j} \muu{-\frac{4(r_{j}/x')^2-1}{4(r_{j}/y')^2-1}\cdot\frac{1}{T}}\cdot \muu{\frac{4(r_{j+1}/x')^2-1}{4(r_{j+1}/y')^2-1}\cdot\frac{1}{T}}&&\text{(property {\bf (3)} for $(R,A,B)$)}\\
	&\le   b_{y,j} &&\text{(by~\eqref{eq:monotonecj} together with $r_j<r_{j+1}$)} \\
	&\le a_{y,j-1} &&\text{(property {\bf (2b)} for $(R,A,B)$)}
	\end{align*}
	
	\paragraph{Verifying condition {\bf (2c)}.}	
	We need to prove
	\begin{align}\label{eq:bpyj1trunc}
	\sum_{y\in [0,r_{j+1}(\sqrt{2}-\psi)]\cap D}~b'_{y,j}\le \frac{\tau}{1-2\tau}\cdot \sum_{y\in(r_{j+1}(\sqrt{2}-\psi),r_{j+1}(\sqrt{2}+\psi))\cap D}b'_{y,j}
	\end{align}
	Note that by property {\bf (2c)} we have 
	\begin{align}\label{eq:byj1trunc}
	\sum_{y\in [0,r_{j+1}(\sqrt{2}-\psi)]\cap D}~b_{y,j+1}\le \frac{\tau}{1-2\tau}\cdot \sum_{y\in(r_{j+1}(\sqrt{2}-\psi),r_{j+1}(\sqrt{2}+\psi))\cap D}b_{y,j+1}
	\end{align}
	Also, recall that by \eqref{eq:h3j3k4h} we have
	\begin{align*}
	b'_{y,j}=b_{y,j+1}\cdot \muu{\frac{4(r_{j+1}/x')^2-1}{4(r_{j+1}/y')^2-1}\cdot\frac{1}{T}}
	\end{align*}
	Now, combing the fact that $\muu{\frac{4(r_{j+1}/x')^2-1}{4(r_{j+1}/y')^2-1}\cdot\frac{1}{T}}$ is increasing in $y'$ with \eqref{eq:byj1trunc}, proves \eqref{eq:bpyj1trunc}.
	
	We have thus shown that $(R', A', B')$ is a valid execution path. Note that $b'_{y,J+1}=b_{y,J+1}$ for all $y\in D$ by definition of $b'$, as required.
\end{proofof}

\section{Feasible LP solutions based on valid execution paths}\label{sec:main-tech-lemma}
First, we state the main result of this section informally below. We refer the reader to Claim~\ref{claim:feasible} for the formal version of this claim. 
\begin{claim}(Informal)
	If the length of a valid execution path is large enough, then the terminal densities must be small. 
\end{claim}
We prove this claim, by arguing that if the terminal densities are not small then there exists a feasible solution to the LP. However, the feasible solution that we construct, has a cost larger than the optimal solution of the LP, which is a contradiction. This implies that we cannot have large terminal densities.

We use Definition~\ref{def:zx}, Definition~\ref{def:fy} and the corresponding notations in the rest of this section. At this point, one should recall that the definition of valid execution paths (Definition~\ref{def:validpath}) is over the continuous densities. Now, we need to present a similar notion for \emph{discretized log-densities}.
\begin{claim}[Discretized execution path]\label{claim:quantizedc}
	 If the $f_{z_i, j}$ variables are defined as per~\eqref{eq:f-def} (based on a zero distance monotone execution path $(R, A, B)$, with $J=|R|$) then
	
	\begin{description}
		\item[(1) Initial densities:] For any integer $i$:
		$		f_{z_i,1}\le \min\left\{ \frac{\cdot z_{i}^2-x^2}{2}+3\delta_z, 1-\frac{x^2}{2}\right\}$.
		\item[(2) Truncation:]  for any $j\in[J]$ and $i\in \{k_j+1,\ldots,I\}$  one has $f_{z_i,j} \le   f_{z_{k_j},j} + \log_{1/\mu}\frac{2-2\tau}{1-2\tau}$.
		
		\item[(3) Locality Sensitive Hashing:] for any $j\in[J]$ and any integer $i\in \{k_j,\ldots,I\}$ one has $f_{z_i,j+1}\le f_{z_i,j}  -\frac{2(z_{k_j}/x)^2-1}{2(z_{k_j}/z_i)^2-1}\cdot \frac{1}{T}\cdot(1-10^{-4})$.

	\end{description}
\end{claim}

\begin{proof}
	For the purposes of the proof it is convenient to introduce an auxiliary definition. For every $j \in \left[J\right]$ 
	and every integer $i \in \{k_j,\ldots, I\}$ define
	
	\begin{align}
	\wt{a}_{z_i,j} &:= \sum_{y\in D_x \cap [z_{i+1},z_{i-1}) } a_{y,j}. \label{eq:tildea}
	\end{align} 
	and 
	\begin{align}
	\wt{b}_{z_i,j} &:= \sum_{y\in D_x \cap [z_{i+1},z_{i-1}) } b_{y,j}.  \label{eq:tildeb}
	\end{align}
	
	Note that with these definitions in place ~\eqref{eq:f-def} is equivalent to
	
	\begin{align}
	f_{z_i,j} &:= \log_{1/\mu} \wt{b}_{z_i,j}.
	\end{align}
	
	We also let $D:=D_x$, omitting the dependence on $x$, to simplify notation. We now prove the properties one by one.
	
	{\bf (1) Initial densities condition:}
	First, note that by the initial densities condition for the execution path $(R, A, B)$ together with the truncation conditions (Definition~\ref{def:validpath}) one has
	\begin{align*}
	\sum_{y'\in[0,y]\cap D}b_{y',1}\le \min\left\{ \muu{\frac{y^2-x^2}{2}},\muu{\frac{1-x^2}{2}}\right\}.
	\end{align*}
	Combining this with~\eqref{eq:tildeb}, we get
	\begin{align*}
	\wt{b}_{z_i,1}&= \sum_{y\in D\cap (z_{i+1},z_{i-1})}b_{y,1} \le  \sum_{y\in[0,z_{i-1}]\cap D}b_{y,1}\\
	&\le \min\left\{ \muu{\frac{z_{i-1}^2-x^2}{2}},\muu{\frac{1-x^2}{2}}\right\}\\
	&= \min\left\{\muu{\frac{(1+\delta_z)^2\cdot z_{i}^2-x^2}{2}}, \muu{1-\frac{x^2}{2}}\right\}\\
	&\le  \min\left\{\muu{\frac{\cdot z_{i}^2-x^2}{2}+3\delta_z}, \muu{1-\frac{x^2}{2}}\right\},
	\end{align*}
	where we used the definition of the grid $Z$, the fact that $\mu=o(1)$ and that for $z_i\ge \sqrt{2}$ the second term is the minimum term.
	
	{\bf (2) Truncation conditions (effect of \textsc{PseudoRandomify}):}
	We have, using~\eqref{eq:tildeb},
	\begin{align}
	\sum_{i=k_j+1}^{I}\wt{b}_{z_i,j}&\le 2\sum_{y\in D\cap (0,z_{k_j+1})}b_{y,j}+\sum_{y\in D\cap (z_{k_j+1},z_{k_j})}b_{y,j}\nonumber\\
	&\le 2 \sum_{y\in D\cap (0,z_{k_j})}b_{y,j}\label{eq:923hg92h3g}\\
	&\le 2 \sum_{y\in D\cap (0,r_j(\sqrt{2}+\psi))}b_{y,j}.\nonumber
	\end{align}
	The last transition uses the fact that by definition of $k_j$ (see~\eqref{eq:zkj}) we have 	$r_j \cdot (\sqrt{2}+ \psi) \in   [z_{k_j} , z_{k_j-1} )$, and in particular, $r_j \cdot (\sqrt{2}+ \psi) \geq   z_{k_j}$.
	
	We now note that since $10\psi\leq \delta_z = 10^{-6}$ by assumption of the claim and $z_{k_j+1}=(1+\delta_z)^{-1}z_{k_j}$, we further have 
	\begin{align*}
	(r_j(\sqrt{2}-\psi),r_j(\sqrt{2}+\psi)) \subset [z_{k_j+1},z_{k_j-1})
	\end{align*}
	which implies
	\begin{align}\label{eq:92h39h23gs}
	\sum_{y\in D\cap (r_j(\sqrt{2}-\psi),r_j(\sqrt{2}+\psi))}b_{y,j} \le \wt{b}_{z_{k_j},j}.
	\end{align}
	At the same time, since $(R,A,B)$ was a valid execution path, then by property {\bf (2c)} in Definition~\ref{def:validpath}, we have
	\begin{align}
	\sum_{y\in D\cap (0,r_j(\sqrt{2}+\psi))}b_{y,j} &\le \left(1+\frac{\tau}{1-2\tau}\right) \sum_{y\in D\cap (r_j(\sqrt{2}-\psi),r_j(\sqrt{2}+\psi))}b_{y,j}\nonumber\\
	&=\frac{1-\tau}{1-2\tau} \sum_{y\in D\cap (r_j(\sqrt{2}-\psi),r_j(\sqrt{2}+\psi))}b_{y,j}.\nonumber
	\end{align}
	
	Substituting the bound above into~\eqref{eq:923hg92h3g} and using~\eqref{eq:92h39h23gs} yields
	\begin{align*}
	\sum_{i=k_j+1}^{I}\wt{b}_{z_i,j} \le \frac{2-2\tau}{1-2\tau} \cdot  \wt{b}_{z_{k_j},j},
	\end{align*}
	establishing the claim.

	{\bf (3) LSH conditions:}
	For all $j\in[J]$ let $c_j:=\sqrt{2}r_j$. One can think of $c_j$ as the distance from a query on the surface of the the $j$-th sphere in the execution path to a `typical' point on the sphere.  Note that ~\eqref{eq:zkj}  defines a rounding of $c_j$'s points on the grid $D$. Specifically, $c_j$ is rounded to $z_{k_j}$'s.

	\begin{claim}\label{claim:approx}
		Let $x':=x+\Delta$ and let $y\in (z_{i+1},z_{i-1}]$, if  $y'=y-\Delta/2$, and $i \in \{k_j,\ldots, I\}$ then we have the following claim. 
		\begin{align*}
		-\frac{4\left(\frac{r_j}{x'}\right)^2-1}{4\left(\frac{r_j}{y'}\right)^2-1}& \le -\frac{2\left(\frac{z_{k_j}}{x}\right)^2-1}{2\left(\frac{z_{k_j}}{z_i}\right)^2-1} (1-10^{-4})
		\end{align*}
	\end{claim}
	We prove this claim in Appendix~\ref{app:Grid}.
	
	By property {\bf (3)} in Definition~\ref{def:validpath}, one has
	\begin{align}\label{eq:tildeLSH}
	a_{y,j} \le {b_{y,j}}\cdot{\muu{-\frac{2(c_j/x')^2-1}{2(c_j/y')^2-1}\cdot \frac{1}{T}}}.
	\end{align}
	Thus, for all $i\in \{k_j,\ldots,I\}$ we have 
	\begin{align*}
	\wt{b}_{z_i,j+1}&=\sum_{y\in D\cap (z_{i+1},z_{i-1})} b_{y,j+1} &&\text{By \eqref{eq:tildeb}}\\
	&\le \sum_{y\in D\cap (z_{i+1},z_{i-1})} a_{y,j}&&\text{Property {\bf (2b)} for $(R,A,B)$}\\ 
	&\le \sum_{y\in D\cap (z_{i+1},z_{i-1})} b_{y,j} \cdot \muu{-\frac{2(c_j/x')^2-1}{2(c_j/y')^2-1}\cdot \frac{1}{T}}&&\text{By \eqref{eq:tildeLSH}}\\
	&\le \wt{b}_{z_i,j} \cdot \muu{-\frac{2(z_{k_j}/x)^2-1}{2(z_{k_j}/z_i)^2-1}\cdot \frac{1}{T}\cdot(1-10^{-4})}&&\text{By \eqref{eq:tildeb} and Claim~\ref{claim:approx}}
	\end{align*}
	This completes the proof of {\bf (3)}.
\end{proof}
\subsection{Construction of a feasible solution}\label{sec:construction}
In this section, we construct a feasible solution to the LP, i.e., $g_{y,j}$'s and $\alpha_j$'s, based on the execution path that we are considering. Later, we show the relation between the cost of this solution and the length of the execution path.

First, letting $J=|R|$, recall that $R=(r_j)_{j=1}^{J}$. Then, for all $s\in [J]$ define $c_s := \sqrt{2}\cdot r_s$ and let $c_0=+\infty$ for convenience. Let $\wt{T}$ be such that 
\begin{align}\frac{1}{T}\cdot (1-10^{-4})=\frac{1}{\wt{T}}\label{eq:TwtT}.\end{align} Let $x'=x+\Delta$. We classify steps $s=1,\ldots, J$ into three types:
\begin{itemize}
	\item We say that a step $s$ is {\bf stationary} if $c_s=c_{s-1}$ (this corresponds to the algorithm performing multiple rounds of hashing on the same sphere).
	\item Otherwise we call step $s$ {\bf minor}, if $\frac{c_s}{c_{s-1}}\ge 1- \frac{1}{\sqrt{T}}$,
	\item and call step $s$ {\bf major}, otherwise.
\end{itemize}  
 Let $R=R_{stat}\cup R_m \cup R_M$ denote the partition of $R$ into stationary, minor and major steps. Let $j_1,\ldots, j_{|R_m\cup R_M|}$ be such that  $z_{j_1}>z_{j_2}>\ldots>z_{j_{|R_m\cup R_M|}}$ are exactly the $c_s$ values corresponding to non-stationary steps, in decreasing order.

Note that by Lemma~\ref{lm:diam} and parameter settings in the algorithm, $c_1\leq R_{max}\sqrt{2}=O(1)$. Since the grid $D$ (see \eqref{eq:dx-def}) contains only elements at least as large as $x+1.5\Delta$, and if we let $x$ to be lower bounded by an absolute constant we have  $|R_M|=O(\sqrt{T})$. The reason is that by the definition above, for any major step $s$, we have $$\frac{c_s}{c_{s-1}}< 1- \frac{1}{\sqrt{T}}.$$

We define the feasible solution $g_{y, j}$'s and $\alpha_j$'s to LP($x, j^*$) as defined in \eqref{def:lp} without the {\bf non-empty range constraint}. 
We construct feasible $g_{y, j}$ and $\alpha_j$ by induction on $j=1,\ldots, j^*$. It will be important that the constructed solutions for $g_{y,j}$'s are non-decreasing in $y$ for $y\in [0, z_j]$ for every $j=1,\ldots, j^*$. 

For the rest of the section, whenever we are working with discrete functions and it is clear from the context, we drop the condition $y\in Z_x$.  

On the other hand, it is more convenient to work with the following formulation of the LP constraints, since we construct the solution in an inductive way.

\begin{center}
	\fbox{
		\begin{minipage}{0.99\textwidth}
			\begin{align*}
			\forall y&: g_{y,1}\le \min\left\{ {\frac{ y^2-x^2}{2}}, {1-\frac{x^2}{2}}\right\}\\
			  \text{~for all~}& j<j^*, y<z_j:\\
			g_{y,{j+1}} &\le \min\left\{ g_{y,{j}}-\frac{2\left(z_j/x\right)^2-1}{2\left(z_j/y\right)^2-1}\cdot \alpha_j,~g_{z_{j+1},{j}}-\frac{2\left(z_j/x\right)^2-1}{2(z_j/z_{j+1})^2-1}\cdot \alpha_j\right\}
			\end{align*}
		\end{minipage}
	}
\end{center}
\noindent{\bf Base:} For all $j\in [j_1]$ such that $y\le z_j$, we set
\begin{align}\label{eq:densayj12}
g_{y,j} := \min\left\{ {\frac{y^2-x^2}{2}}, {1-\frac{x^2}{2}}\right\}.
\end{align}

We let $\alpha_j:= 0$ for all $j\in[j_1-1]$ so that {\bf spherical LSH} constraints of the LP in \eqref{def:lp} are satisfied for $j\in[j_1-1]$. That is, we don't have any progress using spherical LSH, since $\alpha_j=0$ for all $j\in [j_1-1]$. The {\bf truncation} constraints of the LP in~\eqref{def:lp} are satisfied since the rhs of \eqref{eq:densayj12} is non-decreasing in $y$.

\noindent{\bf Inductive step $j_i\rightarrow j_i+1,\ldots, j_{i+1}$:} We let $a:=j_i$ and $b:=j_{i+1}$ to simplify notation. Let $s$ be the first step on sphere $z_a$, i.e., $c_s=z_a$ and $c_{s-1}\ne z_a$. Also, let $N$ be the number of steps that we stay on sphere $z_a$, i.e.,
\begin{align*}
c_s=c_{s+1}=\ldots=c_{s+N-1}=z_a \text{~and~} c_{s+N}\ne z_a.
\end{align*}
Note that steps $s+1,\ldots, s+N-1$ are stationary as per our definitions. 

It is convenient to define a sequence of auxiliary variables in order to handle the sequence of $N-1$ stationary steps (note that $N-1$ could be zero). 
\begin{center}
    \fbox{
      \begin{minipage}{0.99\textwidth}
Upper bounds $h_y^{(q)}, q=0,\ldots, N,$ on density after (possible) stationary steps:
\begin{align}
&\text{For all~} y\le z_a:\text{~~~~~~~~~~~~~~~~~~~~~~~~~~~~~~~~~~~~~~~~~~(Starting density)}\nonumber& \\
&~~~~~~~~~~~~~~~~~~~~~~~~~~~~~~~~~h_{y}^{(0)}:=g_{y, a}\label{eq:h-init-cond}\\
&\text{For all~}q\in[N-1], y\le z_{a}:\text{~~~~~~~~~~~~~~~~~~~~~~~~(Stationary steps)}&\nonumber\\
&~~~~~~~~~~~~~~~~~~~~~~~~~~~~~~~~~h_{y}^{(q)}:=\min\left\{    h_{y}^{(q-1)}-\frac{2(z_a/x)^2-1}{2(z_a/y)^2-1}\frac{1}{\wt{T}},~h_{z_a}^{(q-1)}-\frac{2(z_a/x)^2-1}{\wt{T}}\right\}\label{eq:bLSHtr}\\
&\nonumber\\
&\text{For~} y\le z_a:\text{~~~~~~~~~~~~~~~~~~~~~~~~~~~~~~~~~~~~~~~~~~~(Final density)}&\nonumber\\
&~~~~~~~~~~~~~~~~~~~~~~~~~~~~~~~~~~~~h_y^{(N)}:=h_{y}^{(N-1)}-\frac{2(z_a/x)^2-1}{2(z_a/y)^2-1}\frac{1}{\wt{T}}\label{eq:bN}.
\end{align}
\end{minipage}
}
\end{center}

Equipped with the definitions of $h$ above, we now define $g$ to satisfy the inductive step. First let $\alpha_a:=\frac{N}{\wt{T}}$ and  let $\alpha_{a+s}:=0$ for $s=1,\ldots, N-1$. Then define for all $y\leq z_b$:
\begin{align*}
g_{y, a+1}&:=\min\left\{h_y^{(N)},~h_{z_{a+1}}^{(N)}\right\}
\end{align*}

and for $ j = a+2,\ldots, b$ and all $y\leq z_j$  let 
\begin{align*}
g_{y, j}&:=\min\{g_{y, j-1},g_{z_{j},j-1}\}.
\end{align*} 
We note that this implies for all $y \le z_b$
\begin{align}
g_{y,b}:=\min\left\{ \min_{j\in \{a+1,\ldots, b\}}\left\{    h_{z_j}^{(N)} \right\}, h_y^{(N)} \right\}\label{eq:mina}.
\end{align}
This finished the inductive step. 

Note that in the last step, i.e., when $z_a=z_{j^*-1}$, since we do not have truncation condition for $g_{y,j^*}$'s, we define  
\begin{align}\label{eq:terminalg}
\forall y\le {z_{j^*-1}}:~g_{y,j^*}=h_y^{(N)}.
\end{align}

\subsection{Monotonicity claims}
\begin{claim}[Unique maximum after LSH]\label{claim:monotone}
	For every integer $t\ge1$, $x\in(0,\sqrt{2})$ and any sequence $c_1\ge c_2\ge \ldots\ge c_t\ge x$, such that 
	\begin{equation*}
	f(y)=\frac{y^2-x^2}{2}-\sum_{s=1}^{t} \frac{2(c_s/x)^2-1}{2(c_s/y)^2-1}\cdot \frac{1}{T}
	\end{equation*}
	satisfies $f(\sqrt{2}c_t)>0$, the following conditions hold. There exists $y^*\in (x, \sqrt{2}c_t]$ such that the function 
	satisfies $f(y^*)=0$ is monotone increasing on the interval $[y^*, \eta]$, where $\eta$ is where the (unique) maximum of $f$ on $(y^*, \sqrt{2}c_t]$ happens. 
\end{claim}

\begin{proof}
	We prove that $\frac{\partial^2 f(y)}{\partial y^2}$ is a monotone decreasing function. One should note that
	\begin{align*}
	\frac{\partial^2 f(y)}{\partial y^2}&=1-\sum_{s=1}^{t} \frac{4c_s^2(2c_s^2+3y^2)}{(2c_s^2-y^2)^3}\cdot \frac{2(c_s/x)^2-1}{T}
	\end{align*}
	Now, one can see that $\frac{\partial^2 f(y)}{\partial y^2}$ is a monotone decreasing function in $y$. 	We then note that $f(x)\leq 0$, and the function $f(y)$ has exactly one maximum on $(y^*, 
	\sqrt{2}c_s)$.
\end{proof}
We will need
\begin{claim}[Monotonicity]\label{cl:mina}
For every $i\in [|R|]$ we have
\begin{description}
\item[(a)] If $g_{z_{j_i},j_i} > 0$ then there exists a $y^*\in (x, \sqrt{2})$ such that $g_{y^*,j_i}\ge0$, $g_{y, j_i}\leq 0$ for any $y\in Z_x$ such that $y\leq y^*$,  and $g_{y,j_i}$ is non-decreasing in $y$ for $y\in [y^*, z_{j_i}]$;
\item[(b)] If $h_{z_{j_i}}^{(N-1)} >0$ then there exists a $y^*\in (x, \sqrt{2})$ such that  $h_{y^*}^{(N-1)}\ge0$, $h_y^{(N-1)}\leq 0$ for any $y\in Z_x$ such that $y\leq y^*$ and $h_{y}^{(N-1)}$ is non-decreasing in $y$ for $y\in [y^*, z_{j_i}]$. 
\end{description}
\end{claim}
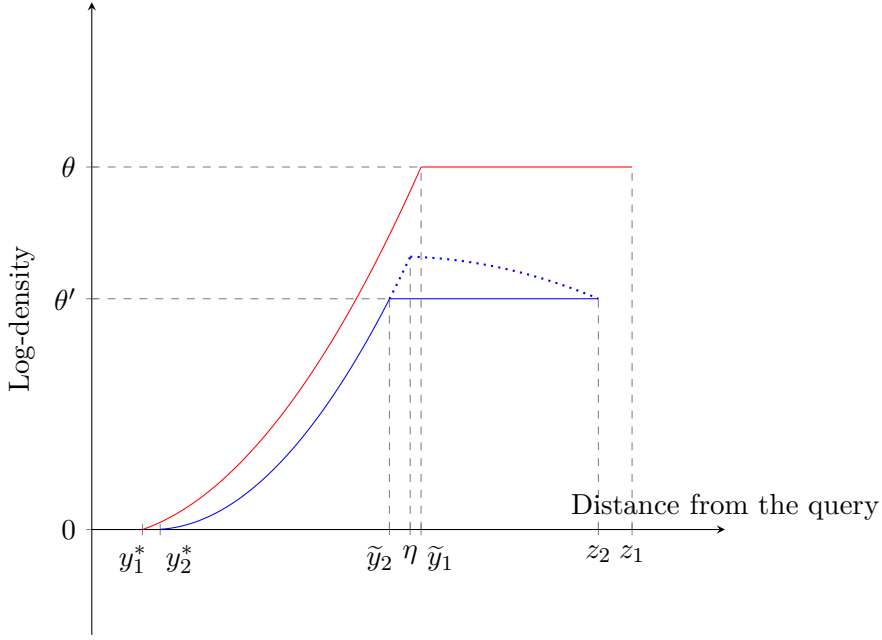
\begin{figure}
	\begin{center}
		\begin{tikzpicture}
		\begin{axis}[
		restrict y to domain=-10:10,
		samples=1000,
		width=10cm, height=10cm,
		ymin=-0.2 ,ymax=1,
		xmin=0, xmax=2.5,
		xtick={0,0.2,0.27,1.175,1.257,1.3,2,2.133},
		xticklabels={$0$,$y^*_1~~$,$~~~~y^*_2$,$\wt{y}_2~~$,$\eta$, $~~~~\wt{y}_1$,$z_2$,$z_1$},
		ytick={0,0.4376652, 0.6875},
		yticklabels={$0$,$\theta'$,$\theta$},
		xlabel={Distance from the query},
		ylabel={Log-density},
		axis x line=center,
		axis y line=left,
		every axis x label/.style={
			at={(ticklabel* cs:1)},
			anchor=south,}
		]

		\addplot+[dashed,gray, const plot, no marks] coordinates { (1.3,0) (1.3 , 0.6875
			) } node[above,pos=.57,black] {};
		
		\addplot+[dashed,gray, const plot, no marks] coordinates { (2.133,0) (2.133 ,0.6875 ) } node[above,pos=.57,black] {};

		\addplot+[dashed,gray, const plot, no marks] coordinates { (2,0) (2 , 0.437665) } node[above,pos=.57,black] {};

		\addplot+[dashed,gray, const plot, no marks] coordinates { (1.175,0) (1.175 , 0.43766) } node[above,pos=.57,black] {};

		\addplot+[dashed,gray, const plot, no marks] coordinates { (1.257,0) (1.257 ,0.51714 ) } node[above,pos=.57,black] {};

		\addplot [
		domain=0:1.175, 
		samples=100, 
		color=gray,
		style=dashed,
		]
		{0.43766};
		
		\addplot [
		domain=0:1.3, 
		samples=100, 
		color=gray,
		style=dashed,
		]
		{0.6875};

		\addplot [
		domain=0.2:1.3, 
		samples=100, 
		color=red,
		]
		{(x^2/2.4-0.04/2.4)};

		\addplot [
		domain=1.3:2.133, 
		samples=100, 
		color=red,
		]
		{  0.6875 };

		\addplot [
		domain=0.27:1.175, 
		samples=100, 
		color=blue,
		]
		{  (x-0.24)^2/2 
		};
		
		\addplot [
		domain=1.175:1.257, 
		samples=100, 
		color=blue, dotted, thick
		]
		{ (x-0.24)^2/2};
		
		\addplot [
		domain=1.257:2, 
		samples=100, 
		color=blue, dotted, thick
		]
		{ 0.51781-(x-1.2)^2/8   };

		\addplot [
		domain=1.175:2, 
		samples=100, 
		color=blue,
		]
		{0.437665};

		\end{axis}
		\end{tikzpicture}
		\caption{An illustration of proof of Claim~\ref{cl:mina}. The red and blue curves represent functions $G_1$ and $G_2$. The dotted part of the blue curve represents $G_1-\hat{q}$ function for interval $[\wt{y}_2,z_2]$, which gets truncated by $\theta'$. }	\label{fig:theta}
	\end{center}
	
\end{figure}
\begin{proof}

Let $$q(y):=\sum_{i=1}^{t}\frac{2(c_s/x)^2-1}{2(c_s/y)^2-1}\frac{1}{T},$$  where $c_1\ge c_2 \ge \ldots \ge c_t\ge z_1\ge x$ for some $z_1\ge x$. And let $y^*_1$ be such that $\frac{(y^*_1)^2-x^2}{2}-q(y^*_1)=0$ and let $\wt{y}_1$ be the smallest value such that $\wt{y}_1\ge y^*_1$ and $\frac{\wt{y}_1^2-x^2}{2}-q(\wt{y}_1)=\theta$ for some $\theta\ge 0$. Now define $G_1(y)$  on $[y^*_1,z_1]$, for some $z_1\ge \wt{y}_1$ as follows
\begin{align}\label{eq:form1}
G_1(y):=\begin{cases}
\frac{y^2-x^2}{2}-q(y)  & y \in [y^*_1,\wt{y}_1)\\
\theta & y\in[\wt{y}_1,z_1]
\end{cases}
\end{align}
 See the red curve in Figure~\ref{fig:theta}.\\
 
 Also, let $\hat{q}(y):=\frac{2(z_1/x)^2-1}{2(z_1/y)^2-1}\frac{1}{T}$. Let $y^*_2\ge y^*_1$ such that $G_1(y^*_2)-\hat{q}(y^*_2)=0$. Now, we define $G_2(y)$ for $y\in [y^*_2,z_2]$ as follows:
\begin{align*}
G_2(y):=\min\left\{  G_1(y)-\hat{q}(y), \theta'   \right\}
\end{align*}
where $\theta':=G_1(z_2)-\hat{q}(z_2)$ and $\theta'\ge 0$ for some $z_2\le z_1$. By the definition of $y^*_2$, function $G_2(y)$ for $y\in [y^*_2,\wt{y}_1]$ is in the form of the function in Claim~\ref{claim:monotone} and thus, it has a unique maximum at some $\eta\in[y^*_2,\wt{y}_1]$. Also, recall that $G_1(y)=\theta$ for $y\in [\wt{y}_1,z_2]$. Also, one should note that since $\hat{q}(y)$ is a monotone increasing function for $y\in (0,\sqrt{2}z_1)$ and hence for $y\in [\wt{y}_1,z_2]$, then $\theta' \le G_2(\wt{y}_1)$ and therefore $\theta'\le G_2(\eta)$. This guarantees that there exist a $\wt{y}_2\in[y^*_2,\eta]$ such that $G_2(\wt{y}_2)=\theta'$. The reason is that $G_2(y)$ is a continuous increasing function for $y\in[\wt{y}_2,\eta]$. So, we have
\begin{align}
G_2(y):=\begin{cases}
\frac{y^2-x^2}{2}-q'(y)  & y \in [y^*_2,\wt{y}_2)\\
\theta' & y\in[\wt{y}_2,z_2]
\end{cases}
\end{align}
where, $q'(y):=q(y)-\hat{q}(y)$. See the blue curve in Figure~\ref{fig:theta}. Now, one can see that by a simple inductive argument starting with the initial densities $$\min\left\{ \frac{y^2-x^2}{2},1-\frac{x^2}{2}\right\}$$ which is in the form of \eqref{eq:form1}, the statement of the claim holds. 
\end{proof}
\subsection{Bounding terminal densities using feasible LP solutions}

\begin{claim}[Feasible LP solution from an execution path]\label{claim:feasible}
If integer $J$ is such that $J> \frac{T}{1-10^{-4}}\mathrm{OPT(LP)}$ then, for all $y\le z_{j^*-1}$ , $f_{y,J+1}< 7\delta_z$ for $j^*= k_{J}+1$ (see Definition~\ref{def:fy} for the definition of $k_J$).
\end{claim}

\begin{proof} We prove the claim by contradiction. Suppose that there exists $z\le z_{j^*-1}$ such that $f_{z,J+1} \ge 7\delta_z$. We define the feasible solution $g_{y, j}$'s and $\alpha_j$'s to LP($x, j^*$) as defined in \eqref{def:lp}. However, the cost of this solution will be more than the optimal cost of the LP, which gives us the contradiction. First, we construct the solution without considering the {\bf non-empty range constraint}. Then, we show that applying $f_{z,J+1} \ge 7\delta_z$ the {\bf non-empty range constraint} is satisfied too.
	
	Let $g_{y, j}$ and $\alpha_j$ be defined as by induction on $j=1,\ldots, j^*$ as above. We prove by induction on $j$ that if there exists a $s$ such that $c_s=z_j$ and $c_{s-1}\ne z_j$ then for all $y\le z_j$,
	\begin{equation}\label{eq:ind-claim}
	f_{y, s}\leq g_{y, j}+X_s,
	\end{equation}
	where we define
	\begin{equation}\label{eq:xt-def}
	X_s:=\sum_{r\in R_m \text{~s.t.~} r\le s} O\left(\frac1{\sqrt{T}}\right)\cdot \frac{1}{\wt{T}}+\sum_{r\in R_M \text{~s.t.~} r\le s} \frac{O(1)}{\wt{T}}+s\cdot \delta+3\delta_z.
	\end{equation}
	for ease of notation, and let
	$\delta=\log_{1/\mu}\frac{2-2\tau}{1-2\tau}=\Theta\left(\frac{1}{\log 1/\mu}\right)$. One should note that the $\delta$ used in this proof is not related to the $\delta$ used in the algorithm. Also, let $c_0=\infty$ and $c_{J+1}=z_{j^*}$, for easing the corner case analysis.

	\noindent{\bf Base:} For all $j\in \{1, 2,\ldots, j_1\}=[j_1]$ and all  $y\le z_j$, in \eqref{eq:densayj12} we did set
	\begin{align}\label{eq:densayj1}
	g_{y,j} := \min\left\{ {\frac{y^2-x^2}{2}}, {1-\frac{x^2}{2}}\right\}.
	\end{align}
	Also, recall that we let $\alpha_j:= 0$ for all $j\in[j_1-1]$ (see {\bf base case} in Section~\ref{sec:construction}). Now, note that we have $f_{y,1}\le g_{y,1}+3\delta_z$ for all $y$ by Claim~\ref{claim:quantizedc}, {\bf (1)} combined with the assumption that $\delta_z\leq 1$. So, the base holds. 
	
	\noindent{\bf Inductive step $j_i\rightarrow j_i+1,\ldots, j_{i+1}$:} We let $a:=j_i$ and $b:=j_{i+1}$ to simplify notation. Let $s$ be the first step on sphere $z_a$: $c_s=z_a$ and $c_{s-1}\ne z_a$. Also, let $N$ be the number of steps that we stay on sphere $z_a$, i.e.,
	\begin{align*}
	c_s=c_{s+1}=\ldots=c_{s+N-1}=z_a \text{~and~} c_{s+N}\ne z_a.
	\end{align*}
	Note that steps $s+1,\ldots, s+N-1$ are stationary as per our definitions. We let $t:=s+N$ for convenience. By the inductive hypothesis for any $y\le z_a$ we have 
	\begin{equation}\label{eq:ind-hyp}
	f_{y, s}\leq g_{y, a}+X_s
	\end{equation}
	We prove that for any $y\le z_b$
	\begin{equation}\label{eq:ind-step}
	f_{y, t}\leq g_{y, b}+X_t.
	\end{equation}

	Let $h_y^{(q)}, q=0,\ldots, N$ and $g_{y, j}, j=a,\ldots, b$,  be defined as above. We now upper bound $f_{y,s+q}$ in terms of $f_{y, s+(q-1)}$.
	We have for all $q\in[N-1]$ and $y\le z_b$:
	\begin{equation}\label{eq:fLSHtr}
	\begin{split}
	f_{y,s+q}&\le \min\left\{    f_{y,s+(q-1)}-\frac{2(z_a/x)^2-1}{2(z_a/y)^2-1}\frac{1}{\wt{T}},~f_{z_a,s+(q-1)}-\frac{2(z_a/x)^2-1}{\wt{T}}+ \delta\right\}\\
	&\le \min\left\{    f_{y,s+(q-1)}-\frac{2(z_a/x)^2-1}{2(z_a/y)^2-1}\frac{1}{\wt{T}},~f_{z_a,s+(q-1)}-\frac{2(z_a/x)^2-1}{\wt{T}}\right\}+ \delta.
	\end{split}
	\end{equation}
	where the first transition is by Claim~\ref{claim:quantizedc}.
	Similarly we have (again by Claim~\ref{claim:quantizedc})
	\begin{equation}\label{eq:fLSHtr2}
	\begin{split}
	f_{y, t}& \le\min\left\{ f_{y,s+(N-1)}-\frac{2(z_a/x)^2-1}{2(z_a/y)^2-1}\frac{1}{\wt{T}},~f_{z_{j_b},s+(N-1)}-\frac{2(z_a/x)^2-1}{2(z_a/z_b)^2-1}\frac{1}{\wt{T}}+\delta\right\}\\
	& \le\min\left\{ f_{y,s+(N-1)}-\frac{2(z_a/x)^2-1}{2(z_a/y)^2-1}\frac{1}{\wt{T}},~f_{z_{j_b},s+(N-1)}-\frac{2(z_a/x)^2-1}{2(z_a/z_b)^2-1}\frac{1}{\wt{T}}\right\}+\delta
	\end{split}
	\end{equation}
	We now note that the recurrence relations~\eqref{eq:bLSHtr} and~\eqref{eq:bN} defining $h^{(q)}_y$ are only different from the above by an additive $\delta$ term, and the initial condition~\eqref{eq:h-init-cond} for $h_{y}^{(0)}$ is only different from the inductive hypothesis~\eqref{eq:ind-hyp} by an additive $X_s$ term.  Combining these observations, we get 
	\begin{equation}\label{eq:230jt23jt}
	\begin{split}
	f_{y,t}&\le \min\left\{h_{y}^{(N)},h_{z_b}^{(N)}\right\}+X_s+\delta\cdot (t-s).
	\end{split}
	\end{equation}
	Now, one can see that we have the following upper bound for $X_s$ for any $s$ using the definition of $X_s$
	\begin{align}
	X_s&=\sum_{r\in R_m \text{~s.t.~} r\le s} O\left(\frac1{\sqrt{T}}\right)\cdot \frac{1}{\wt{T}}+\sum_{r\in R_M \text{~s.t.~} r\le s} \frac{O(1)}{\wt{T}}+s\cdot \delta+3\delta_z\nonumber\\
	&\le O\left(\frac{1}{\sqrt{T}}\right)+3\delta_z \label{eq:Xsorder}
	\end{align}
	since we have at most $O(\sqrt{T})$ major steps, at most $O(T)$ minor steps, and $\delta=\Theta\left(\frac{1}{T^2}\right)$. This implies
	\begin{equation}\label{eq:79}
	\begin{split}
	f_{y,t}&\le \min\left\{h_{y}^{(N)},h_{z_b}^{(N)}\right\}+3\delta_z+O\left(\frac1{\sqrt{T}}\right)\leq h_{y}^{(N)}+4\delta_z.
	\end{split}
	\end{equation}
	Combining this with the assumption that there exists a $z\le z_{j^*-1}$ such that $f_{z,J+1}\geq 7\delta_z$  we have 
	\begin{equation}\label{eq:23ihihfihNCCm}
	h_{y}^{(N)}\geq 0\text{~for all~}y\geq z_{j^*-1},
	\end{equation}
	which we prove below and will be useful whenever we want to invoke Claim~\ref{cl:mina}.
\begin{claim}
		$\forall y \ge z_{j^*-1}$, we have $h_y^{(N)}\ge 0$.
\end{claim}
\begin{proof} $\forall y\ge z_{j^*}:~h_y^{(N)}\ge 0 $.
				Assume that there exists a $z\le z_{j^*-1}$ such that $f_{z,J+1}\ge 7\delta_z$. Now, by the fact that $f_{y,j}$'s are monotone in $j$, and by \eqref{eq:79} we have 
				\begin{align*}
				7\delta_z \le f_{z,J+1} \le f_{z,t}\le h_{z}^{(N)}+4\delta_z, 
				\end{align*}
				which implies
				\begin{align*}
				3\delta_z \le h_z^{(N)}.
				\end{align*}
				On the other hand, by \eqref{eq:bLSHtr} we get
				\begin{align*}
				h_z^{(N-1)} \le h_{z_a}^{(N-1)}.
				\end{align*}
				which implies $h_{z_a}^{(N-1)}\ge 3\delta_z\ge 0$. Thus, by Claim~\ref{cl:mina}, $h_{y}^{(N-1)}$ is non-decreasing in $[z,z_a]$. So, for any $y \in [z,z_a]$, 
				\begin{align}\label{eq:N13dz}
				3\delta_z \le h_y^{(N-1)} 
				\end{align}
				Also, by \eqref{eq:bN} we have
				\begin{align}\label{eq:laststepz}
				h_y^{(N)}=h_y^{(N-1)}-\frac{2(z_a/x)^2-1}{2(z_a/y)^2-1}\frac{1}{\wt{T}}=h_y^{(N-1)}-O\left(\frac{1}{T}\right)
				\end{align}
				Combing \eqref{eq:N13dz} and \eqref{eq:laststepz}, we prove that for $y\ge z_{j^*-1}$:
				\begin{align*}
				h_{y}^{(N)}\ge 2\delta_z\ge 0.
				\end{align*}
				
\end{proof}
	
	The following characterization of $X_t-X_s$ will be useful:
	\begin{equation}\label{eq:xtxs-diff}
	X_t-X_s=\delta \cdot (t-s)+\left\{
	\begin{array}{ll}
	O\left(\frac1{\sqrt{T}}\right)\cdot \frac{1}{\wt{T}}&\text{~if~}t\in R_m\\
	\frac{O(1)}{\wt{T}}&\text{~if~}t\in R_M.
	\end{array}
	\right.
	\end{equation}
	The above follows by \eqref{eq:xt-def} since all steps between $s$ and $t$ are stationary.
	
	We now the upper bound the minimum on the rhs in~\eqref{eq:230jt23jt}.  Recall from \eqref{eq:mina} that
	\begin{align*}
	g_{y,b}:=\min\left\{ \min_{j\in \{a+1,\ldots, b\}}\left\{    h_{z_j}^{(N)} \right\}, h_y^{(N)} \right\}.
	\end{align*}
	Let $y''$ be such that $g_{y, b}=h_{y''}^{(N)}$. We consider two cases, depending on whether $y''=y$. 
	
	\paragraph{Case 1: $y=y''$ (the simple case).}  In that case we have
	\begin{align*}
	f_{y, t}&\le \min\left\{h_{y}^{(N)},h_b^{(N)}\right\}+X_s+\delta\cdot (t-s)&&\text{By \eqref{eq:230jt23jt}}\\
	&= h_{y}^{(N)}+X_s+\delta\cdot (t-s)&&\text{Since $y''=y$}\\
	&= g_{y, b}+X_s+\delta\cdot (t-s)&&\text{Combining $y''=y$ and \eqref{eq:mina}} \\
	&\leq g_{y, b}+X_t,&&\text{By \eqref{eq:xtxs-diff}}
	\end{align*}
	as required.
	
	\paragraph{Case 2: $y\neq y''$ (the main case).}
	\begin{equation}\label{eq:923ht923hg}
	\begin{split}
	f_{y,t}&\le \min\left\{h_{y}^{(N)},h_{z_b}^{(N)}\right\}+X_s+\delta\cdot (t-s)\\
	&\le h_{z_b}^{(N)}+X_s+\delta\cdot (t-s)\\
	&=g_{y, b}+X_s+\delta\cdot (t-s)+(h_{z_b}^{(N)}-g_{y, b}).\\
	\end{split}
	\end{equation}

	In what follows we show that 
	\begin{equation}\label{eq:923hg2g23tccc}
	h_{z_b}^{(N)}-g_{y, b}=\left\{
	\begin{array}{ll}
	O\left(\frac1{\sqrt{T}}\right)\cdot \frac{1}{\wt{T}}&\text{~if~}t\in R_m\\
	\frac{O(1)}{\wt{T}}&\text{~if~}t\in R_M,
	\end{array}\right.
	\end{equation}
	which gives the result once substituted in~\eqref{eq:923ht923hg}, as per~\eqref{eq:xtxs-diff}.

	We now consider two case, depending on whether $t$ is a \emph{minor} or a \emph{major} step. For both steps we use the fact that $y''\ne y$ implies $y''\ge z_b$ (this follows by definition of $y''$ together with~\eqref{eq:mina}).
	
	\paragraph{Minor steps ($t\in R_m$).} In this case we have
	\begin{align*}
	&h_{z_b}^{(N)}-g_{y,b}\\
	=&h_{z_b}^{(N)}-h_{y''}^{(N)}&&\text{By definition of $y''$ }\\
	=&h_{z_b}^{(N-1)}-\frac{2(z_a/x)^2-1}{2(z_a/z_b)^2-1}\cdot \frac{1}{\wt{T}}-h_{y''}^{(N-1)}+\frac{2(z_a/x)^2-1}{2(z_a/y'')^2-1}\cdot \frac{1}{\wt{T}}&&\text{By \eqref{eq:bN}}\\
	\le& \frac{2(z_a/x)^2-1}{2(z_a/y'')^2-1}\cdot \frac{1}{\wt{T}}-\frac{2(z_a/x)^2-1}{2(z_a/z_b)^2-1}\cdot \frac{1}{\wt{T}}.&&\\
	\end{align*}
	The last transition used Claim~\ref{cl:mina}, {\bf (b)}, and the fact that $y'' \ge z_b$: we only need to verify the preconditions of Claim~\ref{cl:mina}, which follows by~\eqref{eq:23ihihfihNCCm} together with the fact that $h^{(N-1)}_y\geq h^{(N)}_y$ for all $y$.
	
	We now bound the rhs of the equation above by
	\begin{align*}
	& \frac{2(z_a/x)^2-1}{2(z_a/y'')^2-1}\cdot \frac{1}{\wt{T}}-\frac{2(z_a/x)^2-1}{2(z_a/z_b)^2-1}\cdot \frac{1}{\wt{T}}&&\\
	\leq &\frac{2(z_a/x)^2-1}{\wt{T}}\left(1-\frac{1}{2(z_a/z_b)^2-1}\right)&&\text{Since $y''\le z_a$}\\
	= &\frac{2(z_a/x)^2-1}{\wt{T}}\left(1-\frac{1}{2(1-1/\sqrt{T})^{-2}-1}\right)&&\text{Since this is a \emph{minor} step}\\
	= &\frac{2(z_a/x)^2-1}{\wt{T}}\cdot O(1/\sqrt{T}) \\
	\le &\frac{2(z_a/\Delta)^2-1}{\wt{T}}\cdot O(1/\sqrt{T}) \\
	= &\frac{1}{\wt{T}}\cdot O\left(\frac{1}{\sqrt{T}}\right),&&\text{Since $z_a\leq R_{max}=O(1)$ and $\Delta=\Omega(1)$} 
	\end{align*}
	
	\paragraph{Major steps ($t\in R_M$).} Now, we consider the case when the step is \emph{major}. 
	\begin{align*}
	&h_{z_b}^{(N)}-g_{z_b, b}\\
	=&h_{z_b}^{(N)}-h_{y''}^{(N)}&&\text{By definition of $y''$ and \eqref{eq:mina}}\\
	=&h_{z_b}^{(N-1)}-\frac{2(z_a/x)^2-1}{2(z_a/z_b)^2-1}\cdot \frac{1}{\wt{T}}-h_{y''}^{(N-1)}+\frac{2(z_a/x)^2-1}{2(z_a/y'')^2-1}\cdot \frac{1}{\wt{T}}&&\text{By \eqref{eq:bN}}\\
	\le& \frac{2(z_a/x)^2-1}{2(z_a/y'')^2-1}\cdot \frac{1}{\wt{T}}-\frac{2(z_a/x)^2-1}{2(z_a/z_b)^2-1}\cdot \frac{1}{\wt{T}}.&&
	\end{align*}
	The last transition used Claim~\ref{cl:mina}, {\bf (b)}, and the fact that $y'' \ge z_b$: we only need to verify the preconditions of Claim~\ref{cl:mina}, which follows by~\eqref{eq:23ihihfihNCCm} together with the fact that $h^{(N-1)}_y\geq h^{(N)}_y$ for all $y$. We now upper bound the rhs of the equation above:
	
	\begin{align*}
	& \frac{2(z_a/x)^2-1}{2(z_a/y'')^2-1}\cdot \frac{1}{\wt{T}}-\frac{2(z_a/x)^2-1}{2(z_a/z_b)^2-1}\cdot \frac{1}{\wt{T}}&&\\
	\leq &\frac{2(z_a/x)^2-1}{\wt{T}}\left(1-\frac{1}{2(z_a/z_b)^2-1}\right)\\
	\le & \frac{2(z_a/x)^2-1}{\wt{T}}\\
	\le & \frac{2(z_a/\Delta)^2-1}{\wt{T}}\\
	=& \frac{O(1)}{\wt{T}}&&\text{Since $z_a\leq R_{max}=O(1)$ and $\Delta=\Omega(1)$} 
	\end{align*}
	This completes the inductive claim and establishes~\eqref{eq:ind-claim} for all $j=1,\ldots, j^*$.

	The only thing we need to verify is that the solution that we presented, satisfies the {\bf non-empty range constraint}. For the sake of this proof, let us define $g_{z_{j^*-1},j^*}$ as follows:
	\begin{align}\label{eq:nemptyconst1}
	g_{z_{j^*-1},j^*} := g_{z_{j^*-1},{j^*-1}}-\frac{2\left(z_{j^*-1}/x\right)^2-1}{2\left(z_{j^*-1}/z_{j^*-1}\right)^2-1}\cdot \alpha_{j^*-1} =g_{z_j,{j}}-\frac{2\left(z_{j^*-1}/x\right)^2-1}{1}\cdot \alpha_{j^*-1}
	\end{align}
	If $s$ is such that $c_s=z_{j^*-1}$ and $c_{s-1}\ne z_{j^*-1}$, then by the discussion above
	\begin{align}\label{eq:nempty-induction}
	f_{y,s}\le g_{y,j^*-1}+X_{s}.
	\end{align}
	and more specifically, when $y=z$ by the assumption we have 
	\begin{align*}
	7\delta_z \le f_{z,J+1} \le g_{z,j^*} + O\left(\frac{1}{\sqrt{T}}\right)+3\delta_z&&\text{By \eqref{eq:Xsorder}}
	\end{align*}
	which implies
	\begin{align}\label{eq:3dz}
	g_{z,j^*} \ge 3\delta_z.
	\end{align}
	Now, we prove that $g_{z_{j^*},j^*}\ge 0$. The same as the discussion above, if we took $N$ steps on sphere $z_{j^*-1}$ then by \eqref{eq:terminalg} we have
	\begin{align*}
	g_{z,j^*} = h_z^{(N)},
	\end{align*}
	where $h$'s are the auxiliary variables defined for sphere $z_{j^*-1}$.
	Now, we also have 
	\begin{align*}
	g_{z,j^*}=h_{z}^{(N)} &\le h_{z}^{(N-1)} &&\text{By \eqref{eq:bN}}\\
	&\le h_{z_{j^*-1}}^{(N-1)} && \text{By \eqref{eq:bLSHtr}}
	\end{align*}
	On the other hand, we have
	\begin{align*}
	h_{z_{j^*-1}}^{(N)}=h_{z_{j^*-1}}^{(N-1)}- \frac{2(z_{j^*-1}/x)^2-1}{1}\frac{1}{\tilde{T}}=h_{z_{j^*-1}}^{(N-1)}- O\left(\frac{1}{T}\right) \ge h_{z_{j^*-1}}^{(N-1)}- \delta_z
	\end{align*}
	Combining these facts we get
	\begin{align*}
	g_{z_{j^*-1},j^*}=h_{z_{j^*-1}}^{(N)}\ge h_{z_{j^*-1}}^{(N-1)}- \delta_z \ge g_{z,j^*}-\delta_z\ge 2\delta_z
	\end{align*}
	where the last inequality is due to \eqref{eq:3dz}. Also, by the construction of the solution and the fact that the function $\frac{2(z/x)^2-1}{2(z/y)^2-1}$ is increasing in $y$, we have
	\begin{align*}
	g_{z_{j^*-1},j^*}-g_{z_{j^*},j^*}&\le \min\left\{\frac{z_{j^*-1}^2-x^2}{2},1-\frac{x^2}{2}\right\}-\min\left\{\frac{z_{j^*}^2-x^2}{2},1-\frac{x^2}{2}\right\}\\
	&\le \frac{(\sqrt{2})^2-\left((1-\delta_z)\sqrt{2}\right)^2}{2}\le 2\delta_z
	\end{align*}
	which implies that $g_{z_{j^*},j^*}\ge 0$ (the non-empty range constraint in the LP~\eqref{def:lp}).
	
	Now, recalling the values of $\alpha_j$'s, one can see the cost of this solution of LP is equal to $\frac{J(1-10^{-4})}{T}$, which is greater than the optimal solution for the LP, which is a contradiction. So, the claim holds.  
\end{proof}

It is important to note that Claim~\ref{claim:feasible} is not universally true for any shift-invariant kernel, even under natural monotonicity assumptions. An example is presented in Section~\ref{sec:tech-overview} in Figure~\ref{fig:goodexample}. We show, however, that our linear programming formulation is indeed a tight relaxation for a wide class of kernels that includes the Gaussian kernel, the exponential kernel as well as any log-convex kernel.

\section{Upper bounding LP value}\label{sec:lp-sol}

The main result of this section is a proof of Lemma~\ref{lm:lp-value-clean} below:

\begin{lem}\label{lm:lp-value-clean}
	For every $x\geq 0, y\geq x$ the value of the LP in~\eqref{def:lp} (restated below as~\eqref{eq:lp-primal})  is upper bounded by $0.1718$. Furthermore, for every $x\in (0, \sqrt{2})$ and $y\geq \sqrt{2}$ the LP value is bounded by $3-2\sqrt{2}<0.1718$, and for every $x\in (0, \sqrt{2})$ and every $y\in [x, \sqrt{2}]$ the LP value is bounded by $\frac{x^2}{2}\left(1-\frac{x^2}{2}\right)$.
\end{lem}

The main result of this section is an upper bound on the value of the LP~\eqref{eq:lp-primal} below. We first derive a dual formulation, then exhibit a feasible dual solution and then verify numerically that the value of dual is bounded by $0.172$ for all values of the input parameters $x$ and $y^*$. 

Fix $x\in [0, \sqrt{2}]$. Let $z_1>z_2>\ldots>z_I$, denote the distances on the grid, and we define $Z:=\{z_1,z_2,\ldots,z_I\}$, we will consider $I$ linear programs, enumerating over all $j^*\in [I]$ such that $z_{j^*}\geq x$.
\begin{align}
\max_{\alpha\ge 0}&~\sum_{j=1}^{j^*-1}  \alpha_j\label{eq:lp-primal}\\
\text{such that}&:\nonumber\\
\forall y\in Z&: g_{y,1}\le
 \min\left\{\frac{y^2-x^2}{2},1-\frac{x^2}{2}\right\}&(r_{y, 0})&~~~\text{Density constraints}\nonumber\\
  \forall j\in[j^*-1], \forall y \in Z \text{ s.t.~} y<z_j&:~ g_{y,j}\le {g_{z_j,j}}&(q_{y, j})&~~~\text{Truncation}\nonumber\\
 & g_{y,{j+1}} \le g_{y,{j}}-\frac{2\left(z_j/x\right)^2-1}{2\left(z_j/y\right)^2-1}\cdot \alpha_j&(r_{y, j})&~~~\text{Spherical LSH}\nonumber\\
 & g_{z_{j^*},j^*}\geq 0&(\eta)\nonumber
\end{align}

The dual of~\eqref{eq:lp-primal} is 
\begin{align}
\min&~\sum_{y\in Z} \left\{\frac{y^2-x^2}{2},1-\frac{x^2}{2}\right\} r_{y, 0}\label{eq:dual}\\
\text{such that}&:\nonumber\\
\forall  j\in [j^*-1], y\in Z, y<z_j&: r_{y, j-1}-r_{y, j}+q_{y, j}=0&(g_{y,j})&~~~\text{Mass transportation}\nonumber\\
\forall  j\in [j^*-1]&: r_{z_j, j-1}-\sum_{x\in Z, x<z_j} q_{x, j}=0&(g_{z_j,j})&~~~\text{Max tracking}\nonumber\\
\forall y\in Z, y<z_{j^*}&: r_{y, j^*-1}=0&(g_{y,j^*})&~~~\text{Sink}\nonumber\\
&-\eta+r_{z_{j^*}, j^*-1}=0&(g_{z_{j^*},j^*})&~~~\text{Terminal flow}\nonumber\\
j\in[j^*-1]&:\sum_{y\in Z: y<z_j} \frac{2\left(z_j/x\right)^2-1}{2\left(z_j/y\right)^2-1}r_{y, j}\geq 1&(\alpha_j)&\nonumber\\
&r_{y, j}, q_{y, j}\geq 0&&\nonumber\\
&\eta\geq 0&&\nonumber
\end{align}

We start by exhibiting a simple feasible solution for the dual that reproduces our result from Section~\ref{sec:data-independent}.

\paragraph{Upper bound of $\frac{x^2}{2}\cdot (1-\frac{x^2}{2})$ for every $x$.} Let $q_{y, j}=0$ for all $y, j$. Let $$r_{z_{j^*}, j}=\left(\frac{x}{y}\right)^2$$ for all $j=0, 1,\ldots, j^*-1$ and let $r_{y, j}=0$ for $y\neq z_{j^*}$ and all $y$. We let $\eta=r_{z_{j^*}, j^*-1}$. We first verify feasibility.
We have for every $j=1,\ldots, j^*-1$
\begin{equation*}
\begin{split}
\sum_{y\in Z: y<z_j} \frac{2\left(z_j/x\right)^2-1}{2\left(z_j/y\right)^2-1}r_{y, j}&=\frac{2\left(z_{j^*}/x\right)^2-1}{2\left(z_{j^*}/y\right)^2-1}r_{z_{j^*}, j^*-1}\\
&\geq \frac{2\left(z_{j^*}/x\right)^2}{2\left(z_{j^*}/y\right)^2}r_{z_{j^*}, j^*-1}\\
&=\left(\frac{y}{x}\right)^2\cdot \left(\frac{x}{y}\right)^2\\
&=1,
\end{split}
\end{equation*}
where we used the fact that $x\leq y$. We thus have a feasible solution. The value of the solution is 
$$
\left(\frac{x}{y}\right)^2\cdot \min\left\{\frac{y^2-x^2}{2},1-\frac{x^2}{2}\right\}\leq \left(\frac{x}{y}\right)^2\cdot \min\left\{\frac{y^2-x^2}{2},1-\frac{x^2}{2}\right\}
$$
When $y\geq \sqrt{2}$, we get, $\left(\frac{x}{y}\right)^2\cdot \left(1-\frac{x^2}{2}\right)$, which is maximized at $y=\sqrt{2}$ and gives $\left(\frac{x}{\sqrt{2}}\right)^2\cdot (1-\frac{x^2}{2})$. Similarly, when $y\leq \sqrt{2}$, we get $$\left(\frac{x}{y}\right)^2\cdot\left( \frac{y^2-x^2}{2}\right)=\frac{x^2}{2}-\frac{x^4}{2y^2},$$ which is again 
maximized when $y=\sqrt{2}$. Thus, we get that the value of the LP in~\eqref{eq:lp-primal} is bounded by 
$$
\frac{x^2}{2}\cdot (1-\frac{x^2}{2}).
$$
and we obtain the exponent of $0.25$. This (almost) recovers the result of Section~\ref{sec:data-independent} . In what follows we obtain a stronger bound of $0.1718$ on the value of the LP in~\eqref{eq:lp-primal}, obtaining our main result on data-dependent KDE.

\paragraph{Upper bound of $0.1718$ on LP value for all $x$.}

We exhibit a feasible solution for the dual in which for every $j<j^*$
\begin{equation}\label{eq:iwrht23r}
\begin{split}
q_{z_{j+1}, j}>0\\
q_{y, j}=0\text{~for all~}y<z_{j+1}\\
\end{split}
\end{equation}
and $q_{y, j^*}=0$ for all $y<z_{j^*}$. We later show numerically that our dual solution is optimal for the Gaussian kernel.

Simplifying~\eqref{eq:dual} under the assumptions from~\eqref{eq:iwrht23r}, we get
\begin{align}
\min&~\sum_{y\in Z} \left\{\frac{y^2-x^2}{2},1-\frac{x^2}{2}\right\} r_{y, 0}\nonumber\\
\text{such that}&:\nonumber\\
\forall  j\in [j^*-1], y\in Z, y<z_{j+1}&: r_{y, j-1}-r_{y, j}=0&(g_{y,j})&\nonumber\\
\forall  j\in [j^*-1]&: r_{z_{j+1}, j-1}-r_{z_{j+1}, j}+q_{z_{j+1}, j}=0&(g_{y,j})&\nonumber\\
\forall  j\in [j^*-1]&: r_{z_j, j-1}-q_{z_{j+1}, j}=0&(g_{z_j,j})&\nonumber\\
\forall y\in Z, y<z_{j^*}&: r_{y, j^*-1}=0&(g_{y,j^*})&\nonumber\\
&-\eta+r_{z_{j^*}, j^*-1}=0&(g_{z_{j^*},j^*})&\nonumber\\
j\in[j^*-1]&:\sum_{y\in Z: y<z_j} \frac{2\left(z_j/x\right)^2-1}{2\left(z_j/y\right)^2-1}r_{y, j}\geq 1&(\alpha_j)&\nonumber\\
&r_{y, j}, q_{y, j}\geq 0&&\nonumber\\
&\eta\geq 0&&\nonumber
\end{align}

Eliminating the $q$ variables from the above for simplicity, we get, making the inequality for the $(\alpha_j)$ constraints an equality (recall that we only need to exhibit a dual feasible solution),
\begin{align}
\min&\sum_{y\in Z} \left\{\frac{y^2-x^2}{2},1-\frac{x^2}{2}\right\} r_{y, 0}\label{eq:dual-s}\\
\text{such that}&:\nonumber\\
\forall  j\in [j^*-1], y\in Z, y<z_{j+1}&: r_{y, j-1}-r_{y, j}=0&(g_{y,j})&\nonumber\\
\forall  j\in [j^*-1]&: r_{z_{j+1}, j-1}=r_{z_{j+1}, j}-r_{z_j, j-1}&(g_{y,j})&\nonumber\\
\forall y\in Z, y<z_{j^*}&: r_{y, j^*-1}=0&(g_{y,j^*})&\nonumber\\
&r_{z_{j^*}, j^*-1}=\eta&(g_{z_{j^*},j^*})&\nonumber\\
j\in[j^*-1]&:\sum_{y\in Z: y<z_j} \frac{2\left(z_j/x\right)^2-1}{2\left(z_j/y\right)^2-1}r_{y, j}=1&(\alpha_j)&\nonumber\\
&r_{y, j}\geq 0&&\nonumber\\
&\eta\geq 0&&\nonumber
\end{align}

\paragraph{Defining a dual feasible solution $r$.} We now derive an expression for a feasible solution $r$. The construction is by {\bf induction}: starting with $j=j^*-1$ as the {\bf base} we define  $r_{y, j}$ variables for $y\in Z, y\leq z_j$ that satisfy dual feasibility. The {\bf base} is provided by
\begin{equation}\label{eq:initial-condition-r}
r_{z_{j^*}, j^*-1}=\eta=\left(\frac{2\left(z_{j^*-1}/x\right)^2-1}{2\left(z_{j^*-1}/z_{j^*}\right)^2-1}\right)^{-1}.
\end{equation}
Note that this fully defines $r_{y,j}$ for $j=j^*-1$, since $r_{y,j^*-1}=0$ for $y<z_{j^*}$.

We now give the {\bf inductive step: $j\to j-1$}. By the inductive hypothesis the variables $r_{y, j}$ that we defined satisfy the $(\alpha_j)$ constraints in the dual~\eqref{eq:dual-s}, which means:
\begin{equation}\label{eq:0923hg9jgggD}
\sum_{i>j} \frac{2\left(z_j/x\right)^2-1}{2\left(z_j/z_i\right)^2-1}r_{z_i, j}=1.
\end{equation}
We will define $r_{y, j-1}$ so that
\begin{equation}\label{eq:0293yt9ut}
\sum_{i>j-1} \frac{2\left(z_{j-1}/x\right)^2-1}{2\left(z_{j-1}/z_i\right)^2-1}r_{z_i, j-1}=1
\end{equation}
and at the same time the $(g_{y, j})$  constraints relating $r_{y, j}$ to $r_{y, j-1}$ are satisfied.

By the first constraint in~\eqref{eq:dual-s} we have $r_{z_i, j-1}=r_{z_i, j}$ for all $i>j+1$, since $y<z_{j+1}$ is equivalent to $i>j+1$. By the second constraint in~\eqref{eq:dual-s} we have
$r_{z_{j+1}, j-1}=r_{z_{j+1}, j}-r_{z_j, j-1}$. Putting these two constraints together, we now find $r_{z_j, j-1}$ and therefore $r_{z_{j+1}, j-1}$. We  rewrite the left hand side of~\eqref{eq:0293yt9ut} as
\begin{equation}
\begin{split}
\sum_{i>j-1} \frac{2\left(z_{j-1}/x\right)^2-1}{2\left(z_{j-1}/z_i\right)^2-1}r_{z_i, j-1}&=\frac{2\left(z_{j-1}/x\right)^2-1}{2\left(z_{j-1}/z_j\right)^2-1}r_{z_j, j-1}+\frac{2\left(z_{j-1}/x\right)^2-1}{2\left(z_{j-1}/z_{j+1}\right)^2-1}(r_{z_{j+1}, j}-r_{z_j, j-1})\\
&+\sum_{i>j+1} \frac{2\left(z_{j-1}/x\right)^2-1}{2\left(z_{j-1}/z_i\right)^2-1}r_{z_i, j}\\
&=\left(\frac{2\left(z_{j-1}/x\right)^2-1}{2\left(z_{j-1}/z_j\right)^2-1}-\frac{2\left(z_{j-1}/x\right)^2-1}{2\left(z_{j-1}/z_{j+1}\right)^2-1}\right)r_{z_j, j-1}\\
&+\sum_{i>j} \frac{2\left(z_{j-1}/x\right)^2-1}{2\left(z_{j-1}/z_i\right)^2-1}r_{z_i, j}\\
\end{split}
\end{equation}
Combining this with~\eqref{eq:0293yt9ut}, we thus get that
\begin{equation}\label{eq:o243hgf}
\begin{split}
r_{z_j, j-1}=\left(\frac{2\left(z_{j-1}/x\right)^2-1}{2\left(z_{j-1}/z_j\right)^2-1}-\frac{2\left(z_{j-1}/x\right)^2-1}{2\left(z_{j-1}/z_{j+1}\right)^2-1}\right)^{-1} \left(1-\sum_{i>j} \frac{2\left(z_{j-1}/x\right)^2-1}{2\left(z_{j-1}/z_i\right)^2-1}r_{z_i, j}\right).
\end{split}
\end{equation}
We now show that $r_{z_j, j-1}\geq 0$. The first multiplier in the expression above is non-negative since $\frac{2\left(z_{j-1}/x\right)^2-1}{2\left(z_{j-1}/z\right)^2-1}$ is increasing in $z$ and $z_j\geq z_{j+1}$.
For the second multiplier we have 
\begin{equation}
\begin{split}
1-\sum_{i>j} \frac{2\left(z_{j-1}/x\right)^2-1}{2\left(z_{j-1}/z_i\right)^2-1}r_{z_i, j}&=1-\sum_{i>j} \frac{2\left(z_{j-1}/x\right)^2-1}{2\left(z_{j-1}/z_i\right)^2-1}r_{z_i, j}\\
&\geq 1-\sum_{i>j} \frac{2\left(z_j/x\right)^2-1}{2\left(z_j/z_i\right)^2-1}r_{z_i, j}\\
&=0.\\
\end{split}
\end{equation}
Here the first transition used~\eqref{eq:0923hg9jgggD} (the inductive hypothesis), and the second transition used the fact that the function $\frac{2\left(z/x\right)^2-1}{2\left(z/y\right)^2-1}$ is non-increasing in $z$ for $x\leq y$.  To summarize, we let $r_{z_j, j-1}$ be defined by~\eqref{eq:o243hgf}. Also, we let 
\begin{equation}
\begin{split}
r_{z_{j+1}, j-1}=r_{z_{j+1}, j}-r_{z_j, j-1}
\end{split}
\end{equation}
and let $r_{z_i, j-1}=r_{z_i, j}$ for $i>j+1$. We verify numerically that $r_{z_{j+1},j-1}\ge0$.

\paragraph{Integral equation representation of the dual solution.} While we do not use the following in our analysis, it is interesting to note that the dual solution that we propose satisfies an integral equation in the limit as the grid size goes to $0$. Let $z_j$  denote a uniform grid with step size $\Delta\to 0$ on the interval $[0, C]$ for some constant $C\geq \sqrt{2}$.
We now rewrite~\eqref{eq:o243hgf} as 
\begin{equation}\label{eq:o24wgihihg}
\begin{split}
\left(\frac{2\left(z_{j-1}/x\right)^2-1}{2\left(z_{j-1}/z_j\right)^2-1}-\frac{2\left(z_{j-1}/x\right)^2-1}{2\left(z_{j-1}/z_{j+1}\right)^2-1}\right)r_{z_j, j-1}=\left(1-\sum_{i>j} \frac{2\left(z_{j-1}/x\right)^2-1}{2\left(z_{j-1}/z_i\right)^2-1}r_{z_i, j}\right).
\end{split}
\end{equation}
Note by the Mean Value Theorem and the fact that the derivative $\left(\frac{2\left(z_{j-1}/x\right)^2-1}{2\left(z_{j-1}/z\right)^2-1}\right)'_z$ is Lipschitz within $[z_{j+1},z_j]$ it follows that
\begin{equation*}
\begin{split}
\left(\frac{2\left(z_{j-1}/x\right)^2-1}{2\left(z_{j-1}/z_j\right)^2-1}-\frac{2\left(z_{j-1}/x\right)^2-1}{2\left(z_{j-1}/z_{j+1}\right)^2-1}\right)&=\left.\left(\frac{2\left(z_{j-1}/x\right)^2-1}{2\left(z_{j-1}/z\right)^2-1}\right)'_z\right|_{z=z_j}\cdot \Delta\cdot (1+O(\Delta)).
\end{split}
\end{equation*}

We thus have that $r_{z_j, j-1}/\Delta$ converges to the solution $g(y)$ to the following integral equation:
\begin{equation}
\left.\left(\frac{2\left(u/x\right)^2-1}{2\left(u/z\right)^2-1}\right)'_z\right|_{z=u}\cdot g(u)=1-\int_{y^*}^u \frac{2\left(u/x\right)^2-1}{2\left(u/r\right)^2-1}g(r)dr
\end{equation}
The initial condition is a point mass at $y^*$. 

\paragraph{Exact solution to the primal when $z_{j^*}\geq \sqrt{2}$.} We note that if $z_{j^*}\geq \sqrt{2}$ an optimal solution to the LP~\eqref{eq:lp-primal} is easy to obtain. The reason is that we can simplify the constraints of LP for band $z_{j^*}$ as follows: for all $j\in[j^*-2]$
\begin{align*}
	g_{z_{j+2},{j+1}} &\le \min\left\{ g_{z_{j+2},{j}}-\frac{2\left(z_j/x\right)^2-1}{2\left(z_j/z_{j+2}\right)^2-1}\cdot \alpha_j,~g_{z_{j+1},{j}}-\frac{2\left(z_j/x\right)^2-1}{2(z_j/z_{j+1})^2-1}\cdot \alpha_j\right\}\\
	&\le g_{z_{j+1},{j}}-\frac{2\left(z_j/x\right)^2-1}{2(z_j/z_{j+1})^2-1}\cdot \alpha_j
\end{align*}	
	and 
	\begin{align*}
	g_{z_{j^*},j^*}\le g_{z_{j^*},j^*-1}-\frac{2\left(z_{j^*-1}/x\right)^2-1}{2\left(z_{j^*-1}/z_{j^*}\right)^2-1}\cdot \alpha_{j^*-1}.
	\end{align*}
	  Now, combining these inequalities with the fact that $g_{z_2,1}\le 1-\frac{x^2}{2}$, one has  
\begin{equation}
\begin{split}
g_{z_{j^*}, j^*}&\leq 1-x^2/2-\sum_{j=1}^{j^*-1} \frac{2\left(z_j/x\right)^2-1}{2\left(z_j/z_{j+1}\right)^2-1}\alpha_j\\
&\leq 1-\frac{x^2}{2}- \frac{1}{1+5\delta_z}\sum_{j=1}^{j^*-1}\left({2\left(z_{j}/x\right)^2-1}\right)\cdot \alpha_j\\
&\leq 1-\frac{x^2}{2}- \frac{1}{1+5\delta_z}\sum_{j=1}^{j^*-1}\left({2\left(z_{j^*}/x\right)^2-1}\right)\cdot \alpha_j\\
&\leq 1-\frac{x^2}{2}- \frac{1}{1+5\delta_z}\left({2\left(\sqrt{2}/x\right)^2-1}\right)\cdot\sum_{j=1}^{j^*-1} \alpha_j,
\end{split}
\end{equation}
where we used the fact that $2\left(z_j/z_{j+1}\right)^2-1=2(1+\delta_z)^2-1\le 1+5\delta_z$ and the function $2\left(z/x\right)^2-1$ is increasing in $z$. 

Letting $\gamma:=\sum_{j=1}^{j^*-1}\alpha_j$ denote the LP objective we need to maximize  $\gamma$ subject to $1-x^2/2- \frac{1}{1+5\delta_z}\left({2\left(\sqrt{2}/x\right)^2-1}\right)\cdot \gamma\geq 0$ (the nonempty range LP constraint), where $y^*=z_{j^*}$. The solution is 
$$
\gamma=\left(1+5\delta_z\right)(1-x^2/2) \left({2\left(\sqrt{2}/x\right)^2-1}\right)^{-1}.
$$
Finally, one has
$$
\text{max}_{x\in [0, \sqrt{2}]} \left(\frac{1-x^2/2}{2(2/x^2)-1}\right)=3-2\sqrt{2}\approx 0.171573,
$$
which is achieved at $x=\sqrt{4-2\sqrt{2}}$. It remains to note that this is achievable by letting $\alpha_{j^*-1}=\gamma$ and letting $\alpha_j=0$ for $j<j^*-1$, when $z_{j^*}=\sqrt{2}$.

\paragraph{Numerical verification for $x\in [0, \sqrt{2}], y\in [0, \sqrt{2}]$.}
Implementing this in Matlab and optimizing over $x$ and $j^*$ (with a uniform grid on $[0, \sqrt{2}]$ consisting of $J=400$ points) yields the exponent of $\approx 0.1716$, achieved at $x\approx 1.0842$ and $z_{j^*}\approx \sqrt{2}$. The Matlab code is given below. Then $0.1718$ is an upper-bound on the optimal cost of LP.  Moreover, for the analysis if we set $\alpha^*=0.172$ (as in Section~\ref{sec:param}) then $\alpha^*(1-10^{-4})$ strictly upper bounds $\text{OPT(LP)}$ (this simplifies the notation in other sections). 
\newpage
\begin{scriptsize}
\begin{verbatim}
J=400;
vmax=0;
xIdxMax=0;
yIdxMax=0;

for xIdx=5:5:J-5,
    for yIdx=xIdx-5:-5:1,
        z=sqrt(2)*(J-(1:J))/J;
        density=zeros(J);
        for j=1:J,
          %% density for exp(-x^2/2)	
          density(j)=min((z(j)^2-z(xIdx)^2)/2, 1-z(xIdx)^2/2);							 
        end;
        r=zeros(J);
        r(yIdx-1)=((2*(z(yIdx-1)/z(xIdx))^2-1)/(2*(z(yIdx-1)/z(yIdx))^2-1))^(-1);        
        for j=yIdx-2:-1:1,
          coeff=zeros(J);
          for i=j:yIdx,
              coeff(i)=(2*(z(j)/z(xIdx))^2-1)/(2*(z(j)/z(i))^2-1);
          end;
          val=0;
          for i=j+1:yIdx-1,
              val=val+coeff(i+1)*r(i);
          end;
	
          r(j)=(coeff(j+1)-coeff(j+2))^(-1)*(1-val);
          r(j+1)=r(j+1)-r(j);
        end;
        val=0;
        for i=1:J,
          val=val+density(i)*r(i);
        end;
        if vmax<val
          vmax=val;
          xIdxMax=xIdx;
          yIdxMax=yIdx;
        end;
    end;
end;

vmax
xIdxMax
yIdxMax
%%%%%%%%%%%%

Matlab output:

vmax = 0.1716

xIdxMax = 95

yIdxMax = 5
%%%%%%%%%%%%
\end{verbatim}
\end{scriptsize}

For other densities replace the density assignment above accordingly. For example, for the $\exp(-||x||_2)$ (exponential kernel,  scaled by $\sqrt{2}$ for convenience) set
\begin{scriptsize}
\begin{verbatim}
          %% density for exp(-|x|/sqrt{2})    
          density(j)=min((z(j)-z(xIdx))/sqrt(2), 1-z(xIdx)/sqrt(2)); 						 
\end{verbatim}
\end{scriptsize}
and for the $\exp\left(-\sqrt{||x||_2}\right)$ kernel (scaled to $\sqrt{2}$ for convenience) set
\begin{scriptsize}
\begin{verbatim}          
          %% density for exp(-(x/\sqrt{2})^{1/2})    
          density(j)=min(sqrt(z(j)/sqrt(2))-sqrt(z(xIdx)/sqrt(2)), 1-sqrt(z(xIdx)/sqrt(2)));	 	 
\end{verbatim}
\end{scriptsize}
respectively.

\newpage

%
\newcommand{\etalchar}[1]{$^{#1}$}

\appendix
\section{Omitted proofs from Section~\ref{sec:prelims}}\label{app:prelim}
\begin{proofof}{Lemma~\ref{lem:projection}}
	Let $x':=||\mathbf{q}'-\mathbf{p}||$. If we consider the plane containing $q,p,o$, then $q'$ also belongs to this plane, since this plane contains $\mathbf{q},o$.\footnote{The cases when $\mathbf{q}'=\mathbf{p}$ or $||\mathbf{p}-\mathbf{q}||=R_1+R_2$ are the cases when the plane is not unique, but the reader should note that these cases correspond to $x'=0 , x'=2R_2$ cases, which are trivial.} Then, without loss of generality we can assume that we are working on $\RR^2$, where $o=(0,0), \mathbf{q}=(R_1,0), \mathbf{q}'=(R_2,0)$. Let $\mathbf{p}=(\alpha,\beta)$ such that 
	\begin{align*}
	x^2&=\left(\alpha-R_1\right)^2+\beta^2=\alpha^2+\beta^2-2\alpha R_1+R_1^2\\
	R_2^2&=\alpha^2+\beta^2\\
	x'^2 &= (\alpha-R_2)^2+\beta^2=\alpha^2+\beta^2-2\alpha R_2 +R_2^2
	\end{align*}
	Therefore, one has
	\begin{align*}
	x^2&=R_2^2-2\alpha R_1 + R_1^2.
	\end{align*}
	Thus,
	\begin{align*}
	x'^2&=R_2^2-2\alpha R_2 + R_2^2\\
	&=2R_2(R_2-\alpha)\\
	&=2R_2\left(R_2- \frac{R_2^2+R_1^2-x^2}{2R_1}\right)\\
	&=\frac{R_2}{R_1}\left(x^2-(R_1-R_2)^2\right),
	\end{align*}
	which proves the claim. One should note that the claim holds for both $R_1\ge R_2$ and $R_1<R_2$.
	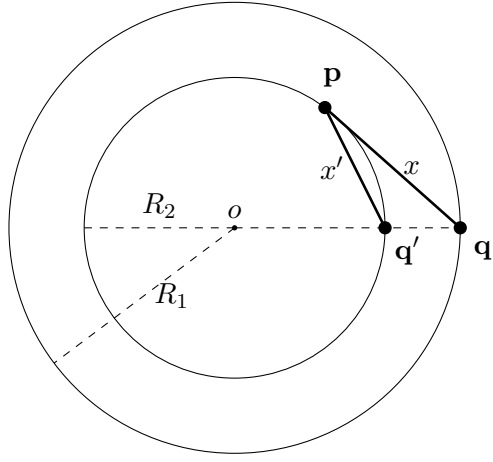
\begin{figure}
		\centering
		\tikzstyle{vertex}=[circle, fill=black!70, minimum size=3,inner sep=1pt]
		\tikzstyle{svertex}=[circle, fill=black!100, minimum size=5,inner sep=1pt]
		\tikzstyle{gvertex}=[circle, fill=green!80, minimum size=7,inner sep=1pt]
		
		\tikzstyle{evertex}=[circle,draw=none, minimum size=25pt,inner sep=1pt]
		\tikzstyle{edge} = [draw,-, color=red!100, very  thick]
		\tikzstyle{bedge} = [draw,-, color=green2!100, very  thick]
		\begin{tikzpicture}[scale=1, auto,swap]
		
		\draw (0,0) circle (2cm);
		\draw (0,0) circle (3cm);

		\fill[fill=black] (0,0) circle (1pt);
		\draw[dashed] (0,0 ) -- node[above]{$R_2$} (-2,0);
		\draw[dashed] (0,0 ) -- node[right]{$R_1$} (-2.4,-1.8);

		\node[svertex](v1) at (3, 0) {};
		\draw (3.3,-0.3) node {{$\mathbf{q}$}};
		\node[svertex](v1) at (2, 0) {};
		\draw (2.3,-0.3) node {{$\mathbf{q}'$}};
		\path[draw, line width=1pt, -, color=black!100, line width=1pt] (3,0) --node[right]{$x$} (1.2,1.6);  
		\path[draw, line width=1pt, -, color=black!100, line width=1pt] (2,0) --node[left]{$x'$} (1.2,1.6);
		\draw[dashed] (0,0 ) -- (3,0);
		\node[svertex](v1) at (1.2, 1.6) {};   
		\draw (1.3,2) node {{$\mathbf{p}$}};
		\draw (0,0)[above] node {{$o$}};
		\end{tikzpicture}
		\caption{Illustration of $x'=\textsc{Project}(x,R_1,R_2)$} \label{fig:projection}
	\end{figure}
\end{proofof}

\begin{proofof}{Claim~\ref{claim:psrs}}
	Now let $\mathbf{q}'$ be the projection of the query on the sphere and let $\mathbf{q}''$ be the antipodal point of $\mathbf{q}'$ on this sphere (see Figure~\ref{fig:epsilon}). By Definition~\ref{def:psrs}, we have 
	\begin{align*}
	\left| \left\{  \mathbf{u}\in P : ||\mathbf{u}-\mathbf{q}'||\le r(\sqrt{2}-\gamma)       \right\}\right| \le \tau\cdot |P|
	\end{align*}
	and 
	\begin{align}\label{eq:epstau}
	\left| \left\{  \mathbf{u}\in P : ||\mathbf{u}-\mathbf{q}''||\le r(\sqrt{2}-\gamma)       \right\}\right| \le \tau\cdot |P|.
	\end{align}
	On the other hand, in Figure~\ref{fig:epsilon} let $a$, $c$ be points at distances  $r(\sqrt{2}-\gamma)$ and $r\sqrt{2}$ respectively from $\mathbf{q}'$ and $\mathbf{d}$ be a point at distance $r(\sqrt{2}-\gamma)$ from $\mathbf{q}''$. Then by Pythagoras theorem, we have 
	\begin{align*}
	||\mathbf{q}'-\mathbf{d}||^2+|\mathbf{q}''-\mathbf{d}||^2=||\mathbf{q}'-\mathbf{q}''||^2,
	\end{align*}
	which implies
	\begin{align*} ||\mathbf{q}'-\mathbf{d}||=r\sqrt{4-\left(\sqrt{2}-\gamma\right)^2}=r\sqrt{2-\gamma^2+2\sqrt{2}\gamma},
	\end{align*}
	since $||\mathbf{q}''-\mathbf{d}||=||\mathbf{q}'-\mathbf{a}||$. 
	Therefore, we have the following
	\begin{align*}
	\left| \left\{  \mathbf{u}\in P : ||\mathbf{u}-\mathbf{q}''||\le r\left(\sqrt{2}-\gamma\right)       \right\}\right|=\left| \left\{  \mathbf{u}\in P : ||\mathbf{u}-\mathbf{q}'||\ge r\left(\sqrt{2-\gamma^2+2\sqrt{2}\gamma}\right)       \right\}\right| \le \tau\cdot |P|.
	\end{align*}
	So one has
	\begin{align*}
	\left| \left\{  \mathbf{u}\in P :||\mathbf{u}-\mathbf{q}'||\in\left(r\left(\sqrt{2}-\gamma\right), r\cdot \sqrt{2-{\gamma^2}+2\sqrt{2}\gamma} \right)    \right\}\right| \ge (1-2\tau)\cdot |P|.
	\end{align*}
	On the other hand, we have
	\begin{align*}
	\left(r\left(\sqrt{2}-\gamma\right), r\cdot \sqrt{2-\gamma^2+2\sqrt{2}\gamma} \right) \subseteq \left(r\left(\sqrt{2}-\gamma\right), r \left(\sqrt{2}+\gamma\right) \right),
	\end{align*}
	and hence
	\begin{align*}
	\left| \left\{  \mathbf{u}\in P :||\mathbf{u}-\mathbf{q}'||\in\left(r\left(\sqrt{2}-\gamma\right), r\cdot \left(\sqrt{2}+\gamma\right)\right)    \right\}\right| \ge (1-2\tau)\cdot |P|.
	\end{align*}
	which proves the second part of the claim. Now, using \eqref{eq:epstau} we have
	\begin{align*}
	\left| \left\{  \mathbf{u}\in P : ||\mathbf{u}-\mathbf{q}'||\le r(\sqrt{2}-\gamma)       \right\}\right| \le \frac{\tau}{1-2\tau}\cdot \left| \left\{  \mathbf{u}\in P :||\mathbf{u}-\mathbf{q}'||\in\left(r\left(\sqrt{2}-\gamma\right), r\cdot \left(\sqrt{2}+\gamma\right)\right)    \right\}\right|.
	\end{align*}
	which proves the first part of the claim. 
	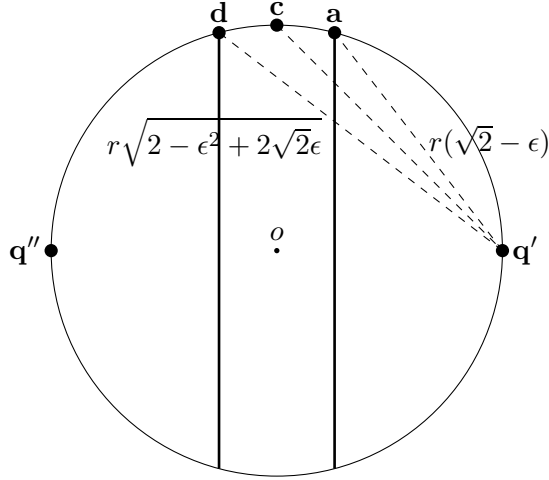
\begin{figure}
		\centering
		\tikzstyle{vertex}=[circle, fill=black!70, minimum size=3,inner sep=1pt]
		\tikzstyle{svertex}=[circle, fill=black!100, minimum size=5,inner sep=1pt]
		\tikzstyle{gvertex}=[circle, fill=green!80, minimum size=7,inner sep=1pt]
		
		\tikzstyle{evertex}=[circle,draw=none, minimum size=25pt,inner sep=1pt]
		\tikzstyle{edge} = [draw,-, color=red!100, very  thick]
		\tikzstyle{bedge} = [draw,-, color=green2!100, very  thick]
		\begin{tikzpicture}[scale=1, auto,swap]
		
		\draw (0,0) circle (3cm);

		\fill[fill=black] (0,0) circle (1pt);
		
		\path[draw, line width=1pt, -, color=black!100, line width=1pt] (0.76811457478,-2.9) -- (0.76811457478,2.9);
		\node[svertex](v1) at (0.76811457478,2.9) {};   
		\draw (0.76811457478,2.9) node[above] {{$\mathbf{a}$}};
		
		\path[draw, line width=1pt, -, color=black!100, line width=1pt] (-0.76811457478,-2.9) -- (-0.76811457478,2.9);
		\node[svertex](v1) at (-0.76811457478,2.9) {};   
		\draw (-0.76811457478,2.9) node[above] {{$\mathbf{d}$}};
		\node[svertex](v1) at (3,0) {};   
		\draw (3,0) node[right] {{$\mathbf{q}'$}};
		\node[svertex](v1) at (-3,0) {};   
		\draw (-3,0) node[left] {{$\mathbf{q}''$}};
		\draw[dashed] (3,0 ) -- node[right] {$r(\sqrt{2}-\epsilon)$}(0.76811457478,2.9);
		\draw[dashed] (3,0 ) -- node[left] {$r\sqrt{2-\epsilon^2+2\sqrt{2}\epsilon}~~~$}(-0.76811457478,2.9);
		\draw[dashed] (0,3 ) -- node[left] {$ $}(3,0);
		\node[svertex](v4) at (0,3) {};   
		\draw (0,3) node[above] {{$\mathbf{c}$}};
		\draw (0,0)[above] node {{$o$}};
		
		\end{tikzpicture}
		\caption{$\epsilon$-neighborhood of orthogonal points in a sphere of radius $r$} \label{fig:epsilon}
	\end{figure}
\end{proofof}

\section{Pseudo-random data sets via Ball carving}
\label{sec:ball_carving}
In this section we provide a simple self-contained proof of the claim that one can efficiently (near-linear time) detect and remove a dense ball on the sphere when it exists that avoids invoking VC-dimension arguments as in \cite{AR15}. 

\begin{lem}\label{lem:partitioning} There is a randomized procedure $\textsc{Certify}_{\epsilon,\tau,\delta}(P)$ that given a set $P\subset \mathcal{S}^{d-1}$ and parameters $\epsilon,\tau,\delta\in(0,\frac{1}{3})$, runs in time $O(\frac{d}{\epsilon^{2}\tau}\log(2n/\delta)\cdot n)$ and with probability $1-\delta$ 
	\begin{enumerate}
		\item either returns a point $p^{*}\in P$ such that $|B(p^{*},\sqrt{2(1-\epsilon^{2})})\cap P|\geq \Omega(\epsilon^{2}) \cdot \tau |P|$ 
		\item or certifies that the set $P$ is \emph{$(\epsilon',\tau)$-pseudo-random} with $\epsilon'=\sqrt{2}(1-\sqrt{1-2\epsilon})=\Theta(\epsilon)$.
	\end{enumerate}
\end{lem}

 \begin{algorithm}[ht]
 	\caption{$\textsc{Certify}_{\epsilon,\tau,\delta}(P)$}
 	\label{alg:certify}
 	\begin{algorithmic}[1]
 		\State {\bfseries Input:} parameters $\epsilon,\tau,\delta\in (0,\frac{1}{3})$ and $P\subset \mathcal{S}^{d-1}$. 
 		\State $\zeta \gets \frac{1}{4}$, $m\gets \lceil\frac{48}{\epsilon^{2}\tau}\log(2n/\delta)\rceil$
 		\State $Q_{m}\gets \{ m \text{  uniform random points with replacement from } P \}$ \Comment{Sub-sampling}
 		\State $p^{*}\gets \arg\max_{p\in P}\{|B(p,\sqrt{2(1-\epsilon^{2})})\cap Q_{m}|\}$
 		\If {$|B(p^{*},\sqrt{2(1-\epsilon^{2})})\cap Q_{m}|\geq (1-\zeta) (3\epsilon^{2})\tau \cdot m$}
 		\State \textbf{return} $p^{*}\in P$ \Comment{Center for a ``dense ball" is found.}
 		\Else
 		\State  {\bf return} $\bot$. \Comment{Set $P$ is $(\Theta(\epsilon),\tau)$-pseudo random}
 		\EndIf
 	\end{algorithmic}
 \end{algorithm}
 
 The partitioning procedure is based on the following lemma adapted from \cite{AR15} showing that in any set that contains a dense ball on the unit-sphere, one can find a point in the dataset  that  captures a large fraction of the points in the dense ball.
 \begin{lem}[Certificate]\label{lem:certification}
 	Let $S'\subset \mathcal{S}^{d-1}$ and  $x^{*}\in \mathcal{S}^{d-1}$ such that  for $\epsilon\in(0,\frac{1}{3})$ and all $x\in S'$, $\|x^{*}-x\|\leq  \sqrt{2(1-2\epsilon)}$. There exists a point $x_{0} \in S'$ such that
 	\begin{equation}
 	\left|\left\{x\in S':\|x-x_{0}\|\leq \sqrt{2(1-\epsilon^{2})}  \right\} \right| \geq (3\epsilon^{2})\cdot  |S|'.
 	\end{equation} 
 \end{lem} 
 The contrapositive  is that if no balls of a certain radius and density exist with points of the dataset as centers, then the dataset is \emph{pseudo-random} with appropriate constants. We can use this lemma to show that either the data set is pseudo-random or we can always find a dense ball and decrease the size of the remaining data set by a non-trivial factor.   The issue that is left to discuss is efficiency of the process.

 By repeatedly applying the lemma and by stopping only when $|P|\leq \frac{1}{\tau}$ we can decompose any set on the sphere in at most $T=O(\frac{\log |P|}{\epsilon^{2}\tau})$ ``dense balls" and a pseudo-random remainder. Let $\chi\in(0,1)$ be a bound on the failure probability and set $\delta=\chi/T$, then  this can be done in time \begin{equation}
 O\left(\frac{d}{\epsilon^{4}\tau^{2}}\log\left(2n\log(n)/\epsilon^{2}\tau\chi\right) \log(n)\cdot n\right).
 \end{equation}

For any point $p\in P$ let $B_{p}:=B(p,\sqrt{2(1-\epsilon^{2})})\cap P$ and  $\mathcal{B}:=\{ B_{p}: p\in P\wedge |B_{p}|\geq 3\epsilon^{2}\tau\}$. If   $Q_{m}$  is a random sample of $m$ points from $P$ with replacement, then  by Chernoff bounds  $\forall B\in \mathcal{B}$  we get $\Pr\left[\left| |B\cap Q_{m}| - \frac{|B|}{|P|}m\right| \geq \zeta  \frac{|B|}{|P|}m \right]\leq 2e^{-\frac{\zeta^{2}}{3}\frac{|B|}{|P|}m}$. 
Setting $\zeta =\frac{1}{4}$ and $m\geq \frac{48}{\epsilon^{2}\tau}\log(2n/\delta)$ and taking union bound over  at most $|\mathcal{B}|\leq |P|$ events, we get that with probability at least $1-\chi$ for all   $B\in\mathcal{B}$    we have $\frac{3}{4}\frac{|B|}{|P|}\leq \frac{|B\cap Q_{m}|}{m}\leq \frac{5}{4}\frac{|B|}{|P|}$.   Conditional on the above event we have that:

\begin{itemize}
	\item  If there exists $p^{*}\in P$ such that $|B_{p^{*}}\cap Q_{m}| \geq \frac{3}{4}(3\epsilon^{2})\tau m$, then $|B_{p^{*}}|\geq \frac{9}{16} (3\epsilon^{2})\tau \cdot |P|$.
	\item If for all $p\in P$, $|B_{p}\cap Q_{m}| <(1-\zeta) (3\epsilon^{2})\tau m$ then $|B_{p}|<3\epsilon^{2}\tau$ for all $p\in P$ and  by Lemma \ref{lem:certification} the set $P$ is $(\epsilon',\tau)$-pseudo random, with $\epsilon'=\sqrt{2}(1-\sqrt{1-2\epsilon})\Rightarrow \epsilon=\frac{\epsilon'}{\sqrt{2}}(1-\frac{\epsilon'}{2\sqrt{2}})$ as $\sqrt{2(1-2\epsilon)}=\sqrt{2(1-\sqrt{2}\epsilon'(1-\frac{\epsilon'}{2\sqrt{2}}))}=\sqrt{(\sqrt{2}-\epsilon')^{2}}=\sqrt{2}-\epsilon'$
\end{itemize}
This shows correctness of the Procedure $\textsc{Certify}_{\epsilon,\tau,\delta}$ (Algorithm \ref{alg:certify}).  The overall cost of this procedure is $O(dmn)=O(\frac{d\log(2n/\chi)}{\epsilon^{2}\tau}n)$ dominated by the cost of finding the ball of radius $\sqrt{2(1-\epsilon^{2})}$ centered at one of the points in $P$ with the most number of points in $Q_{m}$.\qed

To prove Lemma~\ref{lem:certification} we are going to use the following simple lemma.

\begin{prop}[\cite{AR15}] For any set $S\subset \mathcal{S}^{d-1}$ such that there exists $c\in \mathcal{S}^{d-1}$, $\|c-x\|\leq r_{\epsilon}= \sqrt{2(1-2\epsilon)}$ for all $x\in S$, 
\begin{equation}
\frac{1}{|S|^{2}}\sum_{x,y\in S}\langle x,y\rangle \geq \left(1-\frac{r_{\epsilon}^{2}}{2}\right)^{2} = 4\epsilon^{2}
\end{equation} 
\end{prop}
\begin{proof}
Given $x,c\in \mathcal{S}^{d-1}$, $\|x-c\|\leq r_{\epsilon}\Rightarrow \langle x, c\rangle \geq 1- \frac{r_{\epsilon}^{2}}{2}$. Thus, 
\begin{equation}
\sum_{i,j\in S} \langle x_{i},x_{j}\rangle =  \|\sum_{x\in S}x\|^{2}\|c\|^{2}\geq \left|\sum_{x\in S}\langle x, c\rangle\right|^{2} \geq (1- \frac{r_{\epsilon}^{2}}{2})^{2}|S|^{2}
\end{equation}
The proof is concluded by substituting $r_{\epsilon}=\sqrt{2(1-2\epsilon)}$.
\end{proof}
\begin{proof}[Proof of Lemma \ref{lem:certification}] We proceed with a proof by contradiction. Assuming that the statement is not true, then  
\begin{equation}
 \forall y\in S, \ \left|\left\{x\in S:\|x-y\|\leq \sqrt{2(1-\epsilon^{2})}\right\} \right| <  (3\epsilon^{2})\cdot  |S| 
\end{equation}
Moreover, $\|x-y\|  > \sqrt{2(1-\epsilon^{2})}   \Rightarrow \langle x,y\rangle < \epsilon^{2}$. Therefore, we get:
\begin{align}
\frac{1}{|S|^{2}}\sum_{x,y\in S} \langle x, y\rangle  < \left(1-3\epsilon^{2}\right) \epsilon^{2} + 3\epsilon^{2}\cdot 1= (4-3\epsilon^{2})\epsilon^{2}<4\epsilon^{2}
\end{align}
Using Proposition 1 and the hypothesis we arrive at a contradiction.
\end{proof}
\section{Correctness proof of the data-dependent algorithm}\label{sec:dd-kde}
In this section we present the outer algorithms for our approach. The procedure is quite routine and similar to Section~\ref{sec:data-independent}. First, in Algorithm~\ref{alg:DD-KDE-Prep} we present the outer procedure of preprocessing phase. In Algorithm~\ref{alg:DD-KDE-Prep} for any $x\in\{\delta_x,2\delta_x,3\delta_x,\ldots\,\delta_x \lfloor \frac{\sqrt{2}}{\delta_x}\rfloor\}$, we sample the data set with probability $\min\left\{\frac{1}{n}\left(\frac{1}{\mu}\right)^{1-x^2/2},1\right\}$, and then using Algorithm~\ref{alg:DD-KDE-Prep2}, we prepare a data structure that after receiving the query, one can recover any point that is present in the sample and has distance $[x-\delta_x,x)$ from the query using Algorithm~\ref{alg:DD-KDE-Q}, with probability $0.8$ (see Lemma~\ref{lem:tree-correctness} below). 
\begin{lem}\label{lem:tree-correctness}
	Under Assumption~\ref{assump:1}, if $\mathcal{T}=\textsc{PreProcess}(P,x,\mu)$, then for every point $\mathbf{p}\in P$ such that $\mathbf{p}\in v_0.P$, where $v_0$ is the root of tree $\mathcal{T}$ and $||\mathbf{q}-\mathbf{p}||\le x$, one has $p\in \textsc{Query}(\mathbf{q},\mathcal{T},x)$ with probability at least $0.8$.
\end{lem}
\begin{proof}
	By Corollary~\ref{cor:AI}, if $\mathcal{H}$ is a $(\alpha,x,\mu)$-AI hash family then for any point $\mathbf{p}$ such that $||\mathbf{p}-\mathbf{q}||\le x$ 
	\begin{align*}
	\Pr_{h\sim \mathcal{H}}[h(\mathbf{q})=h(\mathbf{p})] \ge \mu^{\alpha}
	\end{align*}
	Now, noting the number of repetitions of the Andoni-Indyk LSH round, i.e., setting of $K_1=100 \left(\frac{1}{\mu}\right)^{\alpha}$ (see line~\ref{line:const-c} of Algorithm~\ref{alg:DD-KDE-Prep2}), with probability at least $0.9$ we know that there exists a hash bucket that both query and point $\mathbf{p}$ are hashed. Now, we prove by induction on depth of the tree, that if $\mathbf{p}$ belongs to the dataset of root of any tree $\mathcal{T}'$ then $\textsc{Query}(\mathbf{q},\mathcal{T}',x)$ recovers it with probability at least 0.9. \\
	
	\noindent{\bf Base:} If the depth of $\mathcal{T}'$ is 1, then $\mathbf{p}\in P_x$ by line~\ref{line:depth1} of Algorithm~\ref{alg:DD-KDE-Query}.\\
	\noindent{\bf Inductive step:} Suppose that $\mathbf{p}\in v.P$ such that $v$ is the root of $\mathcal{T}'$. One should note that $v$ is a pseudo-random sphere. Also, suppose that $\mathbf{q}$ is at distance $R_2$ from the center of this sphere. Then let $x':=\textsc{Project}(x+\Delta,R_2,R)$ and let $x''$ be the smallest element in the grid $W$ which is not less than $x'$. Let $N:=\left\lceil  \frac{100}{G(x''/R,\eta)}   \right\rceil$. Then, by Algorithm~\ref{alg:DD-KDE-spherical} we know that $v$ has $N$ children $u_1,u_2,\ldots,u_N$ such that $u_j.x=x''$ for all $j\in [N]$. If $\mathbf{p}\in u_j.P$ for some $j$ then $\mathbf{p}$ will appear in exactly one of the children of $u_j$, we call this node $u_j(\mathbf{p})$, and if $\mathbf{p}\notin u_j.P$ then $u_j(\mathbf{p})=\perp$. Let $\mathbf{q}'$ be the projection of $\mathbf{q}$ on the sphere. Now, note that for any $j\in [N]$, if $w$  is a child of $u_j$ then
	\begin{align*}
	&\Pr\left[  \textsc{Query}(\mathbf{q},\mathcal{T}_{u_j(\mathbf{p})},x) \text{ will be called}  \text{ and } \mathbf{p} \in u_j(\mathbf{p}).P \text{ and  } \mathbf{p}\in \textsc{Query}(\mathbf{q},\mathcal{T}_{u_j(\mathbf{p})},x) \right]\\
	&=\Pr\left[  \left\langle u_j.g , \frac{\mathbf{q}-o}{||\mathbf{q}-o||} \right\rangle \ge \eta  \text{ and } \mathbf{p} \in u_j.P \text{ and  } \mathbf{p}\in \textsc{Query}(\mathbf{q},\mathcal{T}_{u_j(\mathbf{p})},x) \right]\\
	& = \Pr\left[ \mathbf{p}\in \textsc{Query}(\mathbf{q},\mathcal{T}_{u_j(\mathbf{p})},x) \mid  \left\langle u_j.g , \frac{\mathbf{q}-o}{||\mathbf{q}-o||} \right\rangle \ge \eta  \text{ and } \mathbf{p} \in u_j.P \right]\\
	&~~~~\cdot \Pr\left[    \left\langle u_j.g , \frac{\mathbf{q}-o}{||\mathbf{q}-o||} \right\rangle \ge \eta  \text{ and } \mathbf{p} \in u_j.P \right]\\
	&\ge 0.9\cdot \Pr\left[     \left\langle u_j.g , \frac{\mathbf{q}-o}{||\mathbf{q}-o||} \right\rangle \ge \eta  \text{ and } \mathbf{p} \in u_j.P \right]\\
	&\ge 0.9 \cdot G(||\mathbf{q}'-\mathbf{p}.new||/R,\eta)\\
	&\ge 0.9\cdot G\left(\textsc{Project}(x+\delta,R_2,R)/R,\eta\right)\\
	&\ge 0.9\cdot G(x'/R,\eta).
	\end{align*}
	The first inequality holds  by induction. The second inequality holds by Definition~\ref{def:g-eta}. The third inequality holds since $||\mathbf{q}'-\mathbf{p}.new|| \le \textsc{Project}(x+\delta,R_2,R)$. Also, the last inequality holds since $\Delta \ge \delta$.
	Then, we have
	\begin{align*}
	&\Pr\left[  \textsc{Query}(\mathbf{q},\mathcal{T}_{u_j(\mathbf{p})},x) \text{ will not be called}  \text{ or } \mathbf{p} \notin u_j(\mathbf{p}).P \text{ or  } \mathbf{p}\notin \textsc{Query}(\mathbf{q},\mathcal{T}_{u_j(\mathbf{p})},x) \right]\\&\le 1-0.9\cdot G(x'/R,\eta).
	\end{align*}
	Consequently 
	\begin{align*}
	\Pr\left[     \mathbf{p} \notin \textsc{Query}(\mathbf{q},\mathcal{T}',x)    \right] &\le (1-0.9\cdot G(x'/R,\eta))^N\\
	&\le 0.1,
	\end{align*}
	where the second inequality uses the following fact that since  $x''\geq x'$, we have
	\begin{align*}
	N&= \left\lceil  \frac{100}{G(x''/R,\eta)}   \right\rceil \ge\left\lceil  \frac{100}{G(x'/R,\eta)}   \right\rceil\ge  \frac{100}{G(x'/R,\eta)}.
	\end{align*}
	So the inductive step goes through, and the statement of the lemma holds. Now, by taking the union bound over the failure probability of the Andoni-Indyk round (which succeeds with high probability) and the failure probability of the data dependent part, we succeed by probability at least $1-0.1-0.1= 0.8$.
\end{proof}

For points beyond $\delta_x \lfloor \frac{\sqrt{2}}{\delta_x}\rfloor$ we just sample the data set with rate $\frac{1}{n}$ and just store the sampled set (see line~\ref{line:samplebeyondsqrt2} in Algorithm~\ref{alg:DD-KDE-Prep}). In the query procedure we just scan the sub-sampled data set for recovering these points (see line~\ref{line:samplebeyondsqrt2} of Algorithm~\ref{alg:DD-KDE-Q}). We repeat this procedure $O(\log n)$ times to boost the success probability to high probability. After recovering the sampled points from the various bands using corresponding data structures, Algorithm~\ref{alg:DD-KDE-Q} applies the standard procedure of importance sampling by calculating $Z_{\mu}$. 

\begin{algorithm}
	\caption{\textsc{PreProcess-KDE}: $P$ is the data-set}  
	\label{alg:DD-KDE-Prep} 
	\begin{algorithmic}[1]
		\Procedure{$\textsc{PreProcess-KDE}(P,\mu)$}{}
		\State $\delta_x \gets 10^{-8}$ \label{line:x-grid-def} \Comment{Step size for grid over $x$}
		
		\State $K_1\gets\lceil \frac{C\log n}{\epsilon^2}\cdot \mu^{-4\delta_x}\rceil$\Comment{where $C$ is some large enough constant}
		\For {$k =1,2,\ldots,K_1$}
		
		\For {$j=1,\ldots, \lfloor \sqrt{2}/\delta_x\rfloor$} \Comment{Uniform grid with step size $\delta_x$ over $[0, \sqrt{2}]$}
		\State $x \gets j\cdot \delta_x$ \label{line:xingrid}
		\State $\wt{P}_{k,x} \gets $ sample each point in $P$ with probability $\min\left\{\frac{1}{n}\left(\frac{1}{\mu}\right)^{1-x^2/2},1\right\}$\label{line:wtPkx}
		\For {$i=1,\ldots,10\log n$} \label{line:repeathighprob}
		\State $\mathcal{T}_{x,k,i}\gets \textsc{PreProcess}(\wt{P}_{k,x},x,\mu)$\label{line:TreeDS}
		\EndFor
		\EndFor
		\State $\wt{P}_k \gets$ sample each point in $P$ with probability $\frac{1}{n}$\label{line:samplebeyondsqrt2}
		\State Store $\wt{P}_k$ \Comment{This set will be used to recover points beyond $\delta_x\lfloor \sqrt{2}/\delta_x\rfloor$.}	
		\EndFor
		
		\EndProcedure
		
	\end{algorithmic}
\end{algorithm}
\begin{algorithm}[H]
	\caption{\textsc{Query-KDE}: $\mathbf{q}$ is the query point}  
	\label{alg:DD-KDE-Q} 
	\begin{algorithmic}[1]
		\Procedure{$\textsc{Query-KDE}(\mathbf{q},\mu)$}{}
		\State $\delta_x\gets 10^{-8}$
		\State $C_x \gets \lfloor \frac{\sqrt{2}}{\delta_x}\rfloor$

		\State $K_1\gets\lceil \frac{C\log n}{\epsilon^2}\cdot \mu^{-4\delta_x}\rceil$ \Comment{where $C$ is some large enough constant}
		\State $Z_{\mu}\gets 0$
		\For {$k =1,2,\ldots,K_1$}\label{line:forkK1}
		\State $Z_{\mu,k}, \gets 0$
		\For {$j=1,\ldots, C_x$}
		\State $x \gets j\cdot \delta_x$
		\State $S_x \gets \emptyset$
			\For {$i=1,\ldots,10\log n$}
		\State $\mathcal{T}_{x,k,i} \gets $ the data structure prepared by line~\ref{line:TreeDS} of Algorithm~\ref{alg:DD-KDE-Prep}
		\State $P_{x,k,i}\gets \textsc{Query}(\mathbf{q},\mathcal{T}_{x,k,i},x)$
		\For{$\mathbf{p}\in P_{x,k,i}$}

		\If {$||\mathbf{q}-\mathbf{p}|| \in \left[x-\delta_x,x\right)$}
		\State $S_x \gets S_x \cup \{\mathbf{p}\}$ \label{line:SX}
		\EndIf

		\EndFor
		
		\EndFor	
		\For {$\mathbf{p}\in S_x$}
		\State $\hat{x}\gets ||\mathbf{q}-\mathbf{p}||$
		\State $Z_{\mu,k} \gets Z_{\mu,k} + \left(\mu^{\frac{\hat{x}^2}{2}}\right)\left( \min\left\{\frac{1}{n}\muu{1-\frac{x^2}{2}},1\right\}\right)^{-1}$
		\EndFor
			\EndFor	
		\For {$p\in \wt{P}_k$} \Comment{Importance sampling for points beyond $\delta_xC_x$.} \label{line:beyond2query}
		\If {$||\mathbf{q}-\mathbf{p}||\ge \delta_xC_x$}
		\State $\hat{x} \gets ||\mathbf{q}-\mathbf{p}||$
		\State $Z_{\mu,k}\gets Z_{\mu,k}+n\left(\mu^{\hat{x}^2/2}\right)$
		\EndIf
		\EndFor
		\State $Z_{\mu}\gets Z_{\mu}+\frac{Z_{\mu,k}}{K_1}$
			
		\EndFor
		\State \Return $Z_{\mu}$
	
		\EndProcedure
		
	\end{algorithmic}
\end{algorithm}
Below, we present the proof of correctness for the outer algorithm, which is very similar to the proof in Section~\ref{sec:data-independent}. 

\begin{claim}[Unbiasedness of the estimator]
	The estimator $Z_{\mu,k}$ for any $\mu\ge \mu^*$ and any $k\in[K_1]$(see line~\ref{line:forkK1} of Algorithm~\ref{alg:DD-KDE-Q})
	satisfies the following:
	\begin{align*}
	(1-n^{-9}) n \mu^*\le \mathbb{E}[Z_{\mu,k}] \le n\mu^* .
	\end{align*}
\end{claim}
\begin{proof}
	First note that if a point $\mathbf{p}\in \wt{P}_{k,x}$ for some $k$ and $x$ in line~\ref{line:wtPkx} of Algorithm~\ref{alg:DD-KDE-Prep} is such that $||\mathbf{q}-\mathbf{p}||\in [x-\delta_x,x)$, then since we are preparing $10\log n$ data structures, alongside with Lemma~\ref{lem:tree-correctness} with probability at least $1-n^{-10}$, $\mathbf{p}\in S_x$ (see line~\ref{line:SX} of Algorithm~\ref{alg:DD-KDE-Q}). Taking union bound over all the points, with probability $1-n^{-9}$, any point in distance $[x-\delta_x,x)$ is being sampled with probability $\min\left\{\frac{1}{n}\left(\frac{1}{\mu}\right)^{1-x^2/2},1\right\}$ for any $x\in\{\delta_x,2\delta_x,\ldots\} \cap (0,\sqrt{2})$. We call this event $\mathcal{E}$. 
	Now, since $ p_i\cdot (1-n^{-9})\le \Pr[\chi_i = 1]\le p_i$, we have
	\begin{align*}
	\mathbb{E}[Z_{\mu,k}]=\mathbb{E}\left[\sum_{i=1}^{n}\chi_i \frac{w_i}{p_i}\right]\ge (1-n^{-9})\sum_{i=1}^{n} w_i = (1-n^{-10}) n \mu^*.
	\end{align*}
	and 
	\begin{align*}
	\mathbb{E}[Z_{\mu,k}]\le n\mu^*
	\end{align*}
	where $p_i$ is the probability of sampling $i$'th point, and $\chi_i$ is the indicator for the event that $i$'th point is recovered. 
\end{proof}
We proved that our estimator is unbiased\footnote{Up to some small inverse polynomial error.} for {\bf any choice} of $\mu\ge \mu^*$. Therefore if $\mu\ge4\mu^*$,  by Markov's inequality the estimator outputs a value larger than $\mu$ at most with probability $1/4$. We perform $O(\log n)$ independent estimates, and conclude that $\mu$ is higher than $\mu^*$ if the median of the estimated values is below $\mu$. This estimate is correct with high probability, which suffices to ensure that we find a value of $\mu$ that satisfies $\mu/4< \mu^*\le \mu$ with high probability by starting with $\mu=n^{-\Theta(1)}$ (since our analysis assumes $\mu^*=n^{-\Theta(1)}$) and repeatedly halving our estimate (the number of times that we need to halve the estimate is $O(\log n)$ assuming that $\mu$ is lower bounded by a polynomial in $n$, an assumption that we make).
\begin{claim}\label{claim:outer-thm}
	For $\mu$ such that $\mu/4\le \mu^*\le \mu$, $\textsc{Query-KDE}(\mathbf{q},\mu)$ (Algorithm~\ref{alg:DD-KDE-Q}) returns a $(1\pm \epsilon)$-approximation to $\mu^*$. 
\end{claim}

\begin{proof}
	Also, one should note that $Z_{\mu,k} < n^2\left(\frac{1}{\mu}\right)$ which implies
	\begin{align*}
	\mathbb{E}\left[ Z_{\mu,k}|\mathcal{E}\right]\cdot \Pr[\mathcal{E}] + n^2\left(\frac{1}{\mu}\right) (1-\Pr[\mathcal{E}]) \ge \mathbb{E}[Z_{\mu,k}]
	\end{align*}
	
		So, 
	\begin{align*}
	\mathbb{E}[Z_{\mu,k}]|\mathcal{E}] \ge\left( (1-n^{-10})n\mu^* - \frac{1}{n^2}\frac{1}{\mu}   \right)=n\mu^*-o(1/n^5) 
	\end{align*}
	Also, since $Z_{\mu,k}$ is a non-negative random variable, we have  
	\begin{align*}
	\mathbb{E}\left[Z_{\mu,k}| \mathcal{E} \right]\le 	\frac{\mathbb{E}\left[Z_{\mu,k}\right]}{\Pr[\mathcal{E}]}\le \frac{n\mu^*}{\Pr[\mathcal{E}]}=n\mu^* + o(1/n^5)
	\end{align*}
	
	Also, 
	\begin{align*}
	\mathbb{E}[Z_{\mu,k}^2]&=\mathbb{E}\left[\left(\sum_{i\in [n]}\chi_i \frac{w_i}{p_i}\right)^2\right]\\
	&=\sum_{i\ne j} \mathbb{E}\left[\chi_i\chi_j \frac{w_iw_j}{p_ip_j}\right]+\sum_{i\in [n]}\mathbb{E}\left[\chi_i \frac{w_i^2}{p^2_i}\right]\\
	&\le \sum_{i\ne j}w_iw_j + \sum_{i\in [n]}\frac{w_i^2}{p_i}\mathbb{I}[p_i=1]+\sum_{i\in [n]}\frac{w_i^2}{p_i}\mathbb{I}[p_i\ne 1]\\
	&\le \left(\sum_{i}w_i\right)^2 +\sum_{i}w_i^2+ \max_i\left\{\frac{w_i}{p_i}\mathbb{I}[p_i\ne 1]\right\}\sum_{i\in [n]}w_i\\
	&\le  2n^2{\mu^*}^2+ n^2 \left(\frac{1}{\mu}\right)^{-1+4\delta_x}  \cdot {\mu^*} \le n^2\mu^{2-4\delta_x}
	\end{align*}
	and 
	\begin{align*}
	\mathbb{E}[Z_{\mu,k}^2|\mathcal{E}]\le \frac{\mathbb{E}[Z_{\mu,k}^2]}{\Pr[\mathcal{E}]}\le n^2 \mu^{2-4\delta_x} + o(1/n^5)
	\end{align*}
	So in order to get a $(1\pm \epsilon)$-factor approximation to $n\mu$, with high probability, it suffices to repeat the whole process $K_1=\frac{C\log n}{\epsilon^2}\cdot \mu^{-4\delta_x}$ times (see Algorithm~\ref{alg:DD-KDE-Prep} and Algorithm~\ref{alg:DD-KDE-Q}), where $C$ is a universal constant. 
\end{proof}

\section{Omitted discussion from Section~\ref{sec:pathgeometry}}\label{app:b}

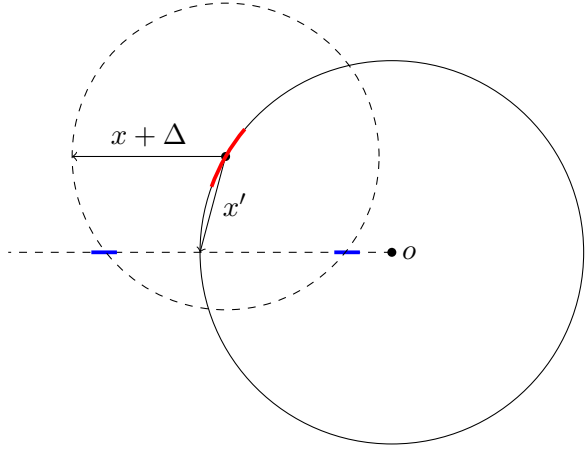
\begin{figure}[H]
	\centering
	\tikzstyle{vertex}=[circle, fill=black!70, minimum size=3,inner sep=1pt]
	\tikzstyle{svertex}=[circle, fill=black!100, minimum size=5,inner sep=1pt]
	\tikzstyle{gvertex}=[circle, fill=green!80, minimum size=7,inner sep=1pt]
	
	\tikzstyle{evertex}=[circle,draw=none, minimum size=25pt,inner sep=1pt]
	\tikzstyle{edge} = [draw,-, color=red!100, very  thick]
	\tikzstyle{bedge} = [draw,-, color=green2!100, very  thick]
	\begin{tikzpicture}[scale=1.7, auto,swap]

	\draw (0,0) circle (1.5cm);

	\fill[fill=black] (0,0) circle (1pt);

	\draw[dashed] (0,0 ) --node[below]{} (-3,0);
	\draw[line width=0.5mm, blue ] (-0.25,0 ) --node[below]{} (-0.45,0);
	
	\draw[line width=0.5mm, blue ] (-2.15,0 ) --node[below]{} (-2.35,0);
	\draw (0,0)[right] node {{$o$}};

	\draw[->] (-1.3,0.75 ) --node[above]{$x+\Delta$} (-2.5,0.75);
	\draw[->] (-1.3,0.75 ) --node[right]{$x'$} (-1.5,0);
	\draw[dashed] (-1.3,0.75) circle (1.2cm);
	\fill[fill=black] (-1.3,0.75) circle (1pt);

	\draw[line width=0.5mm,red] (-1.15,0.964) arc (140:160:1.5cm);

	\end{tikzpicture}
	\caption{Geometric illustration of equation $x'=\textsc{Project}(x+\Delta,\wt{\ell},r)$ when we have access to an approximation of $x'$ (red arc).} 
\end{figure}

Given query $\mathbf{q}$, and a LSH node $v$ with  $(v.x,v.r)=(x'',r)$, we define $b\in\{1,2\}$ which we use in the definition of the path geometry (Definition~\ref{def:pathgeom}). Let $\wt{\ell}:=||\mathbf{q}-o||$, where $o$ is the center of the sphere. Note that, if we solve $x' = \textsc{Project}(x+\Delta,\ell,r)$ for $\ell$ we get the following roots for this equation.
\begin{align}\label{eq:ell1}
\ell_1 = \frac{\sqrt{4r^2((x+\Delta)^2-x'^2)+x'^4}+2r^2-x'^2}{2r}
\end{align}
and 
\begin{align}\label{eq:ell2}
\ell_2 = \frac{-\sqrt{4r^2((x+\Delta)^2-x'^2)+x'^4}+2r^2-x'^2}{2r}
\end{align}
{\bf Stability of $\ell_1$ and $\ell_2$ for small changes of $x'$:} Let $\wt{x}'$ be such that $\wt{x}' \in [x', x'+\delta']$. Since $\delta'=o(1)$, $r=\Theta(1)$ and $x=\Theta(1)$, if we solve equation $ \wt{x}' = \textsc{Project}(x+\Delta,\ell,r)$ for $\ell$ then we get roots $\wt{\ell}_1$ and $\wt{\ell}_2$ such that 
$\wt{\ell}_1\in(\ell_1-\delta'^{1/3},\ell_1+\delta^{1/3})$
and 
$\wt{\ell}_2=(\ell_2-\delta'^{1/3},\ell_2+\delta'^{1/3})$
for large enough $n$, since $\delta'=\exp(-(\log\log n)^C)$ (see line~\ref{line:delta-prime-def} of Algorithm~\ref{alg:DD-KDE-spherical}).  

Suppose that we solve $x'=\textsc{Project}(x+\Delta,\ell,r)$ for $\ell$ for all values of $x'\in [x''-\delta',x'']$, and let $\ell^*_1$ be the largest quantity that we get by \eqref{eq:ell1} and let $\ell^*_2$ be the largest quantity that we get by \eqref{eq:ell2}. More formally,
\begin{align*}
\ell^*_1 :=\max_{x'\in [x''-\delta',x'']} \frac{\sqrt{4r^2((x+\Delta)^2-x'^2)+x'^4}+2r^2-x'^2}{2r}
\end{align*}
and 
\begin{align*}
\ell^*_2 :=\max_{x'\in [x''-\delta',x'']} \frac{-\sqrt{4r^2((x+\Delta)^2-x'^2)+x'^4}+2r^2-x'^2}{2r}.
\end{align*}

Now, if $\wt{\ell} \in [\ell_1^*-\delta'^{1/3},\ell_1^*]$ then we let $b=1$ and $\ell:=\ell^*_1$, and otherwise we let $b=2$ and $\ell:=\ell^*_2$. Note that when $b=2$ it is guaranteed that $\wt{\ell}\in [\ell_2^*-\delta'^{1/3},\ell_2^*]$. One should note that since we define geometry for root to leaf paths, then it is guaranteed that $x'=\textsc{Project}(x+\Delta,\ell,r)$ has at least a real valued solution for $\ell$, because otherwise such a root to leaf path is not possible in the tree that the query explores. Also, note that the maximizations above are over the real values, and we ignore the imaginary solutions.

\section{Omitted claims and proofs from Section~\ref{sec:query}}\label{app:query}

\begin{claim}\label{claim:roundedpsrsph}
	Given query $\mathbf{q}$ and a pseudo random sphere with geometry $(x'',r,b)$ that induces distance $\ell$ let $\mathbf{q}'$ be the projection of $\mathbf{q}$ on the sphere. In that case, if a point $\mathbf{p}.new$ on the sphere is such that $||\mathbf{q}'-\mathbf{p}.new||\in (r(\sqrt{2}-\gamma),r(\sqrt{2}+\gamma))$, then 
	$$||\mathbf{p}-\mathbf{q}|| \in \left(c-r\psi,c+r\psi\right)$$
	where $\psi:=\gamma^{1/3}+\delta'^{1/4}+\delta^{1/4}$, $c:=\sqrt{\ell^2+r^2}$.
\end{claim}
\begin{proof}
	Since $\mathbf{q}$ and the geometry of the sphere induce distance $\ell$, then $||\mathbf{q}-o|| \in [\ell-\delta'^{1/3},\ell]$. Now, suppose that we move $\mathbf{q}$ in the direction of the vector from $o$ to $\mathbf{q}$ and reach a point $\wt{\mathbf{q}}$ such that $||\wt{\mathbf{q}}-o||=\ell$. Then, by assumption
	\begin{align*}
	\textsc{Project}(||\wt{\mathbf{q}}-\mathbf{p}.new||,\ell,r) \in (r(\sqrt{2}-\gamma),r(\sqrt{2}+\gamma)).
	\end{align*}
	Let $\wt{y}:=||\wt{\mathbf{q}}-\mathbf{p}.new||$. Then,
	\begin{align*}
	\frac{r}{\ell}\left(\wt{y}^2-(\ell-r)^2\right) \in \left(r^2(\sqrt{2}-\gamma)^2,r^2(\sqrt{2}+\gamma)^2\right)
	\end{align*}
	which using the definition $c:=\sqrt{r^2+\ell^2}$ (see Figure~\ref{fig:cj}) translates to 
	\begin{align*}
	\wt{y}^2&\in \left( r\ell(2-2\sqrt{2}\gamma + \gamma^2)+(\ell-r)^2,r\ell(2+2\sqrt{2}\gamma + \gamma^2)+(\ell-r)^2\right)\\
	&=\left(c^2-2\sqrt{2}r\ell\gamma +r\ell\gamma^2,c^2+2\sqrt{2}r\ell\gamma +r\ell\gamma^2\right)
	\end{align*} 
	which also translates to 
	\begin{align*}
	\wt{y}\in\left(\sqrt{c^2-2\sqrt{2}r\ell\gamma +r\ell\gamma^2},\sqrt{c^2+2\sqrt{2}r\ell\gamma +r\ell\gamma^2}\right)
	\end{align*}
	Now, noting that $\ell=O(1)$ and $r=\Theta(1)$, for large enough $n$ we get that
	\begin{align*}
	\sqrt{c^2-2\sqrt{2}r\ell\gamma +r\ell\gamma^2} &\ge c - \sqrt{2\sqrt{2}r\ell\gamma -r\ell\gamma^2}\\
	&\ge c-r \gamma^{1/3}.
	\end{align*}
	And Similarly, 
	\begin{align*}
	\sqrt{c^2+2\sqrt{2}r\ell\gamma +r\ell\gamma^2} &\le c - \sqrt{2\sqrt{2}r\ell\gamma +\ell\gamma^2}\\
	&\le c+r \gamma^{1/3}.
	\end{align*}
	So, overall
	\begin{align*}
	\wt{y}\in \left(c-r\gamma^{1/3},c+r\gamma^{1/3}\right)
	\end{align*}
	Noting that $||\mathbf{q}-\wt{\mathbf{q}}|| \le \delta'^{1/3}$ and $||\mathbf{p}-\mathbf{p}.new||\le \delta$, using the triangle inequality, for $y:=||\mathbf{q}-\mathbf{p}||$ we get
	\begin{align*}
	y \in \left(c-r\gamma^{1/3}-\delta'^{1/3}-\delta,c+r\gamma^{1/3}+\delta'^{1/3}+\delta\right)
	\end{align*}
	Again noting that $r=\Theta(1)$ and setting $\psi=\gamma^{1/3}+\delta'^{1/4}+\delta^{1/4}$
	\begin{align*}
	y\in \left(c-r\psi,c+r\psi\right).
	\end{align*}	
	Note that in this proof we did not optimize the inequalities and we were generous in bounding variables for the sake of brevity.
\end{proof}

\begin{claim}\label{cl:collision-prob}
	Let $y$ be such that $y\ge x+\Delta$ for some $x\in (\delta_x,\sqrt{2})$, and $y''$ is such that 
	\begin{align*}
	y'' \in \left[ \textsc{Project}(y-\delta,R_2,R) , \textsc{Project}(y+\delta,R_2,R)\right]
	\end{align*}
	for some $R_2$ and $R$. Let $x'=\textsc{Project}(x+\Delta,R_2,R)$, and let $x''$ be the smallest element in $W_x$ which is not larger than $x'$. Additionally, assume that we have the following properties:
	\begin{description}
		\item[(p1)] $\frac{\delta}{\Delta} = o(1)$
		\item[(p2)] $\frac{\delta'}{\Delta}= o(1)$
		\item[(p3)] $\Delta=\Theta(1)$
		\item[(p4)] $x' \le \frac{8}{5}\cdot R$
	\end{description}
	If $\eta$ is such that $\frac{F(\eta)}{G(x''/R,\eta)}=\muu{\frac{1}{T}}$, then, we have {\bf (a)}
	\begin{align*}
	\frac{G(y''/R,\eta)}{F(\eta)} \le \muu{-(1- o(1))  \frac{4(R/x')^2-1}{4(R/y')^2-1}\cdot \frac{1}{T}}
	\end{align*}
	and Furthermore, {\bf (b)} when $R=O(1)$, then
	\begin{align*}
	\frac{G(y''/R,\eta)}{F(\eta)} \le \muu{-  \frac{4(R/x')^2-1}{4(R/y')^2-1}\cdot \frac{1}{T}}.
	\end{align*}
\end{claim}
\begin{proof}
	
	By assumption we have
	\begin{align}
	\frac{F(\eta)}{G(x''/R,\eta)}=\muu{\frac{1}{T}}.\label{eq:FGT}
	\end{align}
	On the other hand, if we set $s=x''/R$
	\begin{align}
	\frac{F(\eta)}{G(x''/R,\eta)}&=\frac{e^{-(1+o(1))\frac{\eta^2}{2}}}{e^{-(1+o(1))\frac{\eta^2}{2}\cdot \frac{4}{4-s^2}}}\nonumber\\
	&=e^{(1+o(1))\frac{\eta^2}{2}\cdot \frac{s^2}{4-s^2}}\nonumber\\
	&=\exp{ \left(   (1+o(1))\frac{\eta^2}{2}\cdot \frac{x''^2}{4R^2-x''^2}  \right) }.
	\end{align}
	By triangle inequality in Euclidean space (see Figure~\ref{fig:TriangleIneq}) we have,
	
	\begin{figure}
		\centering
		\begin{subfigure}[b]{0.8\textwidth}
			\centering
			\centering
			\tikzstyle{vertex}=[circle, fill=black!70, minimum size=3,inner sep=1pt]
			\tikzstyle{svertex}=[circle, fill=black!100, minimum size=5,inner sep=1pt]
			\tikzstyle{gvertex}=[circle, fill=green!80, minimum size=7,inner sep=1pt]
			
			\tikzstyle{evertex}=[circle,draw=none, minimum size=25pt,inner sep=1pt]
			\tikzstyle{edge} = [draw,-, color=red!100, very  thick]
			\tikzstyle{bedge} = [draw,-, color=green2!100, very  thick]
			\begin{tikzpicture}[line width=0.3mm, scale=1]
			
			\draw (0,0) circle (2cm);

			\fill[fill=black] (0,0) circle (1pt);
			\draw[dashed] (0,0 ) -- node[above]{$R$} (-2,0);

			\node[svertex](v1) at (5, 0) {};
			\draw (5.3,-0.3) node {{$q$}};
			\node[svertex](v1) at (2, 0) {};
			\draw (2.3,-0.3) node {{$q'$}};
			\path[draw, line width=1pt, -, color=black!100, line width=1pt] (5,0) --(1.2,1.6);  
			\path[draw, line width=1pt, -, color=black!100, line width=1pt] (2,0) --node[left]{$x'$} (1.2,1.6);
			\draw[dashed] (2,0 ) --node[below]{$|R_2-R|$} (5,0);
			\node[svertex](v1) at (1.2, 1.6) {};   
			\draw (0,0)[above] node {{$o$}};
			\draw (3,1)[above]node{{$x+\Delta$} };
			\end{tikzpicture}
			\caption{When the query is outside the sphere} 
		\end{subfigure}
		\begin{subfigure}[b]{0.8\textwidth}
			\centering
			
			\centering
			\tikzstyle{vertex}=[circle, fill=black!70, minimum size=3,inner sep=1pt]
			\tikzstyle{svertex}=[circle, fill=black!100, minimum size=5,inner sep=1pt]
			\tikzstyle{gvertex}=[circle, fill=green!80, minimum size=7,inner sep=1pt]
			
			\tikzstyle{evertex}=[circle,draw=none, minimum size=25pt,inner sep=1pt]
			\tikzstyle{edge} = [draw,-, color=red!100, very  thick]
			\tikzstyle{bedge} = [draw,-, color=green2!100, very  thick]
			\begin{tikzpicture}[line width=0.3mm, scale=1.2]
			
			\draw (0,0) circle (2cm);

			\fill[fill=black] (0,0) circle (1pt);
			\draw[dashed] (0,0 ) -- node[above]{$R$} (-2,0);

			\node[svertex](v1) at (0.5, 0) {};
			\draw (0.3,-0.3) node {{$q$}};
			\node[svertex](v1) at (2, 0) {};
			\draw (2.3,-0.3) node {{$q'$}};
			\path[draw, line width=1pt, -, color=black!100, line width=1pt] (0.5,0) --node[left]{$x+\Delta$}(1.2,1.6);  
			\path[draw, line width=1pt, -, color=black!100, line width=1pt] (2,0) --node[left]{$x'$} (1.2,1.6);
			\draw[dashed] (2,0 ) --node[below]{$|R_2-R|$} (0.5,0);
			\node[svertex](v1) at (1.2, 1.6) {};   
			\draw (0,0)[above] node {{$o$}};
			\end{tikzpicture}
			\caption{When the query is inside the sphere} 
		\end{subfigure}
		\caption{Triangle inequality instances for \eqref{eq:tr} }	\label{fig:TriangleIneq}
	\end{figure}
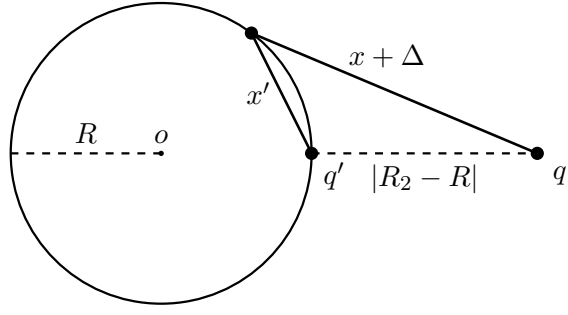
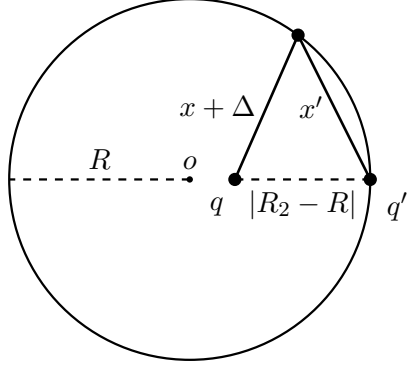
	\begin{align}
	x' &\ge (x + \Delta) - |R_2-R| \label{eq:tr} \\
	&\ge (x+\Delta)-(x+\delta) \nonumber&&\text{By line~\ref{line:xclose} of Algorithm~\ref{alg:DD-KDE-Query}}\\
	&= \Delta-\delta = (1-o(1))\Delta&&\text{By property {\bf (p1)}}. \nonumber
	\end{align}
	Also note that by assumption
	\begin{align*}
	x'' \in \left[ x' ,x' +\delta'\right],
	\end{align*}
	then, since $\frac{\delta'}{\Delta}=o(1)$ by property {\bf (p2)}, we have 
	\begin{align*}
	x''= (1\pm o(1)) \cdot x'.
	\end{align*}
	And by property {\bf (p4)}, we have
	\begin{align*}
	\frac{x''^2}{4R^2-x''^2} = (1\pm o(1))\cdot \frac{x'^2}{4R^2-x'^2}
	\end{align*}
	Therefore
	\begin{align}
	\frac{F(\eta)}{G(x'/R,\eta)}&=\exp{ \left(   (1+o(1))\frac{\eta^2}{2}\cdot \frac{x'^2}{4R^2-x'^2}  \right) }\label{eq:FGxpp}\\
	&=\muu{(1\pm o(1))\frac{1}{T}}\label{eq:FGT2}.
	\end{align}
	On the other hand, if we set $s'=y'/R$, similarly 
	\begin{align}
	\frac{F(\eta)}{G(y'/R,\eta)}&=\frac{e^{-(1+o(1))\frac{\eta^2}{2}}}{e^{-(1+o(1))\frac{\eta^2}{2}\cdot \frac{4}{4-s'^2}}}\nonumber\\
	&=e^{(1+o(1))\frac{\eta^2}{2}\cdot \frac{s'^2}{4-s'^2}}\nonumber\\
	&=\exp{ \left((1+o(1))\frac{\eta^2}{2}\cdot \frac{y'^2}{4R^2-y'^2}\right)  }\label{eq:FGyp}.
	\end{align}
	Thus, by  \eqref{eq:FGxpp}, \eqref{eq:FGT2} and \eqref{eq:FGyp}
	\begin{align}
	\frac{G(y'/R,\eta)}{F(\eta)} &= \muu{-(1\pm o(1))\frac{1}{T} \frac{4-s^2}{s^2}\cdot \frac{s'^2}{4-s'^2}  }\nonumber \\
	& =\muu{-(1\pm o(1)) \frac{1}{T} \frac{4(R/x')^2-1}{4(R/y')^2-1}}\label{eq:Gypp}.
	\end{align}
	
	Note that 
	\begin{align*}
	y'' \in \left[ \textsc{Project}(y-\delta,R_2,R) , \textsc{Project}(y+\delta,R_2,R)\right]
	\end{align*}
	Then since $\frac{\delta'}{\Delta}=o(1)$ by property {\bf (p2)}, we have
	\begin{align*}
	y'' \ge \textsc{Project}(y - \Delta/2,R_2,R)=y',
	\end{align*}
	Now since $G(s,\eta)$ is monotone decreasing in $s$, we have 
	\begin{align}\label{eq:Gppp}
	\frac{G(y'/R,\eta)}{F(\eta)}\ge \frac{G(y''/R,\eta)}{F(\eta)}.
	\end{align}
	Now, by \eqref{eq:Gypp} and \eqref{eq:Gppp}, we have the statement of the first part of the claim.

	For the case when $R=O(1)$, and consequently $R_2=O(1)$ (by the assumption fact that $x\le \sqrt{2}$), by property {\bf (p3)}:
	\begin{align}\label{eq:yppyp}
	y''^2-y'^2 &\ge \left(\textsc{Project}\left(y-\delta,R_2,R\right)\right)^2-\left(\textsc{Project}\left(y-\Delta/2,R_2,R\right)\right)^2\nonumber\\
	&= \frac{R}{R_2}\left(\left(y-\delta\right)^2-(R_2-R)^2\right)-\frac{R}{R_2}\left(\left(y-\Delta/2\right)^2-(R_2-R)^2\right)\nonumber\\
	&=\Omega(1)
	\end{align}
	Then,
	\begin{align*}
	(1-o(1))\cdot \frac{1}{4(R/y'')^2-1} &\ge (1-o(1))\cdot \frac{y'^2}{4R^2-y''^2}&&\text{Since $y''\ge y'$}\\
	&\ge \frac{y'^2}{4R^2-y'^2}&&\text{By \eqref{eq:yppyp}}\\
	&\ge \frac{1}{4(R/y')^2-1}
	\end{align*}
	which implies,
	\begin{align*}
	\muu{-\frac{1}{T} \frac{4(R/x')^2-1}{4(R/y')^2-1}}\ge\muu{-(1\pm o(1)) \frac{1}{T} \frac{4(R/x')^2-1}{4(R/y'')^2-1}}=\frac{G(y''/R,\eta)}{F(\eta)},
	\end{align*}
	which proves the second part of the claim. 
\end{proof}

\begin{claim}\label{cl:pseudorandomify}
Let $V$ denote the output of $\textsc{PseudoRandomify}(v,\gamma)$ on a node $v$ of a recursion tree $\mathcal{T}$ associated with a dataset $P$ of diameter bounded by $D$.  Then for every positive integer $j$, where $R_{min}$ is the parameter from line~\ref{line:d-min} of Algorithm~\ref{alg:pseudorandomify}, the number of sets with diameter at least $(1-\gamma^2/2)^j D$ contained in $V$ is upper bounded by 
$\Lambda^j$  for $\Lambda=O(D\log |P|)/\delta$.
 \end{claim}
 \begin{proof}
 
Note that an input dataset is first partitioned into at most $\lceil R/\delta\rceil=O(D/\delta)$ spherical shells. For each spherical shell one repeatedly removes dense clusters (containing at least a $1/10$ fraction of the current dataset), repeating this process $O(\log |P|)$ times, since at most $10$ clusters are removed before the dataset size decreases by a constant factor. Every such ball has radius smaller than the original dataset by a $(1-\gamma^2/2)$ factor \cite{DBLP:conf/soda/AndoniLRW17}. This gives the claimed bound. 
 \end{proof}
 
 We now give
 
 \begin{proofof}{Lemma~\ref{lm:numleaves-explored}}
The proof is by induction on $(a, b)$, where $a$ is the number $\ell$ of LSH nodes on the path from $v\in \mathcal{T}$ to the closest leaf, $b$ is the number of pseudorandomification nodes on such a path and $r=v.R$ is the radius of the dataset.  We prove that the expected number of nodes in the subtree of such a node $v$ in $\mathcal{T}$ is upper bounded by 
$$
(L\cdot \Lambda)^a\cdot (100/\mu^{1/T})^b\cdot \Lambda^j.
$$
Here $\Lambda=(O(D\log |P|)/\delta)$ is the parameter from Claim~\ref{cl:pseudorandomify}, $j=\log_{\frac{1}{1-\gamma^2/2}} (R_{max}/r)$ is an upper bound on the number of times the radius of the sphere could have shrunk through calls to \textsc{Pseudorandomify} from the largest possible (bounded $R_{max}$) to its current value $r$, and $L=\log_{\frac{1}{1-\gamma^2/2}} (R_{max}/R_{min})$ is the maximum number of times a point can be part of a dataset that \textsc{Pseudorandomify} is called on (since the radius reduces by a factor of $1-\gamma^2/2$ in every such call).

The {\bf base} is provided by the case of $v$ being a leaf. We now give the {\bf inductive step}. First suppose that $u\in \mathcal{T}$ is a pseudorandomification node. Let $x'$ denote the value of rounded projected distance computed in line~\ref{line:x-tilde} of Algorithm~\ref{alg:DD-KDE-Query}. Then Algorithm~\ref{alg:DD-KDE-spherical} generates $\frac{100}{G(x'/R,\eta)}$ Gaussians, and the expected number of Gaussians for which the condition in line~\ref{line:g-capture} is satisfied (i.e. the number of children of $u$ that the query $q$ explores) is exactly $\frac{100 F(\eta)}{G(x'/R,\eta)}$ by definition of $F(\eta)$ (see Lemma~~\ref{lm:f-eta} in Section~\ref{sec:prelims}). We also have $\frac{F(\eta)}{G(x'/R,\eta)}=(1/\mu)^{1/T}$ by setting of parameters in line~\ref{line:1T} of Algorithm~\ref{alg:DD-KDE-Query}. Putting this together with the inductive hypothesis and noting that LSH nodes do not change the radius of the sphere, we get that the expected number of nodes of $\mathcal{T}$ that the query explores is bounded by 

$$
100 \left(1/\mu\right)^{1/T}\cdot (L\cdot \Lambda)^a\cdot (100/\mu^{1/T})^{b-1}\cdot \Lambda^j=(L\cdot \Lambda)^a\cdot (100/\mu^{1/T})^b\cdot \Lambda^j, 
$$
as required.

Now suppose that $u\in \mathcal{T}$ is a pseudorandomification node. Then by Claim~\ref{cl:pseudorandomify} for every $i$ the number of datasets with diameter at least $(1-\gamma^2/2)^i r$ generated by \textsc{Pseudorandomify} is bounded by 
$\Lambda^i$. For every $i=0,\ldots, L$ the number of nodes with radius in $((1-\gamma^2/2)^{i-1} r, (1-\gamma^2/2)^i r]$ that are generated is bounded by $\Lambda^{i-1}$. For such nodes we have by the inductive hypothesis that the expected number of nodes of $\mathcal{T}$ explored in their subtree is upper bounded by 
$$
(L\cdot \Lambda)^{a-1} \left(100/\mu^{1/T}\right)^b\cdot \Lambda^{j-i+1}.
$$
Summing over all $i$ between $1$ and $\log_{\frac{1}{1-\gamma^2/2}} (r/R_{min})$, we get that the total number of nodes that the query is expected to explore in the subtree of $u$ is bounded by 
\begin{equation*}
\begin{split}
\sum_{i=1}^{\log_{\frac{1}{1-\gamma^2/2}} (r/R_{min})} (L\cdot \Lambda)^{a-1} \left(100/\mu^{1/T}\right)^b\cdot \Lambda^{j-i+1}\cdot \Lambda^i&\leq L\cdot (L\cdot \Lambda)^{a-1} \left(100/\mu^{1/T}\right)^b\cdot \Lambda^{j+1}\\
&\leq (L\cdot \Lambda)\cdot (L\cdot \Lambda)^{a-1} \left(100/\mu^{1/T}\right)^b\cdot \Lambda^{j+1}\\
&\leq (L\cdot \Lambda)^a \cdot \left(100/\mu^{1/T}\right)^b\cdot \Lambda^j
\end{split}
\end{equation*}
proving the inductive step.

Substituting $ \alpha^*\cdot T$ as the upper bound on the number of levels in $\mathcal{T}$ as per Algorithm~\ref{alg:DD-KDE-spherical}, we thus get that the number of nodes explored by the query is bounded by 
$$
(L\cdot \Lambda)^T\cdot (100/\mu^{1/T})^{\alpha^*T}\cdot \Lambda^L\leq (100 L\cdot \Lambda)^T\cdot \Lambda^L\cdot (1/\mu)^{\alpha^*}=n^{o(1)}\cdot  (1/\mu)^{\alpha^*}
$$
in expectation. In the last transition we used the fact that 
$$
(100 L\cdot \Lambda)^T\cdot \Lambda^L=(100\cdot \log_{\frac{1}{1-\gamma^2/2}} (R_{max}/R_{min}) \cdot (O(R_{max}\log |P|)/\delta))^{\sqrt{\log n}}\cdot ((O(D\log |P|)/\delta)^{\sqrt{\log n}}=n^{o(1)}
$$
by our setting of parameters since $\gamma=1/\log\log \log n$, $R_{max}=O(1)$, $R_{min}=\Omega(1)$ and $\delta=\exp(-(\log\log n)^{O(1)})$ as per Algorithm~\ref{alg:DD-KDE-spherical} and Algorithm~\ref{alg:pseudorandomify}. And also since we use $100\left(\frac{1}{\mu}\right)^{\alpha}$ Andoni-Indyk hash functions (see Algorithm~\ref{alg:DD-KDE-Prep2}), we get $$n^{o(1)}\cdot \left(\frac{1}{\mu}\right)^{\alpha^*+\alpha}$$
in total.
 \end{proofof} 
 
 \begin{proofof}{Claim~\ref{cl:fg}}
 
By Lemma~\ref{lm:f-eta} and Lemma~\ref{lm:g-eta} and Definition~\ref{def:g-eta} one has
$$
F(\eta)=e^{-(1+o(1))\cdot\frac{\eta^2}{2}} 
$$
and
$$
G(x'/R,\eta)=e^{-(1+o(1))\cdot \frac{2\eta^2(1-\alpha(x'/R))}{2\beta^2(x'/R)}}=e^{-(1+o(1))\cdot \frac{2\eta^2}{2(1+\alpha(x'/R))}},
$$
where $\alpha(x'/R):=1-\frac{(x'/R)^2}{2}$. Using the assumption that $x'>\Delta$ we get that 
\begin{equation*}
G(x'/R, \eta)\leq e^{-(1+o(1))\cdot \frac{\eta^2}{2-\left(\left(\Delta/R\right)^2/2\right)}}.
\end{equation*}
And in particular using the fact that $R\ge \Delta$
\begin{equation*}
\frac{F(\eta)}{G(x'/R, \eta)}\geq e^{-(1+o(1))\cdot\frac{\eta^2}{2}+(1+o(1))\cdot \frac{\eta^2}{2-\left(\left(\Delta/R\right)^2/2\right)}} =(G(x'/R, \eta))^{\Omega(\Delta^2)},
\end{equation*}
or, equivalently, $\frac1{G(x'/R, \eta)}=\left(\frac{F(\eta)}{G(x'/R, \eta)}\right)^{O(1/\Delta^2)}$.  
\end{proofof}
~~~~\newline
\begin{proofof}{Claim~\ref{cl:datasetsizef}}
		For $j^*=k_J+1$, by definition of $f_{z_j,J+1}$ for $i\in \{j^*-1,\ldots,I\}$ and the fact that $$b'_{y,J+1}=\wt{B}_{y,J+1}=\wt{A}_{y,J},$$ we have 
		\begin{align}\label{eq:deffyay}
		f_{z_i,J+1}  
		&=\log_{1/\mu} \left(\sum_{y\in D \cap [z_{i+1},z_{i-1}) } \wt{A}_{y,J}\right) 
		\end{align}
		On the other hand, \eqref{eq:beyondorthogonalpts2} and the fact that $\wt{B}_{y,j}=\mathbb{E}[B_{y,j}]$ we have
		\begin{align*}
		\sum_{y\ge c_J+\psi R_J} \wt{B}_{y, J}\leq \frac{\tau}{1-2\tau}\cdot\sum_{y\in (c_J-\psi R_J, c_J+\psi R_J)} \wt{B}_{y, J}.
		\end{align*}
		Also recall \eqref{eq:wtABp}, where we have
		\begin{align*}
		\wt{A}_{y,J}=\wt{B}_{y,J}\cdot p_{y,J}
		\end{align*}
		where $p_{y,J}$ is a decreasing and non-negative function in $y$ (for the valid range of $y$). This implies that 
		\begin{align*}
		\sum_{y\ge c_J+\psi R_J} \wt{A}_{y, J}\leq \frac{\tau}{1-2\tau}\cdot\sum_{y\in (c_J-\psi R_J, c_J+\psi R_J)} \wt{A}_{y, J}.
		\end{align*} 
		
		Now, we have
		\begin{align*}
		\sum_{y}\wt{A}_{y, J} &= \sum_{y< c_J+\psi R_J} \wt{A}_{y,J} +\sum_{y\ge c_J+\psi R_J} \wt{A}_{y,J} \\
		&\le \sum_{y< c_J+\psi R_J} \wt{A}_{y,J} +\frac{\tau}{1-2\tau}\cdot\sum_{y\in (c_J-\psi R_J, c_J+\psi R_J)} \wt{A}_{y, J}\\
		&\le \sum_{y\le z_{j^*-1}} \muu{f_{y,J+1}} + \muu{f_{z_{j^*-1},J+1}}\\
		 &\le~O(1)\cdot \muu{7\delta_z}=\muu{7\delta_z+o(1)}
		\end{align*}
		where the second inequality is based on Definition~\ref{def:fy}, \eqref{eq:deffyay} and setting of parameters (the fact that $\psi=o(1)$, $\delta_z=\Theta(1)$ and $R_{\mathrm{max}}=O(1)$). The last inequality is by the assumption that $f_{y,J+1}<7\delta_z$ for $y\le z_{j^*-1}$.
\end{proofof}
\section{Proof of Claim~\ref{claim:approx}}\label{app:Grid}

\begin{proofof}{Claim~\ref{claim:approx}}
	We want to prove that 	
	\begin{align*}
	\left(\frac{4\left(\frac{r_j}{x'}\right)^2-1}{4\left(\frac{r_j}{y'}\right)^2-1}\right)\left(\frac{2\left(\frac{z_{k_j}}{z_i}\right)^2-1}{2\left(\frac{z_{k_j}}{x}\right)^2-1}\right)  \ge (1-10^{-4}).
	\end{align*}
	By defining $z:=z_{k_j}$, $s:=z_i$ and $r:=r_j$ for the sake of brevity, the left hand side becomes
	\begin{align*}
	\left(\frac{x^2}{x'^2}\right)\cdot\left(\frac{y'^2}{s^2}\right) \cdot\left(\frac{4r^2-x'^2}{2z^2-x^2}\right)\left(\frac{2z^2-s^2}{4r^2-y'^2}\right).
	\end{align*} 
	We upper-bound each term one by one. 
	\paragraph{First term:}
	Since $x':=x+\Delta$ then
	\begin{align*}
	\frac{x}{x'}=\frac{x}{x+\Delta}=1-\frac{\Delta}{x+\Delta}\ge 1-\frac{\Delta}{\delta_x+\Delta}\ge 1-\frac{\Delta}{\delta_x}=1-10^{-12}
	\end{align*}
	where we used the fact that $x\ge \delta_x$, and the last transition is by the setting of parameters. So, 
	\begin{align*}
	\frac{x^2}{x'^2}\ge 1-10^{-11}
	\end{align*}
	\paragraph{Second term:} Since $y':=y-\Delta/2$ and $y\in\left(s(1+\delta_z)^{-1},s(1+\delta_z)\right)$ then
	\begin{align*}
	\frac{y'}{s}\ge \frac{y-\Delta/2}{y(1+\delta_z)}\ge \frac{1-\delta_z}{1+\delta_z}\ge 1-3\delta_z
	\end{align*}
	where we used the fact that $y\ge \delta_x$ (since $y\ge x$) and also considered that by the parameter setting $\delta_z=10^{-6}$, $\delta_x=10^{-8}$ and $\Delta=10^{-20}$. Consequently, we have
	\begin{align*}
	\frac{y'^2}{s^2}\ge 1+9\delta_z^2-6\delta_z\ge 1-10^{-5}.
	\end{align*}
	
	\paragraph{Third term:} Note that by \eqref{eq:zkj} we have
	\begin{align*}
	r(\sqrt{2}+\psi) \in \left[z, z(1+\delta_z)\right)
	\end{align*}
	which combining with the fact that $\psi=o(1)$ implies
	\begin{align}\label{eq:rzrelation}
	r \in \left[ \frac{z(1-o(1))}{\sqrt{2}} , \frac{z(1+\delta_z)}{\sqrt{2}} \right).
	\end{align}
	Note that we used the fact that $z\ge x$  so $z=\Omega(1)$ (actually we have $z=\Theta(1)$).
	On the other hand, by the bound for the {\bf first term} we have 
	\begin{align*}
	x'^2 \le x^2 \left(1+10^{-10}\right)
	\end{align*}
	Now, we use these tools to bound the third term\footnote{Note that for the sake of brevity we are being generous in bounding terms and the inequalities are not tight}:
	\begin{align*}
	\frac{4r^2-x'^2}{2z^2-x^2}&\ge \frac{2z^2(1-o(1))-x^2(1+10^{-10})}{2z^2-x^2}\\
	&\ge 1-\frac{2o(1)z^2+10^{-10}x^2}{2z^2-x^2}\\
	&\ge  1-\frac{2\times 10^{-10}x^2}{2z^2-x^2}\\
	&\ge 1-10^{-9}
	\end{align*}
	\paragraph{Fourth term:} For the fourth term, actually its easier to upper-bound the inverse of it. First, note that 
	\begin{align*}
	y'=y-\Delta/2 \ge y (1-\delta_z) \ge s(1-\delta_z)(1+\delta_z)^{-1}\ge s(1-10^{-5}).
	\end{align*}
	The first inequality is due to $y\ge x\ge \delta_x= 10^{-8}$ and $\delta_z=10^{-6}$. This also implies that $$y'^2\ge s^2(1-3\times10^{-5}).$$
	
	On the other hand, by \eqref{eq:rzrelation} we have
	\begin{align*}
	2r^2 \le z^2(1+\delta_z)^2 \le z^2(1+10^{-5}).
	\end{align*}
	
	Combining these facts we have
	\begin{align*}
	\frac{4r^2-y'^2}{2z^2-s^2}&\le \frac{2z^2(1+10^{-5})-s^2(1-3\times 10^{-5})}{2z^2-s^2}\\
	&\le 1+10^{-5}+4\times 10^{-5} \frac{s^2}{2z^2-s^2}\\
	&\le 1+5\times 10^{-5}
	\end{align*}
	where the last transition is due to the fact that $s\le z$ (or $z_i\le z_{k_j}$ equivalently). Therefore, we have a lower-bound of $1-5\times 10^{-5}$ for the fourth term. 
	\paragraph{Combining the bounds:} Now, we have:
	\begin{align*}
	\left(\frac{x^2}{x'^2}\right)\cdot\left(\frac{y'^2}{s^2}\right) \cdot\left(\frac{4r^2-x'^2}{2z^2-x^2}\right)\left(\frac{2z^2-s^2}{4r^2-y'^2}\right)&\ge (1-10^{-11})(1-10^{-5})(1-10^{-9})(1-5\times 10^{-5})\\
	&\ge 1-10^{-4}
	\end{align*}
	which proves the claim.
\end{proofof}

\section*{General Kernels}\label{app:generalkernels}
\begin{lem}[Uniqueness of Maximum]\label{lem:general_uniqueness}
Let $f:[a,b]\to \R$ be a three times differentiable function in $(a,b)$ such that:
\begin{itemize}
	\item $f(a)< 0$
	\item $\exists y'\in (a,b]$ such that $f(y')>0$
	\item for all $y\in (a,b)$ it holds  $\frac{d^{3}}{dy^{3}}f(y)\leq 0$.
\end{itemize}Then 
\begin{enumerate}
	\item $\exists y^{*}\in (a,y')$ such that $f(y^{*})=0$.
	\item $\exists \eta \in (y^{*},b]$ such that  $\eta$ is the unique maximum of $f$ in $[a,b]$  and the function is monotone increasing in $[a,\eta]$.
\end{enumerate}
\end{lem}
\begin{proof} We prove the statements in order:
	\begin{enumerate}
		\item Using the first two assumptions and continuity of $f$ (since it is differentiable) we get by the Intermediate Value Theorem that $\exists
		y^{*}\in (a,y')$ such that $f(y^{*})=0$. 
		\item Since the function is defined on a closed interval it attains a maximum. We show that there exists only one maximum. Assume that there exist two local maxima $\eta_{1}<\eta_{2} \in (a,b]$. Then,  there must be a local minimum $\eta_{0} \in (\eta_{1},\eta_{2})$ for which $f''(\eta_{0})>0$. However, this is impossible since $f''(\eta_{1})<0$ and the function $f''$ is non-increasing. Hence, there is exactly one local  maximum $\eta$ in $(a,b]$ and the function is increasing in $[a,\eta]$ (and decreasing in $(\eta,b]$ if $\eta\neq b$).
	\end{enumerate}
	 
\end{proof}
\begin{cor}\label{cor:concave_kernels}
	 Let $\phi:\R_{+}\to \R$ be any function such that $\phi^{'''}(y)\leq 0$. For all $x>0$, $T\geq 1$ and $c_{1}\geq c_{2}\geq \ldots\geq c_{t} >\frac{x}{\sqrt{2}}$  such that $\exists y' \in (x,\sqrt{2}c_{t}]$ with $f(y')>0$ define:
		\begin{equation*}
		f(y):=\left[\phi(y)-\phi(x)\right]-\sum_{s=1}^{t} \frac{2(c_s/x)^2-1}{2(c_s/y)^2-1}\cdot \frac{1}{T}.
		\end{equation*}
Then,  the conclusion of Lemma \ref{lem:general_uniqueness} holds. In particular, it holds  for all $\phi(y)\propto (y)^{p}$ with $p\leq 2$.
\end{cor}
\begin{proof}
	Follows by observing that the second derivative of the summation term is decreasing and that $\phi^{'''}(y)\propto - (2-p)p\cdot (p-1)\frac{1}{y^{2-p}}\leq 0$ for all $p\leq 2$ and $y>0$.
\end{proof}

\begin{claim}[Monotonicity]\label{cl:mina_2}
	For every $i\in [|R|]$ and $\phi:\R_{+}\to \R$ as in Corollary \ref{cor:concave_kernels}   we have
	\begin{description}
		\item[(a)] there exists a $y^*\in (x, \sqrt{2})$ such that $g_{y^*,j_i}\ge0$, $g_{y, j_i}\leq 0$ for any $y\in Z_x$ such that $y\leq y^*$,  and $g_{y,j_i}$ is non-decreasing in $y$ for $y\in [y^*, z_{j_i}]$;
		\item[(b)] there exists a $y^*\in (x, \sqrt{2})$ such that  $h_{y^*}^{(N-1)}\ge0$, $h_y^{(N-1)}\leq 0$ for any $y\in Z_x$ such that $y\leq y^*$ and $h_{y}^{(N-1)}$ is non-decreasing in $y$ for $y\in [y^*, z_{j_i}]$. 
	\end{description}
\end{claim}
\begin{proof}
Let $$q(y):=\sum_{i=1}^{t}\frac{2(c_s/x)^2-1}{2(c_s/y)^2-1}\frac{1}{T},$$  where $c_1\ge c_2 \ge \ldots \ge c_t\ge z_1\ge x$ for some $z_1\ge x$. And let $y^*_1$ be such that $\phi(y^{*}_{1})-\phi(x)-q(y^*_1)=0$ and let $\wt{y}_1$ be the smallest value such that $\wt{y}_1\ge y^*_1$ and $\phi(\wt{y}_1)-\phi(x)-q(\wt{y}_1)=\theta$ for some $\theta\ge 0$. Now define $G_1(y)$  on $[y^*_1,z_1]$, for some $z_1\ge \wt{y}_1$ as follows
\begin{align}
G_1(y):=\begin{cases}
\phi(y)-\phi(x)-q(y)  & y \in [y^*_1,\wt{y}_1)\\
\theta & y\in[\wt{y}_1,z_1]
\end{cases}
\end{align}
See the red curve in Figure~\ref{fig:theta}.\\

Also, let $\hat{q}(y):=\frac{2(z_1/x)^2-1}{2(z_1/y)^2-1}\frac{1}{T}$. Let $y^*_2\ge y^*_1$ such that $G_1(y^*_2)-\hat{q}(y^*_2)=0$. Now, we define $G_2(y)$ for $y\in [y^*_2,z_2]$ as follows:
\begin{align*}
G_2(y):=\min\left\{  G_1(y)-\hat{q}(y), \theta'   \right\}
\end{align*}
where $\theta':=G_1(z_2)-\hat{q}(z_2)$ and $\theta'\ge 0$ for some $z_2\le z_1$. By the definition of $y^*_2$, function $G_2(y)$ for $y\in [y^*_2,\wt{y}_1]$ is in the form of the function in Claim~\ref{claim:monotone} and thus, it has a unique maximum at some $\eta\in[y^*_2,\wt{y}_1]$. Also, recall that $G_1(y)=\theta$ for $y\in [\wt{y}_1,z_2]$. Also, one should note that since $\hat{q}(y)$ is a monotone increasing function for $y\in (0,\sqrt{2}z_1)$ and hence for $y\in [\wt{y}_1,z_2]$, then $\theta' \le G_2(\wt{y}_1)$ and therefore $\theta'\le G_2(\eta)$. This guarantees that there exist a $\wt{y}_2\in[y^*_2,\eta]$ such that $G_2(\wt{y}_2)=\theta'$. The reason is that $G_2(y)$ is a continuous increasing function for $y\in[\wt{y}_2,\eta]$. So, we have
\begin{align}
G_2(y):=\begin{cases}
\phi(y)-\phi(x)-q'(y)  & y \in [y^*_2,\wt{y}_2)\\
\theta' & y\in[\wt{y}_2,z_2]
\end{cases}
\end{align}
where, $q'(y):=q(y)-\hat{q}(y)$. See the blue curve in Figure~\ref{fig:theta}.

\end{proof}

\end{document}